\documentclass[11pt,a4paper]{article}
\pdfoutput=1
\usepackage{jcappub,graphicx,bm}

\makeatletter
\gdef\@fpheader{}
\makeatother

\usepackage[export]{adjustbox}
\newcommand{\eqfig}[1]{\includegraphics[valign=c]{#1}}

\usepackage{multirow}
\newcommand{\tablestyle}{\displaystyle\rule[-20pt]{0pt}{42pt}}

\usepackage{microtype}
\usepackage[defaultlines=2,all]{nowidow}
\newcommand{\vph}[1]{{\vphantom{#1}}}

\newlength\customcitespacing
\setcitestyle{citesep={,\kern-\customcitespacing}}

\usepackage{mathtools}
\newcommand{\coloneq}{\coloneqq}

\newcommand{\pp}{\mathrm{pp}}
\newcommand{\mpl}{m_\mathrm{Pl}} % Planck mass
\newcommand{\M}{\mathcal{M}} % Disformal energy scale
\newcommand{\avg}[1]{{\langle #1 \rangle}} % angled brackets
\newcommand{\w}{\omega}
\newcommand{\vecell}{\hat{\bm\ell}}
\newcommand{\projell}[1]{\vecell\!\hspace{0.5pt}\cdot\!\bm{#1}}
\newcommand{\yl}{\mathfrak{l}}
\newcommand{\ym}{\mathfrak{m}}
\newcommand{\x}{\mathtt{x}}

\newcommand{\GN}{G_\mathrm{N}}
\newcommand{\Geff}{G_{12}}
\newcommand{\gammaeff}{\gamma_{12}}
\newcommand{\gammapeff}{\gamma'\!{}_{12}}
\newcommand{\alphaprime}{\alpha'\!{}}
\newcommand{\comboA}[2]{\mathfrak a_{#1}^{(#2)}}
\newcommand{\comboAp}[2]{{\mathfrak a}_{#1}^{\prime\,(#2)}}
\newcommand{\comboBS}[2]{\tilde{\mathfrak b}_{#1}^{(#2)}}

\newcommand{\K}{\kappa}
\newcommand{\sumKK}{\sum_{\K^\vph{\prime}}\sum_{\K'\neq\K}}

\newcommand{\A}{\mathcal{A}}
\newcommand{\B}{\mathcal{B}}

\newcommand{\epsLadder}{\epsilon_\mathrm{L}} % Ladder
\newcommand{\epsSpin}{\epsilon_\mathrm{S}} % Spin

\newcommand{\dx}{\mathrm{d}}
\newcommand{\diff}[2]{\frac{\dx #1}{\dx #2}}
\newcommand{\covdiff}[2]{\frac{D #1}{D #2}}
\newcommand{\ddiff}[2]{\frac{\dx^2 #1}{\dx #2^2}}
\newcommand{\pdiff}[2]{\frac{\partial #1}{\partial #2}}

\newcommand{\scount}[1]{{\small\textsc{[\MakeLowercase{#1}]}}}
\newcommand{\sector}[1]{{(\mathrm{#1})}}
\newcommand{\scriptsector}[1]{{\scriptscriptstyle\sector{#1}}}
\newcommand{\dip}{\mathrm{d}}
\newcommand{\nondip}{\mathrm{nd}}
\newcommand{\DD}{\mathrm{DD}}
\newcommand{\QD}{\mathrm{QD}}

\title{Spin-orbit effects for compact binaries in scalar-tensor gravity}

\author[a]{Philippe Brax,}
\author[b]{Anne-Christine Davis,}
\author[b]{Scott Melville}
\author[a]{and Leong~Khim~Wong}

\affiliation[a]{%
Universit\'e Paris-Saclay, CNRS, CEA, Institut de Physique Th\'eorique,
\protect\\91191, Gif-sur-Yvette, France}
\affiliation[b]{%
DAMTP, Centre for Mathematical Sciences, University of Cambridge,
\protect\\Wilberforce Road, Cambridge CB3 0WA, U.K.}

\emailAdd{philippe.brax@ipht.fr}
\emailAdd{a.c.davis@damtp.cam.ac.uk}
\emailAdd{scott.melville@damtp.cam.ac.uk}
\emailAdd{leong-khim.wong@ipht.fr}

\abstract{%
Gravitational waves provide us with a new window into our Universe, and have already been used to place strong constrains on the existence of light scalar fields, which are a common feature in many alternative theories of gravity. However, spin effects are still relatively unexplored in this context. In this work, we construct an effective point-particle action for a generic spinning body that can couple both conformally and disformally to a real scalar field, and we show that requiring the existence of a self-consistent solution automatically implies that if a scalar couples to the mass of a body, then it must also couple to its spin. We then use well-established effective field theory techniques to conduct a comprehensive study of spin-orbit effects in binary systems to leading order in the post-Newtonian~(PN) expansion. Focusing on quasicircular nonprecessing binaries for simplicity, we systematically compute all key quantities, including the conservative potential, the orbital binding energy, the radiated power, and the gravitational-wave phase. We show that depending on how strongly each member of the binary couples to the scalar, the spin-orbit effects that are due to a conformal coupling first enter into the phase at either 0.5PN or 1.5PN~order, while those that arise from a disformal coupling start at either 3.5PN or 4.5PN~order. This suppression by additional PN~orders notwithstanding, we find that the disformal spin-orbit terms can actually dominate over their conformal counterparts due to an enhancement by a large prefactor. Accordingly, our results suggest that upcoming gravitational-wave detectors could be sensitive to disformal spin-orbit effects in double neutron star binaries if at least one of the two bodies is~sufficiently~scalarised.
}

\begin{document}
\maketitle
\raggedbottom

% ============================================== %
% Introduction
% ============================================== %
\section{Introduction}
\label{sec:intro}

The successful detection of gravitational waves is a triumph of our time. Observations by the LIGO Scientific and Virgo collaborations continue to fall into excellent agreement with the predictions of general relativity~\cite{Abbott:2016blz, TheLIGOScientific:2016src, LIGOScientific:2018dkp, LIGOScientific:2019fpa, LIGOScientific:2020tif}, and can thus provide powerful constraints on any departure from Einstein's theory. However, fully exploiting this new window into our Universe requires that we develop a more complete understanding of how deviations from general relativity can arise. In this work, we take another step forward in this direction by studying how a real scalar field can couple to both the mass \emph{and the spin} of a compact object. We compute (for the first time) the effect of such a coupling on the phase of the outgoing gravitational-wave signal from a compact binary; thereby providing a novel means with which to search for the imprints of hidden~scalar~degrees~of~freedom.

There are many reasons why one might think to augment general relativity in this~way. On the one hand, light scalar fields lie at the heart of much cosmological model building \cite{Capozziello:2007ec, Capozziello:2011et, Clifton:2011jh, Joyce:2014kja, Bull:2015stt, CANTATA:2021ktz}; motivated primarily by the difficulty in otherwise accounting for the observed, late-time acceleration of our Universe~\cite{Riess:1998cb, Perlmutter:1998np}. This is the well-known cosmological constant problem~\cite{Weinberg:1988cp}. Indeed, because general relativity is only an effective description of gravity at low energies\,---\,requiring as-of-yet unknown new physics to become important at the Planck scale, if not before\,---\,the act of including additional, light degrees of freedom in the effective field theory allows us to capture a greater range of possible effects that could stem~from~new~fundamental~physics. Since the simplest way of going beyond the tensor modes of general relativity is to add a single scalar field, scalar-tensor theories have become the main workhorse for building and testing alternative models of late-time cosmology, where the scalar plays the role of dark energy~\cite{Chiba:1999ka, ArmendarizPicon:2000dh, ArmendarizPicon:2000ah, Boisseau:2000pr, Copeland:2006wr, Bamba:2012cp, Gubitosi:2012hu, Bloomfield:2012ff, Gleyzes:2014rba, Bellini:2014fua}, or even dark matter~\cite{Sin:1992bg, Hu:2000ke, Burgess:2000yq, Bekenstein:2004ne, Hui:2016ltb, Urena-Lopez:2019kud, Hui:2021tkt, Burrage:2018zuj, Brax:2020gqg}. On smaller scales, scalar-tensor theories are also widely used to effect deviations from general relativity in the strong-field regime~\cite{Damour:1993hw, Damour:1996ke, Minamitsuji:2016hkk, Silva:2017uqg, Doneva:2017duq, Andreou:2019ikc, Ventagli:2020rnx, Shao:2017gwu, Zhao:2019suc, Guo:2021leu, Yagi:2021loe, Kanti:1995vq, Kleihaus:2011tg, Pani:2011gy, Yunes:2011we, Maselli:2015tta, Sotiriou:2014pfa, Antoniou:2017hxj, Delgado:2020rev, Sullivan:2020zpf, Doneva:2017bvd, Cunha:2019dwb, Minamitsuji:2019iwp, Herdeiro:2020wei, Berti:2020kgk, East:2021bqk, Yagi:2011xp, Julie:2019sab, Shiralilou:2020gah, Shiralilou:2021mfl, Barausse:2012da, Palenzuela:2013hsa, Shibata:2013pra, Taniguchi:2014fqa, Sennett:2016rwa, Sennett:2017lcx, Khalil:2019wyy, Silva:2020omi}.

By virtue of their interesting phenomenology, the existence of light scalar fields and how they couple to gravity and/or matter have been the subject of extensive study. On theoretical grounds, the most general, causality-preserving coupling between a scalar~$\phi$ and matter with energy-momentum tensor~$T^{\mu\nu}$ is composed of two parts: a \emph{conformal} part and a \emph{disformal} part~\cite{Bekenstein:1992pj}. The conformal part of the coupling, which has the schematic form~${\sim T^{\mu\nu} g_{\mu\nu} \, \phi/\mpl}$, is already tightly constrained in the Solar System by measurements from the Cassini spacecraft~\cite{Bertotti:2003rm} and lunar laser ranging~\cite{Hofmann:2018llr}. The implications of this kind of interaction have also been confronted with experiments in the laboratory~\cite{Adelberger:2009zz, Burrage:2017qrf, Brax:2018iyo, Brax:2018zfb, Berge:2017ovy}, in~other astrophysical settings~\cite{Sakstein:2018fwz, Naik:2019moz, Desmond:2020gzn, Hees:2017aal, Kramer:2006nb, Freire:2012mg, Seymour:2019tir, Shao:2017gwu, Zhao:2019suc, Guo:2021leu}, and on cosmological scales~\cite{Koyama:2015vza, Joyce:2016vqv, Ferreira:2019xrr}. (See also refs.~\cite{Will:2014kxa, Berti:2015itd, Damour:PDGReview} for general reviews.) The disformal part of the coupling, meanwhile, is more challenging to constrain in nonrelativistic settings, as its general form ${\sim  T^{\mu\nu} \partial_\mu \phi \,\partial_\nu \phi / \M^4}$ inevitably results in a suppression by the relevant velocity scale in the problem, and so causes it to vanish around static sources. Nevertheless, constraining this interaction is important, as it arises naturally in a variety of scenarios, including (beyond) Horndeski and general DHOST theories~\cite{Horndeski:1974wa, Deffayet:2009wt, Zumalacarregui:2012us, Bettoni:2013diz, Zumalacarregui:2013pma, Gleyzes:2014dya, Gleyzes:2014qga, Langlois:2015cwa, Crisostomi:2016tcp, Crisostomi:2016czh, BenAchour:2016cay, BenAchour:2016fzp}, in the decoupling limit of massive gravity~\cite{deRham:2010ik, deRham:2010kj, deRham:2014zqa}, in branon models~\cite{Alcaraz:2002iu, Cembranos:2004jp}, and in various brane-world scenarios~\cite{deRham:2010eu, Koivisto:2013fta}. Moreover, a~disformal interaction often appears at a much lower energy scale than its conformal counterpart (i.e.,~${\M \ll \mpl}$), thus making it a (potentially) very powerful probe of the dark sector. Our goal in this work is to complement existing laboratory and astrophysical searches for disformal couplings~\cite{Kaloper:2003yf, Koivisto:2008ak, Zumalacarregui:2010wj, Koivisto:2012za, Brax:2012ie, vandeBruck:2012vq, vandeBruck:2013yxa, Brax:2013nsa, Neveu:2014vua, Sakstein:2014isa, Sakstein:2014aca,  Brax:2014vva, Brax:2015hma, Ip:2015qsa, Sakstein:2015jca, vandeBruck:2015ida, vandeBruck:2016cnh} by examining the impact of such an interaction on the gravitational waves emitted~by~a~compact~binary.

As we enter this age of gravitational-wave astronomy, binary systems are fast becoming one of the most promising probes of light scalar fields. Arguably the most striking example to date is the neutron-star merger event GW170817 \cite{TheLIGOScientific:2017qsa, Goldstein:2017mmi, Monitor:2017mdv, GBM:2017lvd}, whose optical counterpart suggests that photons and gravitational waves travel at the same speed to within one part in~$10^{15}$
(at~least, at LIGO/Virgo frequencies\footnote{Since the speed of gravitational waves, like any coupling in a quantum field theory, depends on the scale at which it is measured, care must be taken when applying this constraint to effective field theories whose cutoff is close to LIGO/Virgo scales ($\sim 100$~Hz), as is often the case for dark energy \cite{deRham:2018red}.});
leading to very stringent constraints on certain scalar-tensor theories~\cite{Creminelli:2017sry, Sakstein:2017xjx, Ezquiaga:2017ekz, Baker:2017hug, Langlois:2017dyl, Heisenberg:2017qka, Akrami:2018yjz, BeltranJimenez:2018ymu, Copeland:2018yuh}.  Propagation effects aside, a light scalar field will also influence the orbital motion of the binary itself, and thereby leave imprints in the precise shape and phase of the gravitational waveform that is produced~\cite{Will:1994fb, Damour:1998jk, Berti:2012bp, Yunes:2013dva, Yunes:2016jcc, Chamberlain:2017fjl, Barack:2018yly, Carson:2019fxr, Perkins:2020tra}. Decoding these imprints is theoretically challenging, but is nevertheless important because it probes the dark sector in a unique regime; far away from what is possible on laboratory, Solar~System, and~cosmological~scales. 

To date, the majority of work on the two-body problem in scalar-tensor gravity has focused on nonspinning systems, with relatively little attention paid to the effect that a scalar would have on a spinning body. Perhaps the main reason for this is that, up until very recently, black holes were thought incapable of coupling to scalars on account of the no-hair theorems~\cite{Thorne:1971, Bekenstein:1971hc, Adler:1978dp, Hawking:1972qk, Zannias:1994jf, Bekenstein:1995un, Saa:1996aw, Sotiriou:2011dz, Hui:2012qt, Graham:2014mda, Chrusciel:2012jk}, while neutron stars (and other bodies composed of matter) are generally expected to spin too slowly for a spin-dependent coupling to leave significant imprints on the gravitational-wave signal~\cite{Berti:2004bd}. Today we know that black holes can indeed couple to a scalar field\,---\,either indirectly through the effect of a time-dependent background~\cite{Jacobson:1999vr, Horbatsch:2011ye, Berti:2013gfa, Wong:2019yoc, Clough:2019jpm, Hui:2019aqm, Bamber:2020bpu, Babichev:2013cya, Kobayashi:2014eva}, or directly via a coupling to a quadratic curvature invariant like the Gauss-Bonnet term \cite{Kanti:1995vq, Kleihaus:2011tg, Pani:2011gy, Yunes:2011we, Maselli:2015tta, Sotiriou:2014pfa, Antoniou:2017hxj, Delgado:2020rev, Sullivan:2020zpf, Doneva:2017bvd, Silva:2017uqg, Cunha:2019dwb, Minamitsuji:2019iwp, Herdeiro:2020wei, Berti:2020kgk, East:2021bqk}.%
\footnote{While we focus on real scalar fields in this work, it is worth noting that massive, complex scalars~\cite{Herdeiro:2014goa, Degollado:2018ypf} and pseudoscalars~\cite{Alexander:2009tp, Yunes:2009hc, Delsate:2018ome} can also produce black hole solutions that circumvent the no-hair theorems. Binary black holes that couple to the latter are studied in~refs.~\cite{Yagi:2012vf, Loutrel:2018ydv, Loutrel:2018rxs}.}
Characterising how a scalar interacts with the spins of these ``hairy'' black holes will undoubtedly give us a better understanding of their behaviour in a binary system, and could be important for establishing more accurate constraints on this class of models. As~for neutron stars, these objects are still expected to have much smaller spins than their black hole cousins, and while it remains unlikely that a \emph{conformal} spin-dependent coupling would leave a significant imprint on the gravitational waveform, it is conceivable that a sufficiently small value of~$\M$ could be enough to boost the size of the \emph{disformal} coupling between the scalar and the spin of the neutron star to a detectable level. This spin-dependent interaction will be especially important when searching for disformally coupled scalar fields in (nearly) circular binaries, like those observed by LIGO and Virgo, as the spin-\emph{independent} effects associated with a disformal coupling are known to vanish for~circular~orbits~\cite{Brax:2018bow, Brax:2019tcy}.

With all of this in mind, the central aim of this paper is to compute the leading spin-orbit effects that would arise when a compact binary is coupled both conformally and disformally to a light scalar field. We focus here on the early portion of the binary's inspiral, which is amenable to analytic methods, and will use a generalisation of the ``nonrelativistic general relativity''\,(NRGR) approach~\cite{Goldberger:2004jt, Goldberger:2007hy, Goldberger:2009qd, Galley:2009px, Porto:2005ac, Porto:2006bt, Porto:2008tb, Porto:2008jj, Levi:2010zu, Levi:2015msa, Foffa:2013qca, Rothstein:2014sra, Porto:2016pyg, Levi:2018nxp} to perform our calculations. When read alongside existing spin-independent results~\cite{Damour:1992we, Huang:2018pbu, Kuntz:2019zef, Mirshekari:2013vb, Lang:2013fna, Lang:2014osa, Sennett:2016klh, Bernard:2018hta, Bernard:2018ivi, Brax:2018bow, Brax:2019tcy}, the spin-orbit results of this paper offer a more complete picture of how these systems behave in scalar-tensor~gravity.

It is worth noting at this stage that some partial spin-orbit results have already been derived in a previous paper of ours~\cite{Brax:2020vgg}, albeit with two key limitations. The first is that ref.~\cite{Brax:2020vgg} was restricted to a purely conservative setting, whereas the results of this paper go as far as to also include the leading spin-orbit effects from a scalar in the radiative sector. The second limitation of ref.~\cite{Brax:2020vgg} is that it applies only to weakly gravitating bodies that are universally coupled to the scalar, and so satisfy the weak equivalence principle. In contrast, the results of this paper remain valid also for strongly gravitating bodies, like black holes and neutron stars, which emphatically do \emph{not} respect the strong equivalence principle~\cite{Nordtvedt:1968qs}, even if the underlying scalar-tensor theory respects the weak equivalence principle at a microscopic level. Both of these advancements were made possible by our use of the NRGR approach, which allows us to work efficiently at the level of the action, rather than laboriously at the level of the equations of motion (as~we~did~in~ref.~\cite{Brax:2020vgg}).

The main conceptual ideas that underpin this NRGR approach are reviewed in section~\ref{sec:review}, where we also review the general pipeline that takes us from a fully covariant, microscopic theory of fields to an effective field theory for the compact binary\,---\,wherein the two extended bodies are replaced by point-particle sources\,---\,and ultimately to a concrete prediction for the gravitational-wave phase (see also figure~\ref{fig:pipeline}). We then work through this sequence of steps in explicit detail across sections~\ref{sec:pp}--\ref{sec:rad}. In section~\ref{sec:pp}, we begin by constructing a general point-particle action for a single spinning body that is conformally and disformally coupled to the scalar field. The most salient details about the body's internal structure are encoded in this action through the values of Wilson coefficients, and while it is possible to match this point-particle action back onto the underlying, microscopic theory to determine what values these coefficients must take to coincide with a particular object (see section~\ref{sec:pp_stt} for details), we will leave these Wilson coefficients as free parameters in this work; thereby keeping our results general enough to describe the behaviour of \emph{any}~spinning~body.

In section~\ref{sec:consv}, we turn to consider a binary system of two such objects. We use Feynman diagrams to compute the spin-orbit part of the conservative two-body potential; working perturbatively to leading order in the post-Newtonian expansion and up to first order in the disformal coupling. For added simplicity, we also assume throughout this paper that any nonlinear self-interactions of the scalar are subleading on the scales of the binary, and can therefore be neglected as a first approximation. This is a valid simplification even for theories that exhibit thin-shell screening~\cite{Khoury:2003aq, Khoury:2003rn, Hinterbichler:2010es, Hinterbichler:2011ca, Brax:2010gi, Zhang:2017srh, Liu:2018sia},%
\footnote{In these models, nonlinearities become important only in the high-density interiors of the two bodies; hence, the primary effect of thin-shell screening can be captured by simply matching the right values for the bodies' Wilson coefficients. See refs.~\cite{Zhang:2017srh, Liu:2018sia} for more details.\looseness=-1}
although models that utilise derivative self-interactions to trigger kinetic or Vainshtein screening~\cite{Nicolis:2008in, Deffayet:2009wt, Babichev:2009ee, Deffayet:2011gz, Babichev:2013usa} are notably outside the purview of our results. (See,~e.g., refs.~\cite{deRham:2012fw, deRham:2012fg, Dar:2018dra, Kuntz:2019plo, Renevey:2021tcz, Bezares:2021dma} for spin-independent progress in this direction.) To~complete our discussion of the binary's conservative sector, we then derive the corresponding equations of motion for this system and identify some of its constants of motion; the most important of which being its orbital binding energy.

In section~\ref{sec:rad}, we move on to the radiative sector, where we use similar diagrammatic techniques to compute the multipole moments responsible for sourcing the outgoing scalar and gravitational waves. Focusing on quasicircular nonprecessing binaries for simplicity, we then combine our result for the total radiated power with our earlier expression for the orbital binding energy to determine the evolution of the binary's orbital phase, and by extension, its gravitational-wave phase. These general results establish what the leading spin-orbit effects would be in any kind of binary system, but for the sake of having a concrete example, we specialise to double neutron stars in section~\ref{sec:obsv}, where we provide estimates for the typical size of these effects. We then argue that a future ground-based detector, like the Einstein Telescope, could be capable of observing the imprints of a disformal spin-orbit interaction if at least one of the neutron stars is sufficiently scalarised. Finally, in section~\ref{sec:discussion}, we conclude with a summary of our main results and identify several key directions for~future~work.

Some of the more technical aspects of this paper are relegated to appendices~\ref{app:ssc}--\ref{app:gw}.
Units in which ${\hbar = c = 1}$ are used throughout and, as is common in the NRGR literature, we define the reduced Planck mass by ${\mpl \equiv (32\pi\GN)^{-1/2}}$. Our metric signature is~$(-,+,+,+)$.

% ============================================== %
% Section 2
% ============================================== %
\section{The effective field theory pipeline}
\label{sec:review}

The decisive property that allows us to make analytic predictions about how compact binaries evolve is the presence of large separations of scales that are inherent during the early inspiral. In the NRGR approach~\cite{Goldberger:2004jt, Goldberger:2007hy, Goldberger:2009qd, Galley:2009px, Porto:2005ac, Porto:2006bt, Porto:2008tb, Porto:2008jj, Levi:2010zu, Levi:2015msa, Foffa:2013qca, Rothstein:2014sra, Porto:2016pyg, Levi:2018nxp}, these separations of scales are used to establish a tower of effective field theories (EFTs) that\,---\,when matched onto another\,---\,provide a coherent description of the system and facilitate the systematic computation of observables, like the phase of the emitted gravitational-wave signal. It is this approach, suitably generalised to scalar-tensor theories, that we shall adopt in this paper.

\begin{figure}
\includegraphics[width=440pt]{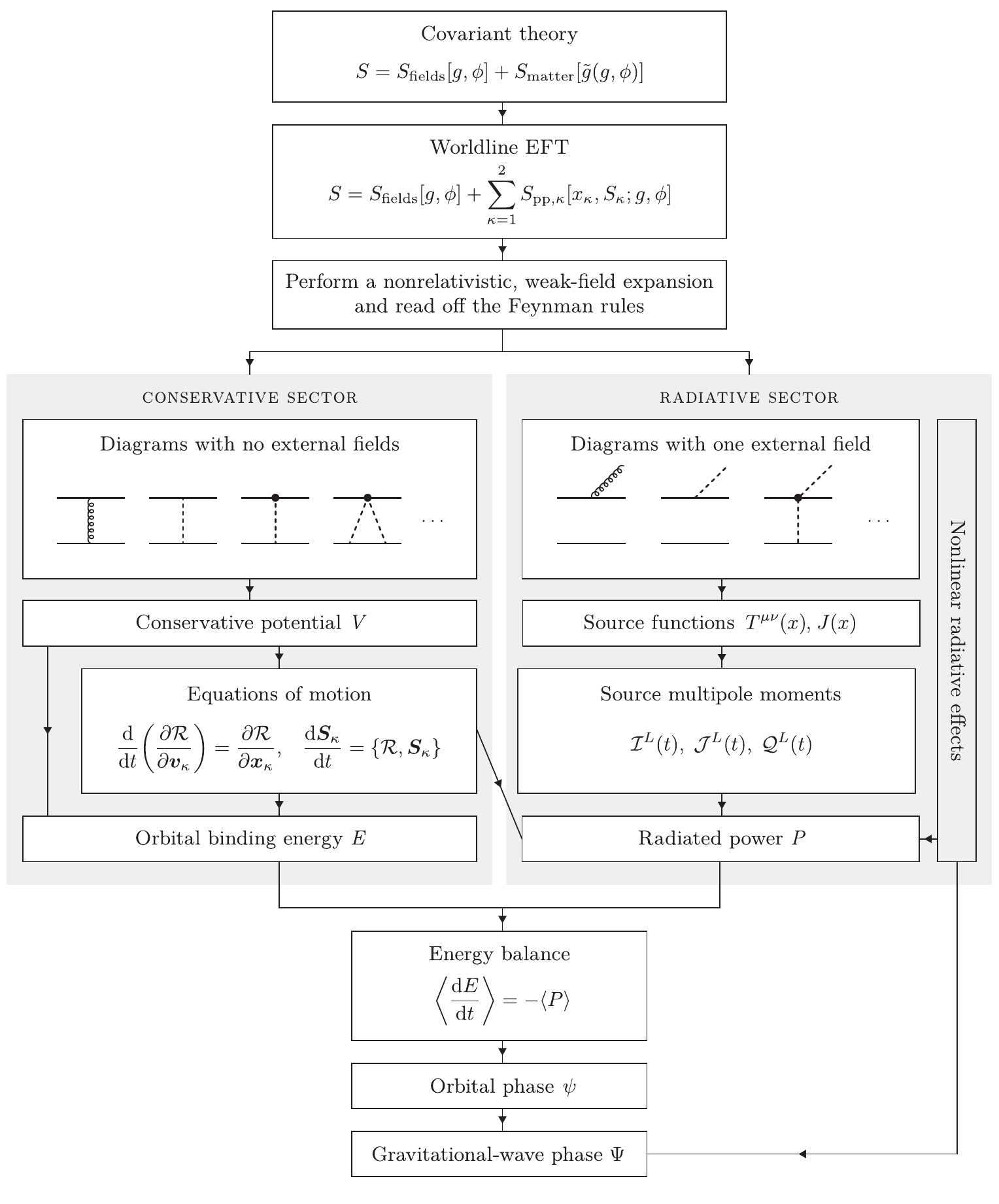}
\caption{The general pipeline for computing the phase of the gravitational-wave signal emitted by a compact binary in the effective field theory approach. This figure makes allusions to the need for additional steps if one wishes to incorporate nonlinearities like tail effects in the radiative sector, but the linear effects that are the focus of this paper will suffice at low orders in the post-Newtonian expansion.\looseness=-1}
\label{fig:pipeline}
\end{figure}

Our goal in this section is to first outline the general sequence of steps that will take us from a given scalar-tensor theory to a concrete result for the gravitational-wave phase; the intention being that having this overview to refer to will help guide our discussion when it comes time to wade through numerous and lengthy calculations. To complement the main text, an illustrated version of this review is also presented in figure~\ref{fig:pipeline}.

\paragraph{Scalar-tensor theories.}
Our very first step, then, is to be precise about the class of scalar-tensor theories under consideration. We will here be interested specifically in metric theories of gravity that can be captured by the general action
\begin{equation}
	S = S_\text{fields}[g,\phi]
	+
	S_\text{matter}[\tilde g(g,\phi)].
\label{eq:review_S_full}
\end{equation}
The kinetic terms for and the interactions between the Einstein-frame metric~$g_{\mu\nu}$ and the scalar field~$\phi$ are contained in the field action~$S_\text{fields}$, while the dynamics of the matter fields and their couplings to $g_{\mu\nu}$ and~$\phi$ are encoded in the matter action~$S_\text{matter}$. Crucially, what makes this a metric theory of gravity is our assumption that the matter fields are coupled minimally and universally to an effective Jordan-frame metric~$\tilde g_{\mu\nu}$, which we take to be~\cite{Bekenstein:1992pj}
\begin{equation}
	\tilde g_{\mu\nu}
	=
	C(\phi ,X) [g_{\mu\nu}
	+
	D(\phi,X) \partial_\mu\phi\,\partial_\nu\phi\,]
	\qquad
	(X \coloneq g^{\mu\nu}\partial_\mu\phi\,\partial_\nu\phi).
\label{eq:review_Jordan_frame}
\end{equation}

Notice that the conformal coupling function~$C$ is dimensionless by construction, and so must depend on $\phi$ and $X$ only via the dimensionless combinations $\phi/\mpl$ and $X/\M^4$, where $\mpl$ is the reduced Planck mass and $\M$ is some strong-coupling scale.%
\footnote{In cosmological applications, one typically sets ${\M^2/\mpl}$ equal to the Hubble constant~$H_0$ so as to produce order-one deviations from general relativity on these scales, but for our purposes it will be more instructive to leave~$\M$ as a free parameter to be constrained by gravitational-wave observations.}
For simplicity, we will suppose that the disformal coupling function~$D$ also depends on~$\phi$ only through these same combinations $\phi/\mpl$ and $X/\M^4$, and that this \emph{dimensionful} function has an overall scaling of the form ${D \propto 1/\M^4}$. 

This kind of scaling automatically tells us that the theory in \eqref{eq:review_S_full} is nonrenormalisable, and so must be viewed as a low-energy EFT of some as-of-yet unknown UV completion. That said, for appropriate choices of the model parameters, this fully covariant EFT can still be valid down to microscopic scales, and so presently encodes more information than is necessary for describing the evolution of astrophysical systems.

\paragraph{Point-particle approximation.}
When modelling the inspiral of a (widely separated) two-body system, this surplus of irrelevant information is discarded by replacing \eqref{eq:review_S_full} with the effective~action
\begin{equation}
	S_\text{eff} = S_\text{fields}[g,\phi]
	+
	\sum_{\K=1}^2 S_{\pp,\K}[x_\K,S_\K; g,\phi],
\label{eq:review_S_eff}
\end{equation}
which is valid on length scales greater than the individual radii of the binary's constituents. In this regime, the two extended bodies (labelled by ${\K \in \{1,2\}}$) have been replaced by effective point particles whose centres of energy travel along the worldlines $x^\mu_\K(\lambda)$. The details about how these point particles move in the bulk of the spacetime and how they interact with the metric and scalar are encoded in the point-particle actions~$S_{\pp,\K}$, while the dynamics of the bulk fields $(g_{\mu\nu},\phi)$ continue to be captured by the same field action~$S_\text{fields}$ as before.

We may regard this coarse-grained description of the system as emerging directly from \eqref{eq:review_S_full} after having integrated out all of the irrelevant short-wavelength degrees of freedom, although in practice it is much simpler to construct these point-particle actions from the bottom~up. In a Lagrangian approach, the relevant objects available for contraction with $g_{\mu\nu}$, $\phi$, and their derivatives are the particle's 4-velocity~${v^\mu}$ and its angular velocity tensor~$\Omega^{\mu\nu}$ (see section~\ref{sec:pp} for details), but it is traditional\,---\,and also more convenient\,---\,to work with the spin tensor~$S_{\mu\nu}$ in place of~$\Omega^{\mu\nu}$. As this spin tensor is the momentum variable conjugate to~$\Omega^{\mu\nu}$, we can pass from one to the other by performing a partial Legendre transform of the Lagrangian~$\mathcal L_\pp(x,\Omega)$ to obtain the point-particle Routhian~$\mathcal R_\pp(x,S)$~\cite{Yee:1993ya, Porto:2006bt, Porto:2008tb}.

Having done so, it follows that the action $S_{\pp,\K}$~(${= \int\dx\lambda\,\mathcal R_{\pp,\K}}$) is composed of an infinite series of terms, which one can order by relevancy, that couple $v^\mu_\K$ and $S^{\mu\nu}_\K$ to the bulk fields. Multiplying most of these terms are Wilson coefficients ${ \{\alpha_\K, \beta_\K, \dots \} }$, whose values encode all of the salient information about the underlying scalar-tensor theory, as well as the internal structure of the extended objects. One can then determine these values by matching appropriate observables computed within this point-particle EFT to results obtained in the full theory~\eqref{eq:review_S_full}, but we shall make no attempt to do so here and will simply leave these Wilson coefficients as free parameters. The advantage in doing so is that our results will be applicable to a broad class of models.

\paragraph{Potential and radiation modes.}
Two more simplifications are necessary to render the problem analytically tractable. We first make the weak-field approximation by writing
\begin{equation}
	(g_{\mu\nu},\phi) = (\eta_{\mu\nu},\phi_0)
	+
	(h_{\mu\nu}/\mpl,\varphi),
\label{eq:review_weak_field_approximation}
\end{equation}
where the part $(h_{\mu\nu},\varphi)$ that is sourced by the binary is to be treated as a weak perturbation about some fixed background $(\eta_{\mu\nu},\phi_0)$. Implicit in this choice of background is the assumption that the binary's immediate neighbourhood is mostly just empty space, albeit permeated by some ambient value~$\phi_0$ of the scalar that we take to be a constant.%
\footnote{If the scalar is a candidate for dark energy or fuzzy dark matter, this background value~$\phi_0$ will exhibit some mild time dependence, although the large hierarchy between the typical timescales in a binary and those of cosmology mean that any effects resulting from this time-dependent background are usually small~\cite{Horbatsch:2011ye, Kuntz:2019zef, Wong:2019yoc, Blas:2016ddr, Blas:2019hxz}, and so will be neglected here.}

We then split the field perturbations further into different parts based on their kinematics. In pure general relativity, the metric perturbation (or ``graviton'') $h_{\mu\nu}$ splits into two parts~\cite{Goldberger:2004jt}: there are \emph{potential modes} that are always off shell and are responsible for mediating the attractive forces holding the two-body system together, and there are \emph{radiation modes} that can go on shell and transport energy and momentum away from the binary. The former varies on length scales on the order of the distance~$r$ between the two bodies, while the latter varies on length scales~$\sim r/v$, where $v$ is the binary's orbital velocity. If we now suppose that the scalar perturbation~$\varphi$ is sufficiently light that it is effectively massless on length scales $\lesssim r/v$, then it too admits this simple decomposition into potential and radiation modes. We shall make this assumption throughout, and so~are~justified~in~writing
\looseness=-1
\begin{equation}
	(h_{\mu\nu},\varphi)
	=
	(\bar h_{\mu\nu},\bar\varphi)
	+
	(\hat h_{\mu\nu},\hat\varphi),
\label{eq:review_potential_radiation_split}
\end{equation}
where $(\hat h_{\mu\nu},\hat\varphi)$ denote the potential modes, while $(\bar h_{\mu\nu},\bar\varphi)$ are the radiation modes.

That the radiation modes have wavelengths~$\lambda$~$(\sim r/v)$ much larger than the characteristic size~$r$ of the binary during the early inspiral allows us to ``zoom out'' on this system even further by integrating out the potential modes. The result is a new~EFT,
\begin{equation}
	S'_\text{eff}
	=
	-i\log\int D[\hat h,\hat\varphi]\,
	\exp(iS_\text{eff}),
\label{eq:review_S_eff'_integral}
\end{equation}
whose action can be organised into a conservative and radiative sector:
\begin{equation}
	S'_\text{eff}
	=
	S_\text{con}[x_\K,S_\K]
	+
	S_\text{rad}[x_\K,S_\K;\,\bar h,\bar\varphi].
\label{eq:review_S_eff'}
\end{equation} 
Most of the hard work in the NRGR approach is in the evaluation of~\eqref{eq:review_S_eff'_integral}. It goes without saying that, for astrophysical binaries, we are ultimately interested only in the classical limit of this result, but framing the problem in this kind of quantum field theoretic language allows us to use the machinery of Feynman diagrams to our advantage when computing \eqref{eq:review_S_eff'} systematically to some prescribed order in the post-Newtonian~(PN) expansion; i.e., to some order in~$v$. For classical processes, only the tree-level Feynman diagrams are~needed~\cite{Goldberger:2004jt}.

\paragraph{Conservative sector.}
The conservative sector of~\eqref{eq:review_S_eff'} encodes information about the orbital dynamics of the binary in the absence of any outgoing radiation, and may be written as the integral of the two-body Routhian
\begin{equation}
	\mathcal R
	=
	-\sum_\K^\vph{\K} m_\K^{} \sqrt{1-\bm v_\K^2}
	-
	V(x_\K,v_\K,S_\K)
\label{eq:review_def_Routhian}
\end{equation}
 over the coordinate time~$t$ (i.e., ${S_\text{con} = \int\dx t\, \mathcal R}$). The conservative potential~$V$ arises from summing over all Feynman diagrams involving internal potential lines only and no external~ones:\looseness=-1%
\begin{equation}
	-\int\dx t\;V
	\eqfig{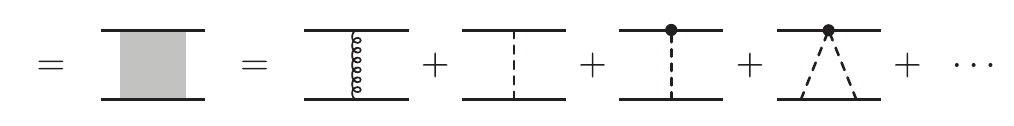}.
\label{eq:review_def_V}
\end{equation}
The explicit calculations leading to this potential are discussed further in section~\ref{sec:consv}.

Because the Routhian in \eqref{eq:review_def_Routhian} has been obtained by Legendre transforming the particles' rotational degrees of freedom, it behaves like a Lagrangian from the point of view of the position variables, but behaves like a Hamiltonian with respect to the spin variables. The equations of motion therefore follow from a mixture of Euler--Lagrange and Hamilton equations; namely,%
\footnote{When working in harmonic coordinates to 2PN order or higher, the conservative potential~$V$ will also generally depend on the accelerations~$\bm a_\K$ and their higher time derivatives~\cite{Damour:1985mt}. However, appropriate field redefinitions can always be made to replace these by functions that depend on $\bm x_\K$ and $\bm v_\K$ only~\cite{Damour:1990jh}, thus ensuring that $\bm x_\K$ satisfies a second-order equation of motion. One also encounters time derivatives of~$\bm S_\K$ in the potential at sufficiently high PN~orders, but in this case, the appropriate field redefinitions cannot be made when working with a Routhian (in fact, the spin equations of motion no longer follow from a Poisson bracket), and so one is compelled to revert back to a Lagrangian of first-order form involving both $\Omega$ and~$S$~\cite{Levi:2014sba, Levi:2015msa}. That said, we will only consider spin effects to leading PN~order in this paper, for which the Routhian approach will suffice, and will in fact be the most convenient.}
\looseness=-1
\begin{equation}
	\diff{}{t}\bigg( \pdiff{\mathcal R}{\bm v_\K} \bigg)
	=
	\pdiff{\mathcal R}{\bm x_\K},
	\quad
	\diff{\bm S_\K}{t}
	=
	\{ \bm S_\K, \mathcal R \}.
\label{eq:review_consv_eom}
\end{equation}
It is worth highlighting that these equations are expressed in terms of the 3-vectors $\bm x_\K(t)$ and~$\bm S_\K(t)$, rather than the worldline coordinates $x_\K^\mu(\lambda)$ and spin tensor~$S^{\mu\nu}_\K(\lambda)$ that we started with in~\eqref{eq:review_S_eff}. The gauge-fixing procedure that removes the unphysical degrees of freedom in $x_\K^\mu(\lambda)$ and $S^{\mu\nu}_\K(\lambda)$ is discussed in detail in sections~\ref{sec:pp} and~\ref{sec:consv}.

\paragraph{Radiative sector.}
Turning now to the radiative sector, we find that the remaining terms in $S_\text{eff}'$ naturally group themselves into two categories:
\begin{equation}
	S_\text{rad} = S_\text{fields}[\eta + \bar h/\mpl, \phi_0 + \bar\varphi]
	+
	S_\text{binary}[\mathcal M^L(x_\K,S_\K);\bar h,\bar\varphi].
\label{eq:review_S_rad}
\end{equation}
The first term gives us the propagators for and the interactions between the radiation modes, while the second accounts for how these fields are sourced by the binary. As we discussed previously, the radiation modes vary on length scales much greater than the orbital separation between the two bodies, and so are unable to resolve either of them individually\,---\,instead perceiving the entire two-body system as behaving like a single point particle. Mathematically, this notion is reflected in the way the interactions are organised in $S_\text{binary}$; namely, in terms of multipole moments $\mathcal M^L = \{\mathcal I^L,\mathcal J^L,\mathcal Q^L,\dots\}$ for the binary as a whole.

To go from \eqref{eq:review_S_eff} to the multipole moments in \eqref{eq:review_S_rad}, one proceeds as follows. First, we integrate out the potential modes while holding the radiation modes fixed to determine the source functions~$J(x)$ and $T^{\mu\nu}(x)$. In the scalar sector, we sum over all Feynman diagrams with one external radiation-mode scalar to~obtain
\begin{equation}
	\int\dx^4x\, J(x) \bar\varphi(x)
	\eqfig{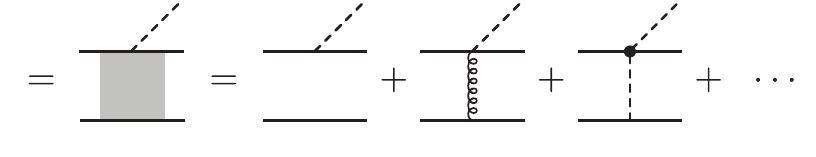},
\label{eq:review_def_J}
\end{equation}
from which $J(x)$ can be deduced. (Any internal lines in these diagrams correspond to potential modes.) Similarly, the binary's energy-momentum tensor~$T^{\mu\nu}(x)$ follows from summing over all Feynman diagrams with one external radiation-mode graviton:
\begin{equation}
	\frac{1}{2\mpl}\int\dx^4x\, T^{\mu\nu}(x) \bar h_{\mu\nu}(x)
	\eqfig{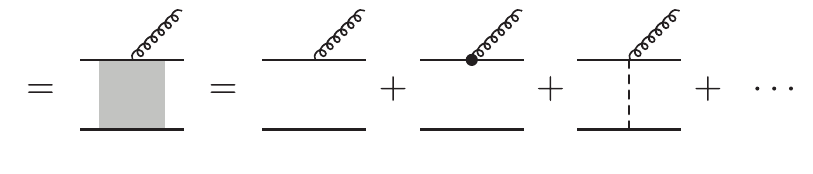}.
\label{eq:review_def_T}
\end{equation}

Next, we perform a Taylor expansion of the radiation fields about the binary's centre of energy, which we may take to coincide with the origin of our coordinate system without any loss of generality. Then regrouping terms into symmetric and trace-free~(STF) operators, the result in the scalar sector~is%
\footnote{We are employing standard multi-index notation:~we write ${\mathcal Q^L \equiv \mathcal Q^{i_1 \cdots\, i_\ell}}$ to denote any tensor with $\ell$ spatial indices, while any vector repeated $\ell$ times is abbreviated to read ${\bm x^L \equiv \bm x^{i_1} \cdots\, \bm x^{i_\ell}}$, and likewise ${\partial_L \equiv \partial_{i_1} \!\cdots \partial_{i_\ell}}$. Lastly, angled brackets around indices instruct us to keep only the STF part of that tensor.\looseness=-1}
\begin{equation}
	\int\dx^4x\, J(x)\,\bar\varphi(x)
	=
	\int\dx^4x\, J(x)
	\bigg(
		\sum_{\ell=0}^\infty
		\frac{1}{\ell!} \bm x^L \partial_L \bar\varphi(t,\bm 0)
	\bigg)
	=
	\sum_{\ell=0}^\infty\frac{1}{\ell!}\int\dx t\,
	\mathcal Q^L(t)\partial_L\bar\varphi(t,\bm 0),
\label{eq:review_Q_multipole_expansion}
\end{equation}
where the binary's scalar multipole moments $\mathcal Q^L(t)$ are given by~\cite{Ross:2012fc}
\begin{equation}
	\mathcal Q^L(t)
	=
	\sum_{p=0}^\infty \frac{(2\ell+1)!!}{(2p)!!(2\ell+2p+1)!!}
	\left( \diff{}{t} \right)^{2p}\!\int\dx^3x\,
	J(t,\bm x)\, \bm x^{2p} \bm x^\avg{L}.
\label{eq:review_def_Q}
\end{equation}

Identical steps can then be taken in the gravitational sector to obtain analogous formulae relating $T^{\mu\nu}(x)$ to the binary's mass-type and current-type multipoles, $\mathcal I^L(t)$ and~$\mathcal J^L(t)$; see, e.g., eqs.~(7.44) and~(7.45) of ref.~\cite{Porto:2016pyg} for the explicit expressions. With these multipole moments in hand, it is then a straightforward exercise to solve the wave equations (as derived from~$S_\text{rad}$) for $\bar\varphi$ and $\bar h_{\mu\nu}$ to determine the rate at which energy is carried off to infinity. Ignoring nonlinear effects like tail terms, the power radiated away into scalar waves is~\cite{Ross:2012fc}
\begin{subequations}
\label{eq:review_rad_power}
\begin{equation}
	P_\phi
	=
	\sum_{\ell=0}^\infty
	\frac{ \langle ( {}^{(\ell+1)}\!\mathcal Q^L )^2 \rangle }%
		{4\pi\ell!(2\ell+1)!!},
\label{eq:review_rad_power_scalar}
\end{equation}
while the power radiated into gravitational waves is~\cite{Ross:2012fc, Thorne:1980ru}
\begin{equation}
	P_g =
	\sum_{\ell=2}^\infty
	\frac{\GN}{\ell!(2\ell+1)!!}
	\bigg( \frac{ (\ell+2)(\ell+1) }{\ell(\ell-1)}
	\langle ( {}^{(\ell+1)}\!\mathcal I^L )^2 \rangle
	+
	\frac{ 4\ell(\ell+2) }{(\ell+1)(\ell-1)}
	\langle ( {}^{(\ell+1)}\!\mathcal J^L )^2 \rangle
	\bigg).
\label{eq:review_rad_power_gw}
\end{equation}
\end{subequations}
In writing these formulae, we have used the short hand ${{}^{(\ell)}\! X \equiv \dx^\ell X/\dx t^\ell}$ to denote the action of multiple time derivatives, and note also that the angled brackets around the multipoles denote a time average over several orbital periods. These multipole moments, and the corresponding power that is radiated away, are discussed in more detail in section~\ref{sec:rad}.
\looseness=-1

\paragraph{Phase evolution.}
The two key outputs from the above calculations\,---\,the binding energy~$E$ from the conservative sector, which follows from the equations of motion in~\eqref{eq:review_consv_eom}, and the total power ${P = P_\phi + P_g}$ computed in the radiative sector\,---\,can now be combined into the balance equation
\begin{equation}
	\bigg\langle \diff{E}{t} \bigg\rangle
	=
	-\avg{P}
\label{eq:review_balance}
\end{equation}
to tell us how the binary's orbit evolves as it emits gravitational and scalar waves. For circular nonprecessing orbits, which will be the targets of our focus in this paper, both $E$ and $P$ depend on time only through the orbital frequency~$\Omega$, and so \eqref{eq:review_balance} can in this case be recast into a differential equation for the binary's orbital phase ${\psi = \int\dx t\,\Omega}$.

After solving this equation (a task we undertake towards the end of section~\ref{sec:rad}), all that remains is to relate the orbital phase~$\psi$ to the gravitational-wave phase~$\Psi$. At the detector, the gravitational-wave signal can be decomposed into spin-weighted spherical harmonics labelled by the familiar integers $(\yl,\ym)$, and the phase of the $(\yl,\ym)$ mode is simply given by ${\Psi_{\yl\ym} = \ym\psi}$, up to corrections from nonlinear effects like tail terms~\cite{Blanchet:2013haa, Sennett:2016klh}, which we neglect. Any scalar waves that reach the detector can also be decomposed into spherical harmonics, and the corresponding $(\yl,\ym)$ modes will possess the same phases~$\ym\psi$ up to nonlinear corrections. However, in viable scalar-tensor theories, for which the strength of the scalar-matter couplings in the weak-field regime are strongly constrained, these scalar waves leave a much smaller imprint on the detector than their gravitational counterparts~\cite{Damour:1998jk, Will:1994fb}. It~will therefore suffice to focus on just the gravitational-wave signal in what follows.

% ============================================== %
% Section 3
% ============================================== %
\section{Spinning point particles in scalar-tensor theories}
\label{sec:pp}

Having reviewed the general EFT pipeline, we are now in a better position to begin undertaking detailed calculations. The aim of this section is to construct a general point-particle action for spinning bodies that are coupled to both a metric~$g_{\mu\nu}$ and a real scalar~$\phi$.

We do so in three stages. We start off in section~\ref{sec:pp_gr} by first describing the phase space of one such particle, before reviewing the simple arguments that lead to a first-order Lagrangian that minimally couples it to general relativity. This Lagrangian is then generalised to include additional couplings to the scalar~$\phi$ in section~\ref{sec:pp_stt}. Finally, in section~\ref{sec:pp_Routhian}, we perform the requisite Legendre transformation that turns this Lagrangian into a Routhian, as will be needed in the next step of our EFT pipeline.

% ============================================== %
\subsection{Covariant degrees of freedom}
\label{sec:pp_gr}

In any generally covariant theory, a spinning point particle can be described by a worldline $x^\mu(\lambda)$ and an orthonormal tetrad~$e^\mu_A(\lambda)$. The first specifies for us the trajectory along which this particle's centre of energy travels, while the latter may be regarded as the Jacobian ${e^\mu_A \equiv \partial x^\mu/\partial y^A}$ for transforming between a general coordinate chart~$x^\mu$ and the particle's body-fixed frame~$y^A$. This transformation encodes information about the intrinsic rotation of the particle, which proceeds with an angular velocity given by ${\Omega^{\mu\nu} \coloneq \eta^{AB} e^\mu_A D e^{\nu \vph{\mu}}_B/D\lambda}$, where ${D/D\lambda\equiv \dot x^\mu\nabla_\mu}$ is the covariant derivative along the tangent $\dot x^\mu$~($\equiv \dx x^\mu/\dx\lambda$) to the worldline. If we now introduce the conjugate momentum variables
\begin{equation}
	p_\mu
	=
	\bigg(\frac{\delta S_\pp}{\delta\dot x^\mu}\bigg)_\Omega
	\quad\text{and}\quad
	S_{\mu\nu}
	=
	\bigg(\frac{\delta S_\pp}{\delta\Omega^{\mu\nu}}\bigg)_{\dot x},
\end{equation}
we may easily write down the action for this point particle in first-order form:
\begin{equation}
	S_\pp
	=
	\int \mathcal L_\pp\,\dx\lambda,
	\quad
	\mathcal L_\pp
	=
	p_\mu \dot x^\mu
	+
	\frac{1}{2} S_{\mu\nu}\Omega^{\mu\nu}
	-
	\mathcal H_\pp.
\label{eq:pp_general_action}
\end{equation}

Before we can specify the exact form of the Hamiltonian~$\mathcal H_\pp$, a few more words on the phase-space variables are in order. To start with, note that the transformation from a general set of coordinates~$x^\mu$ to the body-fixed coordinates~$y^A$ can be effected in two stages: one first performs a rescaling of the metric to go into the locally flat frame, followed by a Lorentz transformation to go into the body-fixed frame. Mathematically, we write ${e^\mu_A = \Lambda^a{}^\vph{\mu}_A e^\mu_{a\vph{A}} }$, where $e^\mu_a$ is the vierbein that takes us into the locally flat frame, while $\Lambda^a{}_A$ is the appropriate Lorentz transformation. The usefulness of this decomposition is that it separates $e^\mu_A$ into a part that depends only on the particle's translational degrees of freedom (the vierbein $e^\mu_a$) and a part that depends only on its rotational degrees of freedom (the Lorentz matrix $\Lambda^a{}_A$). Substituting this into our definition of $\Omega^{\mu\nu}$ then reveals that
\begin{equation}
	S_{\mu\nu}\Omega^{\mu\nu}
	=
	S_{ab} (\Omega^{ab}_\Lambda +  \w_\mu^{ab} \dot x^\mu),
\label{eq:pp_SOmega}
\end{equation}
where ${\Omega^{ab}_\Lambda = \eta^{AB} \Lambda^a{}_A \dot\Lambda^b{}_B}$ is the angular velocity relative to the locally flat frame (which is emphatically different from ${\Omega^{ab} = e^a_\mu e^b_\nu \Omega^{\mu\nu}}$),
while
${S_{ab} = e^\mu_{a \vph{bb}} e^{\nu \vph{\mu}}_b S_{\mu\nu} }$
is the corresponding spin tensor in this frame. Crucially, because the ``kinetic term'' $S_{ab}\Omega^{ab}_\Lambda$ for the rotational degrees of freedom is independent of the metric, we see that a minimal coupling between gravity and spin appears solely through an interaction term involving the spin connection
\begin{equation}
	\w_\mu^{ab}
	\coloneq
	g^{\rho\sigma} e^b_\sigma \nabla_\mu^\vph{b} e^{a \vph{b}}_\rho.
\end{equation}

Let us now count the total number of degrees of freedom. The generalised coordinates $(x^\mu,\Lambda^a{}_A)$ and their conjugate momenta $(p_\mu, S_{ab})$ together give us a total of 20 phase-space variables: the worldline coordinates~$x^\mu$ and the linear momentum~$p_\mu$ contain four degrees of freedom each, the Lorentz matrix is constrained by its defining property ${\Lambda^a{}_A \Lambda^b{}_B \eta^{AB} = \eta^{ab}}$ and so has six degrees of freedom, while the spin tensor~$S_{ab}$ is antisymmetric by construction and so carries another six degrees of freedom. This is ultimately eight too many (only~12~phase-space variables, or six generalised coordinates, are needed to uniquely describe a spinning point particle\,---\,three coordinates for its position and three angles to describe its orientation); hence, we must impose a commensurate number of constraints. This can be accomplished by~choosing~\cite{Steinhoff:2009ei}
\begin{equation}
	\mathcal H_\pp
	=
	\frac{e}{2m}(p^2 + m^2)
	+
	e\chi_a ({\textstyle\sqrt{-p^2}}\Lambda^a{}_0 - p^a)
	+
	e\,\xi^a S_{ab} p^b,
\label{eq:pp_H}
\end{equation}
where the fields~$e(\lambda)$, $\chi_a(\lambda)$, and $\xi^a(\lambda)$ serve as Lagrange multipliers.

Starting from the right, we see that extremising the action with respect to~$\xi^a$ enforces the constraint ${ S_{ab} p^b \approx 0}$~\cite{Tulczyjew:1959ssc}.%
\footnote{We use the ``$\approx$'' symbol to denote a weak equality in the sense of Dirac~\cite{Dirac:1964lqm}.}%
\,This constraint, which is known as the covariant spin supplementary condition~(SSC), removes the three unphysical degrees of freedom contained in the spin tensor~$S_{ab}$, and removes no more than three because $S_{ab} p^b$ is trivially orthogonal to~$p^a$. Next, extremising with respect to $\chi_a$ imposes the conjugate constraint~${\Lambda^a{}_0 \approx p^a/\sqrt{-p^2}}$~\cite{Hanson:1974qy}, which also removes just three degrees of freedom, since ${(\sqrt{-p^2}\Lambda^a{}_0 - p^a)}$ is orthogonal to ${(\sqrt{-p^2}\Lambda^a{}_0 + p^a)}$. In physical terms, what this constraint on $\Lambda^a{}_0$ amounts to is a gauge fixing of the (superfluous) boost degrees of freedom, which thereby sets the timelike vector~$e^\mu_{A=0}$ parallel to the particle's 4-momentum~$p^\mu$. The choice of SSC, meanwhile, essentially determines what we mean by the ``centre of energy'' of this spinning object~\cite{Pryce:1948pf}. We will work exclusively with these gauge choices in what follows, although it is worth mentioning that other choices are certainly possible~\cite{Pryce:1948pf, Porto:2005ac, Levi:2015msa}.

Returning to \eqref{eq:pp_H}, we now see that extremising the action with respect to the einbein~$e$ enforces the mass-shell constraint ${p^2 + m^2 \approx 0}$. Because rotational energy necessarily gravitates, the particle's Arnowitt-Deser-Misner (ADM) mass~$m$ is generically a function of the spin's absolute magnitude; i.e., ${m^2\equiv f(S^2)}$ with ${S^2 = S_{ab} S^{ab}/2}$~\cite{Hanson:1974qy, Lorentsen:1997wt, Steinhoff:2015ksa}. Specifying an exact form for this functional dependence will, however, turn out to be unnecessary. As we show in appendix~\ref{app:ssc}, the equations of motion descending from this action guarantee that $S^2$ is conserved; hence, $m$~is also a constant of the motion regardless of how it depends on~$S^2$.%
\footnote{See, e.g., ref.~\cite{Goldberger:2020fot} for a point-particle action that conserves neither $p^2$ nor $S^2$. Such an action is useful for incorporating dissipative effects, such as the absorptive nature of a black hole's horizon.}

The reader keeping score should have counted a total of seven independent constraints imposed thus far, which means that there is still one more to go. This last constraint is also enforced by the einbein, albeit more subtly, as its presence renders the action~$S_\pp$ reparametrisation invariant. Said in other words, the transformation ${\lambda \mapsto \lambda'}$ and ${e(\lambda)\,\dx\lambda \mapsto e'(\lambda')\,\dx\lambda'}$ is a gauge symmetry of the point-particle action.%
\footnote{That the einbein~$e$ is playing this dual role as both a Lagrange multiplier and a Stueckelberg field is ultimately tied to the fact that, in the Hamiltonian description of this problem, the mass-shell constraint ${p^2 + m^2 \approx 0}$ is first class, whereas the spin constraints ${S_{ab} p^b \approx 0}$ and ${\sqrt{-p^2}\Lambda^a{}_0 - p^a \approx 0}$ are second class~\cite{Hanson:1974qy, Lorentsen:1997wt}.\looseness=-1}
Fixing a gauge for the worldline parameter~$\lambda$ therefore removes the last remaining unphysical degree of freedom, contained in~$x^\mu(\lambda)$. For post-Newtonian applications, the most natural gauge choice would be to set ${\lambda=x^0}$ (as we do in section~\ref{sec:consv}), while for fully relativistic problems, one might instead prefer to work with the condition ${\dot x_\mu \dot x^\mu = -1}$, which sets~$\lambda$ equal to the proper time.

As a final remark, let us also note that this reparametrisation invariance of the action is the reason why the Hamiltonian in \eqref{eq:pp_H} is made up purely of constraint terms\,---\,the canonical Hamiltonian of a generally covariant system typically vanishes~\cite{Henneaux:1992ig}.

% ============================================== %
\subsection{Scalar couplings and strong-gravity effects}
\label{sec:pp_stt}

Up to higher-order terms that account for its finite size~\cite{Goldberger:2004jt, Porto:2005ac, Levi:2015msa, Bini:2012gu, Endlich:2015mke, Steinhoff:2016rfi, Kol:2011vg, Hui:2020xxx, Goldberger:2005cd, Goldberger:2020fot}, the action that we have just written down can be used to describe the behaviour of any extended, spinning object in general relativity. Moreover, because the weak equivalence principle holds in the Jordan frame by construction, this same action will also describe how \emph{weakly} gravitating objects behave in the scalar-tensor theories of~\eqref{eq:review_S_full} once we replace $g_{\mu\nu}$ by the Jordan-frame metric~$\tilde g_{\mu\nu}$. (Strongly gravitating objects will be discussed shortly.) To make this manifest, let us affix tildes to everything, such that the Lagrangian in the Jordan~frame~reads
\looseness=-1
\begin{align}
	\mathcal L_\pp
	&=
	\tilde p_\mu \dot x^\mu
	+
	\frac{1}{2} \tilde S_{ab}
	(\tilde \Omega_\Lambda^{ab} + \tilde\w_\mu^{ab}\dot x^\mu)
	-
	\frac{\tilde e}{2m}(\tilde p^2 + m^2)
	-
	\tilde e\tilde\chi_a
	({\textstyle\sqrt{-\tilde p^2}}\tilde\Lambda^a{}_0 - \tilde p^a)
	-
	\tilde e\,\tilde\xi^a \tilde S_{ab} \tilde p^b.
\label{eq:pp_L_Jordan}
\end{align}

\paragraph{Disformal transformations.}
Going back to the Einstein frame by way of \eqref{eq:review_Jordan_frame} will now generate direct couplings between this point particle and the scalar~$\phi$. At the level of the vierbeins, this transformation reads
\begin{equation}
	\tilde e_\mu^a = A(\phi,X)[ e_\mu^a + B(\phi,X) \nabla^a\phi\nabla_\mu\phi],
\label{eq:pp_disformal_transformation}
\end{equation}
where $A$ and $B$ are related to the coupling functions $C$ and $D$ in \eqref{eq:review_Jordan_frame} by the equations ${C=A^2}$ and ${D = B + XB^2}$. Accordingly, the conformal coupling function~$A$ is dimensionless, while the disformal coupling function~$B$ has an overall scaling of the form~${B \propto 1/\M^4}$.

In addition to \eqref{eq:pp_disformal_transformation}, we will also need expressions for $\tilde e^\mu_a$, $\tilde g^{\mu\nu}$, and $\tilde \w_\mu^{ab}$ in terms of their Einstein-frame counterparts. The first appears in the Lagrangian when projecting the 4-momentum onto the locally flat frame, ${\tilde p^a \equiv \eta^{ab}\tilde e^\mu_b \tilde p_\mu}$; the second in the definition of the inner product ${\tilde p^2 \equiv \tilde g^{\mu\nu} \tilde p_\mu \tilde p_\nu}$; and the third in the spin-gravity coupling $\tilde S_{ab} \tilde\w_\mu^{ab} \dot x^\mu$. For simplicity, we will here work only to leading order in the disformal coupling, and so will systematically discard any term that scales like $\sim (\nabla\phi\nabla\phi/\M^4)^{1+p}(\phi/\mpl)^q$ with $p+q \geq 1$. Since $A$ and $B$ depend on the scalar only via the dimensionless combinations $\phi/\mpl$ and $X/\M^4$ (recall our discussion in section~\ref{sec:review}), what this means in practice is that we will neglect any and all terms involving two or more powers of~$B$, terms where $B$ is multiplied by one or more powers of~$\nabla_\mu A$, and terms involving derivatives of~$B$. With this in mind, it is straightforward to~show~that 
\looseness=-1 
\begin{subequations}
\label{eq:pp_more_disformal_transformations}
\begin{align}
	\tilde e^\mu_a
	&=
	A^{-1}(e^\mu_a - B\nabla^\mu\phi\nabla_a\phi),
	\\
	\tilde g^{\mu\nu}
	&=
	A^{-2}(g^{\mu\nu} - 2B \nabla^\mu\phi \nabla^\nu\phi),
	\\
	\tilde\w_\mu^{ab}
	&=
	\w_\mu^{ab} - 2 e_\mu^{[a} \nabla^{b]}\log A
	-
	2B \nabla^{[a}\phi \nabla^{b]}\nabla_\mu\phi
\end{align}
\end{subequations}
at the order to which we are working, and consequently
\begin{align}
	\mathcal L_\pp
	=&\;
	\tilde p_\mu \dot x^\mu
	+
	\frac{1}{2}\tilde S_{ab}
	\big(
		\tilde\Omega_\Lambda^{ab}
		+
		\w_\mu^{ab} \dot x^\mu
		-
		2\dot x^\mu e_\mu^{a} \nabla^{b}\log A
		-
		2B \dot x^\mu \nabla^{a}\phi \nabla^{b}\nabla_\mu\phi
	\big)
	\nonumber\\[-3pt]&
	-
	\frac{\tilde e}{2mA^2}
	\big[ \tilde p^2 - 2B(\tilde p\cdot\!\nabla\phi)^2 + A^2m^2 \big]
	-
	\tilde e A^{-1}\tilde\xi^a \tilde S_{ab}
	[\tilde p^b - B(\tilde p\cdot\!\nabla\phi)\nabla^b\phi]
	\nonumber\\[2pt]&
	-
	\tilde e A^{-1}\tilde\chi_a
	\big\{
		\sqrt{-[\tilde p - B(\tilde p\cdot\!\nabla\phi)\nabla\phi]^2}
		\,\tilde \Lambda^a{}_0
		-
		[\tilde p^a - B(\tilde p\cdot\!\nabla\phi)\nabla^a\phi]
	\big\}
\label{eq:pp_L_Einstein_raw}
\end{align}
when written in the Einstein frame.

It is no surprise that this frame transformation has rendered the point-particle action considerably more complex, but we will now show that much of this complexity can be removed by appropriate redefinitions of the phase-space variables ${ ( x^\mu, \tilde p_\mu, \tilde\Lambda^a{}_A, \tilde S_{ab} ) }$ and the Lagrange multipliers ${ (\tilde e, \tilde\chi_a, \tilde\xi^a) }$. Indeed, we affixed tildes onto these quantities for exactly this reason. All of them are naturally defined without any reference to a metric, and so remain unchanged under frame transformations\,---\,the tildes that we have introduced instead serve to distinguish them from a new set of variables (without tildes) that we will now introduce.

\paragraph{Field redefinitions.}
To motivate these field redefinitions, consider what would happen if we worked with the Lagrangian in \eqref{eq:pp_L_Einstein_raw} as is. Extremising the corresponding action with respect to $\tilde\xi^a$ gives us an SSC that reads ${\tilde S_{ab}[\tilde p^b - B(\tilde p\cdot\!\nabla\phi)\nabla^b\phi] \approx 0}$, which implies that a term that is proportional to $\tilde S_{ab} \tilde p^b$ in the equations of motion is secretly suppressed by one extra power of the disformal coupling, since $\tilde S_{ab} \tilde p^b$ gets replaced by $B \tilde S_{ab} (\tilde p\cdot\!\nabla\phi)\nabla^b\phi$ after we impose the SSC. There is nothing wrong with this per se, but this kind of mixing between sectors with different powers of $B$ can be quite cumbersome to keep track of, especially at high post-Newtonian orders, and so it would be desirable if we could find a field redefinition ${ (\tilde p_\mu, \tilde S_{ab}, \,\dots) \mapsto (p_\mu, S_{ab}, \,\dots) }$ such that the new variables satisfy the usual covariant~SSC, ${S_{ab} p^b \approx 0}$, as in general relativity. Additionally, we would also like for this redefinition to be such that the new momentum variable~$p_\mu$ is proportional to the tangent vector~$\dot x^\mu$, up to corrections from the spin; thereby guaranteeing that no further mixing can arise between the different sectors when we Legendre transform to the Routhian in the next subsection. Finally, we will also require that this field redefinition be such that $\Lambda^a{}_A$ satisfies the usual conjugate~constraint,~${\sqrt{-p^2}\Lambda^a{}_0 - p^a \approx 0}$. It is worth emphasising now that these field redefinitions are essentially just gauge transformations of the phase-space variables; hence, although they will invariably affect the explicit form of gauge-dependent results like the equations of motion, they will necessarily have no effect on gauge-invariant quantities like the gravitational-wave phase.

Mostly by trial and error, we have found that the desired outcomes above can all be achieved (at least, at the order to which we are working) by making the transformations
\begin{subequations}
\label{eq:pp_field_redefinitions}
\begin{align}
	\tilde p_\mu
	&=
	p_\mu + 2B(p\cdot\!\nabla\phi)\nabla_\mu\phi,
	\\
	\tilde\Lambda^a{}_A
	&=
	L^a{}_b \Lambda^b{}_A,
	\allowdisplaybreaks\\
	\tilde S_{ab}
	&=
	L_a{}^c L_b{}^d S_{cd},
	\allowdisplaybreaks\\
	\tilde\chi_a &= A^{-1}
	\bigg(
		L_a{}^b\chi_b
		+
		\frac{ B(p\cdot\!\nabla\phi)^2\chi_a }%
			{(-p)^2}
	\bigg),
	\allowdisplaybreaks\\
	\tilde\xi^a
	&= A^{-1}
	\bigg(
		\xi^a
		+
		\frac{2B(p\cdot\!\nabla\phi)p_b\, \xi^{[a}\nabla^{b]}\phi}%
			{(-p^2)}
	\bigg),
	\\
	\tilde e &= A^2 e.
\end{align}
\end{subequations}
The matrix $L^a{}_b$ that appears in the second, third, and fourth lines is an infinitesimal Lorentz transformation%
\footnote{One can see this easily by verifying that $L^a{}_b$ satisfies the constraint ${L^a{}_c L^b{}_d \,\eta_{ab} = \eta_{cd}}$ at the order in~$B$ to which we are working. This property is essential as it guarantees that both the old and the new Lorentz matrices satisfy the requisite orthonormality constraint, ${\tilde\Lambda^a{}_A \tilde\Lambda^b{}_B \eta^{AB} = \Lambda^a{}_A \Lambda^b{}_B \eta^{AB} = \eta^{ab}}$.}
given by ${L^a{}_b = \eta^{ac} e_{c \vph{b}}^\mu e_b^{\nu \vph{\mu}} L_{\mu\nu}}$, where ${L_{\mu\nu} = g_{\mu\nu} + \Theta_{\mu\nu}}$ and
\begin{equation}
	\Theta_{\mu\nu}
	=
	\frac{2B(p\cdot\!\nabla\phi) \, p_{[\mu\!}\nabla_{\nu]}\phi }{(-p^2)}.
\label{eq:pp_field_redefinitions_Theta}
\end{equation}
The angular velocity tensor can then be shown to transform as
\begin{equation}
	\tilde\Omega^{\mu\nu}
	=
	L^\mu{}_\rho L^\nu{}_\sigma \Omega^{\rho\sigma}
	+
	L^\mu{}_\rho \covdiff{}{\lambda} L^{\nu\rho}
	=
	L^\mu{}_\rho L^\nu{}_\sigma \Omega^{\rho\sigma}
	-
	\covdiff{}{\lambda} \Theta^{\mu\nu}
\end{equation}
under the action of \eqref{eq:pp_field_redefinitions}, and after putting everything together, we find that the Lagrangian in terms of these new variables is
\begin{align}
	\mathcal L_\pp
	&=
	[p_\mu + 2B(p\cdot\!\nabla\phi)\nabla_\mu\phi]\, \dot x^\mu
	-
	\frac{e}{2m}
	\big[ p^2 + 2B(p\cdot\!\nabla\phi)^2 + A^2m^2 \big]
	\nonumber\\&\quad
	+
	\frac{1}{2} S_{ab}
	\big(
		\Omega_\Lambda^{ab}
		+
		\w_\mu^{ab}\dot x^\mu
	\big)
	-
	S_{\mu\nu}
	(
		\dot x^{\mu}\nabla^{\nu}\log A
		+
		B \dot x^\rho \nabla^{\mu}\phi \nabla^{\nu}\nabla_\rho\phi
	)
	\nonumber\\&\quad
	-
	\frac{S_{\mu\nu}}{2}\covdiff{\Theta^{\mu\nu}}{\lambda}
	-
	e\xi^a S_{ab} p^b
	-
	e\chi_a
	( {\textstyle\sqrt{-p^2}}\Lambda^a{}_0 - p^a).
\end{align}

One last field redefinition can be made to further simplify the term involving $D\Theta^{\mu\nu}\!/D\lambda$. Looking at \eqref{eq:pp_field_redefinitions_Theta}, we see that the application of the product rule will produce two categories of terms: those that are proportional to $D p_\alpha/D\lambda$, and those that are not. The terms in the former category, which prima facie lead to third-order equations of motion for the worldline, turn out to be redundant operators that can be order-reduced by an appropriate redefinition of the worldline coordinates, ${ x^\mu \mapsto x^\mu + \delta x^\mu }$~\cite{Damour:1990jh}. To see how this works in our case, first take ${\delta x^\mu = B S^{\rho\sigma} z^\mu_{\rho\sigma}}$, where $z^\mu_{\rho\sigma}$ is some object to be determined. The net effect of this shift in the worldline is to produce a corresponding shift in the action, given by $\delta S_\pp = \int\dx\lambda\, B S^{\rho\sigma} z^\mu_{\rho\sigma} \mathcal E_\mu + \mathcal O(B^2)$, where ${\mathcal E_\mu = \delta S_\pp/\delta x^\mu}$ is the equation of motion that follows from extremising the (unshifted) action with respect to~$x^\mu$. At the order to which we are working, only the general relativistic part of $\mathcal E_\mu$ contributes, since there is already one explicit power of $B$ in~$\delta S_\pp$. This general relativistic part is given by the Mathisson--Papapetrou--Dixon equations \cite{Mathisson:1937zz, Papapetrou:1951pa, Dixon:1970zza, Dixon:1970zz, Dixon:1974}, and reads
${ \mathcal E^\text{(GR)}_\mu = -Dp_\mu/D\lambda - R_{\mu\nu\rho\sigma}\dot x^\nu S^{\rho\sigma}/2 }$ 
[see~also~\eqref{eq:app_ssc_eom_p}]. It~now follows that if we choose $z^\mu_{\rho\sigma}$ in exactly the right way, then the term $\int\dx\lambda\, B S^{\rho\sigma} z^\mu_{\rho\sigma} (-Dp_\mu/D\lambda)$ that arises from this shift of the worldline coordinates can be made to exactly cancel the terms in $-\int\dx\lambda\, (S_{\mu\nu}/2)D\Theta^{\mu\nu}\!/D\lambda$ that are proportional to $Dp_\alpha/D\lambda$. Left behind in their place is a new contribution to the action, $\int\dx\lambda\, B S^{\rho\sigma}z^\mu_{\rho\sigma}(- R_{\mu\nu\alpha\beta}\dot x^\nu S^{\alpha\beta}/2)$, which cannot be eliminated any further, but since this term is quadratic in the spins, it will play no role in the spin-orbit effects that are the main subject of this paper, and so will~be~neglected~henceforth.

All that remains after this procedure is the second category of terms in $D\Theta^{\mu\nu}\!/D\lambda$ that are \emph{not} proportional~to~$Dp_\alpha/D\lambda$. Written out explicitly, we have that
\begin{align}
	\mathcal L_\pp
	=\;&
	[p_\mu + 2B(p\cdot\!\nabla\phi)\nabla_\mu\phi]\, \dot x^\mu
	-
	\frac{e}{2m}
	\big[ p^2 + 2B(p\cdot\!\nabla\phi)^2 + A^2m^2 \big]
	\nonumber\\&
	+
	\frac{1}{2} S_{ab}
	\big(
		\Omega_\Lambda^{ab}
		+
		\w_\mu^{ab}\dot x^\mu
	\big)
	-
	S_{\mu\nu}
	(
		\dot x^{\mu}\nabla^{\nu}\log A
		+
		B \dot x^\rho \nabla^{\mu}\phi \nabla^{\nu}\nabla_\rho\phi
	)
	\nonumber\\&
	+
	\frac{p^\sigma p^\mu}{p^2}
	BS_{\mu\nu} \dot x^\rho
		\nabla_\rho(\nabla_\sigma\phi \nabla^\nu\phi)
	-
	e\xi^a S_{ab} p^b
	-
	e\chi_a
	( {\textstyle\sqrt{-p^2}}\Lambda^a{}_0 - p^a).
\label{eq:pp_L_weak}
\end{align}

\paragraph{Strong-gravity effects.}
We asserted earlier that the Jordan-frame Lagrangian in \eqref{eq:pp_L_Jordan} was valid only for weakly gravitating objects, and since \eqref{eq:pp_L_weak} follows directly from \eqref{eq:pp_L_Jordan}~after a series of smooth transformations, this Einstein-frame Lagrangian must be subject to the same limitations as well. That said, very few modifications will turn out to be necessary to render this Lagrangian capable of also describing strongly gravitating objects.
\looseness=-1

To see why, consider what would change were we to construct this point-particle Lagrangian in a different fashion. Rather than generate couplings to the scalar by way of the frame transformation in~\eqref{eq:pp_disformal_transformation}, we could have also elected to build this Lagrangian from the bottom up by simply writing down all possible contractions between the phase-space variables, the fields $(g_{\mu\nu},\phi)$, and their derivatives. Up to redundant operators, which can be removed by appropriate field redefinitions~\cite{Damour:1998jk, Goldberger:2007hy}, one finds that the most relevant terms arising from this more general procedure are exactly those in~\eqref{eq:pp_L_weak}; the~only difference being that this bottom-up approach does not specify a~priori what the values of the Wilson coefficients multiplying each of these terms ought to be. Thus, the result of this~construction~is
\begin{align}
	\mathcal L_\pp
	&=
	[p_\mu + 2\B(p\cdot\!\nabla\phi)\nabla_\mu\phi]\, \dot x^\mu
	-
	\frac{e}{2m}
	\big[ p^2 + 2\B(p\cdot\!\nabla\phi)^2 + \A^2m^2 \big]
	\nonumber\\&\quad
	+
	\frac{1}{2} S_{ab}
	\big(
		\Omega_\Lambda^{ab}
		+
		\w_\mu^{ab}\dot x^\mu
	\big)
	-
	S_{\mu\nu}
	(
		\dot x^{\mu}\nabla^{\nu}\log\widetilde\A
		+
		\widetilde\B \dot x^\rho \nabla^{\mu}\phi
		\nabla^{\nu}\nabla_\rho\phi
	)
	\nonumber\\&\quad
	+
	\frac{p^\sigma p^\mu}{p^2}
	(
		\acute\B S_{\mu\nu} \dot x^\rho
		\nabla_\rho\nabla_\sigma\phi \nabla^\nu\phi
		+
		\grave\B S_{\mu\nu} \dot x^\rho
		\nabla_\sigma\phi \nabla_\rho\nabla^\nu\phi
	)
	\nonumber\\&\quad
	-
	e\xi^a S_{ab} p^b
	-
	e\chi_a
	( {\textstyle\sqrt{-p^2}}\Lambda^a{}_0 - p^a),
\label{eq:pp_L_final}
\end{align}
which is highly reminiscent of~\eqref{eq:pp_L_weak}, except that the coupling functions~$(A,B)$ have now been promoted to a set of arbitrary functions $(\A,\widetilde\A,\B,\widetilde\B, \acute\B, \grave\B)$. As these must reduce back to $(A,B)$ in the limit of a weakly gravitating object, the $\A$-type functions are all dimensionless, while the $\B$-type functions are all proportional to~$1/\M^4$.
Notice also that each function appears in this Lagrangian only once, with the exception of the function~$\B$, which appears twice in the first line. No loss of generality is incurred, however, because field redefinitions similar to those in~\eqref{eq:pp_field_redefinitions} can always be made to put the Lagrangian into such a form. In fact, recall that one of the criteria we demanded of the transformations in~\eqref{eq:pp_field_redefinitions} was that the new momentum variable~$p_\mu$ should be proportional to the tangent vector~$\dot x^\mu$ (up to corrections from the spin), and indeed, this is guaranteed only if the two disformal spin-independent terms in~\eqref{eq:pp_L_final} depend on the~same~function~$\B$.

Each of these arbitrary functions ${\mathcal F \in \{ \A, \B, \dots \}}$ should be viewed as a formal power series of the form
\begin{equation}
	\mathcal F
	=
	\sum_{p=0}^\infty \sum_{q=0}^\infty
	\frac{ \mathcal F^{(p,q)} }{p!(2q)!}
	\bigg( \frac{\phi-\phi_0}{\mpl}\bigg)^p
	\bigg( \frac{\nabla^\mu\phi\nabla_{\!\mu}\phi}{\M^4}\bigg)^q,
\label{eq:pp_formal_power_series}
\end{equation}
where $\phi_0$ is the ambient value of the scalar in the absence of this body [cf.~\eqref{eq:review_weak_field_approximation}], and $\mathcal F^{(p,q)}$ are the aforementioned Wilson coefficients. That these coefficients now appear as free parameters in the Lagrangian is exactly what allows it to be general enough to describe the behaviour of any extended, spinning object within the class of models in~\eqref{eq:review_S_full}, up to subleading effects associated with the set of higher-order terms that we have been systematically discarding (namely, higher-order disformal interactions, spin effects of quadratic order or higher, and finite-size effects like tidal~deformations%
\footnote{Dipolar tidal effects due to a scalar field are discussed in ref.~\cite{Bernard:2019yfz}.}%
). If desired, one can then specialise to a specific object, like a black hole or a neutron star, by performing a number of~matching~calculations.

In most of this paper, these Wilson coefficients will be left unspecified for the sake of generality, although it will still be instructive to briefly discuss how this kind of matching is done. Consider the coefficient $\mathcal A^{(1,0)}$, which is more commonly denoted by ${\alpha = -2\mathcal A^{(1,0)}}$, as an example. Aside from the mass parameter~$m$, this is the only Wilson coefficient that contributes to the dynamics of a binary system at Newtonian order, where it is responsible for setting the overall strength of the force that the scalar mediates between the two bodies. For the purposes of a matching calculation, however, it will suffice to consider a much simpler scenario in which a single body just remains at rest in what is otherwise empty space. At distances~$r$ much greater than the size of this body (but much smaller than the Compton wavelength of the scalar, which we are assuming is very light), the point-particle theory predicts that the surrounding scalar-field profile should be given by ${\phi = \phi_0 + \alpha m/(8\pi\mpl r) + \mathcal O(1/r^2)}$. The value of this \emph{effective coupling strength}~$\alpha$ can then be determined for a given body by matching this result onto the one obtained by solving the field equations of the full~theory~in~\eqref{eq:review_S_full}. 

Working perturbatively in powers of ${s \sim \GN m/R}$, Damour and Esposito--Far\`{e}se~\cite{Damour:1992we} showed that the effective coupling strength for a (fluid)~body of radius~$R$ is given schematically by ${\alpha \sim \alpha_\text{weak}(1 + a_1 s + a_2 s^2 + \cdots)}$.%
\footnote{See also ref.~\cite{Kuntz:2019zef} for a nice discussion in terms of bare and renormalised couplings.}
The overall prefactor ${\alpha_\text{weak} = -2\mpl \,\dx A/\dx \phi|_{\phi = \phi_0}}$ is~the value of this coupling in the limit of negligible self-gravity (${s\ll 1}$), and is in~complete agreement with the predictions of our Lagrangian for weakly gravitating objects in~\eqref{eq:pp_L_weak}. For larger values of~$s$, the microscopic details of the fluid begin to affect the overall strength with which this body couples to the scalar, and this information is encoded in the coefficients~$(a_1, a_2, \dots)$. How exactly these coefficients depend on the body's equation of state and on the parameters of the underlying theory are details that we will not care to go into here, but we would be remiss not to highlight the existence of a certain class of scalar-tensor theories for which the infinite series of self-gravity contributions ${(1 + a_1 s + a_2 s^2 + \cdots)}$ can compensate for a vanishingly small $\alpha_\text{weak}$ to give a (relatively) large coupling strength~$\alpha$. This phenomenon, known as spontaneous scalarisation~\cite{Damour:1993hw, Damour:1996ke, Minamitsuji:2016hkk, Silva:2017uqg, Doneva:2017duq, Andreou:2019ikc, Ventagli:2020rnx}, is particularly interesting from an observational standpoint, as it would allow for neutron stars in a given mass range to couple strongly to the scalar field, and thus exhibit substantial deviations from general relativity in the strong-field regime. Meanwhile, the smallness of $\alpha_\text{weak}$ ensures that these models remain indistinguishable from general relativity in the weak-field regime, where it is already tightly constrained by Solar~System~tests~\cite{Bertotti:2003rm, Hofmann:2018llr}.%
\footnote{The simplest models that exhibit spontaneous scalarisation also lead to a cosmological history that is inconsistent with observations~\cite{Damour:1992kf, Damour:1993id, Anderson:2016aoi, Alby:2017dzl, Anson:2019uto, Franchini:2019npi}, although recent proposals~\cite{Anson:2019ebp, Antoniou:2020nax} have been put forward to resolve this issue. In any case, we shall not dwell too much on these cosmological concerns here, as our results also apply to other classes of models that do not possess this issue.}

We could also tune the Wilson coefficients in~\eqref{eq:pp_L_final} to describe a black hole, although in this case we know that\,---\,for almost all choices of the field action~$S_\text{fields}$\,---\,no-hair theorems preclude the possibility of a nontrivial scalar-field profile~\cite{Thorne:1971, Bekenstein:1971hc, Adler:1978dp, Hawking:1972qk, Zannias:1994jf, Bekenstein:1995un, Saa:1996aw, Sotiriou:2011dz, Hui:2012qt, Graham:2014mda, Chrusciel:2012jk}; hence, black holes will typically have ${\alpha=0}$. Exceptionally, one can circumvent these restrictions to find solutions with scalar ``hair'' if the ambient field value~$\phi_0$ is allowed to vary with time~\cite{Jacobson:1999vr, Horbatsch:2011ye, Berti:2013gfa, Wong:2019yoc, Clough:2019jpm, Hui:2019aqm, Bamber:2020bpu, Babichev:2013cya, Kobayashi:2014eva}, or if the field action is chosen to include a coupling between the scalar and the Gauss--Bonnet invariant~\cite{Kanti:1995vq, Kleihaus:2011tg, Pani:2011gy, Yunes:2011we, Maselli:2015tta, Sotiriou:2014pfa, Antoniou:2017hxj, Delgado:2020rev, Sullivan:2020zpf, Doneva:2017bvd, Silva:2017uqg, Cunha:2019dwb, Minamitsuji:2019iwp, Herdeiro:2020wei, Berti:2020kgk, East:2021bqk}, and indeed~${\alpha \neq 0}$ in~these~cases.%
\footnote{Our Lagrangian is valid only for real scalar fields varying on length scales much larger than the size of the compact object, and so cannot be used to describe some known hairy black hole solutions involving massive, complex scalars~\cite{Herdeiro:2014goa, Degollado:2018ypf} or pseudoscalars~\cite{Alexander:2009tp, Yunes:2009hc, Delsate:2018ome}. A point-particle Lagrangian for the latter was recently constructed in ref.~\cite{Loutrel:2018ydv}.}

% ============================================== %
\subsection{The point-particle Routhian}
\label{sec:pp_Routhian}

The first-order formalism that we have been using thus far has proven itself ideal for building Lagrangians from the bottom up, as it makes the symmetries and the constraint structure of the problem manifest. Having completed this task, it is no longer advantageous to continue working with the full set of phase-space variables. In preparation for the next section, we shall now transform~\eqref{eq:pp_L_final} into a Routhian that depends only on the worldline coordinates~$x^\mu$, the tangent vector~$\dot x^\mu$, and the spin~tensor~$S_{ab}$.

Our first step is to ``integrate out'' the momentum~$p_\mu$, which at the classical level is tantamount to solving the corresponding equation of motion (${\delta S_\pp/\delta p_\mu = 0}$) and substituting the solution back into the action. It turns out that the algebra simplifies dramatically if we can ignore the constraint on~$\Lambda^a{}_0$, and so we shall set ${\chi_a = 0}$ for the time being. Having done~so, extremising the action with respect to~$p_\mu$ yields
\begin{align}
	p_\mu 
	=&\;
	me^{-1}\dot x_\mu
	-
	2\B[(p-me^{-1}\dot x)\cdot\!\nabla\phi]\nabla_\mu\phi
	+
	m S_{\mu\nu}\xi^\nu
	\nonumber\\&
	+
	\bigg(
		\frac{\delta_\mu^{(\sigma} p^{\alpha)}_\vph{\mu} }{p^2}
		-
		\frac{p_\mu p^\sigma p^\alpha }{p^4}
	\bigg)
	2S_{\alpha\nu} me^{-1}\dot x^\rho 	
	(\acute\B \nabla_\rho\nabla_\sigma\phi\nabla^\nu\phi
	+
	\grave\B \nabla_\sigma\phi\nabla_\rho\nabla^\nu\phi),
\label{eq:pp_Routhian_eom_x_raw}
\end{align}
which we can solve order by order in the spin and the disformal coupling to get
\begin{align}
	p_\mu
	=&\;
	me^{-1}\dot x_\mu + m S_{\mu\nu} \xi^\nu
	+
	2\B m (S_{\rho\sigma}\xi^\rho\nabla^\sigma\phi)\nabla_\mu\phi
	\nonumber\\&
	+
	\bigg(
		\frac{\delta_\mu^{(\sigma} \dot x^{\alpha)}_\vph{\mu} }{\dot x^2}
		-
		\frac{\dot x_\mu \dot x^\sigma \dot x^\alpha }{\dot x^4}
	\bigg)
	2S_{\alpha\nu} \dot x^\rho 	
	(\acute\B \nabla_\rho\nabla_\sigma\phi\nabla^\nu\phi
	+
	\grave\B \nabla_\sigma\phi\nabla_\rho\nabla^\nu\phi).
\label{eq:pp_Routhian_eom_x_raw_sol}	
\end{align}
Substituting this back into \eqref{eq:pp_L_final} and discarding all of the higher-order terms as per usual, we then find
\begin{align}
	\mathcal L_\pp
	=&\;
	\frac{m}{2e}[\dot x^2 + 2\B(\dot x\cdot\!\nabla\phi)^2 - \A^2 e^2]
	+
	\frac{1}{2}S_{ab}
	(\Omega_\Lambda^{ab}+\w_\mu^{ab}\dot x^\mu)
	-
	m \xi^a S_{ab} \dot x^b
	-
	S_{\mu\nu}
	\dot x^{\mu}\nabla^{\nu}\log\widetilde\A
\nonumber\\&
	-
	\widetilde\B S_{\mu\nu}
	\dot x^\rho\nabla^{\mu}\phi \nabla^{\nu}\nabla_\rho\phi
	+
	\frac{\dot x^\sigma \dot x^\mu}{\dot x^2}
	S_{\mu\nu} \dot x^\rho
	(\acute\B \nabla_\rho\nabla_\sigma\phi\nabla^\nu\phi
	+
	\grave\B \nabla_\sigma\phi\nabla_\rho\nabla^\nu\phi).
\end{align}

Notice that the einbein~$e$ now appears nonlinearly and no longer acts as a Lagrange multiplier in this new version of the Lagrangian. This is just as well, because it will allow us to integrate out the einbein in the same way as we did for the momentum. The~solution to the equation ${\delta S_\pp/\delta e = 0}$~is
\begin{equation}
	e
	=
	\A^{-1}\sqrt{-\dot x^2 - 2\B(\dot x\cdot\!\nabla\phi)^2}
	\simeq
	\A^{-1}\sqrt{-\dot x^2}
	\bigg(
		1 + \frac{\B(\dot x\cdot\!\nabla\phi)^2}{\dot x^2}
	\bigg),
\end{equation}
and its substitution back into the action gives us
\begin{align}
	\mathcal L_\pp
	=&
	-m\A
	\sqrt{-\dot x^2}
	\bigg(
		1 + \frac{\B(\dot x\cdot\!\nabla\phi)^2}{\dot x^2}
	\bigg)
	+
	\frac{1}{2}S_{ab}
	(\Omega_\Lambda^{ab}+\w_\mu^{ab}\dot x^\mu)
	-
	m \xi^a S_{ab} \dot x^b
	-
	S_{\mu\nu}
	\dot x^{\mu}\nabla^{\nu}\log\widetilde\A
\nonumber\\&
	-
	\widetilde\B S_{\mu\nu}
	\dot x^\rho \nabla^{\mu}\phi \nabla^{\nu}\nabla_\rho\phi
	+
	\frac{\dot x^\sigma \dot x^\mu}{\dot x^2}
	S_{\mu\nu} \dot x^\rho
	(\acute\B \nabla_\rho\nabla_\sigma\phi\nabla^\nu\phi
	+
	\grave\B \nabla_\sigma\phi\nabla_\rho\nabla^\nu\phi).
\label{eq:pp_L_intermediate}
\end{align}

We could now proceed to integrate out the spin tensor~$S_{ab}$ and the vector~$\xi^a$ in a similar fashion to obtain a Lagrangian of second-order form that depends only on~$x^\mu$,~$\dot x^\mu$, and~$\Omega_\Lambda^{ab}$, but as we discussed already, it is traditional and also more convenient to work with the spin, rather than the angular velocity, when modelling the inspiral of a two-body system. This motivates elimimating $\Omega_\Lambda^{ab}$ from~\eqref{eq:pp_L_intermediate} by performing a partial Legendre transform on just the spin variables. The result is the point-particle~Routhian~\cite{Yee:1993ya, Porto:2006bt, Porto:2008tb}
\begin{equation}
	\mathcal R_\pp
	=
	\mathcal L_\pp - \frac{1}{2} S_{ab}\Omega_\Lambda^{ab}.
\label{eq:pp_def_Routhian}
\end{equation}

Variation of this Routhian with respect to~$\xi^a$ will then enforce the covariant SSC, which up to $\mathcal O(S^2)$ corrections now reads\looseness=-1
\begin{equation}
	S_{ab} \dot x^b \approx 0.
\label{eq:pp_ssc}
\end{equation}
The fact that the equation ${\delta S_\pp/\delta\xi^a = 0}$ is independent of~$\xi^a$ implies that, unlike the einbein, this Lagrange multiplier cannot be integrated out. (Note~that we are now defining ${S_\pp \!=\! \int\dx\lambda\,\mathcal R_\pp}$.) Nevertheless, we would still like to eliminate this vector from the Routhian in~\eqref{eq:pp_def_Routhian}, and one way of doing so is to hand-pick a solution for $\xi^a$ that automatically preserves the SSC under time evolution~\cite{Porto:2016pyg}.

Because the task of solving the equation ${D(S_{ab} \dot x^b)/D\lambda=0}$ for~$\xi^a$ is a straightforward but lengthy one, these ancillary details have been relegated to appendix~\ref{app:ssc}. Here, we shall simply quote the end result, which~is~[cf.~\eqref{eq:app_ssc_sol_xi}]
\begin{equation}
	m\xi^a = \nabla^a\log(\tilde\A/\A) 
	+
	\frac{\dot x^\mu \dot x^\nu}{\dot x^2}
	\big[
		(2\B-\widetilde\B-\acute\B)
		\nabla_\mu\nabla_\nu\phi\nabla^a\phi
		+
		(\widetilde\B-\grave\B)
		\nabla_\mu\phi\nabla^a\nabla_\nu\phi
	\big].
\label{eq:pp_xi_solution}
\end{equation} 
Substituting this into \eqref{eq:pp_def_Routhian} then gives us our final expression for the point-particle Routhian:\looseness=-1
\begin{align}
	\mathcal R_\pp
	=&
	-m\A
	\sqrt{-\dot x^2}
	\bigg(
		1 + \frac{\B(\dot x\cdot\!\nabla\phi)^2}{\dot x^2}
	\bigg)
	+
	\frac{1}{2}S_{ab\,} \w_\mu^{ab}\dot x^\mu
	-
	S_{\mu\nu}
	\dot x^{\mu}\nabla^{\nu}\log\A
	\nonumber\\&
	-
	\widetilde\B S_{\mu\nu}
	\dot x^\rho \nabla^{\mu}\phi \nabla^{\nu}\nabla_\rho\phi
	+
	\frac{\dot x^\sigma \dot x^\mu}{\dot x^2}
	S_{\mu\nu}\dot x^\rho
	\big[
	 (2\B - \widetilde\B) \nabla_\rho\!\nabla_\sigma\phi\nabla^\nu\phi
	 +
	 \widetilde\B \nabla_\rho\phi\nabla_\sigma\nabla^\nu\phi
	\big].
\label{eq:pp_Routhian}
\end{align}

At this stage, we should recall that we previously set ${\chi_a = 0}$, and so must now return to examine what would change were we to relax this condition. There are two ways to see that nothing changes. First, notice that the Lorentz matrix~$\Lambda^a{}_A$ is a cyclic coordinate, as it does not appear explicitly in the Lagrangian in~\eqref{eq:pp_L_final}, except in the constraint term involving~$\chi_a$. Consequently, the equations of motion for $x^\mu$ and $S_{ab}$, which are ultimately the only ones we care about, are the same whether or not we impose any constraint on~$\Lambda^a{}_A$. The other way to reach this conclusion is to note that had we kept track of the Lagrange multiplier~$\chi_a$ in the above calculations, we would eventually want to eliminate it from the Routhian in the same way as we did with~$\xi^a$. The solution to the consistency condition ${D(\sqrt{-p^2}\Lambda^a{}_0 -p^a)/D\lambda=0}$ is simply ${\chi_a = 0}$~[cf.~\eqref{eq:app_ssc_sol_chi}].

By virtue of the partial Legendre transform in~\eqref{eq:pp_def_Routhian}, this Routhian will present as either a Lagrangian or a Hamiltonian depending on the context, and as such, the equations of motion for this particle derive from a mixture of Euler--Lagrange and Hamilton equations:
\begin{equation}
	\frac{\delta}{\delta x^\mu}\int\dx\lambda\,\mathcal R_\pp = 0,
	\quad
	\diff{S_{ab}}{\lambda} = \{ S_{ab}, \mathcal R_\pp\}.
\label{eq:pp_eom}
\end{equation} 
The Poisson brackets for the spin equation follow directly from the structure of the kinetic term~$S_{ab}\Omega^{ab}_\Lambda$~\cite{Steinhoff:2015ksa, Levi:2015msa}, and are given by
\begin{equation}
	\{ S_{ab}, x^\mu \} = 0,
	\quad
	\{ S_{ab}, S_{cd} \}
	=
	- \eta_{ac \vph{d}} S_{bd} + \eta_{ad} S_{bc}
	- \eta_{bd} S_{ac \vph{d}} + \eta_{bc} S_{ad}.
\label{eq:pp_S_PB}
\end{equation}

Notice that because these equations are expressed in terms of the covariant spin tensor~$S_{ab}$, one still has to impose the SSC in~\eqref{eq:pp_ssc} by hand \emph{after} all of the functional derivatives and Poisson brackets have been evaluated. This last step may seem rather unsatisfactory within the framework of an EFT, and while it is indeed possible to remove the unphysical degrees of freedom in~$S_{ab}$ already at the level of the action, one must either bear the cost of a more complicated Dirac algebra or otherwise offset it by working with a different~SSC. More details on these alternative approaches can be found in refs.~\cite{Porto:2005ac, Porto:2008tb, Levi:2014sba, Levi:2015msa}, but in what follows, we shall proceed with imposing \eqref{eq:pp_ssc} by hand as described above. This approach, as advocated by Porto and Rothstein~\cite{Porto:2008tb}, will be the most convenient for our purposes, as we will only be computing spin-orbit effects to leading~post-Newtonian~order.

One final remark about~\eqref{eq:pp_Routhian} is necessary. Observe that while we started off with a Lagrangian in~\eqref{eq:pp_L_final} involving six arbitrary functions of the scalar field, only three have survived in the Routhian of~\eqref{eq:pp_Routhian}\,---\,$\A$, $\B$, and $\widetilde\B$. Because the remaining three functions were eliminated at the point when we substituted the solution for $\xi^a$ in \eqref{eq:pp_xi_solution} back into the action, we deduce that the preservation of the SSC under time evolution establishes new relations between the different Wilson coefficients, which a priori appeared to be independent. The general pattern is easy enough to discern: it is those coefficients that multiply operators proportional to the SSC that are not independent, but whose values end up being fixed in terms of the others. This has particularly important ramifications for the conformal sector of the theory, which we now see is parametrised by the single function~$\A$. Accordingly, we conclude that if a scalar is conformally coupled to the mass of a spinning body, then it must also couple \emph{with the same strength} to its spin.
\looseness=-1

% ============================================== %
% Section 4
% ============================================== %
\section{Conservative potential and the binding energy}
\label{sec:consv}

Armed with a general point-particle action for spinning bodies, we now turn to consider the two-body problem. The~complete effective action for this system is given by~[cf.~\eqref{eq:review_S_eff}]
\begin{equation}
	S_\text{eff} = S_\text{fields}[g,\phi]
	+
	\sum_{\K=1}^2 S_{\pp,\K}[x_\K,S_\K; g,\phi],
\label{eq:consv_S_eff}
\end{equation}
where the label $\K \in \{1,2\}$ distinguishes between the two constituents of the binary. As for the fields, we shall assume that any nonlinear self-interactions of the scalar and any of its non\-minimal couplings to the spacetime curvature are subleading on the scales that we are interested in, and so will take as our fiducial model general relativity plus a massless~scalar:
\begin{equation}
\label{eq:consv_S_fields_raw}
	S_\text{fields} = \int\dx^4x\sqrt{-g}
	\bigg(
		2\mpl^2R
		-
		\frac{1}{2}\, g^{\mu\nu}\partial_\mu\phi\,\partial_\nu\phi
		+
		\cdots
	\bigg).
\end{equation}

The key output of the numerous calculations in this section is the binding energy for binary systems in circular nonprecessing orbits, which we shall derive in four stages. In~section~\ref{sec:consv_rules}, we first determine the Feynman rules that will allow us to integrate out the potential modes from the effective action. We use these rules to calculate the conservative potential in section~\ref{sec:consv_V}, and then derive the corresponding equations of motion and a number of conserved quantities, like the binding energy, in section~\ref{sec:consv_eom}. Finally, after putting ourselves into the centre-of-mass frame, we specialise to circular nonprecessing orbits in~section~\ref{sec:consv_circles}.
\looseness=-1

% ============================================== %
\subsection{Feynman rules and power counting}
\label{sec:consv_rules}

Far away from either member of this isolated binary, the scalar and gravitational fields that these bodies source will only effect weak perturbations about an otherwise flat spacetime. For this reason, we may perform a weak-field expansion as described in~\eqref{eq:review_weak_field_approximation}, and then organise the resulting terms based on how many powers of $h_{\mu\nu}$ and $\varphi$ they contain.

\paragraph{Propagators.}
After also including an appropriate gauge-fixing term~$S_\text{gf}$ to impose the de~Donder gauge,%
\footnote{One should actually gauge fix the potential and radiation modes of $h_{\mu\nu}$ separately~\cite{Goldberger:2004jt, Kuntz:2019zef}, but at the PN order to which are working, this technical subtlety can be glossed over without affecting the final result.}
the field action reads
\begin{equation}
	S_\text{fields} + S_\text{gf}
	=
	\frac{1}{2}
	\int\dx^4x\,
	\bigg(
		h_{\mu\nu} P^{\mu\nu\rho\sigma}\Box h_{\rho\sigma}
		+
		\varphi\Box\varphi
		+
		\cdots
	\bigg),
\label{eq:consv_S_fields}
\end{equation}
where ${\Box \coloneq \eta^{\mu\nu}\partial_\mu\partial_\nu}$ is the wave operator on flat space and
${P^{\mu\nu\rho\sigma} \coloneq \eta^{\mu(\rho}\eta^{\sigma)\nu} - \eta^{\mu\nu}\eta^{\rho\sigma}/2 }$.
Meanwhile, the ellipsis  above alludes to an infinite series of higher-order interactions of the form $h^{n-2}(\partial h)^2$ or $h^{n-2}(\partial\varphi)^2$ with ${n \geq 3}$ (see refs.~\cite{Goldberger:2004jt, Huang:2018pbu, Kuntz:2019zef} for more details), although these will play no role in the spin-orbit effects to be calculated below. At leading PN order, all that is required are the propagators for~these~fields.

Because they are being sourced by objects moving nonrelativistically, it is useful to decompose these fields further into potential and radiation modes, as in~\eqref{eq:review_potential_radiation_split}. The potential modes ${\hat\varrho = (\hat h_{\mu\nu},\hat\varphi)}$ are those whose spatial and temporal derivatives scale nonuniformly with the orbital velocity~$v$, such that ${\partial_i\hat\varrho/\hat\varrho \sim 1/r}$ while ${\partial_t\hat\varrho/\hat\varrho \sim v/r}$~\cite{Goldberger:2004jt}, where $r$ is the binary's orbital separation. This scaling suggests that the quadratic terms involving time derivatives in~\eqref{eq:consv_S_fields} should also be treated perturbatively as interaction terms; hence, the Feynman propagators for the potential modes are just the Green functions to the Poisson equation~\cite{Goldberger:2004jt, Kuntz:2019zef},
\looseness=-1
\begin{subequations}
\label{eq:consv_propagators_potential_modes}%
\begin{align}
	\langle T\hat h_{\mu\nu}(x) \hat h_{\rho\sigma}(x')\rangle
	&=
	-iP_{\mu\nu\rho\sigma} \int\frac{\dx^4p}{(2\pi)^4}
	\frac{e^{ip\cdot(x-x')}}{\bm p^2}
	=
	-iP_{\mu\nu\rho\sigma} \frac{\delta(t-t')}{4\pi|\bm x - \bm x'|},
	\\
	\langle T\hat\varphi(x) \hat\varphi(x')\rangle
	&=
	-i \int\frac{\dx^4p}{(2\pi)^4}
	\frac{e^{ip\cdot(x-x')}}{\bm p^2}
	=
	-i \frac{\delta(t-t')}{4\pi|\bm x - \bm x'|}.
\end{align}
\end{subequations}

In contrast, the derivatives of the radiation modes~${\bar\varrho=(\bar h_{\mu\nu},\bar\varphi)}$ must scale uniformly with~$v$ as $\partial_i\bar\varrho/\bar\varrho \sim \partial_t\bar\varrho/\bar\varrho \sim v/r$, since these modes can go on shell to transport energy and momentum away from the binary. The corresponding  propagators in this case are the Green functions to the wave equation~\cite{Goldberger:2004jt, Kuntz:2019zef},
\begin{subequations}
\label{eq:consv_propagators_radiation_modes}
\begin{align}
	\langle T\bar h_{\mu\nu}(x) \bar h_{\rho\sigma}(x')\rangle
	&=
	-iP_{\mu\nu\rho\sigma} \int\frac{\dx^4p}{(2\pi)^4}
	\frac{e^{ip\cdot(x-x')}}{p^2 + i\epsilon},
	\\
	\langle T\bar\varphi(x) \bar\varphi(x')\rangle
	&=
	-i \int\frac{\dx^4p}{(2\pi)^4}
	\frac{e^{ip\cdot(x-x')}}{p^2 + i\epsilon}.
\end{align}
\end{subequations}
Although not used explicitly in what follows, these radiation-mode propagators are essential to section~\ref{sec:rad} as they underpin the master formulae in \eqref{eq:review_rad_power} for the radiated power.%
\footnote{The computation of real-time quantities like the outgoing waveforms and radiation-reaction forces actually require the use of causal propagators and the in--in formalism, but the Feynman propagators in \eqref{eq:consv_propagators_radiation_modes} will suffice when working with time-averaged quantities, as we do here. See refs.~\cite{Galley:2009px, Galley:2015kus, Foffa:2021pkg} for more~details.\looseness=-1}

\paragraph{Worldline vertices.}
All of the ways in which these fields can be sourced by the binary are encoded in the point-particle actions~$S_{\pp,\K}$. To go from the covariant formulation in~\eqref{eq:pp_Routhian} to a nonrelativistic one that is appropriate for  the inspiral of a two-body system, we now gauge fix the worldline parameters to be equal to the coordinate time~$t$ [i.e., we take ${x_\K^0(\lambda) = \lambda = t}$] and will define ${v_\K^\mu \coloneq \dx x^\mu_\K/\dx t = (1,\bm v_\K)}$ as the corresponding gauge-fixed version of the tangent vector~$\dot x^\mu$. At the same time, we will also perform a weak-field expansion as per~\eqref{eq:review_weak_field_approximation}, which allows us to write
\begin{equation}
	\A_\K = 1 - \alpha_\K \bigg(\frac{\varphi}{2\mpl}\bigg)
	- \frac{\alphaprime_\K}{2} \bigg(\frac{\varphi}{2\mpl}\bigg)^{\!2}
	+ \cdots,
	\quad
	\B_\K = \frac{\beta_\K}{\M^4} + \cdots,
	\quad
	\widetilde\B_\K = \frac{\tilde\beta_\K}{\M^4} + \cdots.
\end{equation}
Factors of 2 in the denominator have been included to render our definition of $\alpha_\K$ consistent with that of ref.~\cite{Kuntz:2019zef}, which uses a different convention for the Planck mass.\looseness=-1

Notice that each body has its own set of Wilson coefficients ${ \{ \alpha_\K, \alphaprime_\K, \beta_\K, \tilde\beta_\K, \dots \} }$ for characterising how strongly it interacts with the scalar field. Allowing for these couplings to be body-dependent is important if we are to construct accurate waveform models for binaries with strongly gravitating objects, like black holes and neutron stars, since (as we discussed previously) strong-gravity effects generally lead to a violation of the strong equivalence principle~\cite{Nordtvedt:1968qs}, even if the underlying scalar-tensor theory respects the weak equivalence principle at a microscopic level. Notice also that we have normalised the conformal coupling functions $\A_\K$ to be equal to~$1$ when ${\varphi = 0}$. This can be done without loss of generality by absorbing any overall factors into the definitions of the mass parameters~$m_\K$.

What we now have are point-particle actions that are each composed of a kinetic term and an infinite series of interaction vertices which couple the fields to the worldlines:
\begin{equation}
	S_{\pp,\K} = -m_\K^\vph{2} \int\dx t\,\sqrt{1-\bm v_\K^2}
	\;\;+\;
	\bigg(
		\eqfig{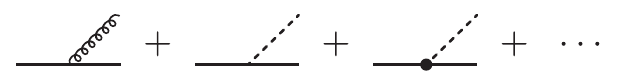}
	\bigg).
\end{equation}
Explicit expressions for the four worldline vertices most relevant to this work are presented here in table~\ref{table:feyn_1} alongside their diagrammatic representations, while a longer list (containing all of the other vertices that will feature in our discussion) can be found in table~\ref{table:feyn_2} of appendix~\ref{app:feyn}. As~$h_{\mu\nu}$ and $\varphi$ have yet to be decomposed into potential and radiation modes, these tables allow us to read off the appropriate Feynman rules for both the conservative and radiative~sectors.

\begin{table}
\centering
\begin{tabular}{|@{\hspace{10pt}} l @{\hspace{25pt}} l @{\hspace{25pt}} l @{\hspace{10pt}}|}
\hline
Diagram & Interaction vertex & Scaling \\
\hline
% Conformal [O]
\includegraphics[valign=c]{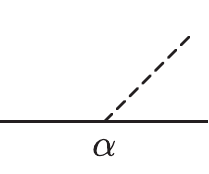}
&
$\tablestyle
\frac{\alpha_\K m_\K}{2\mpl}
\int\dx t\: \varphi$
&
$\sqrt{L}$
\\
% Conformal [SO]
\includegraphics[valign=c]{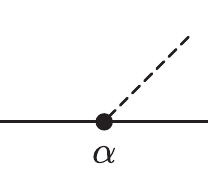}
&
$\tablestyle
\frac{\alpha_\K}{2\mpl}
\int\dx t\:
(S_\K^{i0} - S_\K^{ij}v_\K^j)\,\partial_i\varphi $
&
$\sqrt{L}v^2\epsSpin$
\\
% Disformal [O]
\includegraphics[valign=c]{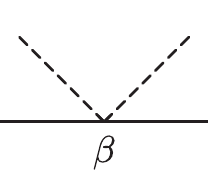}
&
$\tablestyle
\frac{\beta_\K m_\K}{\M^4} \int\dx t\:
\dot\varphi^2$
&
$\tablestyle
\epsLadder e^2$
\\
% Disformal [SO]
\includegraphics[valign=c]{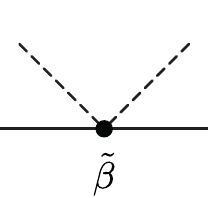}
&
$\tablestyle
\frac{\tilde\beta_\K}{\M^4}
\int\dx t\: S^{ij}_\K
(\partial_i\dot\varphi)\,\partial_j\varphi $
&
$\epsLadder\epsSpin$
\\[25pt]
\hline
\end{tabular}
\caption{Feynman rules for the four worldline vertices most relevant to this work. The scalar field should be understood as being evaluated along the worldline; i.e., ${\varphi \equiv \varphi(t,\bm x_\K)}$, and note that we write $\dot\varphi \equiv v_\K^\mu\partial_\mu\varphi$ as shorthand for denoting a total time derivative. Diagrammatically, each factor of~$\varphi$ is represented by a dashed line, while the worldline itself is drawn as a solid line. Vertices without black dots are spin-independent, while those with a black dot are coupled to one power of the spin. The rightmost column lists the power-counting rules for how each vertex scales with the EFT's expansion parameters, assuming all factors of~$\varphi$ are taken to be potential modes. For each factor of $\varphi$ that is instead taken to be a radiation mode, simply include an extra factor of $\sqrt{v}$.}
\label{table:feyn_1}
\end{table}

\paragraph{Power counting.}
One of the key advantages of the EFT approach is manifest power counting at the level of the action~\cite{Goldberger:2004jt}, which allows us to determine ahead of any detailed calculation the order at which a given Feynman diagram will contribute. For this particular purpose, it~will be convenient to suppose that the binary's two constituents have comparable masses~$m$ and spins~$S$, and that their Wilson coefficients $\{ \alpha_\K, \alphaprime_\K, \beta_\K, \tilde\beta_\K, \dots \}$ are all of order~one. We must emphasise that this is done purely for the sake of simplicity, and that the actual quantitative results to follow will hold for arbitrary values of these parameters, as long as we remain within the EFT's regime of validity (to be discussed shortly). 

Combining these parameters with the binary's orbital velocity~$v$ and separation~$r$, as well as with the mass scales ${\mpl \equiv (32\pi\GN)^{-1/2}}$ and $\M$, now allows us to identify three dimensionless parameters about which to organise a perturbative expansion. These are the orbital velocity~$v$, the ladder parameter ${\epsLadder \sim mv^2/(2\pi \M^4 r^3)}$, and the ratio ${\epsSpin \sim S/L}$, where ${L \sim mrv}$ is the binary's orbital~angular~momentum. (It is worth noting here that in the standard post-Newtonian literature, one typically assumes when power counting that the binary is composed of two maximally rotating black holes, as this sets ${\epsSpin \sim v}$~\cite{Porto:2005ac}. Making this identification does have the benefit of reducing the number of independent expansion parameters, but we shall not do so here, as treating $\epsSpin$ independently of~$v$ will help us easily distinguish between spin-independent and spin-dependent effects.)

The rightmost columns of tables~\ref{table:feyn_1} and~\ref{table:feyn_2} reveal how each worldline vertex scales with our three expansion~parameters. To arrive at these power-counting rules, we have made use of the virial relation ${v^2 \sim \GN m/r}$ to show that ${m/\mpl \sim \sqrt{Lv}}$~\cite{Goldberger:2004jt}, and have also taken ${\int\dx t \sim T \sim r/v}$, since the orbital period~$T$ is the most relevant timescale in the problem. Because ${\int\dx t \,\delta(t) = 1}$ by definition, it now follows that ${\delta(t) \sim 1/T}$, and thus we can deduce from \eqref{eq:consv_propagators_potential_modes} that the potential modes scale like ${\hat\varrho \sim 1/\sqrt{Tr} \sim \sqrt{v}/r}$ when they appear as internal lines in a Feynman diagram. In contrast, the radiation modes can be shown to scale like~${\bar\varrho \sim v/r}$~\cite{Goldberger:2004jt, Kuntz:2019zef}.  Finally, any explicit derivatives in tables~\ref{table:feyn_1} and~\ref{table:feyn_2} are taken to scale as ${(\partial_t,\partial_i) \sim (v/r, 1/r)}$, regardless of whether they act on a potential or radiation mode. The reason this does not conflict with our earlier discussion above \eqref{eq:consv_propagators_radiation_modes} is that, when computing a Feynman diagram that is \emph{linear} in the radiation modes, any derivatives acting on these modes should first be removed via integration by parts before one can read off the contribution to~$J(x)$ or $T^{\mu\nu}(x)$ [cf~\eqref{eq:review_def_J} and~\eqref{eq:review_def_T}], which are the main quantities of interest in the radiative sector. Having performed this integration by parts, all derivatives are left acting only on the potential modes. 

Any quantity~$Y$ built from these Feynman diagrams is thus an infinite series of terms that each scale homogeneously with the expansion parameters as $\sim [Y] v^a_\vph{S} \epsSpin^b \epsLadder^c$, where $[Y]$ is some overall (possibly dimensionful) factor, while $a$, $b$, and $c$ are three nonnegative integers. In what follows, we will find it useful to split any such quantity of interest into different parts based on the values of these integers. First expanding in powers of $\epsSpin$, we write
\begin{subequations}
\label{eq:consv_decomposition}
\begin{equation}
	Y
	=
	Y\scount{o}
	+
	Y\scount{so}
	+
	\mathcal O(\epsSpin^2),
\label{eq:consv_decomposition_spin}
\end{equation}
where the orbital part~(O) is spin-independent, while the spin-orbit part~(SO) is linear in the spins. Any term involving two or more powers of $\epsSpin$ will be neglected. Next, at each order in the spin, we perform a \emph{ladder expansion} in powers of~$\epsLadder$ to~get
\looseness=-1
\begin{equation}
	Y\scount{x}
	=
	Y^\sector{C}\scount{x}
	+
	Y^\sector{D}\scount{x}
	+
	\mathcal O(\epsLadder^2),
\label{eq:consv_decomposition_ladder}
\end{equation}
\end{subequations}
where~X is a placeholder for either O or~SO. The conformal part~(C) is the set of all terms that are independent of $\epsLadder$ (this includes the terms from pure general relativity, as well as the terms due to a conformal coupling with the scalar), while the leading disformal part~(D) is composed of all terms that are linear in~$\epsLadder$. As with the spin, any term that is of quadratic order or higher in~$\epsLadder$ will be neglected. Lastly, each of the terms in \eqref{eq:consv_decomposition_ladder} admits a further expansion in powers of the velocity~$v$. We will usually work only to leading order in this post-Newtonian~(PN) expansion, although certain intermediate results will be needed to next-to-leading (or so-called 1PN) order.

\paragraph{Regime of validity.}
The conditions ${v \ll 1}$, ${\epsSpin \ll 1}$, and ${\epsLadder \ll 1}$ are clearly sufficient to ensure that we remain within the regime of validity of this perturbative EFT, but unlike the first two, it turns out that the last condition on $\epsLadder$ is not actually necessary.

Now is a good time to mention that the two powers of the orbital eccentricity~$e$, which appear in the third row of table~\ref{table:feyn_1}, do not arise automatically from the power-counting rules, but have been included by hand to reflect the results of more detailed calculations, which show that disformal spin-independent effects actually vanish in the limit of a circular orbit~\cite{Brax:2018bow, Brax:2019tcy}. With this factor included, we see that $\epsLadder$ is always accompanied by either $e^2$ or $\epsSpin$ when it appears in a Feynman diagram; hence, the necessary and sufficient conditions that justify truncating some quantity to linear order in $\epsLadder$ are ${\epsLadder e^2 \ll 1}$~and~${\epsLadder\epsSpin \ll 1}$.
\looseness=-1

These conditions assume that the two bodies' Wilson coefficients are all of order one, but it is a straightforward task to generalise them to account for arbitrary values of the parameters. The condition ${\epsLadder\epsSpin \ll 1}$, for instance, becomes
\begin{equation}
	\frac{v}{2\pi\M^4 r^4} \text{max}(\tilde\beta_1 S_1,\tilde\beta_2 S_2)
	\ll 1
\label{eq:consv_boundary_spin_ladder}
\end{equation}
if we allow for arbitrary spins~$S_\K$ and arbitrary coefficients~$\tilde\beta_\K$, while still assuming that the two masses $m_\K$ are comparable. For more general expressions, and for further details on the validity of the ladder expansion, we refer the interested reader to refs.~\cite{Davis:2019ltc, Brax:2020vgg}.

At the same time, an important detail that should not go unmentioned here is that while the perturbative expansion in \eqref{eq:consv_decomposition_ladder} clearly breaks down for sufficiently large values of~$\epsLadder$, the point-particle EFT itself remains valid as long as the formal power series expansion in~\eqref{eq:pp_formal_power_series} holds. This is the case whenever ${\varphi/\mpl \sim v^2/\sqrt{L}}$ and ${X/\M^4 \sim \epsLadder/L}$ are both small; hence, the EFT continues to be predictive even in a regime where $\epsLadder$ is large enough that ${\epsLadder e^2 \gg 1}$ and/or ${\epsLadder\epsSpin \gg 1}$, but small enough that ${\epsLadder/L \ll 1}$. (Note that ${L \gg 1}$ for astrophysical binaries~\cite{Goldberger:2004jt}.) The only difficulty with studying this regime is that a certain class of ``ladder diagrams'' must be resummed, but this has already been shown to be possible, at least for binaries of nonspinning objects~\cite{Davis:2019ltc, Davis:2021oce}. We expect that similar techniques can also be applied to study the spinning case, but we leave this task to the future. In what follows, we will focus solely on the perturbative regime and will truncate our results to~linear~order~in~$\epsLadder$.
\looseness=-1

% ============================================== %
\subsection{Conservative potential} 
\label{sec:consv_V}

Following \eqref{eq:review_def_V}, we can now construct the conservative potential~$V$ up to the order prescribed in~\eqref{eq:consv_decomposition} by summing over all relevant Feynman diagrams with no external~fields.

As the spin-independent part of this potential has already been derived several times in the literature, we feel it sufficient to simply quote the end result here. In the conformal sector, the result up to 1PN order in the de~Donder gauge is~\cite{Damour:1992we, Huang:2018pbu, Kuntz:2019zef}
\begin{subequations}
\label{eq:consv_V}
\begin{align}
	V^\sector{C}\scount{o}
	=&
	-
	\frac{\Geff m_1 m_2}{r}
	\bigg(
		1
		-
		\frac{\Geff m}{2r} (2\gammapeff - 1)
		\nonumber\\&
		+
		\frac{1}{2}
		\big[
			\bm v_1^2 + \bm v_2^2
			-
			3 \bm v_1\cdot\bm v_2
			-
			(\bm n\cdot\bm v_1)(\bm n\cdot\bm v_2)
			+
			2\gammaeff \bm v^2
		\big]
	\bigg),
\label{eq:consv_V_O_C}
\end{align}
where ${\bm v = \bm v_1 - \bm v_2}$ is the relative velocity between the two bodies, ${\bm r = \bm x_1 - \bm x_2}$ is their relative displacement, and ${\bm n = \bm r/r}$ is the unit vector pointing in this direction. From a diagrammatic viewpoint, the leading $1/r$ term in this potential can be attributed to $t$-channel exchanges of potential-mode gravitons and scalars between the two worldlines, as depicted in figures~\ref{fig:feyn_V_O}(a) and~\ref{fig:feyn_V_O}(b), respectively. These two interactions combined give rise to an inverse-square-law force whose overall strength is set by the effective gravitational constant $\Geff \coloneq \GN(1+2\alpha_1\alpha_2)$. The~remaining terms in \eqref{eq:consv_V_O_C} are the 1PN corrections, which stem from a total of 11 diagrams (see, e.g., figure~7 of ref.~\cite{Kuntz:2019zef}) and can be seen to depend on two constants, $\gammaeff$~and~$\gammapeff$, that are built from combinations of the bodies' parameters. These symbols are defined in table~\ref{table:definitions}, as are all of the other variables and parameter combinations used in~this~work.

\begin{figure}
\centering\includegraphics{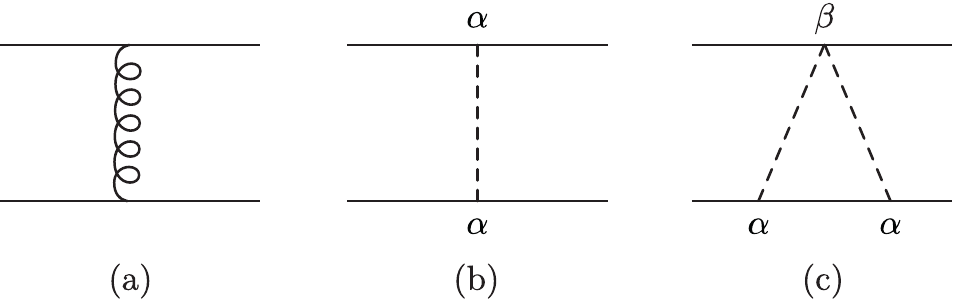}
\caption{Feynman diagrams contributing to the spin-independent potential at leading PN~order. The exchange of a~(a)~graviton and (b)~scalar between the two worldlines is responsible for the leading $1/r$~term in the conformal sector, while the diagram in~(c) gives rise to the leading term in the disformal sector. The mirror inverse of this third diagram, wherein the $\beta$~vertex is attached to the bottom worldline, is included implicitly as we do not distinguish between the two solid lines.}
\label{fig:feyn_V_O}
\end{figure}

\begin{table}[t]
\centering
\def\arraystretch{1.3}
\begin{tabular}{|cl|cl|}
\hline
Symbol & Definition & Symbol & Definition \\
\hline
\multicolumn{2}{|l|}{Relative coordinates}		& \multirow{2}{*}{$\gammaeff$}
												& \multirow{2}{*}{$\displaystyle 1 -
													\frac{4\alpha_1\alpha_2}{1+2\alpha_1\alpha_2}$} \\
$\bm r$	& $\bm x_1 - \bm x_2$ 					&& \\
$\bm n$ 	& $\bm r/r \quad (r=|\bm r|)$     	& \multirow{2}{*}{$\gammapeff$}
												& \multirow{2}{*}{$\displaystyle 1 -
														\frac{2(m \alpha_1^2\alpha_2^2
														+ m_1^\vph{2}\alpha_1^2\alphaprime_2^\vph{2}
														+ m_2^\vph{2}\alpha_2^2\alphaprime_1^\vph{2})%
														}{ m (1+2\alpha_1\alpha_2)^2}$}\\
$\bm v$	& $\bm v_1 - \bm v_2$ 					&&\\
$\bm a$	& $\bm a_1 - \bm a_2$					& \multicolumn{2}{l|}{Radiative sector}\\
\multicolumn{2}{|l|}{Mass combinations}			& $\Delta\alpha$ & $\alpha_1 - \alpha_2$\\
$m$			& $m_1 + m_2$ 						& \multirow{2}{*}{$\comboA{\pm}{\ell}$}
												& \multirow{2}{*}{$\displaystyle\frac{
												\alpha_1^\vph{\ell} m_1^\vph{\ell} m_2^{\ell}
												\pm (-1)^\ell
												\alpha_2^\vph{\ell} m_2^\vph{\ell} m_1^{\ell}
												}{ m^{\ell+1} }$}\\
$\Delta m$	& $m_1 - m_2$						&&\\
$\nu$		& $m_1 m_2 / m^2$ 					& \multirow{2}{*}{$\comboAp{\pm}{\ell}$}
												& \multirow{2}{*}{$\displaystyle\frac{
												\alphaprime_1^\vph{2}\alpha_2^\vph{\ell} m_1^\vph{\ell} m_2^{\ell}
												\pm (-1)^\ell
												\alphaprime_2^\vph{2}\alpha_1^\vph{\ell} m_2^\vph{\ell} m_1^{\ell}
												}{ m^{\ell+1} }$}\\
\multicolumn{2}{|l|}{Spin combinations}			&&\\
$S$			& $S_1 + S_2$ 						& \multirow{2}{*}{$\comboBS{\pm}{\ell}$}
												& \multirow{2}{*}{$\displaystyle\frac{
												\tilde\beta_1^\vph{\ell}\alpha_2^\vph{\ell}
												m_1^\vph{\ell} m_2^{\ell}
												\pm (-1)^\ell
												\tilde\beta_2^\vph{\ell}\alpha_1^\vph{\ell}
												m_2^\vph{\ell} m_1^{\ell}
												}{ m^{\ell+1} } $}\\
$\Sigma$	& $m \, (S_2/m_2 - S_1/m_1)$ 		&&\\
$S_\pm$		& $(m_2/m_1)S_1 \pm (m_1/m_2)S_2$~~	& \multirow{2}{*}{$\comboA{\text{NLO}}{1}$}
												& \multirow{2}{*}{$\displaystyle
												  \frac{ 2\comboAp{-}{2} - \comboA{-}{2} }{1+2\alpha_1\alpha_2}
												  - \frac{3}{5}\comboA{+}{3} $}\\
$\chi_\K^{}$ & $S_\K^{}/(\GN m_\K^2)$ &&\\
\multicolumn{2}{|l|}{Conservative sector}		& \multirow{2}{*}{$\zeta$}
												& \multirow{2}{*}{$\displaystyle
												  1 + \frac{1}{3}\bigg(\frac{ \comboA{+}{2} }{\nu}\bigg)^2 $}\\
 $\Geff$ 	& $\GN(1+2\alpha_1\alpha_2)$ 		&&\\[5pt]
\hline
\end{tabular}
\caption{Definitions for all of the variables and parameter combinations used in this work. Both the antisymmetric tensor~$S_\K^{ij}$ and the 3-vector~$S_\K^i$ encode the same information about the bodies' spins, and one can easily switch between the two by using the definition ${S_\K^{ij} \equiv \epsilon^{ijk} S_\K^k}$. It follows that $S$, $\Sigma$, and $S_\pm$ can also be expressed either as antisymmetric tensors or 3-vectors, since they are just linear combinations of the individual spins. Which form we decide to use is ultimately based~on~convenience.}
\label{table:definitions}
\end{table}

The leading contribution to the disformal spin-independent sector was recently calculated in~ref.~\cite{Brax:2019tcy}, albeit only for the special case in which ${\alpha_1 = \alpha_2}$ and ${\beta_1 = \beta_2}$. It is nevertheless straightforward to repeat the calculation with arbitrary coefficients, in which case one finds
\begin{equation}
	V^\sector{D}\scount{o}
	=
	- 
	\frac{\GN m_1 m_2}{2\pi\M^4 r^4}
	(m_1^\vph{2}\alpha_1^2\beta_2^\vph{2} 
	+
	m_2^\vph{2}\alpha_2^2\beta_1^\vph{2})
	(\bm n \cdot \bm v)^2.
\label{eq:consv_V_O_D}
\end{equation} 
The Feynman diagram responsible for this disformal interaction is shown in figure~\ref{fig:feyn_V_O}(c).

\begin{figure}
\centering\includegraphics{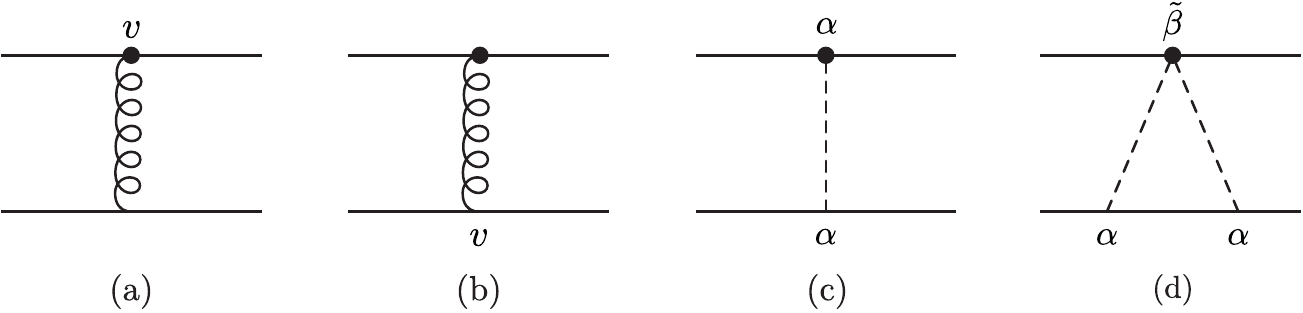}
\caption{Feynman diagrams contributing to the (a--c) conformal and (d)~disformal parts of the spin-orbit potential at leading PN~order. The mirror inverses of these diagrams are all included implicitly as we do not distinguish between the two solid lines.}
\label{fig:feyn_V_SO}
\end{figure}

Now turning our attention to the spin-orbit part of the potential, we find that
\begin{align}
	V^\sector{C}\scount{so}
	&=
	\frac{\GN m_2}{r^2} n^i
	\big[
		2 S_1^{ij} v^j 
		+
		(1 + 2\alpha_1\alpha_2)
		(S^{i0}_1 - S_1^{ij} v_1^j)
	\big]
	+
	(1\leftrightarrow 2),
	\label{eq:consv_V_SO_C}
	\\
	V^\sector{D}\scount{so}
	&=
	\frac{\GN m_2^2 \alpha_2^2 }{2\pi\M^4 r^5} 
	\tilde\beta_1^\vph{j} S^{ij}_1 n^i v^j
	+
	(1\leftrightarrow 2)
\label{eq:consv_V_SO_D}
\end{align}
\end{subequations}
at leading order in~$v$, and note that both $\bm n$ and $\bm v$ swap sign under the interchange of the body labels ${1 \leftrightarrow 2}$. The conformal terms in \eqref{eq:consv_V_SO_C} are the result of three Feynman diagrams, as shown in figures~\ref{fig:feyn_V_SO}(a) to~\ref{fig:feyn_V_SO}(c). The first two are the usual contributions from general relativity, which have previously been calculated in ref.~\cite{Porto:2005ac}, while the third diagram in figure~\ref{fig:feyn_V_SO}(c) accounts for the extra effect from the scalar field, whose evaluation is novel to this work. Likewise, the evaluation of figure~\ref{fig:feyn_V_SO}(d), which leads to the disformal terms in \eqref{eq:consv_V_SO_D}, is also a novel result. The details of these algebraically lengthy calculations are presented in~appendix~\ref{app:feyn}.

It is now worth highlighting that these results are written in terms of the spin components $S_\K^{i0 \vph{j}}$ and~$S_\K^{ij}$. In this ${3+1}$~decomposition, our index notation can no longer distinguish between spin tensors defined in different frames; hence, we should clarify that we will always be working with the spin tensor as defined in the locally flat frame, until stated otherwise. The covariant SSC in \eqref{eq:pp_ssc} can be~rewritten~as
\begin{equation}
	S_\K^{i0 \vph{j}} \approx S_\K^{ij} v_\K^j
\label{eq:consv_ssc}
\end{equation}
in this notation, up to $\mathcal O(v^2)$ corrections that come from projecting the tangent vector~$v^\mu_\K$ onto the locally flat frame~\cite{Levi:2014sba, Porto:2016pyg}.%
\footnote{The vierbein that effects this transformation is ${e_\mu^a = \delta_\mu^a + \mathcal O(v^2)}$, where $\delta^a_\mu$ is the Kronecker delta.}
These corrections need not concern us at the order to which we are working, however. 

% ============================================== %
\subsection{Equations of motion}
\label{sec:consv_eom}

Potential in hand, the equations of motion for this two-body system now follow from the Euler--Lagrange and Hamilton equations in~\eqref{eq:review_consv_eom}. The former tells us that the worldlines trace out trajectories governed by the equation ${\ddot{\bm x}_\K = \bm a_\K}$, where the acceleration vector $\bm a_\K$ specifies the total force per unit mass acting on the $\K$th body. Splitting it up into its four constituent parts as per~\eqref{eq:consv_decomposition}, we find that the acceleration of the first body~is
\begin{subequations}
\label{eq_consv_eom_a1}
\begin{align}
% Conformal [O]
	a_1^\sector{C}\scount{o}^i
	=&
	-\frac{\Geff m_2}{r^2}
	\bigg(
		1
		-
		\frac{\Geff}{r}[2m(\gammaeff + \gammapeff) + m_1]
		+
		\gammaeff\bm v^2
		-
		2\bm v_1\cdot\bm v_2
		\nonumber\\&
		+
		\bm v_2^2
		-
		\frac{3}{2}(\bm n\cdot \bm v_2)^2
	\bigg) n^i
	+
	\frac{\Geff m_2}{r^2}
	\big[
		(2\gammaeff + 1)\bm n\cdot\bm v
		+
		\bm n\cdot\bm v_1
	\big]
	v^i,
\label{eq_consv_eom_a1_O_C}
\allowdisplaybreaks\\[3pt]
% Disformal [O]
	a_1^\sector{D}\scount{o}^i
	=&\:
	\frac{2\GN m_2
		(\beta_2^\vph{2} \alpha_1^2 m_1^\vph{2}
		+
		\beta_1^\vph{2} \alpha_2^2 m_2^\vph{2})%
	}{2\pi\M^4 r^5}
	\left(
		3(\bm n\cdot\bm v)^2 - \bm v^2
		+
		\frac{\Geff m}{r}
	\right)
	n^i,
\label{eq_consv_eom_a1_O_D}
\allowdisplaybreaks\\[3pt]
% Conformal [SO]
	a_1^\sector{C}\scount{so}^i
	=&\:
	\frac{3\GN m_2}{r^3}
	\Big(
		2 n^{\langle i}n^{j\rangle} v^k (\hat S_1^{jk} + \hat S_2^{jk})
		+
		[(1-2\alpha_1\alpha_2) \hat S_1^{ij} + 2\hat S_2^{ij}]
		n^{\langle j}n^{k\rangle} v^k
	\Big),
\label{eq_consv_eom_a1_SO_C}
\allowdisplaybreaks\\[3pt]
% Disformal [SO]
	a_1^\sector{D}\scount{so}^i
	=&
	-\frac{18\GN}{2\pi \M^4 r^6}\frac{1}{m_1}
	\big(
		m_2^{2\vph{j}}\alpha_2^{2\vph{j}}\tilde\beta_1^\vph{j} S_1^{jk}
		+
		m_1^{2\vph{j}}\alpha_1^{2\vph{j}}\tilde\beta_2^\vph{j} S_2^{jk}
	\big)\,
	 n^{\langle i}n^{j\rangle} n^{\langle k}n^{p\rangle} v^p
\label{eq_consv_eom_a1_SO_D}
\end{align}
\end{subequations}
to the order at which we are working, and note that we write ${\hat S_\K \equiv S_\K/m_\K}$ as shorthand. The acceleration of the second body can then be obtained by simply interchanging the labels ${1 \leftrightarrow 2}$, and recall that the vectors $\bm n$ and $\bm v$ swap sign under this interchange. 

To obtain the spin equations of motion, it is useful to first introduce the spin vector~$S_\K^i$ as an alternative but equivalent way of encoding the same information as the antisymmetric tensor~$S_\K^{ij}$. The two are related by the identity ${ S_\K^{ij} \equiv \epsilon^{ijk} S_\K^{k} }$, and in what follows, we shall often switch between them based on whichever proves more convenient. In terms of these spin vectors, the Poisson brackets in~\eqref{eq:pp_S_PB} read
\begin{equation}
	\{ S_{\K\vph{l}}^{i \vph{j}},  x_{\K'}^j \}
	=
	\{ S_{\K\vph{l}}^{i0 \vph{j}}, x_{\K'}^j \}
	= 0,
	\quad
	\{ S^{i \vph{j}}_{\K \vph{l}}, S_{\K'}^{j \vph{i}} \}
	=
	- \delta_{\K\K'}^\vph{j}
	\epsilon^{ijk}_\vph{\K'} S_{\K\vph{l}}^{k \vph{j}},
	\quad
	\{ S^{i\vph{j}}_{\K\vph{'}}, S_{\K'}^{j0} \}
	=
	- \delta_{\K\K'}^\vph{j}
	\epsilon^{ijk}_\vph{\K'} S_{\K\vph{l}}^{k0 \vph{j}},
\end{equation}
and Hamilton's equation then tells us that
\begin{align}
	\diff{\bm S_1}{t}
	=&\:
	\frac{\GN m_2}{r^2}
	\big[
		2(\bm n \times \bm v)\times \bm S_1
		+
		\bm v_1 \times (\bm n \times \bm S_1)
	\big]
	\nonumber\\&
	+
	\frac{2\GN m_2 \alpha_1\alpha_2}{r^2}
	\bm v_1 \times (\bm n \times \bm S_1)
	+
	\frac{\GN m_2^2 \alpha_2^2}{2\pi\M^4 r^5}
	\tilde\beta_1 (\bm n\times \bm v)\times \bm S_1.
\label{eq:consv_eom_spin_loc}
\end{align}
The last term above is the leading contribution in the disformal sector, while the rest make up the leading contribution in the conformal sector. The equation governing the evolution of~$\bm S_2$ follows after interchanging the labels~${1\leftrightarrow 2}$.

Worth reiterating is the requirement that the SSC in \eqref{eq:consv_ssc} be imposed only at the level of the equations of motion, and not beforehand. Having done so, this explains why the conservative potential in~\eqref{eq:consv_V} depends explicitly on both $S_\K^{i0 \vph{j}}$ and $S_\K^{ij}$, whereas the equations of motion in \eqref{eq_consv_eom_a1} and \eqref{eq:consv_eom_spin_loc} depend only on the latter (or its equivalent, $S_\K^i$).

\paragraph{Constants of motion.}
These equations fully specify the conservative dynamics of our two-body system at the order to which are working, but are difficult to solve as is, and so motivate us to look for a number of conserved quantities.

To begin with, the fact that the two-body Routhian%
\footnote{Or, more fundamentally, the two-body Lagrangian ${\mathcal L = \mathcal R + \sum_\K \Omega_{\Lambda,\K}^{ab} S^\vph{b}_{\K,ab}/2}$.}
is invariant under spatial translations tells us via Noether's theorem that the total linear momentum
\begin{equation}
	P^i
	=
	\sum_\K \pdiff{\mathcal R}{v_\K^i}
	=
	\bigg(
		1 + \frac{\bm v_1^2}{2}
		-
		\frac{\Geff m_2}{2r}
	\bigg)
	m_1^\vph{i} v_1^i
	-
	\frac{\Geff m_1 m_2}{2r}
	(\bm n \cdot \bm v_1)\, n^i
	-
	\frac{\Geff m_2}{r^2} S_1^{ij} n^j_\vph{1}
	+
	(1\leftrightarrow 2)
\label{eq:consv_P}
\end{equation}
is a constant of motion. Notice that this momentum receives no contribution from the disformal terms in \eqref{eq:consv_V_O_D} and \eqref{eq:consv_V_SO_D} because any term in $\mathcal R$ that depends on $\bm v_1$ and $\bm v_2$ only via the relative velocity ${\bm v}$~(${=\bm v_1 - \bm v_2}$) cancels itself out once we sum over~$\K$. 

Similarly, rotational invariance leads to a conserved total angular momentum
\begin{equation}
	J^i = L^i + S^i.
\end{equation}
The binary's orbital angular momentum is given by ${ L^i = \sum_\K m_\K (\bm x_\K \times \bm v_\K)^i }$ at leading order, while ${S^i_\vph{\K} = \sum_\K S_\K^i}$ is the sum of the two individual spin vectors.%
\footnote{At higher orders in the PN expansion, $L^i$ will also depend on the spin variables~\cite{Kidder:1992fr, Kidder:1995zr, Faye:2006gx}, but these corrections will be of no concern to us here.}

Another important symmetry of the Lagrangian is its invariance under Lorentz boosts, which although no longer manifest, is guaranteed by the fact that our EFT in \eqref{eq:consv_S_eff} is globally Poincar\'{e} invariant. Consequently, the vector~\cite{deAndrade:2000gf}
\begin{equation}
	K^i \coloneq G^i - P^i t
\end{equation}
is conserved order by order in the PN expansion, and since $P^i$ is itself a conserved quantity, differentiation with respect to time tells us that ${\dx G^i/\dx t = P^i}$. The vector $G^i$ can thus be interpreted as (being proportional to) the position of the binary's centre of mass. In our case, it is easy enough to integrate~\eqref{eq:consv_P} directly to find
\begin{equation}
	G^i =
	\left(
		1 + \frac{\bm v_1^2}{2}
		-
		\frac{\Geff m_2}{2r}
	\right) 
	m_1^\vph{j} x_1^{i \vph{j}}
	+
	S_1^{ij} v_1^j
	+
	(1\leftrightarrow 2),
\label{eq:consv_G}
\end{equation}
up to an irrelevant additive constant. In arriving at this result, we made use of the equation of motion in~\eqref{eq_consv_eom_a1_O_C}, but were free to ignore the other three lines in~\eqref{eq_consv_eom_a1} as they do not contribute at the order to which we are working. Similarly, the spin can be held constant when going from \eqref{eq:consv_P} to \eqref{eq:consv_G}, or vice versa, as the time derivative in \eqref{eq:consv_eom_spin_loc} leads to spin-orbit terms that contribute only at next-to-leading PN~order.

Combining \eqref{eq:consv_G} with our definition for the relative displacement, ${\bm r= \bm x_1 - \bm x_2}$, gives us two equations that can be solved simultaneously for $\bm x_1$ and $\bm x_2$ as functions of $\bm r$ and~$\bm G$; thereby allowing us to disentangle the relative motion of the binary from the motion of its centre of mass. Indeed, the latter can be ignored completely by putting ourselves into the centre-of-mass~(CM) frame, wherein ${G^i = 0}$ and ${P^i = \dx G^i/\dx t = 0}$. Having done so, we find that the expression for~$\bm x_1$ in this frame is
\begin{subequations}
\label{eq:consv_to_CM_frame}	
\begin{equation}
	x^i_1 = \frac{m_2}{m} r^i
	+
	\frac{\Delta m \hspace{0.5pt} \nu}{2m}
	\left(
		\bm v^2 - \frac{\Geff m}{r}
	\right) r^i
	+
	\frac{\nu}{m} \Sigma^{ij} v^j.
\label{eq:consv_x1_to_CM_frame}
\end{equation}
The mass difference~$\Delta m$, the symmetric mass ratio~$\nu$, and the spin parameter~$\Sigma$ are all defined as in table~\ref{table:definitions}. The expression for $\bm x_2$ then follows after interchanging the labels ${1\leftrightarrow 2}$, and note that $\bm r$, $\bm v$, $\Delta m$, and $\Sigma$ all swap sign under this interchange.

Now combining the condition ${P^i = 0}$ with the definition~${\bm v = \bm v_1 - \bm v_2}$ gives us another two equations that can be solved simultaneously for $\bm v_1$ and $\bm v_2$. At the order to which we are working, the result for $\bm v_1$ in the CM frame~is
\begin{equation}
	v^i_1 = \frac{m_2}{m} v^i
	+
	\frac{\Delta m \hspace{0.5pt} \nu}{2m}
	\bigg[
		\bigg(
			\bm v^2 - \frac{\Geff m}{r}
		\bigg)
		v^i
		-
		\frac{\Geff m}{r}(\bm n\cdot\bm v)\, n^i
	\bigg] 
	-
	\frac{\Geff \nu}{r^2} \Sigma^{ij} n^j,
\label{eq:consv_v1_to_CM_frame}
\end{equation}
\end{subequations}
while the result for $\bm v_2$ follows after interchanging the labels~${1 \leftrightarrow 2}$. As a consistency check, note that \eqref{eq:consv_v1_to_CM_frame} can also be obtained by differentiating \eqref{eq:consv_x1_to_CM_frame} with respect to time and~then using the equations of motion.

The last constant of motion we shall consider is the binary's total energy, which can be determined by constructing the two-body Hamiltonian~$\mathcal H$. Having already performed a partial Legendre transform on the spin variables to get to the Routhian, all that remains is to Legendre transform the position variables; hence,
\begin{equation}
	\mathcal H = \sum_\K v_\K^i \pdiff{\mathcal R}{v_\K^i} - \mathcal R.
\end{equation}
Subtracting the rest masses of the two bodies from this Hamiltonian gives us the orbital binding energy ${E \coloneq \mathcal H - m_1 - m_2}$, and it is convenient to further define ${\mathcal{E} \coloneq E/(m\nu)}$ as the orbital binding energy per unit reduced mass. Split into its four constituent parts as per~\eqref{eq:consv_decomposition}, we~have~that
\begin{subequations}
\label{eq:consv_energy}
\begin{align}
% Conformal [O]
	\mathcal{E}^\sector{C}\scount{o}
	=&\;
	\frac{1}{2}\bm v^2
	+
	\frac{3}{8}(1-3\nu)\bm v^4
	-
	\frac{\Geff m}{r}
	\bigg(
		1
		-
		\frac{\Geff m}{2r} (2\gammapeff - 1)
		\nonumber\\&
		-
		\frac{1}{2}(2\gammaeff + 1 + \nu)\,\bm v^2
		-
		\frac{1}{2}\nu(\bm n\cdot\bm v)^2 
	\bigg),
\allowdisplaybreaks\\
% Disformal [O]
	\mathcal{E}^\sector{D}\scount{o}
	=&\;
	\frac{\GN m}{2\pi\M^4 r^4}
	(m_1^\vph{2}\alpha_1^2 \beta_2^\vph{2} 
	+
	m_2^\vph{2}\alpha_2^2\beta_1^\vph{2})
	(\bm n\cdot\bm v)^2,
\allowdisplaybreaks\\
% Conformal [SO]
	\mathcal{E}^\sector{C}\scount{so}
	=&\;
	\frac{\Geff}{r^2}
	n^i v^j S_+^{ij},
% Disformal [SO]
	\\
	\mathcal{E}^\sector{D}\scount{so}
	=&\;
	0
\end{align}
\end{subequations}
when expressed in the CM frame. The disformal spin-orbit contribution vanishes at the order to which we are working because \eqref{eq:consv_V_SO_D} is linear in the velocities, and so cancels itself out in the Legendre transform. Meanwhile, the conformal spin-orbit contribution is nonvanishing only because of the terms in \eqref{eq:consv_V_SO_C} that are proportional to $S_\K^{i0}$, and we reiterate yet again that the SSC should be imposed only at the end of our~manipulations.

\paragraph{Relative acceleration.}
Given the transformation rule in~\eqref{eq:consv_to_CM_frame}, we can now simplify the equations of motion by putting them into the CM frame. For the worldlines, we find that the displacement vector~${\bm r = \bm x_1 - \bm x_2}$ evolves according to the equation ${\ddot{\bm r} = \bm a}$, where ${\bm a = \bm a_1 - \bm a_2}$ is the relative acceleration between the two bodies. Explicitly, we have that
\begin{subequations}
\label{eq:consv_eom_a}
\begin{align}
% Conformal [O]
	a^\sector{C}\scount{o}^i
	=&
	-\frac{\Geff m}{r^2}
	\bigg(
		1
		-
		\frac{2\Geff m}{r}(\gammaeff+\gammapeff+\nu)
		+
		(3\nu + \gammaeff)\,\bm v^2
		\nonumber\\&
		-
		\frac{3\nu}{2} (\bm n\cdot\bm v)^2
	\bigg)
	n^i
	+
	\frac{2\Geff m}{r^2}(1+\gammaeff - \nu)(\bm n\cdot \bm v)\, v^i,
\allowdisplaybreaks\\[3pt]
% Disformal [O]
	a^\sector{D}\scount{o}^i
	=&\;
	\frac{2\GN m
		(m_1^\vph{2} \alpha_1^2 \beta_2^\vph{2} 
		+
		 m_2^\vph{2} \alpha_2^2 \beta_1^\vph{2})%
	}{2\pi\M^4 r^5}
	\left(
		3(\bm n\cdot\bm v)^2 - \bm v^2
		+
		\frac{\Geff m}{r}
	\right)
	n^i,
\allowdisplaybreaks\\[3pt]
% Conformal [SO]
	a^\sector{C}\scount{so}^i
	=&\;
	\frac{3\GN}{r^3}
	\Big(
		2 n^{\langle i}n^{j\rangle} v^k
		(S_+^{jk} + S^{jk})
		+
		[(1-2\alpha_1\alpha_2)S_+^{ij} + 2S^{ij}]
		n^{\langle j}n^{k\rangle} v^k
	\Big),
\allowdisplaybreaks\\[3pt]
\label{eq:consv_eom_a_SO_C}
% Disformal [SO]
	a^\sector{D}\scount{so}^i
	=&
	-\frac{9\GN m}{2\pi\M^4 r^6}
	\big[
		(\alpha_2^2\tilde\beta_1^\vph{2} + \alpha_1^2\tilde\beta_2^\vph{2})
		S_+^{jk}
		+
		(\alpha_2^2\tilde\beta_1^\vph{2} - \alpha_1^2\tilde\beta_2^\vph{2})
		S_-^{jk}
	\big]
	n^{\langle i}n^{j\rangle} n^{\langle k}n^{p\rangle} v^p.
% For spacing purposes
\vphantom{\sum_0}
\end{align}
\end{subequations}
The spin combinations $S_+$ and $S_-$ are defined as in table~\ref{table:definitions}.

\paragraph{Spin precession.}
Applying~\eqref{eq:consv_to_CM_frame} to \eqref{eq:consv_eom_spin_loc} allows us to replace $\bm v_1$ by ${m_2 \bm v/m + \mathcal O(v^3)}$, but the result is still not in a useful form because it does not preserve the magnitude of the spin vector. One can see this easily by taking the inner product of \eqref{eq:consv_eom_spin_loc} with $2\bm S_1$ to find that ${2\bm S_1 \cdot \dx\bm S_1/\dx t = \dx \bm S_1^2/\dx t \neq 0}$. However, recall from our discussion in section~\ref{sec:pp} that the magnitude of the spin tensor \emph{is} conserved [i.e., ${D(S_{\K,ab}^\vph{b} S_\K^{ab})/D\lambda = 0}$]; hence, it must be that the nonconservation of $\bm S_\K^2$ is simply a coordinate artefact.

Indeed, a simple Lorentz boost in the direction of the body's motion is all that is required to put \eqref{eq:consv_eom_spin_loc} into the form of a precession equation. The spin vector $\bm S_1^\text{cmv}$ defined in this \emph{comoving frame} is related to the spin in the locally flat frame by~\cite{Porto:2005ac, Porto:2008tb}
\begin{equation}
	\bm S_1^\text{cmv}
	=
	\bm S_1^\vph{c}
	+
	\frac{1}{2} \bm v_1^\vph{c} \times
	(\bm v_1^\vph{c} \times \bm S_1^\vph{c})
	+
	\mathcal O(v^4).
\label{eq:consv_def_spin_cmv}
\end{equation} 
Making this transformation, using the equation of motion for~$\bm a_1$, and then using \eqref{eq:consv_to_CM_frame} to replace $\bm v_1$ by $m_2 \bm v/m$, we find that \eqref{eq:consv_eom_spin_loc} ultimately becomes
\begin{equation}
	\diff{}{t}\bm S_1^\text{cmv}
	=
	\frac{\GN}{r^3}
	\bigg( \frac{m_2}{m_1} \bigg)
	\bigg(
		\frac{2 m_1}{m_2} + \frac{3}{2}
		-
		\alpha_1\alpha_2
		+
		\frac{m \alpha_2^2\tilde\beta_1^\vph{2}}{2\pi\M^4 r^3}
	\bigg)\,
	\bm L \times \bm S_1^\text{cmv},
\label{eq:consv_eom_spin_cmv}
\end{equation}
where ${\bm L = m\nu \bm r \times \bm v}$ is the orbital angular momentum vector in the CM~frame. As usual, the equation of motion for $\bm S_2^\text{cmv}$ follows after interchanging the labels ${1 \leftrightarrow 2}$.

The fact that the equations of motion automatically conserve the magnitude of $\bm S_\K^\text{cmv}$ makes this spin vector a much more natural variable to work with, and so we shall now switch from working with the spin in the locally flat frame to working with the spin in the comoving frame.%
\footnote{A natural question that arises in retrospect is whether we could have formulated everything in terms of this comoving spin variable from the outset. We ultimately decided on the approach presented herein (which is a generalisation of the Routhian approach by Porto and Rothstein~\cite{Porto:2008jj}) as we found it to be the most convenient for our purposes. Nevertheless, as we mentioned below~\eqref{eq:pp_S_PB}, there are certainly other ways in which we could have arrived at the same result (up to gauge transformations). For instance, Levi and Steinhoff~\cite{Levi:2015msa} have shown how one can implement the SSC directly at the level of the action by introducing what are essentially Stueckelberg fields associated with the internal $\text{SO}(3)$ symmetry of the particle's body-fixed frame, followed by a subsequent transformation of the phase-space variables. Note that this is in stark contrast to the method of this paper, wherein the SSC can be imposed only after the equations of motion are obtained via~\eqref{eq:review_consv_eom}. The specific SSC used in ref.~\cite{Levi:2015msa} is a generalisation of the Pryce--Newton--Wigner~SSC~\cite{Pryce:1948pf, Newton:1949ssc}, and it turns out that the transformed spin variable that satisfies this SSC is closely related to the comoving spin variable defined herein. (To see this, compare~\eqref{eq:consv_def_spin_cmv} with eqs.~(57)--(59) of ref.~\cite{Porto:2008jj}.) Alternatively, section~11 of ref.~\cite{Blanchet:2013haa} (see~also~ref.~\cite{Brax:2020vgg}) describes how one can work with the comoving spin from the outset, albeit at the level of the equations~of~motion.}
This switch is certainly beneficial on a conceptual level, but because the difference between $\bm S_\K^\vph{c}$ and $\bm S_\K^\text{cmv}$ is of order~$v^2$ [see~\eqref{eq:consv_def_spin_cmv}], very little actually changes on a practical level; at least, at the order to which are working. Apart from \eqref{eq:consv_eom_spin_loc}, which has now been superseded by~\eqref{eq:consv_eom_spin_cmv}, all of the other results in this section remain the same regardless of whether they are written in terms of $\bm S_\K^\vph{c}$ or $\bm S_\K^\text{cmv}$. For this reason, we shall further drop the superscript ``cmv'' in what follows to declutter our notation, and so the symbol $\bm S_\K^\vph{c}$ will henceforth refer to the spin of the $\K$th body as defined in the comoving frame.

% ============================================== %
\subsection{Circular nonprecessing orbits}
\label{sec:consv_circles}

To conclude this section, we now specialise to binary systems in circular nonprecessing orbits. A binary is said to be nonprecessing if the spins of its two constituents are either aligned or anti-aligned with the orbital angular momentum. In this particular configuration, \eqref{eq:consv_eom_spin_cmv} tells us that the spin vectors~$\bm S_\K$ are individually conserved, and since the total angular momentum~$\bm J$ is necessarily conserved (when neglecting radiation), it follows that the orbital angular momentum~$\bm L$ is also~conserved. This makes it possible for us to find a solution to~\eqref{eq:consv_eom_a} that describes a~circular~orbit.

Specifically, what we are looking for is a solution that satisfies the conditions ${\bm n\cdot \bm v = 0}$ and ${\bm a = -\Omega^2\bm r}$. We also have that ${\bm v^2 = r^2\Omega^2}$ by definition, and this can be used to eliminate all instances of $\bm v^2$ from \eqref{eq:consv_eom_a}; thus leaving us with an implicit equation that relates the orbital frequency~$\Omega$ to the relative separation~$r$. Order by order in our three expansion parameters, we find that the solution to this implicit equation~is~[note that ${S_+ = S + (\Delta m/m)\Sigma}$]
\begin{align}
	\Omega^{2}
	=&\:
	\frac{\Geff m}{r^3}
	\bigg[
		1
		-
		(2\gammapeff + \gammaeff - \nu) \bigg(\frac{\Geff m}{r}\bigg)
		\nonumber\\&
		-
		\bigg(
			\frac{ \projell{S} }{\Geff m^2}
			\frac{5+2\alpha_1\alpha_2}{1+2\alpha_1\alpha_2}
			+
			\frac{ \projell{\Sigma} }{\Geff m^2}
			\frac{3+2\alpha_1\alpha_2}{1+2\alpha_1\alpha_2}
			\frac{\Delta m}{m}
		\bigg)
		\bigg( \frac{\Geff m}{r} \bigg)^{3/2}
	\nonumber\\&
		-
		\frac{m}{2\pi\M^4 (\Geff m)^3}
		\sum_{\sigma=\pm}
		\frac{ 2\projell{S}_\sigma }{\Geff m^2}
		\frac{\alpha_2^2\tilde\beta_1^\vph{2} + 	
				\sigma\alpha_1^2\tilde\beta_2^\vph{2}}%
			{1+2\alpha_1\alpha_2}
		\bigg( \frac{\Geff m}{r} \bigg)^{9/2}\,
	\bigg],
\label{eq:consv_circles_eom_1}
\end{align}
where $\vecell$ is the unit vector in the direction of~$\bm L$. Notice that there are no disformal spin-independent terms in the above result, as these vanish for circular orbits~\cite{Brax:2018bow, Brax:2019tcy}. 

A more useful result to have is the inverse of this relation, which reads
\begin{align}
	\frac{\Geff m}{r}
	=
	\x\,&
	\bigg[
		1
		+
		\frac{1}{3} (2\gammapeff + \gammaeff - \nu) \,\x
		+
		\frac{1}{3}
		\bigg(
			\frac{ \projell{S} }{\Geff m^2}
			\frac{5+2\alpha_1\alpha_2}{1+2\alpha_1\alpha_2}
			+
			\frac{ \projell{\Sigma} }{\Geff m^2}
			\frac{3+2\alpha_1\alpha_2}{1+2\alpha_1\alpha_2}
			\frac{\Delta m}{m}
		\bigg)
		\,\x^{3/2}
		\nonumber\\&
		+
		\frac{m}{2\pi\M^4 (\Geff m)^3}
		\sum_{\sigma=\pm}
		\frac{2}{3}
		\frac{ \projell{S}_\sigma }{\Geff m^2}
		\frac{\alpha_2^2\tilde\beta_1^\vph{2} + 	
				\sigma\alpha_1^2\tilde\beta_2^\vph{2}}%
				{1+2\alpha_1\alpha_2}
		\,\x^{9/2}
	\bigg]
\label{eq:consv_circles_eom_2}
\end{align}
when written in terms of the post-Newtonian variable
\begin{equation}
	\x \coloneq (\Geff m \Omega)^{2/3}.
\label{eq:consv_def_x}
\end{equation}
For circular nonprecessing orbits, performing a post-Newtonian expansion is tantamount to expanding in powers of this single variable~$\x$, since we have that ${\x \sim \GN m/r \sim v^2}$ on account of the virial theorem. Thus, the order at which a given term contributes can be easily ascertained by simply counting powers of~$\x$.

Now imposing the conditions for a circular orbit on \eqref{eq:consv_energy}, and then using \eqref{eq:consv_circles_eom_2} and \eqref{eq:consv_def_x} to eliminate $r$ and $\Omega$ in favour of $\x$, we finally arrive at the main result of this section. Split into its four constituent parts as per~\eqref{eq:consv_decomposition}, the binding energy for a binary in a circular nonprecessing orbit~is
\begin{subequations}
\label{eq:consv_circles_energy}
\begin{align}
% Conformal [O]
	\mathcal{E}^\sector{C}\scount{o}
	&=
	-\frac{1}{2}\,\x
	+
	\frac{1}{24} (9 + 8\gammaeff - 8\gammapeff + \nu)\,\x^2,
\label{eq:consv_circles_energy_O_C}	
\\
% Disformal [O]
	\mathcal{E}^\sector{D}\scount{o} &= 0,
\allowdisplaybreaks\\
% Conformal [SO]
	\mathcal{E}^\sector{C}\scount{so}
	&=
	- \frac{1}{3}
	\bigg(
		\frac{\projell{S}}{\Geff m^2}
		\frac{7-2\alpha_1\alpha_2}{1+2\alpha_1\alpha_2}
		+
		\frac{ \projell{\Sigma} }{\Geff m^2}
		\frac{ 3-2\alpha_1\alpha_2 }{1+2\alpha_1\alpha_2}
		\frac{\Delta m}{m}
	\bigg)
	\,\x^{5/2},
\label{eq:consv_circles_energy_SO_C}	
\\
% Disformal [SO]
	\mathcal{E}^\sector{D}\scount{so}
	&=
	-
	\frac{m}{2\pi\M^4(\Geff m)^3}
	\sum_{\sigma=\pm}
	\frac{4}{3}
	\frac{ \projell{S}_\sigma }{\Geff m^2}
	\frac{\alpha_2^2\tilde\beta_1^\vph{2}
		+ \sigma
		\alpha_1^2\tilde\beta_2^\vph{2}}%
		{1+2\alpha_1\alpha_2}
	\,\x^{11/2}.
\label{eq:consv_circles_energy_SO_D}	
\end{align}
\end{subequations}
Notice as before that there are no disformal spin-independent terms in this result, as $\mathcal{E}^\sector{D}\scount{o}$ vanishes identically for circular orbits~\cite{Brax:2018bow, Brax:2019tcy}. Our results for $\mathcal{E}^\sector{C}\scount{so}$ and $\mathcal{E}^\sector{D}\scount{so}$, meanwhile, are each given to leading PN~order, while the result for $\mathcal{E}^\sector{C}\scount{o}$ is accurate to next-to-leading~PN~order.

As a general rule, the relative importance of each of these terms can now be estimated by simply comparing their leading powers of~$\x$. Case in point, we find that the conformal spin-orbit term in \eqref{eq:consv_circles_energy_SO_C} is smaller than the leading spin-independent term in~\eqref{eq:consv_circles_energy_O_C} by one and a half powers of~$\x$, and conventionally, we say that the former starts at 1.5PN order relative to the latter. The disformal spin-orbit term in~\eqref{eq:consv_circles_energy_SO_D}, meanwhile, starts at relative $4.5$PN order, although its contribution to the binding energy is not actually as small as this naive power counting would suggest. This is due to the size of the overall~prefactor
\begin{equation}
	\frac{m}{2\pi\M^4(\Geff m)^3}
	\sim
	\bigg(
		\frac{r_V}{r_S}
	\bigg)^3,
\label{eq:consv_Vainshtein_ratio}
\end{equation}
which we may interpret as being the (cubed) ratio of the binary's Vainshtein radius~$r_V$ to its Schwarzschild radius~$r_S$. For suitably small values of the coupling scale~$\M$, this ratio can be large enough to counteract the suppression from the additional powers of~$\x$; thus allowing \eqref{eq:consv_circles_energy_SO_D} to impart a greater contribution to the binding energy than~\eqref{eq:consv_circles_energy_SO_C}.%
\footnote{Our power-counting scheme can be used to verify that this inverted hierarchy does not invalidate our perturbative expansion. We find that \eqref{eq:consv_circles_energy_SO_C} is suppressed by $v^2\epsSpin$ relative to~\eqref{eq:consv_circles_energy_O_C}, while \eqref{eq:consv_circles_energy_SO_D} is suppressed by $\epsLadder\epsSpin$ relative to~\eqref{eq:consv_circles_energy_O_C}; hence, the disformal spin-orbit interaction can be larger than its conformal counterpart whenever $\M$ is small enough that ${v^2\epsSpin \ll \epsLadder\epsSpin }$. This is still within the regime of validity of our perturbative EFT as long as ${\epsLadder\epsSpin \ll 1}$; or, in other words, as long as the disformal spin-orbit term is smaller than the leading spin-independent term.}
In fact, this is precisely why we are keeping track of this disformal spin-orbit interaction.

To close this section, one final remark about the binding energy is worth making. Notice that its general expression for arbitrary orbits in \eqref{eq:consv_energy} has a vanishing disformal spin-orbit part, but $\mathcal{E}^\sector{D}\scount{so}$ is nonvanishing in~\eqref{eq:consv_circles_energy_SO_D}. This difference can be traced back to the leading $\bm v^2/2$~${(\equiv r^2 \Omega^2/2)}$ term in~\eqref{eq:consv_energy}, which gives rise to the extra disformal terms in question upon use of \eqref{eq:consv_circles_eom_2} and \eqref{eq:consv_def_x}; i.e., upon use of the equations of motion. Physically, we may interpret this as saying that the disformal spin-orbit interaction behaves similarly to a magnetic field. While it does not do any work, it can still lead to bound orbits, and a stronger disformal interaction leads to an orbit that is more~tightly~bound.

% ============================================== %
% Section 5
% ============================================== %
\section{Radiated power and the gravitational-wave phase}
\label{sec:rad}
 
An isolated two-body system eventually coalesces as it emits gravitational and scalar waves. In this section, our goal is to characterise this outgoing radiation and to quantify its impact on the evolution of the binary. We begin by first computing the power radiated into scalar waves in section~\ref{sec:rad_scalar}, before turning to a discussion of the accompanying gravitational waves in section~\ref{sec:rad_gw}. In section~\ref{sec:rad_phase}, we then combine these results into a balance law to determine how the phase of the orbit evolves with time. This ultimately feeds into our result for the gravitational-wave phase, which is what our detectors can measure.

Throughout this discussion, our focus will be centred purely on the linear part of the radiative sector, and we will therefore neglect all subleading contributions from nonlinear effects like tail terms. Moreover, as we did in the conservative sector, the key results in this section will be limited to (quasi)circular nonprecessing orbits~for~simplicity.

% ============================================== %
\subsection{Scalar radiation}
\label{sec:rad_scalar}

\begin{figure}
\centering
\includegraphics{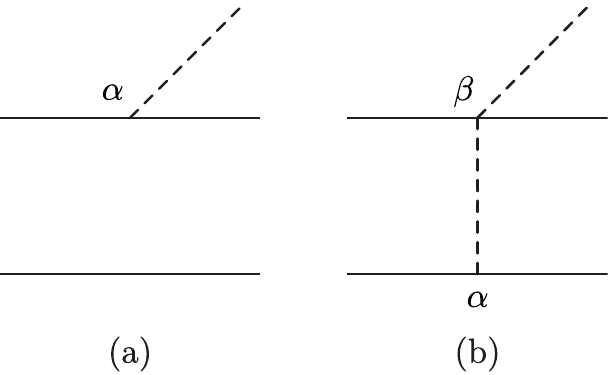}
\caption{Feynman diagrams contributing to scalar radiation at leading PN order.  Diagrams~(a) and~(b) depict the leading spin-independent terms in the conformal and disformal sector, respectively. The external field lines correspond to radiation modes, while any internal field lines correspond to potential modes.~The mirror inverses of these diagrams are all included implicitly as we do not distinguish between the two solid lines.}
\label{fig:feyn_J_O}
\end{figure}

Our starting point for the study of scalar waves is the source function~$J(x)$, which can be obtained by summing over all Feynman diagrams with one external radiation-mode scalar~$\bar\varphi$ [see~\eqref{eq:review_def_J}]. At the order to which we are working, the spin-independent part of $J(x)$ turns out to be localised entirely along the two worldlines, meaning
\begin{equation}
	J\scount{o}
	=
	\sum_\K q_\K(t)\,
	\delta^{(3)}\bm( \bm x - \bm x_\K(t) \bm).
\end{equation}
The outgoing scalar waves can thus be understood as being sourced by the periodic motion of a set of scalar~charges
\begin{align}
	q_\K(t)
	=
	\frac{\alpha_\K m_\K}{2\mpl}
	\sum_{\K' \neq \K}
	\bigg[
		1 - \frac{\bm v_\K^2}{2}
		-
		\frac{\GN m_{\K'}}{r}
		\bigg(
			1
			-
			\frac{2\alphaprime_{\K\vph{'}} \alpha_{\K'}}{\alpha_\K}
		\bigg)
		-
		\frac{\alpha_{\K'}m_{\K'}}{\alpha_\K m_\K}
		\frac{\beta_{\K}m_{\K}}{2\pi\M^4}
		\ddiff{}{t}\!\left(\frac{1}{r}\right)
	\bigg].
\label{eq:rad_J_O_charge}
\end{align}
None of this is new.~The first three terms\,---\,which together make up the conformal part of this result\,---\,have been taken from ref.~\cite{Kuntz:2019zef}, while a straightforward generalisation of the result in ref.~\cite{Brax:2019tcy} gives us the remaining disformal term. (To~clarify, the result in ref.~\cite{Brax:2019tcy} is limited to systems with ${\alpha_1 = \alpha_2}$ and ${\beta_1 = \beta_2}$, but it is easy enough to repeat the calculation with arbitrary Wilson coefficients.) In the language of Feynman diagrams, the leading conformal term in~\eqref{eq:rad_J_O_charge} can be seen to arise from figure~\ref{fig:feyn_J_O}(a), while the leading disformal term comes from figure~\ref{fig:feyn_J_O}(b). Not shown here are the four other diagrams (see, e.g., figure~8~of~ref.~\cite{Kuntz:2019zef}) that are responsible for the 1PN corrections in the conformal~sector.

\begin{figure}
\centering\includegraphics{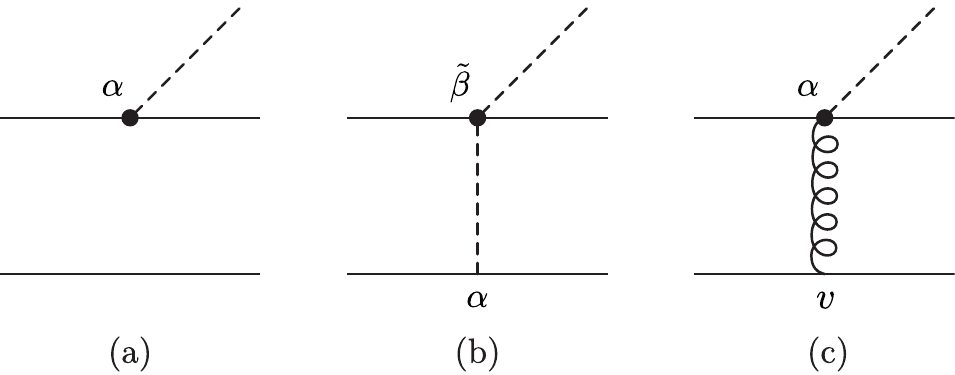}
\caption{Feynman diagrams contributing to scalar radiation in the spin-orbit sector. Diagrams~(a) and~(b) depict the leading conformal and disformal terms, respectively, while diagram~(c) is an example of a conformal term that contributes at next-to-leading PN order. The external field lines correspond to radiation modes, while any internal field lines correspond to potential modes. The mirror inverses of these diagrams are all included implicitly as we do not distinguish between the two~solid~lines.\looseness=-1}
\label{fig:feyn_J_SO}
\end{figure}

Now turning our attention to the spin-orbit part of~$J(x)$, power counting reveals that the leading conformal and disformal terms are given by figures~\ref{fig:feyn_J_SO}(a) and~\ref{fig:feyn_J_SO}(b), respectively. Notice, however, that the diagram in figure~\ref{fig:feyn_J_SO}(a) is proportional to ${S_\K^{i0 \vph{(j}} - S_\K^{ij} v_\K^j}$ (see~table~\ref{table:feyn_1}), and so vanishes when we impose the covariant SSC in~\eqref{eq:consv_ssc}. As the relevant observables in the radiative sector do not involve any Poisson brackets or functional derivatives, we are free to impose this constraint already at the level of the action; hence, the spin-orbit part of~$J(x)$ is purely disformal at the order to which we are working. As it so happens, this part is also localised entirely along the two worldlines, as we can write
\begin{equation}
	J\scount{so}
	=
	- \sum_\K q_\K^i(t) \partial_i
	\delta^{(3)}\bm( \bm x - \bm x_\K(t) \bm).
\end{equation}
The vectors $q_\K^i(t)$ may be interpreted as scalar dipole moments that are both spin and velocity dependent. Explicitly, we have that
\begin{equation}
	q_\K^i(t)
	=
 	- \sum_{\K'\neq \K}
	\frac{\alpha_{\K'}m_{\K'}}{2\mpl}
	\frac{3\tilde\beta_\K }{2\pi\M^4 r^3}
	S_{\K \vph{'}}^{ij \vph{\rangle}}
	n_{\K\K'}^{\langle j} n_{\K\K'}^{k\rangle} 
	v_{\K\K'}^{k\vph{\langle}}
\label{eq:rad_J_SO_dipole},
\end{equation}
having defined ${\bm n_{\K\K'} = (\bm x_{\K\vph{'}} - \bm x_{\K'})/r}$ and ${\bm v_{\K\K'} = \bm v_{\K\vph{'}} - \bm v_{\K'}}$ for the sake of brevity. The evaluation of figure~\ref{fig:feyn_J_SO}(b), which is what leads to this result, is novel to this work and is discussed in more detail in appendix~\ref{app:feyn}.

\paragraph{Scalar multipole moments.}
With $J(x)$ in hand, we can now use \eqref{eq:review_def_Q} to determine the scalar multipole moments~$\mathcal Q^L$ for the binary as a whole. Split into their four constituent parts as per~\eqref{eq:consv_decomposition}, we find that
\begin{subequations}
\label{eq:rad_Q_off_shell}
\begin{align}
	\mathcal Q^L_\sector{C}\scount{o}
	&=
	\sum_\K \frac{\alpha_\K m_\K}{2\mpl} x_\K^\avg{L},
	\\
	\mathcal Q^L_\sector{D}\scount{o}
	&=
	-\sumKK \frac{\alpha_{\K'} m_{\K'}}{2\mpl}
	\frac{\beta_\K m_\K}{2\pi\M^4}
	\,\ddiff{}{t}\bigg(\frac{1}{r}\bigg)
	\, x_\K^\avg{L},
	\allowdisplaybreaks\\
	\mathcal Q^L_\sector{C}\scount{so}
	&=
	0,
	\label{eq:rad_Q_off_shell_CSO}
	\\
	\mathcal Q^L_\sector{D}\scount{so}
	&=
	- \sumKK \frac{\alpha_{\K'}m_{\K'}}{2\mpl}
	\frac{3\ell\tilde\beta_\K}{2\pi\M^4 r^3}
	x_{\K\vph{'}}^{\langle i_2 \cdots\, i_\ell}
	S_{\K \vph{'}}^{i_1\rangle\, j}
	n_{\K\K'}^{\langle j} n_{\K\K'}^{k\rangle} 
	v_{\K\K'}^{k\vph{\langle}},
\end{align}
\end{subequations}
where the result in each line is given to leading order in~$v$.%
\footnote{Only the ${p=0}$ terms in \eqref{eq:review_def_Q} are needed at this order.}
These expressions should be regarded as being ``off shell'' in the sense that they currently make no assumptions about the binary's orbit, apart from requiring that ${|\bm x_\K| \ll r/v}$, as is needed to establish the validity of the multipole expansion in~\eqref{eq:review_Q_multipole_expansion}.

What we will do now is put these results ``on shell'' by going into the CM frame and substituting in the solutions to the equations of motion. For circular nonprecessing orbits, we~find that these scalar multipole moments can all be cast into the general form
\begin{equation}
	\mathcal Q^L
	=
	\frac{m}{2\mpl}
	\bigg( \frac{\Geff m}{\x} \bigg)^{\!\ell}
	Q_\ell(\x)\,
	n^\avg{L},
\label{eq:rad_Q_on_shell}
\end{equation}
although it must be emphasised that this expression holds only at the order to which we are working. The dimensionless functions~$Q_\ell(\x)$ are~given~by
\begin{subequations}
\label{eq:rad_Q_ell}
\begin{align}
% Conformal [O]
	Q_\ell^\sector{C}\scount{o}
	&=
	\comboA{+}{\ell},
	\label{eq:rad_Q_ell_CO}
\\
% Disformal [O]	
	Q_\ell^\sector{D}\scount{o}
	&=
	0,
	\label{eq:rad_Q_ell_DO}
\allowdisplaybreaks\\
% Conformal [SO]
	Q_\ell^\sector{C}\scount{so}
	&=
	\bigg[
		\frac{\projell{\Sigma}}{\Geff m^2}
		\bigg(
			\nu\comboA{+}{\ell-1}
			-
			\frac{3+2\alpha_1\alpha_2}{1+2\alpha_1\alpha_2}
			\frac{\Delta m}{m}
			\frac{ \comboA{+}{\ell} }{3}
		\bigg)
		-
		\frac{\projell{S}}{\Geff m^2}
		\frac{5+2\alpha_1\alpha_2}{1+2\alpha_1\alpha_2}
		\frac{ \comboA{+}{\ell} }{3}
	\bigg]
	\ell\x^{3/2},
\label{eq:rad_Q_ell_CSO}
\\
% Disformal [SO]
	Q_\ell^\sector{D}\scount{so}
	&=
	\frac{m}{2\pi\M^4 (\Geff m)^3}
	\sum_{\sigma=\pm}
	\frac{2}{3}
	\frac{\projell{S}_\sigma}{\Geff m^2}
	\bigg(
		\frac{3}{4}\comboBS{-\sigma}{\ell-1}
		-
		\frac{\alpha_2^2\tilde\beta_1^\vph{2} + 	
				\sigma\alpha_1^2\tilde\beta_2^\vph{2}}%
			{1+2\alpha_1\alpha_2}
		\comboA{+}{\ell}
	\bigg)
	\ell\x^{9/2}
\label{eq:rad_Q_ell_DSO}
\end{align}
\end{subequations} 
when split into their four constituent parts, with $\comboA{\pm}{\ell}$ and $\comboBS{\pm}{\ell}$ defined as in table~\ref{table:definitions}. The result in each line is accurate to leading order in~$v$, with the exception of~\eqref{eq:rad_Q_ell_DO}, which vanishes identically for circular~orbits~\cite{Brax:2019tcy}.

Two other observations about these results are now worth making. First, notice that while the conformal spin-orbit part~$\mathcal Q^L_\scriptsector{C}\scount{so}$ is zero off~shell [see~\eqref{eq:rad_Q_off_shell_CSO}], it is nonzero on~shell. The first term in \eqref{eq:rad_Q_ell_CSO} arises when we use \eqref{eq:consv_to_CM_frame} to go into the CM~frame. This transformation introduces a new term into $\mathcal Q^L_\scriptsector{C}\scount{o}$ that is linear in~$\Sigma$, which must then be reclassified as a new term in~$\mathcal Q^L_\scriptsector{C}\scount{so}$. The subsequent use of the equations of motion in \eqref{eq:consv_circles_eom_2} and~\eqref{eq:consv_def_x} causes more terms to ``leak'' from the spin-independent sector into the spin-orbit sector, and is responsible for the remainder of \eqref{eq:rad_Q_ell_CSO}, as well as the second term in~\eqref{eq:rad_Q_ell_DSO}.

That $\mathcal Q^L_\scriptsector{C}\scount{so}$ vanishes off~shell is, of course, merely an artefact of working at leading PN order. It is nonzero already at next-to-leading order, but for our purposes, these corrections can be safely neglected, since their contribution to $\mathcal Q^L$ is smaller than that of~\eqref{eq:rad_Q_ell_CSO}. To~verify that this is the case, consider just one out of the many Feynman diagrams contributing to $\mathcal Q^L_\scriptsector{C}\scount{so}$ at this order. An example is shown in figure~\ref{fig:feyn_J_SO}(c). Power counting tells us that this diagram scales as
${ (\sqrt{L} v) \, (v^3\epsSpin) \, \sqrt{v} }$\,---\,the first two factors in parentheses come from table~\ref{table:feyn_2}, while the extra factor of~$\sqrt{v}$ accounts for the fact that the outgoing scalar is a radiation mode. Now dividing this by ${\int\dx^4x\,\bar\varphi \sim (r^4/v)(v/r)}$ and then multiplying by ${\int\dx^3x\,\bm x^\avg{L} \sim r^{\ell+3}}$ to get its contribution to~$\mathcal Q^L$ [cf.~\eqref{eq:review_Q_multipole_expansion} and~\eqref{eq:review_def_Q}], we see that figure~\ref{fig:feyn_J_SO}(c) results in a term that scales as $\sqrt{Lv}  r^\ell v^4 \epsSpin$. This should be contrasted with the term in \eqref{eq:rad_Q_ell_CSO}, whose contribution to $\mathcal Q^L$ scales as $\sqrt{Lv} r^\ell v^2 \epsSpin$, which is indeed one~PN~order~lower.
\looseness=-1

Our second observation pertains to the order at which different multipoles contribute to the power~$P_\phi$. In appendix~\ref{app:Q}, we show that the master formula in~\eqref{eq:review_rad_power_scalar} simplifies~to
\begin{equation}
	P_\phi
	=
	\sum_{\ell=0}^\infty
	\frac{8\pi N_\ell}{\Geff(1+2\alpha_1\alpha_2)}
	Q_\ell^2(\x) \, \x^{\ell+3} 
\label{eq:rad_power_scalar_on_shell}
\end{equation}
when the on-shell expression in \eqref{eq:rad_Q_on_shell} holds. The numerical coefficient~$N_\ell$ can be evaluated on a case-by-case basis [see~\eqref{eq:app_Q_Nl}], and we find that ${N_0 = 0}$, ${N_1 = 1/12\pi}$, and ${N_2 = 4/15\pi}$ for the lowest few values of~$\ell$. Written in this way, \eqref{eq:rad_power_scalar_on_shell} clearly shows that the higher multipole moments are suppressed by increasingly many powers of $\x$~(${\sim v^2}$), and so contribute less and less to the final result. Consequently, it will generally suffice to keep only the contribution from the dipole moment when working to leading PN~order.

There is, however, an important exception to this rule that arises when the binary's constituents have comparable coupling strengths~$\alpha_\K$, as this results in a dipole moment whose spin-independent part vanishes at leading PN~order. Said differently, the definitions in table~\ref{table:definitions} can be used to show that ${ Q^\scriptsector{C}_{\ell=1}\scount{o} = \Delta\alpha\nu }$, where ${\Delta\alpha \coloneq \alpha_1 - \alpha_2}$, and we must be able to account for the possibility that ${\Delta\alpha}$ can vanish if our results are to be as general as possible. To that end, we will now determine which terms must be included in our calculation for~$P_\phi$ when~${\Delta\alpha = 0}$ (or~is otherwise suitably small) by making a few power-counting arguments. In~the spin-independent sector, a cursory glance at~\eqref{eq:rad_Q_ell} and~\eqref{eq:rad_power_scalar_on_shell} reveals that the dipole moment contributes a term to $P_\phi\scount{o}$ that is usually proportional to~$\x^4$ (plus higher-order corrections), but that becomes proportional to~$\x^6$ when~${\Delta\alpha = 0}$, since ${Q^\scriptsector{C}_{\ell=1}\scount{o} \sim \mathcal O(\x)}$ in this case. The size of $\Delta\alpha$ has no effect on the starting PN order of the other multipole moments, however, and so the quadrupole moment\,---\,whose contribution to $P_\phi\scount{o}$ is always proportional to~$\x^5$\,---\,takes over as the dominant term when~${\Delta\alpha = 0}$. All of the higher multipoles remain suppressed by additional powers of~$\x$, and thus can always be neglected when working to leading~order.
\looseness=-1

The situation in the spin-orbit sector is rather different, on account of $P_\phi\scount{so}$ being built from cross terms of the form $Q_\ell\scount{o}Q_\ell\scount{so}\,\x^{\ell+3}$. From~\eqref{eq:rad_Q_ell} and~\eqref{eq:rad_power_scalar_on_shell}, we see that the dipole moment contributes a term in $P_\phi\!{}^\scriptsector{C}\scount{so}$ that is usually proportional to
${
	Q^\scriptsector{C}_{\ell=1}\scount{o}
	Q^\scriptsector{C}_{\ell=1}\scount{so}
	\,\x^{4}
	\sim
	\x^{11/2}
}$,
but that becomes proportional to $\x^{13/2}$ when ${\Delta\alpha = 0}$ for the same as reason as before; namely, because ${Q^\scriptsector{C}_{\ell=1}\scount{o} \sim \mathcal O(\x)}$ in this limit. Meanwhile, the quadrupole moment provides a contribution that is also proportional to~$\x^{13/2}$ regardless of the size of~$\Delta\alpha$; hence, both must be included when $\Delta\alpha$ is suitably small, but only the former is needed otherwise. The same is true in the disformal spin-orbit sector as well, except that the dipolar contribution now scales as
${
	Q^\scriptsector{C}_{\ell=1}\scount{o}
	Q^\scriptsector{D}_{\ell=1}\scount{so}
	\,\x^{4}
	\sim
	\x^{17/2}
}$
or~$\x^{19/2}$ depending on the size of~$\Delta\alpha$, while the quadrupolar contribution is always proportional to~$\x^{19/2}$. All of the higher multipoles are consistently suppressed by additional powers of~$\x$ and can therefore~be~neglected.

This simple power-counting exercise has identified which terms are needed for a complete solution to $P_\phi$ at the order to which are working, and we now see that in addition to the leading-order expressions in \eqref{eq:rad_Q_ell}, we will also require $\mathcal Q^i_\scriptsector{C}\scount{o}$ at next-to-leading PN~order. Indeed, it is precisely for this reason that we have kept the conformal spin-independent results in section~\ref{sec:consv} and in~\eqref{eq:rad_J_O_charge} accurate to 1PN~order.

Substituting \eqref{eq:rad_J_O_charge} into \eqref{eq:review_def_Q} as before, but now keeping the 1PN corrections, we find that the conformal spin-independent part of the dipole moment~is
\begin{align}
	\mathcal Q^i_\sector{C}\scount{o}
	=&
	\sumKK
	\frac{\alpha_\K m_\K}{2\mpl}
	\bigg\{
		\bigg[
			1 - \frac{\bm v_\K^2}{2}
			-
			\frac{\GN m_{\K'}}{r}
			\bigg(
				1 - \frac{2\alphaprime_{\K\vph{'}}\alpha_{\K'}}{\alpha_\K}
			\bigg)
		\bigg]
		x_\K^i
		\nonumber\\&
		+
		\frac{1}{10}
		\big[
			2(\bm a_\K^\vph{i} \cdot \bm x_\K^\vph{i} + \bm v_\K^2)\, x_\K^i
			+
			4(\bm x_\K^\vph{i} \cdot \bm v_\K^\vph{i})\, v_\K^i
			+
			|\bm x_\K^\vph{i}|^2 a_\K^i
		\big]
	\bigg\}.
\label{eq:rad_Q_CO_dipole_1PN}
\end{align}
Even with these 1PN corrections included, \eqref{eq:rad_Q_CO_dipole_1PN} can still be cast into the general form of~\eqref{eq:rad_Q_on_shell}  thanks to the vanishing of the $v_\K^i$ term and the fact that ${a_\K^i \propto n^i}$ when evaluated on shell for circular nonprecessing orbits. The same cannot be said of the higher multipoles, but these do not contribute at the order to which we are working. For the dipole moment, we~find~that
\begin{equation}
	Q^\text{(C)}_{\ell=1}\scount{o}
	=
	\Delta\alpha\nu
	\bigg(
		1 - \frac{1}{3}(2\gammapeff + \gammaeff - \nu)\,\x\!
	\bigg)
	+
	\comboA{\text{NLO}}{1}\,\x
\label{eq:rad_Q_1_NLO}
\end{equation}
to 1PN order after using \eqref{eq:consv_to_CM_frame}, \eqref{eq:consv_circles_eom_2}, and \eqref{eq:consv_def_x}; with~$\comboA{\text{NLO}}{1}$ defined as in~table~\ref{table:definitions}. 
\looseness=-1

\paragraph{Radiated power.}
The power radiated into scalar waves can now be computed explicitly. Following ref.~\cite{Sennett:2016klh}, we will account for the possibility that $\Delta\alpha$ can vanish by splitting the power into a dipolar~($\dip$) and nondipolar~($\nondip$) part:
\begin{equation}
	P_\phi = P_{\phi,\dip} + P_{\phi,\nondip}.
\label{eq:rad_power_dipolar_split}
\end{equation}
The nondipolar part is defined as the set of terms that survive in the limit ${\Delta\alpha \to 0}$, while the dipolar part consists of the remaining terms that are all proportional to at least one power of~$\Delta\alpha$. It is worth noting that this terminology\,---\,whilst conventional\,---\,is a bit of a misnomer, as $P_{\phi,\dip}$ and $P_{\phi,\nondip}$ can both receive contributions from multipole moments of any order. In this context, whether a term is deemed to be ``dipolar'' or ``nondipolar'' is simply a statement of how it behaves when~${\alpha_1 = \alpha_2}$.

Having split $P_\phi$ into its dipolar and nondipolar part, we now use \eqref{eq:consv_decomposition_spin} to further subdivide each piece into a spin-independent and spin-orbit part. A subsequent decomposition into a conformal and disformal part \`{a}~la~\eqref{eq:consv_decomposition_ladder} will also be made in the spin-orbit sector, but this will no longer be necessary in the spin-independent sector, as the vanishing of \eqref{eq:rad_Q_ell_DO} renders it purely conformal in the case of a circular nonprecessing orbit. The end result is thus a division of $P_\phi$ into six qualitatively distinct parts:
\begin{equation}
	P_\phi
	=
	\big(
		P_{\phi,\dip}^\vph{(}\scount{o}
		+
		P_{\phi,\dip}^\sector{C}\scount{so}
		+
		P_{\phi,\dip}^\sector{D}\scount{so}
	\big)
	+
	\big(
		P_{\phi,\nondip}^\vph{(}\scount{o}
		+
		P_{\phi,\nondip}^\sector{C}\scount{so}
		+
		P_{\phi,\nondip}^\sector{D}\scount{so}
	\big).
\end{equation}

Expressions for each will now be determined to leading PN order. We do so by substituting \eqref{eq:rad_Q_ell} and \eqref{eq:rad_Q_1_NLO} into~\eqref{eq:rad_power_scalar_on_shell} while keeping only the leading-order contributions, as identified by our earlier power-counting arguments. The result in the spin-independent sector~is
\begin{subequations}
\label{eq:rad_power_phi}	
\begin{align}
% Conformal [O] dip
	P_{\phi,\dip}\scount{o}
	&=
	\frac{2\Delta\alpha^2\nu^2}{3\Geff(1+2\alpha_1\alpha_2)} 
	\,\x^4,
\label{eq:rad_power_phi_O_dip}
	\\
% Conformal [O] nondip
	P_{\phi,\nondip}\scount{o}
	&=
	\frac{32 (\comboA{+}{2})^{2\vph{)}}_\vph{+} }%
		{15 \Geff(1+2\alpha_1\alpha_2)}
	\,\x^5.
\label{eq:rad_power_phi_O_nondip}
\end{align}
These expressions are very well known (see, e.g., refs.~\cite{Damour:1992we, Kuntz:2019zef, Sennett:2016klh}), but the following spin-orbit results are novel to this work. In the conformal spin-orbit sector, we find that
\begin{align}
% Conformal [SO] dip
	P_{\phi,\dip}^\sector{C}\scount{so}
	=&\:
	\frac{2\Delta\alpha^2\nu^2}{3\Geff(1+2\alpha_1\alpha_2)}
	\frac{2}{3}
	\bigg[
		{-}
		\frac{ \projell{S} }{\Geff m^2}
		\frac{5+2\alpha_1\alpha_2}{1+2\alpha_1\alpha_2}
		\nonumber\\[-3pt]&
		+
		\frac{ \projell{\Sigma} }{\Geff m^2}
		\bigg(
			\frac{3 \comboA{+}{0} }{\Delta\alpha}
			-
			\frac{3+2\alpha_1\alpha_2}{1+2\alpha_1\alpha_2}
			\frac{\Delta m}{m}
		\bigg)
	\bigg]
	\,\x^{11/2},
\label{eq:rad_power_phi_SO_C_dip}	
\\[3pt]
% Conformal [SO] nondip
	P_{\phi,\nondip}^\sector{C}\scount{so}
	=&\:
	\frac{32 (\comboA{+}{2})^{2\vph{)}}_\vph{+} }%
		{15 \Geff(1+2\alpha_1\alpha_2)}
	\frac{4}{3}
	\bigg[
		{-}
		\frac{ \projell{S} }{\Geff m^2}
		\frac{5+2\alpha_1\alpha_2}{1+2\alpha_1\alpha_2}
		\nonumber\\[-3pt]&
		+
		\frac{ \projell{\Sigma} }{\Geff m^2}
		\bigg(
			\frac{ 15\nu \comboA{+}{0}\comboA{\text{NLO}}{1} }%
				{ 32 (\comboA{+}{2})^{2\vph{)}}_\vph{+} }
			-
			\frac{3+2\alpha_1\alpha_2}{1+2\alpha_1\alpha_2}
			\frac{\Delta m}{m}
		\bigg)
	\bigg]
	\,\x^{13/2}
\label{eq:rad_power_phi_SO_C_nondip}
\end{align}
at leading PN order, while in the disformal spin-orbit sector, the leading terms~are
\label{eq:rad_power_phi_SO_D}
\begin{align}
% Disformal [SO] dip
	P_{\phi,\dip}^\sector{D}\scount{so}
	=&\:
	\frac{m}{2\pi\M^4(\Geff m)^3}
	\frac{ 2\Delta\alpha^2\nu^2 }{3\Geff (1+2\alpha_1\alpha_2)}
	\sum_{\sigma=\pm}
	\frac{ \projell{S}_\sigma }{\Geff m^2}
	\nonumber\\[-5pt]&
	\times\!
	\bigg(
		\frac{\comboBS{-\sigma}{0}}{\Delta\alpha\nu}
		-
		\frac{4}{3}
		\frac{\alpha_2^2\tilde\beta_1^\vph{2} + 	
			\sigma\alpha_1^2\tilde\beta_2^\vph{2}}%
		{1+2\alpha_1\alpha_2}
	\bigg)
	\,\x^{17/2},
\label{eq:rad_power_phi_SO_D_dip}
\\
% Disformal [SO] nondip
	P_{\phi,\nondip}^\sector{D}\scount{so}
	=&\:
	\frac{m}{2\pi\M^4(\Geff m)^3}
	\frac{32 (\comboA{+}{2})^{2\vph{)}}_\vph{+} }%
		{15 \Geff(1+2\alpha_1\alpha_2)}
	\sum_{\sigma=\pm}
	\frac{ \projell{S}_\sigma }{\Geff m^2}
	\nonumber\\[-5pt]&
	\times\!
	\bigg(
		\frac{ 5\comboA{\text{NLO}}{1} \comboBS{-\sigma}{0} }%
			{ 16 (\comboA{+}{2})^{2\vph{)}}_\vph{+} }
		+
		\frac{ 2\comboA{+}{2} \comboBS{-\sigma}{1} }%
			{ (\comboA{+}{2})^{2\vph{)}}_\vph{+} }
		-
		\frac{8}{3}
		\frac{\alpha_2^2\tilde\beta_1^\vph{2} + 	
				\sigma\alpha_1^2\tilde\beta_2^\vph{2}}%
			{1+2\alpha_1\alpha_2}
	\bigg)
	\,\x^{19/2}.
\label{eq:rad_power_phi_SO_D_nondip}
\end{align}
\end{subequations}

% ============================================== %
\subsection{Gravitational radiation}
\label{sec:rad_gw}

Besides opening a new channel for radiation, the presence of a light scalar field also causes the amount of power~$P_g$ that is radiated into gravitational waves to deviate from the value predicted by general relativity. This occurs for two reasons. First, the gravitational multipole moments, $\mathcal I^L$ and~$\mathcal J^L$, inevitably receive new contributions from Feynman diagrams that involve the exchange of one or more potential-mode scalars [see~\eqref{eq:review_def_T}]. Second, the scalar's impact on the binary's conservative dynamics also affects the final result for~$P_g$, as the solutions to the equations of motion in section~\ref{sec:consv} are needed when putting $\mathcal I^L$ and~$\mathcal J^L$ on~shell. As it turns out, the former contribution is irrelevant at the order to which we are working, and so we need only concern ourselves with the latter.

Power-counting arguments will be used to substantiate this claim in due course, but we will simply take it as a given for now. If so, then the leading-order (off-shell) expressions~for $\mathcal I^L$~and~$\mathcal J^L$ are the same as in general relativity, and it has already been shown~that~\cite{Blanchet:2013haa}
\begin{subequations}
\label{eq:rad_IJ_O}
\begin{align}
	\mathcal I^L\scount{o}
	&=
	m\nu \sigma_\ell r^\avg{L}, 
	\\
	\mathcal J^L\scount{o}
	&=
	m\nu \sigma_{\ell+1} 
	\epsilon^{jk\langle i_1} r^{i_2 \cdots\, i_\ell\rangle j} v^k
\end{align}
\end{subequations}
at leading order in the spin-independent sector, with ${\sigma_\ell \coloneq [m_2^{\ell-1} + (-1)_\vph{2}^{\ell\vph{-1}} m_1^{\ell-1}]/m^{\ell-1}_\vph{2} }$. In the spin-orbit sector, the leading terms are~\cite{Blanchet:2013haa}\looseness=-1%
\begin{subequations}
\label{eq:rad_IJ_SO}
\begin{align}
	\mathcal I^L\scount{so}
	=&\:
	\frac{2\ell\nu}{\ell+1}
	\big[
		\ell
		r^{\langle i_2 \cdots\, i_\ell}
		\big(
			\sigma_\ell S^{i_1\rangle j}
		 	-
		 	\sigma_{\ell+1}\Sigma^{i_1\rangle j}
		\big)
		v^j
		\nonumber\\[-4pt]&
		-
		(\ell-1)\,
		v^{\langle i_2} n^{i_3 \cdots\, i_\ell}
		\big(
			\sigma_\ell S^{i_1\rangle j}
			-
			\sigma_{\ell+1}\Sigma^{i_1\rangle j}
		\big)
		r^j
	\big], 
	\\
	\mathcal J^L\scount{so}
	=&\:
	\frac{1}{2}(\ell+1)\nu 
	\big(
		\sigma_{\ell-1} S^{\langle i_1}
		-
		\sigma_\ell \Sigma^{\langle i_1}
	\big)
	r^{i_2 \cdots\, i_\ell \rangle}.
\end{align}
\end{subequations}
It should be noted that these results are not completely general, as they have already been restricted to the CM~frame. Nevertheless, they remain valid for our purposes because the leading spin-independent and spin-orbit terms in \eqref{eq:consv_to_CM_frame}, which are responsible for effecting this transformation, are also the same as in general relativity.

The power radiated into gravitational waves can now be computed by substituting \eqref{eq:rad_IJ_O} and \eqref{eq:rad_IJ_SO} into the master formula in~\eqref{eq:review_rad_power_gw}. Only the quadrupole moments are needed at leading order,%
\footnote{More precisely, only the mass-type quadrupole moment~$\mathcal I^{ij}\scount{o}$ is needed to calculate~$P_g\scount{o}$ at leading order, but both the mass-type and current-type moments are needed to determine~$P_g\scount{so}$, which is built from the two cross~terms
${}^{(3)}\mathcal I^{ij}\scount{o}{}^{(3)}\mathcal I^{ij}\scount{so}$
and
${}^{(3)}\!\mathcal J^{ij}\scount{o}{}^{(3)}\!\mathcal J^{ij}\scount{so}$.
This is because $\mathcal J^L\scount{o}$ is suppressed by one power of~$v$ relative to $\mathcal I^L\scount{o}$, but conversely, $\mathcal I^L\scount{so}$ is suppressed by one power of~$v$ relative to~$\mathcal J^L\scount{so}$.}
and for circular nonprecessing orbits we find that
\begin{subequations}
\label{eq:rad_power_gw_off_shell}
\begin{align}
	P_g\scount{o}
	&=
	\frac{32}{5}\GN m^2\nu^2 r^4 \Omega^6,
	\\
	P_g\scount{so}
	&=
	\frac{8}{15}\GN \nu^2 r^4\Omega^7
	\big(
		32 m\,\projell{S}
		+
		33 \Delta m \,\projell{\Sigma}
	\big),
\end{align}
\end{subequations}
prior to using the equations of motion. To elaborate, we have already imposed the kinematic conditions ${\bm n \cdot \bm v = 0}$ and ${\bm a = -\Omega^2\bm r}$ to obtain the above result, but have yet to make use of the equation of motion in~\eqref{eq:consv_circles_eom_2}, which establishes a dynamical relation between $r$ and~$\Omega$. Doing so gives us our final result, which~reads
\begin{subequations}
\label{eq:rad_power_gw}
\begin{align}
	P_g\scount{o}
	&=
	\frac{32\nu^2}{5\Geff (1+2\alpha_1\alpha_2)} \,\x^5,
\label{eq:rad_power_gw_O}
\\
	P_g^\sector{C}\scount{so}
	&=
	\frac{32\nu^2}{5 \Geff (1+2\alpha_1\alpha_2)}
	\frac{4}{3}
	\bigg(
		{-}
		\frac{ \projell{S} }{\Geff m^2}
		\frac{3-2\alpha_1\alpha_2}{1+2\alpha_1\alpha_2}
		-
		\frac{ \projell{\Sigma} }{\Geff m^2}
		\frac{15-34\alpha_1\alpha_2}{16+32\alpha_1\alpha_2}
		\frac{\Delta m}{m}
	\bigg)
	\,\x^{13/2},
\label{eq:rad_power_gw_SO_C}
\\[-8pt]
	P_g^\sector{D}\scount{so}
	&=
	\frac{32\nu^2}{5 \Geff (1+2\alpha_1\alpha_2)}
	\frac{m}{2\pi\M^4(\Geff m)^3}
	\sum_{\sigma = \pm}
	\frac{ \projell{S}_\sigma }{\Geff m^2}
	\bigg(
		{-}\frac{8}{3}
		\frac{\alpha_2^2\tilde\beta_1^\vph{2} + 	
				\sigma\alpha_1^2\tilde\beta_2^\vph{2}}%
			{1+2\alpha_1\alpha_2}
	\bigg)
	\,\x^{19/2}.
\label{eq:rad_power_gw_SO_D}
\end{align}
\end{subequations}
We saw previously when computing $P_\phi$ that the use of \eqref{eq:consv_circles_eom_2} invariably causes terms to leak from the spin-independent sector into both the conformal and disformal spin-orbit sectors, and this explains why there is now a nonzero disformal spin-orbit part in~\eqref{eq:rad_power_gw_SO_D}, despite the fact that~\eqref{eq:rad_power_gw_off_shell} is purely~conformal. 

Having determined each part of~$P_g$ to leading PN~order, we are now in a position to verify our earlier claim that~the scalar-induced corrections to $\mathcal I^L$ and~$\mathcal J^L$ contribute only at next-to-leading PN~order. To start with, recall from our discussion in section~\ref{sec:review} that the key quantity of interest in the gravitational sector is the binary's energy-momentum tensor~$T^{\mu\nu}(x)$. Not unlike how $J(x)$ is related to $\mathcal Q^L$ by the master formula in~\eqref{eq:review_def_Q}, analogous formulae exist for relating $T^{\mu\nu}(x)$ to the multipole moments $\mathcal I^L$~and~$\mathcal J^L$; see, e.g., eqs.~(7.44) and~(7.45) of ref.~\cite{Porto:2016pyg} for the explicit expressions. Having said that, the only thing we need to know here is that these master formulae treat the components $T^{00}$, $T^{0i}$, and $T^{ij}$ on unequal footing by virtue of $\mathcal I^L$~and~$\mathcal J^L$ being $\text{SO}(3)$~tensors, rather than Lorentz tensors. We must therefore power count separately for each of $T^{00}$, $T^{0i}$, and $T^{ij}$, but since the three cases are qualitatively very similar, we shall present just the first case with $T^{00}$ below.
\looseness=-1

\begin{figure}
\centering\includegraphics{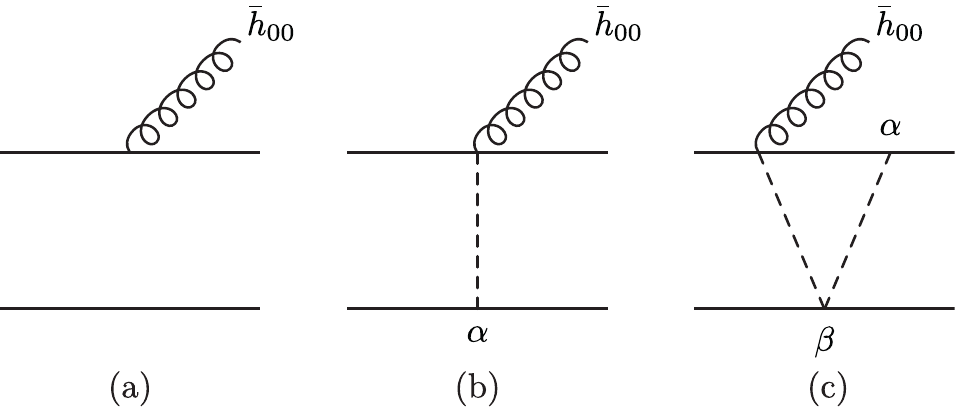}
\caption{Examples of Feynman diagrams contributing to gravitational radiation in the spin-independent sector. The external field lines correspond to radiation modes, while any internal field lines correspond to potential modes. The mirror inverses of these diagrams are all included implicitly as we do not distinguish between the two~solid~lines.}
\label{fig:feyn_T_O}
\end{figure}

Drawn in figure~\ref{fig:feyn_T_O} are a number of Feynman diagrams that contribute to $T^{00}$ according to the definition in~\eqref{eq:review_def_T}. Figure~\ref{fig:feyn_T_O}(a) is the leading term in this series, while figure~\ref{fig:feyn_T_O}(b) is just one out of several diagrams responsible for the leading correction due to a scalar in the conformal spin-independent sector. Our power-counting rules tell~us~that\looseness=-1
\begin{equation}
	\frac{\text{figure~\ref*{fig:feyn_T_O}(b)}}%
		 {\text{figure~\ref*{fig:feyn_T_O}(a)}}
	\sim
	\frac{ (\sqrt{L})(v^2)\sqrt{v} }%
		 { (\sqrt{L})\sqrt{v} }
	\sim
	v^2,
\end{equation}
where each factor in parentheses comes from either table~\ref{table:feyn_1} or table~\ref{table:feyn_2}, while the extra factors of $\sqrt{v}$ account for the fact that the outgoing graviton is a radiation mode. Clearly, figure~\ref{fig:feyn_T_O}(b) is suppressed by one PN~order relative to figure~\ref{fig:feyn_T_O}(a), and will therefore contribute terms to $\mathcal I^L$~and~$\mathcal J^L$ that are one PN order higher than those in~\eqref{eq:rad_IJ_O}. Its contribution to $P_g\scount{o}$ is thus also one PN~order higher than the leading term in \eqref{eq:rad_power_gw_O}, and can therefore be neglected. Meanwhile, figure~\ref{fig:feyn_T_O}(c) is one out of several diagrams responsible for the leading correction due to a scalar in the disformal spin-independent sector, but these disformal terms all vanish identically for circular orbits~\cite{Brax:2019tcy}, and so need~not~concern~us~here.

\begin{figure}
\centering\includegraphics{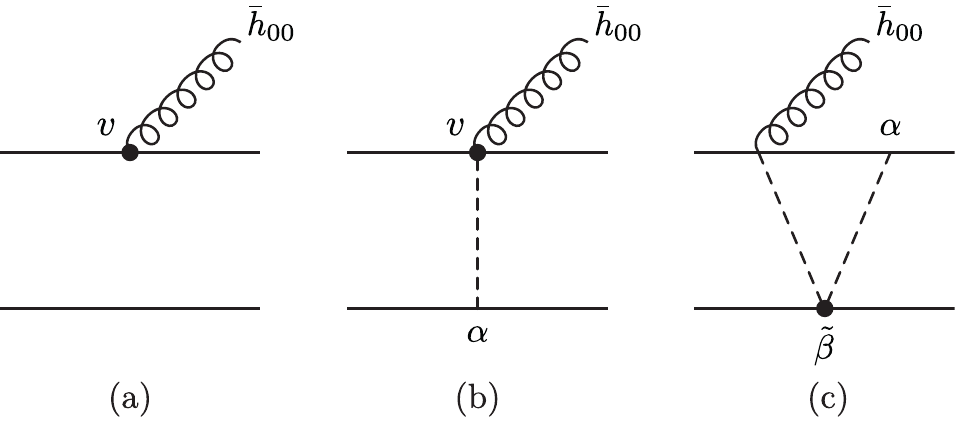}
\caption{Examples of Feynman diagrams contributing to gravitational radiation in the spin-orbit sector. The external field lines correspond to radiation modes, while any internal field lines correspond to potential modes. The mirror inverses of these diagrams are all included implicitly as we do not distinguish between the two~solid~lines.}
\label{fig:feyn_T_SO}
\end{figure}

The leading diagram contributing to~$T^{00}$ in the spin-orbit sector is drawn in figure~\ref{fig:feyn_T_SO}(a), while figures~\ref{fig:feyn_T_SO}(b) and \ref{fig:feyn_T_SO}(c) constitute examples of diagrams responsible for the leading correction from a scalar to its conformal and disformal spin-orbit part, respectively. Our power-counting rules tell us that figure~\ref{fig:feyn_T_SO}(b) is similarly suppressed by one PN~order relative to figure~\ref{fig:feyn_T_SO}(a), and will thus contribute terms to~$P_g\scount{so}$ that are one PN~order higher than those in~\eqref{eq:rad_power_gw_SO_C}. As for the leading disformal spin-orbit correction, we find that
\begin{equation}
	\frac{\text{figure~\ref*{fig:feyn_T_SO}(c)}}%
		 {\text{figure~\ref*{fig:feyn_T_SO}(a)}} 
	\sim
	\frac{ (\sqrt{L})(\epsLadder\epsSpin)(v^2)\sqrt{v} }%
		 { (\sqrt{L}v^2\epsSpin)\sqrt{v} }	
	\sim
	\epsLadder.
\end{equation}
This indicates that figure~\ref{fig:feyn_T_SO}(c) introduces terms into $P_g\scount{so}$ that are suppressed by one power of $\epsLadder$ relative to those in \eqref{eq:rad_power_gw_SO_C}, and which therefore scale as ${\mpl^2 \epsSpin \epsLadder v^{12}}$ overall. In contrast, the leading disformal terms in \eqref{eq:rad_power_gw_SO_D} can be shown to scale as~${\mpl^2 \epsSpin \epsLadder v^{10}}$.

One encounters a similar state of affairs when repeating the argument for  $T^{0i}$ and~$T^{ij}$, and this allows us to conclude that the scalar-induced corrections to~$\mathcal I^L$ and~$\mathcal J^L$ contribute to $P_g$ only at next-to-leading PN~order. Such corrections will not be needed in~what~follows.
\looseness=-1

% ============================================== %
\subsection{Balance law and phase evolution}
\label{sec:rad_phase}
 
Our result for the total power $P$~(${ = P_\phi + P_g }$) can now be used to determine how the binary evolves as it radiates energy into gravitational and scalar waves. The orbit's response to this energy loss is approximately adiabatic (i.e.,~${\dot\Omega \ll \Omega^2}$) during the early inspiral, and is therefore well described by the balance law in~\eqref{eq:review_balance}. For quasicircular nonprecessing orbits, both the binding energy~$E$ and the total power~$P$ depend implicitly on time only through the variable~$\x$; hence, the chain rule can be used to recast~\eqref{eq:review_balance}~into
\looseness=-1%
\begin{equation}
	\diff{t}{\x} = -\frac{E'}{P},
\label{eq:rad_dtdx}
\end{equation}
where~${E' \equiv \dx E/\dx\x}$. Combining this with the definition ${\Omega \equiv \dx\psi/\dx t}$ for the orbital phase~$\psi$ then gives us
\begin{equation}
	\diff{\psi}{\x} = -\frac{\x^{3/2}}{\Geff m}\frac{E'}{P}.
\label{eq:rad_phase_master_formula}
%For spacing
\vphantom{\sum_|}
\end{equation}

There are a number of ways in which we could proceed to solve this set of differential equations~\cite{Buonanno:2009zt}. The particular approach that we shall adopt here goes by the name of TaylorT2, and entails re-expanding the ratio $E'/P$ in powers of~$\x$ to obtain a Laurent series that we can then integrate analytically term by term. What results from this procedure is a pair of parametric equations, ${t \equiv t(\x)}$ and~${\psi \equiv \psi(\x)}$, which together describe how the orbital phase evolves as a function of time. The evolution of the orbital separation~$r$ then follows immediately from our solution for ${t\equiv t(\x)}$ and the equation for ${r\equiv r(\x)}$ in~\eqref{eq:consv_circles_eom_2}.

When solving these equations, care must be taken to distinguish between two substantially different scenarios for the binary's inspiral. The first possibility is that the binary radiates energy predominantly into dipolar scalar waves, which will be the case whenever $\Delta\alpha$ is large enough that \eqref{eq:rad_power_phi_O_dip} constitutes the leading term in the total power~$P$. Alternatively, if $\Delta\alpha$ is suitably small relative to~$\x$, then most of the outgoing radiation will be quadrupolar, and $P$ is instead dominated by the sum of \eqref{eq:rad_power_phi_O_nondip} and~\eqref{eq:rad_power_gw_O}. The precise condition that delineates the former \emph{dipole-driven} regime from the latter \emph{quadrupole-driven} regime can be determined by comparing the relative sizes of these different terms, and we see that
\begin{equation}
	\x \ll \frac{5\Delta\alpha^2}{48\zeta}
\label{eq:rad_DD_QD_boundary}
\end{equation}
if the system is dipole driven, while $\x$ must be much greater than~$5\Delta\alpha^2/48\zeta$ if the system is~quadrupole driven. (The constant~$\zeta$ is~defined as in table~\ref{table:definitions}.) The fact that the leading term in~$P$ changes depending on which regime we are in leads to markedly different expansions for the ratio~$E'/P$, which must be treated separately.

\paragraph{Dipole-driven regime.}
We shall begin with the mathematically simpler case of a dipole-driven~(DD) inspiral, for which the leading term in~$P$ is given by~\eqref{eq:rad_power_phi_O_dip}. As the remaining terms are all proportional to higher powers of~$\x$, the ratio $E'/P$ can be re-expanded as a Laurent series without any obstruction.

To proceed, it proves convenient to introduce the dimensionless function
\begin{equation}
	\rho(\x) \coloneq - \frac{1}{\Geff m}\frac{E'}{P},
\label{eq:rad_def_rho}
\end{equation}
which is proportional to the rhs's of both \eqref{eq:rad_dtdx} and~\eqref{eq:rad_phase_master_formula}. Now re-expanding this function in powers of~$\x$ and then truncating to the order at which we are working, we find that the leading term in the spin-independent sector is
\begin{equation}
	\rho_\DD\scount{o}
	=
	\frac{1+2\alpha_1\alpha_2}{2\Delta\alpha^2\nu}
	\frac{3}{2} \,\x^{-4},
\label{eq:rad_DD_rho_O}
\end{equation}
in agreement with ref.~\cite{Sennett:2016klh}. In fact, $\rho_\DD\scount{o}$ has been determined up to relative 2PN order in ref.~\cite{Sennett:2016klh} (albeit in very different, Jordan-frame centric notation), but these higher-order corrections will be of no concern to us here. Instead, what we are interested in are the additional contributions from spin-orbit interactions, and we find that%
\footnote{To arrive at these results, we made use of the general expression
\begin{equation*}
	\rho_\DD\scount{so}
	=
	- \frac{1}{\Geff m}
	\bigg(
		\frac{E'\scount{so}}{P\scount{o}}
		-
		\frac{ E'\scount{o}P\scount{so} }{ P\scount{o}^2 }
	\bigg),
\end{equation*} 
which we then truncated to leading order in~$\x$ after separating it into its conformal and disformal parts.}
\begin{subequations}
\label{eq:rad_DD_rho_SO}
\begin{align}
	\rho_\DD^\sector{C}\scount{so}
	&=
	\mathcal{S}_\DD^\sector{C}
	\, \x^{-5/2},
\label{eq:rad_DD_rho_SO_C}	
\\
	\rho_\DD^\sector{D}\scount{so}
	&=
	\mathcal{S}_\DD^\sector{D}\, \x^{1/2}
\label{eq:rad_DD_rho_SO_D}
\end{align}
\end{subequations}
at leading PN~order in the conformal and disformal sector, respectively. The two spin-orbit coefficients $\mathcal{S}_{\scriptscriptstyle\DD}^\scriptsector{C}$ and $\mathcal{S}_{\scriptscriptstyle\DD}^\scriptsector{D}$ have rather lengthy expressions that can be found in table~\ref{table:coefficients}.
\looseness=-1

\begin{table}
\centering
\def\arraystretch{1.4}
\begin{tabular}{| @{\hspace{10pt}} l @{\hspace{15pt}} l @{\hspace{10pt}} |}
\hline
% ------ DIPOLE-DRIVE REGIME ------ %
\multicolumn{2}{|c|}{\small Dipole-driven regime}\\[1pt]
\hline
% Conformal [SO]
$\tablestyle \mathcal{S}_\DD^\sector{C}$
&
$
\tablestyle
	\frac{1+2\alpha_1\alpha_2}{2\Delta\alpha^2\nu}
	\frac{3}{2}
	\bigg[
		\frac{ \projell{S} }{\Geff m^2}
		\frac{15-2\alpha_1\alpha_2}{1+2\alpha_1\alpha_2}
		-
		\frac{\projell\Sigma}{\Geff m^2}
		\bigg(
			\frac{ 2\comboA{+}{0} }{\Delta\alpha}
			-
			\frac{7-2\alpha_1\alpha_2}{1+2\alpha_1\alpha_2}
			\frac{\Delta m}{m}
		\bigg)
	\bigg]
$
\\
% Disformal [SO]
$\tablestyle \mathcal{S}_\DD^\sector{D}$
&
$
\tablestyle
	\frac{m}{2\pi\M^4(\Geff m)^3}
	\frac{1+2\alpha_1\alpha_2}{2\Delta\alpha^2\nu}
	\frac{3}{2}
	\sum_{\sigma = \pm}
	\frac{ \projell{S}_\sigma}{\Geff m^2}
	\bigg(
		{-}
		\frac{ \comboBS{-\sigma}{0} }{\Delta\alpha\nu}
		+
		\frac{16 (\alpha_2^2\tilde\beta_1^\vph{2}
			  +\sigma\alpha_1^2\tilde\beta_2^\vph{2} ) }%
			{1+2\alpha_1\alpha_2}
	\bigg)
$
\\[17pt]
\hline
% ------ QUADRUPOLE-DRIVE REGIME ------ %
\multicolumn{2}{|c|}{\small Quadrupole-driven regime}\\[2pt]
\hline
% Conformal [SO] nondip
$\begin{aligned}
\tablestyle
	\mathcal{S}_{\QD,\nondip}^\sector{C}
\\[5pt]~
\end{aligned}$
&
$\begin{aligned}
\tablestyle
	&
	\frac{1+2\alpha_1\alpha_2}{32\zeta\nu}
	\frac{5}{6}
	\bigg\{
		\frac{ \projell{S} }{\Geff m^2}
		\bigg(
			\frac{55-2\alpha_1\alpha_2}
				{1+2\alpha_1\alpha_2}
			-
			\frac{8}{\zeta}
		\bigg)
		\nonumber\\[-10pt]&
		-
		\frac{\projell\Sigma}{\Geff m^2}
		\bigg[
			\frac{ 5\comboA{+}{0}\comboA{\text{NLO}}{1} }
				{8\zeta\nu}
			-
			\bigg(
				\frac{27-2\alpha_1\alpha_2}
					{1+2\alpha_1\alpha_2}
				-
				\frac{33}{4\zeta}
			\bigg)
			\frac{\Delta m}{m}
		\bigg]
	\bigg\}
\end{aligned}$
\\[35pt]
% Conformal [SO] dip
$\begin{aligned}
\tablestyle
	\mathcal{S}_{\QD,\dip}^\sector{C}
\\[5pt]~
\end{aligned}$
&
$\begin{aligned}
\tablestyle
	&
	\frac{1+2\alpha_1\alpha_2}{32\zeta\nu}
	\frac{25\Delta\alpha^2}{336\zeta}
	\frac{7}{6}
	\bigg\{
		{-}
		\frac{ \projell{S} }{\Geff m^2}
		\bigg(
			\frac{65+2\alpha_1\alpha_2}{1+2\alpha_1\alpha_2}
			-
			\frac{48}{3\zeta}
		\bigg)
		\nonumber\\[-10pt]&
		-
		\frac{\projell{\Sigma}}{\Geff m^2}
		\bigg[
			\frac{ 6\comboA{+}{0} }{\Delta\alpha}
			-
			\frac{ 5\comboA{+}{0}\comboA{\text{NLO}}{1} }
				{4\zeta\nu}
			+
			\bigg(
				\frac{33 + 2\alpha_1\alpha_2}{1+2\alpha_1\alpha_2}
				-
				\frac{33}{2\zeta}
			\bigg)
			\frac{\Delta m}{m}
			\bigg)
		\bigg]
	\bigg\}
\end{aligned}$
\\[33pt]
% Disformal [SO] nondip
$\begin{aligned}
\tablestyle
	\mathcal{S}_{\QD,\nondip}^\sector{D}
\\[5pt]~
\end{aligned}$
&
$\begin{aligned}
\tablestyle
	&
	\frac{m}{2\pi\M^4(\Geff m)^3}
	\frac{1+2\alpha_1\alpha_2}{32\zeta\nu}
	\frac{5}{6}
	\sum_{\sigma=\pm}
	\frac{ \projell{S}_\sigma }{\Geff m^2}
	\nonumber\\[-10pt]&
	\times\!
	\bigg(
		\!-
		\frac{ 5\comboA{\text{NLO}}{1}\comboBS{-\sigma}{0} }
			{16\zeta\nu^2}
		-
		\frac{2\comboA{+}{2}\comboBS{-\sigma}{1}}
			{\zeta\nu^2}
		+
		\frac{52( \alpha_2^2\tilde\beta_1^\vph{2}
			+ \sigma\alpha_1^2\tilde\beta_2^\vph{2} ) }%
			{ 1 + 2\alpha_1\alpha_2 }
	\bigg)
\end{aligned}$
\\[33pt]
% Disformal [SO] dip
$\begin{aligned}
\tablestyle
	\mathcal{S}_{\QD,\dip}^\sector{D}
\\[5pt]~
\end{aligned}$
&
$\begin{aligned}
\tablestyle
	&
	\frac{m}{2\pi\M^4(\Geff m)^3}
	\frac{1+2\alpha_1\alpha_2}{32\zeta\nu}
	\frac{25\Delta\alpha^2}{336\zeta}
	\frac{7}{6}
	\sum_{\sigma=\pm}
	\frac{ \projell{S}_\sigma }{\Geff m^2}
	\nonumber\\[-10pt]&
	\times\!
	\bigg(
		\!-
		\frac{3\comboBS{-\sigma\,}{0}}{\Delta\alpha\nu}
		+
		\frac{ 5\comboA{\text{NLO}}{1}\comboBS{-\sigma}{0} }
			{8\zeta\nu^2}
		+
		\frac{4\comboA{+}{2}\comboBS{-\sigma}{1}}
			{\zeta\nu^2}
		-
		\frac{56( \alpha_2^2\tilde\beta_1^\vph{2}
			+ \sigma\alpha_1^2\tilde\beta_2^\vph{2} ) }%
			{ 1 + 2\alpha_1\alpha_2 }
	\bigg)
\end{aligned}$
\\[37pt]
\hline
\end{tabular}
\caption{Explicit expressions for the spin-orbit coefficients that appear in our solutions to the orbital and gravitational-wave phase. All of the symbols used above are defined in table~\ref{table:definitions}.}
\label{table:coefficients}
\end{table}

It is now a straightforward task to integrate \eqref{eq:rad_phase_master_formula} term by term in order to obtain the orbital phase~$\psi$. The end result is schematically of the form
\begin{equation}
	\psi^\vph{()}_\DD
	=
	\psi_0^\vph{()}
	+
	\psi_\DD^\vph{()}\scount{o}
	+
	\psi_\DD^\sector{C}\scount{so}
	+
	\psi_\DD^\sector{D}\scount{so},
\end{equation}
where $\psi_0$ is some integration constant to be fixed by initial conditions, while the other three terms denote the spin-independent, conformal spin-orbit, and disformal spin-orbit parts of the phase, respectively. At leading PN~order in each sector, we find
\begin{subequations}
\label{eq:rad_DD_phase}
\begin{align}
	\psi_\DD^\vph{()}\scount{o}
	&=
	- \frac{1+2\alpha_1\alpha_2}{2\Delta\alpha^2\nu}
	\,\x^{-3/2},
\label{eq:rad_DD_phase_O}
\allowdisplaybreaks\\
	\psi_\DD^\sector{C}\scount{so}
	&=
	\mathcal{S}_\DD^\sector{C}\log\x,
\label{eq:rad_DD_phase_SO_C}
\\
	\psi_\DD^\sector{D}\scount{so}
	&=
	\frac{1}{3}
	\mathcal{S}_\DD^\sector{D}
	\,\x^3.
\label{eq:rad_DD_phase_SO_D}
\end{align}
\end{subequations}
The solution for ${t\equiv t(\x)}$ follows from a similar calculation, and is presented in~appendix~\ref{app:gw}. 
\looseness=-1

As we did with the binding energy in section~\ref{sec:consv_circles}, we can now determine the relative importance of each term in $\psi$ by simply comparing their leading powers of~$\x$. We find that the conformal spin-orbit term in \eqref{eq:rad_DD_phase_SO_C} starts at 1.5PN order relative to the leading spin-independent term in~\eqref{eq:rad_DD_phase_O}, while the disformal spin-orbit term in~\eqref{eq:rad_DD_phase_SO_D} starts at relative 4.5PN~order, although it is enhanced  by the large prefactor $m/[2\pi\M^4(\Geff m)^3]$, which enters via the definition of the spin-orbit coefficient~$\mathcal{S}_{\scriptscriptstyle\DD}^\scriptsector{D}$ (see~table~\ref{table:coefficients}). Per our discussion below~\eqref{eq:consv_Vainshtein_ratio}, it follows that for sufficiently small values of~$\M$, the disformal term in \eqref{eq:rad_DD_phase_SO_D} could have a significantly larger impact on the orbital phase than its conformal counterpart, despite the suppression by three additional powers of~$\x$. This same enhancement of the disformal sector is also present in quadrupole-driven systems, as we shall~see.

\paragraph{Quadrupole-driven regime.}
A binary is said to be in a quadrupole-driven~(QD) inspiral whenever the sum of~\eqref{eq:rad_power_phi_O_nondip} and~\eqref{eq:rad_power_gw_O} serves as the leading term in~$P$. In this regime, the re-expansion of $E'/P$ requires more thoughtful consideration because of the scalar dipole term in~\eqref{eq:rad_power_phi_O_dip}, which\,---\,on account of the smallness of~$\Delta\alpha$ [see~our discussion below~\eqref{eq:rad_DD_QD_boundary}]\,---\,provides only a subleading correction to the overall energy loss, despite being proportional to one less power of~$\x$ than the leading quadrupole term.

Following ref.~\cite{Sennett:2016klh}, we will account for this complication by splitting the total power into a dipolar and non\-dipolar part, as we did for~$P_\phi$ in~\eqref{eq:rad_power_dipolar_split}. This allows us to write
\begin{equation}
	\frac{E'}{P}
	\simeq
	\frac{E'}{P_\nondip}
	\bigg(
		1 - \frac{P_\dip}{P_\nondip}
	\bigg)
\label{eq:rad_QD_first_order_approximation}
\end{equation}
after expanding to first order in the small ratio ${P_\dip/P_\nondip \sim \mathcal O(\Delta\alpha^2/\x)}$, where
\begin{equation}
	P_\dip = P_{\phi,\dip}
	\quad\text{and}\quad
	P_\nondip = P_{\phi,\nondip} + P_g.
\end{equation}
The approximation in \eqref{eq:rad_QD_first_order_approximation} naturally causes $\rho(\x)$ to split into a dipolar and nondipolar part as well, and this motivates writing ${\rho_\QD(\x) = \rho_{\QD,\nondip}(\x) + \rho_{\QD,\dip}(\x) }$, with
\begin{subequations}
\begin{align}
	\rho_{\QD,\nondip}(\x)
	&=
	-\frac{1}{\Geff m}\frac{E'}{P_\nondip},
\label{eq:rad_QD_rho_nondip}
\\
	\rho_{\QD,\dip}(\x)
	&=
	\frac{1}{\Geff m}\frac{E'}{P_\nondip}\frac{P_\dip}{P_\nondip}.
\label{eq:rad_QD_rho_dip}
\end{align}
\end{subequations}

Both parts of $\rho_\QD$ can now be re-expanded in powers of $\x$ without any further obstructions, and we find that
\begin{subequations}
\label{eq:rad_QD_rho}
\begin{align}
% [O]	
	&&
	\rho_{\QD,\nondip}^\vph{()}\scount{o}
	&=
	\frac{1+2\alpha_1\alpha_2}{32\zeta\nu}
	\frac{5}{2}\x^{-5},
&\qquad
	\rho_{\QD,\dip}^\vph{()}\scount{o}
	&=
	-\frac{1+2\alpha_1\alpha_2}{32\zeta\nu}
	\frac{25\Delta\alpha^2}{336\zeta}
	\frac{7}{2}\x^{-6},
&
\label{eq:rad_QD_rho_O}
\\
% Conformal [SO]	
	&&
	\rho_{\QD,\nondip}^\sector{C}\scount{so}
	&=
	\mathcal{S}_{\QD,\nondip}^\sector{C}
	\,\x^{-7/2},
	&
	\rho_{\QD,\dip}^\sector{C}\scount{so}
	&=
	\mathcal{S}_{\QD,\dip}^\sector{C}
	\,\x^{-9/2},
&
\label{eq:rad_QD_rho_SO_C}
\\[3pt]
% Disformal [SO]	
	&&
	\rho_{\QD,\nondip}^\sector{D}\scount{so}
	&=
	\mathcal{S}_{\QD,\nondip}^\sector{D}
	\,\x^{-1/2},
	&
	\rho_{\QD,\dip}^\sector{D}\scount{so}
	&=
	\mathcal{S}_{\QD,\dip}^\sector{D}
	\,\x^{-3/2},
&
\label{eq:rad_QD_rho_SO_D}
\end{align}
\end{subequations}
with the result in each sector given to leading order in~$\x$. As was the case in the dipole-driven regime, the leading spin-independent terms in \eqref{eq:rad_QD_rho_O} are found to be in complete agreement with the results of ref.~\cite{Sennett:2016klh}, while the leading spin-orbit terms in \eqref{eq:rad_QD_rho_SO_C} and~\eqref{eq:rad_QD_rho_SO_D} are novel to this work.%
\footnote{To elaborate further, we obtained the spin-orbit results in \eqref{eq:rad_QD_rho} by starting with the general~expressions
\begin{align*}
	\rho_{\QD,\nondip}\scount{so}
	&=
	-\frac{1}{\Geff m}
	\bigg(
		\frac{E'\scount{so}}{P_\nondip\scount{o}}
		-
		\frac{E'\scount{so} P_\nondip\scount{so}}{P_\nondip\scount{o}^2}
	\bigg),
	\\
	\rho_{\QD,\dip}\scount{so}
	&=
	\frac{1}{\Geff m}
	\bigg[
		\bigg(
			\frac{E'\scount{so}}{P_\nondip\scount{o}}
			-
			\frac{E'\scount{o} P_\nondip\scount{so}}{P_\nondip\scount{o}^2}
		\bigg)
		\frac{P_\dip\scount{o}}{P_\nondip\scount{o}}
		+
		\frac{E'\scount{o}}{P_\nondip\scount{o}}
		\bigg(
			\frac{P_\dip\scount{so}}{P_\nondip\scount{o}}
		-
		\frac{P_\dip\scount{o} P_\nondip\scount{so}}%
			{P_\nondip\scount{o}^2}
		\bigg)
	\bigg],
\end{align*}
which we then truncated to leading order in~$\x$ after separating them into conformal and disformal~parts.\looseness=-1}
The four spin-orbit coefficients appearing in these results have rather lengthy expressions that can be found in~table~\ref{table:coefficients}.

We can now integrate \eqref{eq:rad_phase_master_formula} to obtain the orbital phase~$\psi$, which schematically reads%
\begin{align}
	\psi^\vph{()}_\QD
	=
	\psi_0^\vph{()}
	+
	\big(
		&
		\psi_{\QD,\nondip}^\vph{()}\scount{o}
		+
		\psi_{\QD,\nondip}^\sector{C}\scount{so}
		+
		\psi_{\QD,\nondip}^\sector{D}\scount{so}
	\big)
	\nonumber\\
	+\:
	\big(
		&
		\psi_{\QD,\dip}^\vph{()}\scount{o}
		\:+\:
		\psi_{\QD,\dip}^\sector{C}\scount{so}
		\:+\:
		\psi_{\QD,\dip}^\sector{D}\scount{so}
	\big).
\end{align}
The integration constant~$\psi_0$ is to be fixed by initial conditions, while the remaining terms correspond to a subdivision of the phase into dipolar and nondipolar, spin-independent and spin-orbit, and conformal and disformal parts. At leading PN~order in each sector, we~find
\looseness=-1
\begin{subequations}
\label{eq:rad_QD_phase}
\begin{align}
% [O]	
	&&
	\psi_{\QD,\nondip}^\vph{()}\scount{o}
	&=
	-\frac{1+2\alpha_1\alpha_2}{32\zeta\nu}
	\x^{-5/2},
&\qquad
	\psi_{\QD,\dip}^\vph{()}\scount{o}
	&=
	\frac{1+2\alpha_1\alpha_2}{32\zeta\nu}
	\frac{25\Delta\alpha^2}{336\zeta}
	\x^{-7/2},
&
\label{eq:rad_QD_phase_O}
\\
% Conformal [SO]	
	&&
	\psi_{\QD,\nondip}^\sector{C}\scount{so}
	&=
	-\mathcal{S}_{\QD,\nondip}^\sector{C}
	\,\x^{-1},
	&
	\psi_{\QD,\dip}^\sector{C}\scount{so}
	&=
	-\frac{1}{2}\mathcal{S}_{\QD,\dip}^\sector{C}
	\,\x^{-2},
&
\label{eq:rad_QD_phase_SO_C}
\\[3pt]
% Disformal [SO]	
	&&
	\psi_{\QD,\nondip}^\sector{D}\scount{so}
	&=
	\frac{1}{2}
	\mathcal{S}_{\QD,\nondip}^\sector{D}
	\,\x^{2},
	&
	\psi_{\QD,\dip}^\sector{D}\scount{so}
	&=
	\mathcal{S}_{\QD,\dip}^\sector{D}
	\,\x.
&
\label{eq:rad_QD_phase_SO_D}
\end{align}
\end{subequations}
The solution for ${t\equiv t(\x)}$ follows from a similar calculation, and is presented in~appendix~\ref{app:gw}. 
\looseness=-1

Comparing the different powers of~$\x$ in~\eqref{eq:rad_QD_phase} gives us an estimate for the relative importance of each term. In the spin-independent sector~[see~\eqref{eq:rad_QD_phase_O}], it is well known that the term proportional to~$\x^{-5/2}$, which arises due to quadrupole radiation, is~responsible for the dominant contribution to the phase, while the dipolar term\,---\,despite appearing at relative~${-1}$PN order\,---\,provides only a subleading correction to~$\psi$ on account of the smallness of~$\Delta\alpha$. This much is guaranteed by definition, as a binary is said to be in a quadrupole-driven inspiral if and only if ${\Delta\alpha \ll \x^{1/2}}$~[or more precisely, if ${5\Delta\alpha^2/48\zeta \ll \x}$;~cf.~\eqref{eq:rad_DD_QD_boundary}]. 
\looseness=-1

Interestingly, what occurs in the spin-orbit sector can be rather different. To illustrate this, we find it helpful to distinguish between two extreme cases: when $\alpha_1$ and~$\alpha_2$ are comparable to one another, such that ${\alpha_1 \approx \alpha_2}$ and ${0 \leq \Delta\alpha \ll \alpha_1}$; and when $\alpha_1$ and $\alpha_2$ are vastly different, such that ${\alpha_1 \approx \Delta\alpha}$ while ${\alpha_2 \approx 0}$. Consider the first case with~${\alpha_1\approx\alpha_2}$, and take the conformal spin-orbit sector in~\eqref{eq:rad_QD_phase_SO_C} to start with. Its nondipolar part appears at $1.5$PN order relative to the leading term in~$\psi_{\QD,\nondip}\scount{o}$, while its dipolar part starts earlier at relative $0.5$PN order, but is suppressed due to the smallness of~$\Delta\alpha$. Crucially, however, we see from table~\ref{table:coefficients} that it is suppressed by only one power of~$\Delta\alpha$,%
\footnote{While $\mathcal{S}^\scriptsector{C}_{\scriptscriptstyle\QD,\dip}$ has an overall factor of $\Delta\alpha^2$ out front, the first term in square brackets is proportional to~$1/\Delta\alpha$.\looseness=-1}
unlike $\psi_{\QD,\dip}\scount{o}$, which is suppressed by two. What this means is that the dipolar spin-orbit term
(${
	\psi^\scriptsector{C}_{\scriptscriptstyle\QD,\dip}\scount{so}
	\sim
	\Delta\alpha/\x^2
}$)
can actually be larger than its nondipolar counterpart
(${
	\psi^\scriptsector{C}_{\scriptscriptstyle\QD,\dip}\scount{so}
	\sim
	1/\x
}$)
if $\Delta\alpha$ satisfies the inequality ${\x \lesssim \Delta\alpha \ll \x^{1/2}}$. The value of~$\x$ will continue to increase towards unity as the binary evolves, and the two terms become comparable to one another when~${\x \sim \Delta\alpha}$. Subsequently, the nondipolar part takes over as the larger contribution when~${\Delta\alpha \lesssim \x}$. The same conclusion holds also for the disformal spin-orbit sector in~\eqref{eq:rad_QD_phase_SO_D}, whose dipolar and nondipolar parts start at relative $3.5$PN and relative $4.5$PN orders, respectively.

Now consider the second case with ${\alpha_1\approx\Delta\alpha}$ and ${\alpha_2 \approx 0}$. The dipolar coefficients $\mathcal{S}^\scriptsector{C}_{\scriptscriptstyle\QD,\dip}$ and~$\mathcal{S}^\scriptsector{D}_{\scriptscriptstyle\QD,\dip}$ go from being suppressed by one power of~$\Delta\alpha$ to being suppressed by two as a result of $\comboA{\pm}{\ell}$, $\comboA{\text{NLO}}{1}$, and $\comboBS{\pm}{\ell}$ themselves becoming newly proportional to~$\Delta\alpha$ (cf.~table~\ref{table:definitions}). Whether these dipolar terms can ever be larger than their nondipolar counterparts is now a question of what happens to $\mathcal{S}^\scriptsector{C}_{\scriptscriptstyle\QD,\nondip}$ and~$\mathcal{S}^\scriptsector{D}_{\scriptscriptstyle\QD,\nondip}$. In~the conformal sector, the coefficient $\mathcal{S}^\scriptsector{C}_{\scriptscriptstyle\QD,\nondip}$ remains of order one, and thus the dipolar term
(${
	\psi^\scriptsector{C}_{\scriptscriptstyle\QD,\dip}\scount{so}
	\sim
	\Delta\alpha^2/\x^2
}$)
is always subleading to its nondipolar counterpart
(${
	\psi^\scriptsector{C}_{\scriptscriptstyle\QD,\nondip}\scount{so}
	\sim
	1/\x 
}$),
since we must have ${\Delta\alpha \ll \x^{1/2}}$ in the quadrupole-driven regime. On the other hand, $\mathcal{S}^\scriptsector{D}_{\scriptscriptstyle\QD,\nondip}$ goes from being of order one to being of order~$\Delta\alpha^2$ when we set ${\alpha_1 \approx \Delta\alpha}$ and ${\alpha_2 \approx 0}$; hence, the dipolar term
(${
	\psi^\scriptsector{D}_{\scriptscriptstyle\QD,\dip}\scount{so}
	\sim
	\Delta\alpha^2 \x
}$)
in the disformal spin-orbit sector is always larger than its nondipolar counterpart
(${
	\psi^\scriptsector{D}_{\scriptscriptstyle\QD,\nondip}\scount{so}
	\sim
	\Delta\alpha^2 \x^2
}$).
This difference can be attributed to the fact that $\mathcal{S}^\scriptsector{C}_{\scriptscriptstyle\QD,\nondip}$ contains the usual spin-orbit contributions from general relativity in addition to those from a scalar field, whereas $\mathcal{S}^\scriptsector{D}_{\scriptscriptstyle\QD,\nondip}$ is induced purely by the scalar. All of this serves to illustrate how varied the evolution of~$\psi$ can be in a scalar-tensor theory when spin-orbit effects~are~taken~into~account.%
\footnote{It goes without saying that these conclusions were drawn by simply counting powers of~$\Delta\alpha$. Nevertheless, we do not expect them to change on a qualitative level were we to undertake a more deliberate calculation that accounts for all of the different numerical factors in table~\ref{table:coefficients}.}

As a last remark on~\eqref{eq:rad_QD_phase}, it is worth reiterating what was said earlier in our discussion below \eqref{eq:consv_Vainshtein_ratio} and \eqref{eq:rad_DD_phase}; namely, that although the disformal spin-orbit sector is suppressed by $3$PN orders relative to the conformal spin-orbit sector, the former can still have a greater impact on the orbital phase as a result of being multiplied by the large prefactor~${m/[2\pi\M^4(\Geff m)^3]}$. We~will see this explicitly in the next section when we discuss observational~prospects, but for now, there is one final step to take in our EFT~pipeline.

\paragraph{Gravitational-wave phase.}
What we ultimately measure is the gravitational-wave signal from an inspiraling binary, and as we discussed towards the end of section~\ref{sec:review}, the phase~$\Psi$ of this gravitational wave derives directly from the binary's orbital phase~$\psi$. Specifically, after decomposing the signal into spin-weighted spherical harmonics, one finds that the phase of the ${(\yl,\ym})$~mode is given by ${\Psi_{\yl\ym} = \ym \psi}$, up to higher-order corrections that we shall neglect~\cite{Blanchet:2013haa, Sennett:2016klh}. The pair of parametric equations ${\Psi_{\yl\ym} \equiv \ym\psi(\x)}$ and ${t \equiv t(\x)}$ therefore provides us with our final result for the evolution of the gravitational-wave phase.

This time-domain solution will suffice for our purposes in the next section, although it is worth noting that frequency-domain waveforms are much more convenient for a matched-filtering analysis. If
${
	h_{\yl\ym}(t) = A_{\yl\ym}(t) e^{-i\ym\psi(t)}
}$
is the waveform for a particular oscillatory mode in the time domain, its counterpart in the frequency domain is given by the~Fourier~transform
\begin{equation}
	\tilde h_{\yl\ym}(f)
	=
	\int_{-\infty}^{\infty}\dx t\;e^{2i\pi ft} h_{\yl\ym}(t).
\label{eq:rad_freq_domain}
\end{equation}
Because these gravitational waves are rapidly oscillating but slowly chirping signals, the integral in~\eqref{eq:rad_freq_domain} can be evaluated analytically within the stationary-phase approximation. The result is of the general form
${
	\tilde h_{\yl\ym}(f)
	=
	\tilde A_{\yl\ym}(f)
	e^{-i\tilde\Psi_{\yl\ym}(f)-i\pi/4}
}$,
where the corresponding phase in the frequency domain is given by~\cite{Sennett:2016klh}
\begin{equation}
	\tilde\Psi_{\yl\ym}(f)
	=
	\ym
	\bigg(
		 \psi(\x) - \frac{\x^{3/2}}{\Geff m} \,t(\x)
	\bigg)_{\x=\x_f},
	\quad
	\x_f = ( 2\pi\Geff m f/\ym)^{2/3}.
\label{eq:rad_gw_phase_Fourier}
\end{equation}
Explicit expressions for the leading spin-independent, conformal spin-orbit, and disformal spin-orbit parts of $\tilde\Psi_{\yl\ym}(f)$ are presented in appendix~\ref{app:gw}. Written in this form (which is known as the TaylorF2~approximant~\cite{Buonanno:2009zt}), our results are now ready to be used in model-independent frameworks for testing gravity, like the parametrised~post-Einsteinian~(PPE) or generalised inspiral-merger-ringdown~(gIMR) formalisms~\cite{Yunes:2009ke, Tahura:2018zuq, TheLIGOScientific:2016src}, or can otherwise be incorporated directly into dedicated waveform models for scalar-tensor~theories~\cite{Sennett:2016klh}.

% ============================================== %
% Section 6
% ============================================== %
\section{Double neutron stars and observational prospects}
\label{sec:obsv}

Mention has already been made of the large prefactor $m/[2\pi\M^4(\Geff m)^3]$, which can cause the disformal spin-orbit terms in the waveform to be enhanced relative to those in the conformal spin-orbit sector. Our goal in this penultimate section is to be more quantitative about the extent of this enhancement, and to investigate if gravitational-wave detectors are capable of observing these interactions as a result. To be clear, gravitational waves are by no means the only avenue for probing these kinds of effects, and need not even be best suited to the task~\cite{Damour:1998jk}\,---\,they are simply the most natural option to consider here, given the collection of results derived in this~paper. (We comment briefly on other potential probes in the next section.) The question of how well scalar-tensor theories can be constrained by gravitational waves is nevertheless an interesting and timely one, and so warrants some investigation.

We will be interested primarily in the prospect of probing \emph{disformal} effects, since the smoking gun for a conformal coupling between a binary and a light scalar field is already well understood: it is the emission of scalar dipole radiation, which has been discussed extensively in, e.g., refs.~\cite{Will:1994fb, Damour:1998jk, Berti:2012bp, Yunes:2013dva, Yunes:2016jcc, Chamberlain:2017fjl, Barack:2018yly, Carson:2019fxr, Perkins:2020tra, Shao:2017gwu, Zhao:2019suc, Guo:2021leu}. As these studies emphasise, a robust assessment of detectability requires deliberate consideration of many factors, like the various degeneracies between parameters, expected detection rates for different binary populations, the signal-to-noise ratio of individual events, and so on. A~fully fledged Bayesian approach of this kind, however, is well beyond the scope of this already very lengthy paper; hence, as a precursor to a more in-depth analysis in the future, we shall here employ a more rudimentary diagnostic for assessing detectability. Let $\mathcal N$ be the total number of gravitational-wave cycles that accumulate in the detector as a signal passes through. Following the logic of ref.~\cite{Will:1994fb}, we deem a particular effect or interaction to be undetectable if its contribution to~$\mathcal N$\,---\,integrated over the entire time that the signal spends in the detector's sensitivity band\,---\,is less than~$1$, and conversely, we say that it is (potentially) detectable if its contribution to~$\mathcal N$~is~greater~than~$1$.

\paragraph{Double neutron stars.}
As for the question of which systems are most relevant to our discussion, the choice to focus on double neutron star binaries is essentially forced upon us by two determining factors. The first is the fact that, for the class of scalar-tensor theories considered in this paper, disformal effects enter into the gravitational waveform only if both constituents of the binary are composed of matter. The second is our restriction to circular orbits, which\,---\,as far as gravitational-wave sources are concerned\,---\,is an assumption that is generally valid only for binaries approaching the end of their inspiral~\cite{Peters:1964:ee, Krolak:1987ee, Cardoso:2020iji}. Taken together, these considerations naturally direct our attention towards double neutron star binaries, which are the only systems (that we know of) whose constituents are both composed of matter and whose inspiral comes to an end within the sensitivity band of a gravitational-wave detector; namely, one~that~is~ground-based.

In this frequency window, the evolution of the binary can be split broadly into three phases: what we will call the perturbative inspiral phase, the nonperturbative inspiral phase, and the merger phase. Mathematically, we write
\begin{equation}
	\mathcal N
	=
	\mathcal N_\text{pert}
	+
	\mathcal N_\text{nonpert}
	+
	\mathcal N_\text{merger}.
\label{eq:obsv_N}
\end{equation}
The first of these phases corresponds to the early portion of the inspiral, during which all three expansion parameters\,---\,$v$, $\epsSpin$, and $\epsLadder\epsSpin$\,---\,remain small.%
\footnote{For noncircular orbits, we would also require ${\epsLadder e^2 \ll 1}$; recall our discussion in section~\ref{sec:consv_rules}.
\looseness=-1}
The second phase has no analogue in general relativity, and moreover is present only if the disformal coupling scale~$\M$ is below some critical value. This is because the ladder parameter $\epsLadder$~[$\sim mv^2/(2\pi\M^4 r^3)$] grows as the inspiral progresses, and for sufficiently small values of~$\M$, there  will come a time when the combination~$\epsLadder\epsSpin$ is large enough to invalidate our perturbative expansion. As we discussed towards the end of section~\ref{sec:consv_rules}, the point-particle approximation of the two bodies remains valid in this nonperturbative regime, but maintaining an accurate description of the binary's dynamics requires a certain class of ``ladder diagrams'' to be resummed. Finally, the merger phase begins when the two neutron stars come into sufficiently close contact that their orbit becomes unstable, prompting them to plunge into one another~\cite{Radice:2020ddv, Baiotti:2016qnr, Faber:2012rw}.

It will be useful in what follows to know approximately when a binary transitions from one phase to the next. To that end, let $f_\text{nonpert}$ and $f_\text{contact}$ denote the gravitational-wave frequencies that mark the beginnings of the nonperturbative inspiral phase and merger phase, respectively. The latter invariably depends on the parameters of the system and on the specific details of the neutron stars' equation of state~\cite{Radice:2020ddv}, but as a rough guide, we shall here adopt a fixed value for~${f_\text{contact} \sim 1~\text{kHz}}$. Meanwhile, we can obtain an estimate for~$f_\text{nonpert}$ by appealing to the fact that the combination $\epsLadder\epsSpin$ is small\,---\,or more precisely, that the condition in~\eqref{eq:consv_boundary_spin_ladder} holds\,---\,at all frequencies below this threshold. If we further focus on the dominant $(2,2)$ mode of the gravitational-wave signal for simplicity, we can make the identification~${\Omega = \pi f}$, which allows~\eqref{eq:consv_boundary_spin_ladder} to be rewritten~as
\begin{equation}
	\frac{m}{2\pi\M^4(\Geff m)^3}
	\frac{\max(\tilde\beta_1\chi_1, \tilde\beta_2\chi_2)}{1+2\alpha_1\alpha_2} 
	(\Geff m\pi f)^3
	\ll 1,
\end{equation}
after also defining ${\chi_\K \coloneq \projell{S}_\K/(\GN m_\K^2)}$ and making use of~\eqref{eq:consv_circles_eom_2} and~\eqref{eq:consv_def_x}. Moving forward, we shall replace the vague condition ``${\ll 1}$'' by the inequality ``${ < \epsilon }$'' to be more concrete, and will take ${\epsilon = 0.1}$ to be our criterion for what we consider small. The quantity~$f_\text{nonpert}$ is~then defined as the frequency that saturates this inequality, and we~have~that
\begin{equation}
	f_\text{nonpert}
	=
	\bigg|
		\frac{2\epsilon\M^4(1+2\alpha_1\alpha_2)}%
			{m\pi^2 \max(\tilde\beta_1\chi_1, \tilde\beta_2\chi_2) }
	\bigg|^{1/3}.
\label{eq:obsv_f_nonpert}
\end{equation}
It now follows that a neutron star binary undergoes all three phases as described below~\eqref{eq:obsv_N} if ${ f_\text{nonpert} < f_\text{contact} }$, but if ${ f_\text{nonpert} > f_\text{contact} }$, then the binary transitions directly from the perturbative part of its inspiral into the~merger~phase.

\paragraph{Dephasing.}
The results of section~\ref{sec:rad} enable us to compute only the number of gravitational-wave cycles that accumulate during the early, perturbative part of the binary's inspiral, but this will nevertheless suffice for obtaining a rough estimate of detectability. The reason being: an effect that contributes more than one cycle to~$\mathcal N_\text{pert}$ necessarily also contributes more than one cycle to~$\mathcal N$. We can thus revise our criterion for detectability and will now deem an effect to be (potentially) detectable if its contribution to~$\mathcal N_\text{pert}$ is~greater~than~1.

Focusing on the dominant $(2,2)$ mode of the gravitational-wave signal, we have that
\begin{equation}
	\mathcal N_\text{pert}
	=
	\frac{\Psi_{2,2}(\x_2) - \Psi_{2,2}(\x_1)}{2\pi},
\label{eq:obsv_N_pert}
\end{equation}
where the initial time ${t_1 \equiv t(\x_1)}$ is the time at which the signal enters the detector's sensitivity band, while the final time ${t_2 \equiv t(\x_2)}$ marks the end of the perturbative inspiral phase. For the former, we take ${\x_1 = (\Geff m \pi f_1)^{2/3}}$ [cf.~\eqref{eq:rad_gw_phase_Fourier}], and following ref.~\cite{Shao:2017gwu}, we consider three possible values for~$f_1$:
\begin{equation}
	f_1 \in \{ 10~\text{Hz},\, 5~\text{Hz},\, 1~\text{Hz} \}.
\label{eq:obsv_f1}
\end{equation}
The largest of these frequencies marks the beginning of the sensitivity band for Advanced LIGO~\cite{Martynov:2016fzi}, while the remaining two are the projections for future, third-generation ground-based detectors like Cosmic Explorer~(CE) and the Einstein Telescope~(ET)~\cite{Hild:2010id}. As for the end of the perturbative inspiral phase, we take ${\x_2 = (\Geff m \pi f_2)^{2/3}}$ with
\begin{equation}
	f_2 = \min(f_\text{nonpert}, f_\text{contact}).
\end{equation}
Note also that ${\mathcal N_\text{pert} = 0}$ if ${f_2 \leq f_1}$; in which case, the perturbative part of the inspiral occurs outside the detector's window.

Having specified the appropriate frequency band, we can now write down an explicit expression for~$\mathcal N_\text{pert}$. Current observations of binary pulsars already place strong constraints on the size of~$\Delta\alpha$~\cite{Shao:2017gwu, Zhao:2019suc, Guo:2021leu}, which limit the inspiral of a binary neutron star to be quadrupole driven by the time it enters the LIGO/CE/ET band, even if it did undergo a previous dipole-driven phase at lower frequencies. Accordingly, only the formula for~$\psi_\QD$ in~\eqref{eq:rad_QD_phase} is needed to evaluate~\eqref{eq:obsv_N_pert}. The result can be split into four parts,
\begin{equation}
	\mathcal N_\text{pert}^{}
	=
	\mathcal N^\sector{GR}_\text{pert}
	+
	\delta\mathcal N_\text{pert}^{}\scount{o}
	+
	\delta\mathcal N^\sector{C}_\text{pert}\scount{so}
	+
	\delta\mathcal N^\sector{D}_\text{pert}\scount{so},
\label{eq:obsv_N_pert_split}
\end{equation}
where the first term is the prediction of general relativity, while the remaining three terms specify the amount of dephasing that accrues in the signal due to the presence of the scalar. Expressions accurate to leading PN~order will suffice for an adequate estimate of the size of each of these terms, and it therefore suffices to write
\begin{subequations}
\label{eq:obsv_N_pert_explicit}
\begin{align}
% GR prediction
	\mathcal N_\text{pert}^\sector{GR}
	=&\:
	\bigg\{
		{-} \frac{1}{32\nu}
		(\GN m \pi f)^{-5/3}
	\bigg\}
	\Big._{f_1}^{f_2},
\allowdisplaybreaks\\[5pt]
% [O]
	\delta\mathcal N_\text{pert}^{}\scount{o}
	=&\:
	\bigg\{
		\frac{1+2\alpha_1\alpha_2}{32\zeta\nu}\frac{25\Delta\alpha^2}{336\zeta}
		(\Geff m \pi f)^{-7/3}
		\nonumber\\&
		-
		\frac{1}{32\nu}
		\bigg[
			\frac{1+2\alpha_1\alpha_2}{\zeta}
			-
			\bigg( \frac{\Geff}{\GN} \bigg)^{\! 5/3\,}
		\bigg]
		(\Geff m \pi f)^{-5/3}
	\bigg\}
	\Big._{f_1}^{f_2},
\label{eq:obsv_dN_pert_O}
\allowdisplaybreaks\\[5pt]
% Conformal [SO]
	\delta\mathcal N^\sector{C}_\text{pert}\scount{so}
	=&\:
	\bigg\{
		{-}\frac{1}{2}
		\mathcal{S}^\sector{C}_{\QD,\dip}
		(\Geff m \pi f)^{-4/3}
		-
		\bigg[
			\mathcal{S}^\sector{C}_{\QD,\nondip}
			-
			\frac{1}{32\nu}
			\bigg(
				\frac{235}{6}
				\frac{\projell{S}}{\GN m^2}
				\nonumber\\&
				+
				\frac{125}{8}
				\frac{\Delta m}{m}
				\frac{\projell{\Sigma}}{\GN m^2}
			\bigg)
			\bigg( \frac{\Geff}{\GN} \bigg)^{\! 2/3 \,}
		\bigg]
		(\Geff m \pi f)^{-2/3}
	\bigg\}
	\Big._{f_1}^{f_2},
\label{eq:obsv_dN_pert_SO_C}
\\[5pt]
% Disformal [SO]
	\delta\mathcal N^\sector{D}_\text{pert}\scount{so}
	=&\:
	\bigg\{
		\mathcal{S}^\sector{D}_{\QD,\dip}
		(\Geff m \pi f)^{2/3}
		+
		\frac{1}{2}
		\mathcal{S}^\sector{D}_{\QD,\nondip}
		(\Geff m \pi f)^{4/3}
	\bigg\}
	\Big._{f_1}^{f_2}.
\label{eq:obsv_dN_pert_SO_D}
\end{align} 
\end{subequations}
For each sector in~$\delta\mathcal N_\text{pert}$, we have included both the leading dipolar part and the leading nondipolar part of the result, as either one can provide the dominant contribution, depending on the relative sizes of~$\Delta\alpha$ and~$\x \equiv (\Geff m\pi f)^{2/3}$. (Recall our discussion in section~\ref{sec:rad_phase}.) Note also that in~\eqref{eq:obsv_dN_pert_O} and~\eqref{eq:obsv_dN_pert_SO_C}, the terms proportional to powers of~$\Geff/\GN$ serve to remove the parts of $(1+2\alpha_1\alpha_2)/\zeta$ and $\mathcal{S}^\scriptsector{C}_{\scriptscriptstyle\QD,\nondip}$ that overlap with general relativity, since these are already included in~$\mathcal N^\scriptsector{GR}_\text{pert}$; even if not explicitly~written~down.

\paragraph{Numerical estimates.}
\begin{table}
\small\centering
\def\arraystretch{1.5}
\begin{tabular}{|l l l l |}
\hline
& $m_\K$ & $\alpha_\K$ & $\chi_\K$ \\[2pt]
\hline
Weakly scalarised, slowly spinning (W)
& $1.25~M_\odot$
& $5\times 10^{-4}$
& $2.0\times 10^{-5}$

\\
Strongly scalarised, slowly spinning (S)
& $1.70~M_\odot$
& $7\times 10^{-3}$
& $2.0\times 10^{-5}$
\\
Strongly scalarised, rapidly spinning (S${}^{\textstyle\ast}$)
& $1.70~M_\odot$
& $7\times 10^{-3}$
& $1.4\times 10^{-2}$~
\\[2pt]
\hline
\end{tabular}
\caption{Fiducial parameters for several types of neutron stars.}
\label{table:neutron_star_parameters}
\end{table}
We now evaluate~\eqref{eq:obsv_N_pert_explicit} using a selection of parameter values that are typical of neutron stars, as shown in table~\ref{table:neutron_star_parameters}. To explore an array of different possibilities, we have chosen to distinguish between neutron stars that are either weakly or strongly scalarised, and between neutron stars that are either slowly or rapidly rotating. (We leave a more exhaustive exploration of parameter space to future work.) Recalling our discussion at the end of section~\ref{sec:pp_stt}, we define a ``weakly scalarised'' neutron star as one whose coupling strength~$\alpha_\K$ is equal to~$\alpha_\text{weak}$ up to an order-one factor, while a ``strongly scalarised'' neutron star is one that has undergone spontaneous scalarisation. Of course, this latter possibility occurs only in a certain class of scalar-tensor theories~\cite{Damour:1993hw, Damour:1996ke, Minamitsuji:2016hkk, Silva:2017uqg, Doneva:2017duq, Andreou:2019ikc, Ventagli:2020rnx, Doneva:2017bvd, Cunha:2019dwb, Minamitsuji:2019iwp, Herdeiro:2020wei, Berti:2020kgk, East:2021bqk}, and only if the mass of the body lies in a suitable range~\cite{Shao:2017gwu, Zhao:2019suc, Guo:2021leu, Yagi:2021loe}. (Theories that allow for spontaneous scalarisation also lead to a related phenomenon known as dynamical scalarisation~\cite{Barausse:2012da, Palenzuela:2013hsa, Shibata:2013pra, Taniguchi:2014fqa, Sennett:2016rwa, Sennett:2017lcx, Khalil:2019wyy, Silva:2020omi}, but we have neglected this added complication here for the sake of simplicity.)

\begin{figure}[t]
\centering\includegraphics[width=\textwidth]{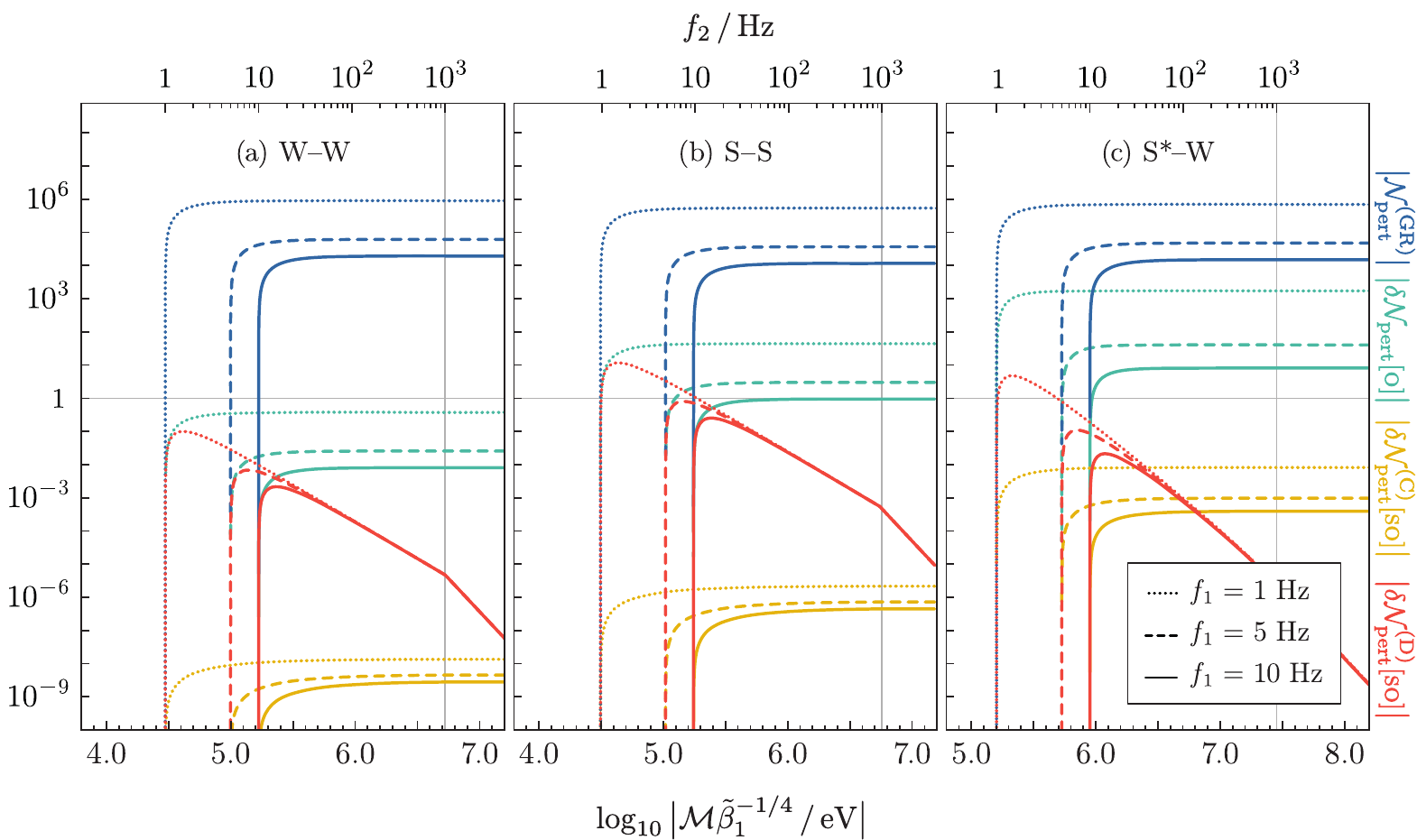}
\caption{Number of gravitational-wave cycles~$\mathcal N_\text{pert}$ that accumulate within the frequency band ${f \in [f_1,f_2]}$ during the inspiral of a double neutron star binary; shown here for different values of the disformal coupling scale~$\M$. The parameter values used for each type of neutron star (labelled W, S, or~S${}^{\textstyle\ast}$) are listed in table~\ref{table:neutron_star_parameters}.
The contributions to~$\mathcal N_\text{pert}$ are split into four parts: $\mathcal N_\text{pert}^\scriptsector{GR}$ is the prediction of general relativity, while $\delta\mathcal N_\text{pert}^{}\scount{o}$, $\delta\mathcal N_\text{pert}^\scriptsector{C}\scount{so}$, and $\delta\mathcal N_\text{pert}^\scriptsector{D}\scount{so}$ are the corrections induced by a spin-independent, conformal spin-orbit, and disformal spin-orbit interaction between the binary and a light scalar field, respectively. In decreasing order of magnitude, the three values used for the lower frequency bound~$f_1$ mark the beginning of the sensitivity bands for Advanced LIGO, Cosmic Explorer, and the Einstein Telescope. Meanwhile, the upper frequency bound~$f_2$, which marks the end of the perturbative inspiral phase, depends on the value of~$\M$. For values of~$\M$ above a certain threshold (shown in each panel as a vertical grey line), the binary transitions directly from its perturbative inspiral phase into the merger phase, in which case we fix ${f_2 = f_\text{contact}}$, with~${ f_\text{contact} \sim 1~\text{kHz} }$. For values of~$\M$ below this threshold, the binary first transitions into a nonperturbative inspiral phase, and we instead set ${f_2 = f_\text{nonpert}}$, with $f_\text{nonpert}$ given by~\eqref{eq:obsv_f_nonpert}. The corresponding numerical values of $f_2$ in this regime are shown along the top axis, and note that ${\mathcal N_\text{pert} = 0}$ when ${ f_2 \leq f_1}$. Also drawn in each panel is a horizontal grey line, which marks the threshold for at least one gravitational-wave cycle to be accumulated. As a rough diagnostic, we deem a particular interaction to be detectable if, for a given value of~$\M$, the point on its corresponding curve lies above this horizontal~line.
\looseness=-1}
\label{fig:dephasing}
\end{figure}

The corresponding values of~$\alpha_\K$ in table~\ref{table:neutron_star_parameters} have been lifted from refs.~\cite{Shao:2017gwu, Zhao:2019suc, Guo:2021leu}, and are consistent with current constraints from binary pulsars and the gravitational-wave event GW170817.%
\footnote{Strictly speaking, the upper bounds on~$|\alpha_\K|$ in refs.~\cite{Shao:2017gwu, Zhao:2019suc, Guo:2021leu} apply only to the specific theory by Damour and Esposito-Far\`{e}se~\cite{Damour:1993hw, Damour:1996ke}, but for the sake of illustration, we shall here assume that these values are also representative of the bounds on other models. We further note that our definition for~$\alpha_\K$ differs from that of refs.~\cite{Damour:1993hw, Damour:1996ke} by a factor of~$\sqrt{2}$; namely, $\alpha_\K^\text{(theirs)} = -\sqrt{2}\alpha_\K^\text{(ours)}$. The minus sign is irrelevant, however, as the effective action in~\eqref{eq:review_S_eff} is invariant under the transformation~${ (\phi,\alpha_\K) \to (-\phi,-\alpha_\K) }$.
\looseness=-1}
The values of~$\chi_\K$, meanwhile, were obtained by making use of the double pulsar PSR~J0737-3039A/B~\cite{Kramer:2006nb}, which is presently the only binary neutron star system for which the spins of both components are known. Following ref.~\cite{Shao:2017gwu}, we use a combination of their measured spins and spin-down rates to infer the would-be values of $\chi_\K$ when the binary enters the LIGO/CE/ET band, about 86~Myr from now. The three types of neutron stars in table~\ref{table:neutron_star_parameters} are then used to construct three different types of binary systems, and for each one, we plot the magnitudes of $\mathcal N_\text{pert}^\scriptsector{GR}$, $\delta\mathcal N_\text{pert}^{}\scount{o}$, $\delta\mathcal N_\text{pert}^\scriptsector{C}\scount{so}$, and $\delta\mathcal N_\text{pert}^\scriptsector{D}\scount{so}$ as~functions~of~$\M$~in~figure~\ref{fig:dephasing}.

The different curves have qualitatively similar shapes across all three panels that are easy enough to explain. Beginning with $\delta\mathcal N_\text{pert}^\scriptsector{D}\scount{so}$, we see from~\eqref{eq:obsv_dN_pert_SO_D} that this quantity is proportional to positive powers of~$f$, and thus has its value determined predominantly by the higher frequency~$f_2$. A vertical grey line in each panel marks the critical value of~$\M$ above which ${f_\text{nonpert} > f_\text{contact}}$, and in so doing, naturally divides the parameter space into two halves. On the right, we have that ${f_2 = f_\text{contact}}$, and thus ${\delta\mathcal N_\text{pert}^\scriptsector{D}\scount{so} \propto 1/\M^4}$, since its dependence on~$\M$ in this region stems solely from the disformal spin-orbit coefficients in table~\ref{table:coefficients}. The slope of this curve then changes abruptly when we cross over to the left of the vertical line, where ${f_2 = f_\text{nonpert}}$, on account of $f_\text{nonpert}$ itself being a function of~$\M$ [see~\eqref{eq:obsv_f_nonpert}]. Depending on whether its dipolar or nondipolar part dominates, $\delta\mathcal N_\text{pert}^\scriptsector{D}\scount{so}$ is now proportional to either $1/\M^{28/9}$ or~$1/\M^{20/9}$ (or more generally, a linear combination of the two). Its magnitude continues to rise as we move to even lower values of~$\M$, until $f_2 $ approaches~$f_1$, at which point $\delta\mathcal N_\text{pert}^\scriptsector{D}\scount{so}$ must necessarily vanish. As for the other three terms in~$\mathcal N_\text{pert}$, we see from~\eqref{eq:obsv_N_pert_explicit} that they are all proportional to negative powers of~$f$, and thus have their values determined predominantly by the lower frequency~$f_1$. They are mostly independent of~$\M$ as a result, except for when the diminishing difference between $f_2$~and~$f_1$~similarly~causes~them~to~vanish.

Having understood these qualitative features, we can now comment on the actual magnitudes of these quantities. To start with, consider the binary in figure~\ref{fig:dephasing}(a), which consists of two identical neutron stars that are both weakly scalarised and slowly rotating. For all three values of~$f_1$, we see that not one of 
$\delta\mathcal N_\text{pert}^{}\scount{o}$, $\delta\mathcal N_\text{pert}^\scriptsector{C}\scount{so}$, or $\delta\mathcal N_\text{pert}^\scriptsector{D}\scount{so}$ results in a dephasing that is large enough to be detectable; indicating that both current and future ground-based detectors will be unable to distinguish this kind of system from its counterpart in general relativity. Naturally, this begs the question as to what is actually needed for there to be a discernible effect from the scalar field. Being interested particularly in the disformal spin-orbit sector, we shall here consider the requirements for $\delta\mathcal N_\text{pert}^\scriptsector{D}\scount{so}$ to surpass the threshold value~of~1.

A~brief inspection of table~\ref{table:coefficients} tells us that we can boost the magnitude of the spin-orbit coefficients $\mathcal{S}^\scriptsector{D}_{\scriptscriptstyle\QD,\dip}$ and $\mathcal{S}^\scriptsector{D}_{\scriptscriptstyle\QD,\nondip}$ by increasing either the magnitudes of~$\chi_\K$ or~$\alpha_\K$. We find when repeating the calculation with larger spins up to the Kerr bound (i.e., for all values of ${|\chi_\K| < 1}$), however, that $\delta\mathcal N_\text{pert}^\scriptsector{D}\scount{so}$ consistently remains smaller than~1. This might seem surprising at first glance, given that in figure~\ref{fig:dephasing}(a), $\delta\mathcal N_\text{pert}^\scriptsector{D}\scount{so}$ already achieves a maximum value of ${\sim 0.1}$ despite the minuscule value of ${\chi_1 = \chi_2 = 2.0 \times 10^{-5}}$. The reason is simply that $f_\text{nonpert}$ also varies with~$\chi_\K$ [see~\eqref{eq:obsv_f_nonpert}]. Accordingly, while a larger spin does increase the overall size of the effect, there is a concomitant reduction in the value of~$f_\text{nonpert}$, which shortens the amount of time that the binary spends in the perturbative part of its inspiral, and so lowers the number of gravitational-wave cycles that can accumulate~during~this~phase. (Of~course, our perturbative results do not preclude the possibility of a large dephasing during the nonperturbative part of the inspiral, although we shall refrain from making any speculations about this regime until a concrete calculation has been achieved.)

The other option to boost $\delta\mathcal N_\text{pert}^\scriptsector{D}\scount{so}$ is to increase the value of~$\alpha_\K$, which is possible if at least one of the neutron stars has undergone spontaneous scalarisation. Shown in figure~\ref{fig:dephasing}(b) are the number of gravitational-wave cycles that would accumulate in the detector if the binary were composed of two identical, strongly scalarised and slowly rotating neutron stars. Meanwhile, figure~\ref{fig:dephasing}(c) shows the corresponding values for an asymmetric binary involving one strongly scalarised, rapidly rotating neutron star and one weakly scalarised, slowly rotating neutron star.%
\footnote{For the asymmetric binary, we take ${\K=1}$ to correspond to the more massive, strongly scalarised star. Moreover, to properly evaluate each part of $\mathcal N_\text{pert}$ in this case, we must also specify appropriate values for the Wilson coefficients $\tilde\beta_2$ and $\alphaprime_\K$. This was not necessary for binaries with identical constituents, since ${\tilde\beta_2 = \tilde\beta_1}$ in that case, while $\alphaprime_\K$ (which enters via the definition of~$\comboA{\text{NLO}}{1}$) is always multiplied by something that vanishes. For the asymmetric binary, we find that the curves in figure~\ref{fig:dephasing}(c) are mostly insensitive to the precise value of $\tilde\beta_2$ and $\alphaprime_\K$ as long as ${\tilde\beta_2/\tilde\beta_1 \lesssim \mathcal O(10)}$, ${\alphaprime\!{}^{}_1 \lesssim \mathcal O(100)}$,~and~${\alphaprime\!{}^{}_2 \lesssim \mathcal O(10)}$.\looseness=-1}
For this latter case, our choice to furnish the more massive object with the larger spin is informed by the standard picture of binary neutron star formation~\cite{Lorimer:2008se, Faber:2012rw}. In~short, the more massive constituent of what is initially a main-sequence binary is the first to undergo core collapse and leave behind a neutron star. This neutron star then gains mass and angular momentum when its companion evolves off the main sequence and overflows its Roche lobe. Eventually, the companion also collapses to form a neutron star, but the preceding period of mass transfer guarantees that this secondary neutron star is always lighter and more slowly rotating than the primary.

As promised, the larger value of~$\alpha_\K$ does indeed lead to a larger value of~${\delta\mathcal N_\text{pert}^\scriptsector{D}\scount{so}}$. While the amount of dephasing is still too small to be detectable by current gravitational-wave detectors, figure~\ref{fig:dephasing} suggests that:
\begin{quotation}
\noindent
A future detector like the Einstein Telescope will be sensitive enough to probe disformal spin-orbit effects in a narrow range of values for~${\M \tilde\beta_1\!{}^{-1/4}}$ if at least one of the binary's constituents is strongly scalarised.	
\end{quotation}
A comparison between figures~\ref{fig:dephasing}(b) and \ref{fig:dephasing}(c) further reveals that while the magnitudes of the spins have little impact on the maximum value of~${\delta\mathcal N_\text{pert}^\scriptsector{D}\scount{so}}$ (for reasons already discussed), their impact on the value of $f_\text{nonpert}$ leads to a discernible shift in the position of this maximum. Case in point, the small and large values of $\chi_\K$ in table~\ref{table:neutron_star_parameters} result in a sensitivity window to~${\M \tilde\beta_1\!{}^{-1/4}}$ that is peaked around 40~keV~and~200~keV,~respectively.

As a final remark on figure~\ref{fig:dephasing}, it is worth noting that while $\delta\mathcal N_\text{pert}^\scriptsector{D}\scount{so}$ can become greater than~1 for suitably small values of~$\M$, $\delta\mathcal N_\text{pert}^\scriptsector{C}\scount{so}$ remains significantly less than~1 in all three panels. When combined with the fact that the values of $\chi_\K$ in table~\ref{table:neutron_star_parameters} are meant to be representative of typical neutron star spins, while the values of $\alpha_\K$ are the largest possible given existing constraints, these results indicate that both current and upcoming ground-based detectors will be insensitive to the effects of a scalar-induced, conformal spin-orbit interaction in (the perturbative part of) a binary neutron star inspiral. This conclusion is in complete agreement with the expectations we laid out in section~\ref{sec:intro}. Note, however, that the magnitudes of $\chi_\K$ and $\alpha_\K$ could be much larger in binary systems of scalarised black holes; hence, our perturbative results for the conformal spin-orbit sector may still play an important role in the modelling of such systems. We will comment briefly on this scenario~in~section~\ref{sec:discussion_future}.

\paragraph{Caveats.}
For now, let us close this section by highlighting several of its limitations. First,~it almost goes without saying that the small selection of values in table~\ref{table:neutron_star_parameters} leaves most of parameter space unexplored. Because different models lead to different predictions for how the Wilson coefficients depend on the properties of the neutron star and the underlying parameters of the scalar-tensor theory, it is difficult to perform a meaningful exploration of parameter space without also being explicit about the specific details of the microscopic physics. In lieu of this, we have elected to simply use the small number of illustrative values in table~\ref{table:neutron_star_parameters} to make some generic statements about the detectability of disformal spin-orbit effects. Our expectation is that these conclusions are fairly representative of a broad class of models, but it is certainly possible that some models could give rise to larger effects that can already be constrained by current-generation detectors, and conversely, there will surely be others whose imprints are too small to be observable by any planned detector. It would therefore be valuable in the future to map our general results for the gravitational-wave phase onto specific models, so as to establish more~definitive~constraints.

Second, recall that in addition to the masses, spins, and Wilson coefficients of the two bodies, our formula for $\mathcal N_\text{pert}$ also takes as input a value for the small parameter~$\epsilon$ [see~\eqref{eq:obsv_f_nonpert}], which sets the criterion for when we think our perturbative ladder expansion breaks down. The results in figure~\ref{fig:dephasing} assume a value of~${\epsilon = 0.1}$, and while we believe this to be a fairly conservative choice, it should be noted that our conclusions turn out to be quite sensitive to this particular parameter. A value of ${\epsilon = 0.5}$, for instance, would mean that even current-generation detectors can be sensitive to disformal spin-orbit effects in strongly scalarised binaries of the kind in figure~\ref{fig:dephasing}(b), while an even larger value of~${\epsilon = 1}$ would allow for a future detector like the Einstein Telescope to see the imprints of a disformal spin-orbit interaction in weakly scalarised binaries of the kind in figure~\ref{fig:dephasing}(a). Going in the opposite direction, we find that for ${\epsilon \lesssim 0.01}$, all three of the ground-based detectors considered in this section would be blind to disformal effects during the perturbative part of the binary's inspiral. It would therefore be instructive to extend our calculations to higher orders in the disformal coupling in the future, so that we can refine our estimate for when our perturbative expansion breaks down; or, better yet, one could consider exploring how to resum the relevant set of ladder diagrams, so as to obtain a reliable prediction for both the perturbative and nonperturbative parts of the binary's inspiral. 
\looseness=-1

Finally, it is worth reiterating that our rudimentary criterion for detectability in this section does not account for several important factors, like the inevitable degeneracies between parameters, expected detection rates, signal-to-noise ratios, and so on. Indeed, we know that in the spin-independent sector, the green lines in figures~\ref{fig:dephasing}(a) and~\ref{fig:dephasing}(b) overestimate how likely we are to detect the presence of a scalar in systems with~${\Delta\alpha = 0}$, as the absence of scalar dipole radiation means that~$\delta\mathcal N_\text{pert}\scount{o}$ is degenerate with a rescaling of the total mass at leading PN order. One therefore has to rely on the corrections from higher-PN-order terms to break this degeneracy, but these higher-order terms are naturally harder to constrain, given that they are suppressed by one or more powers of~$v^2$. For~$\delta\mathcal N_\text{pert}^\scriptsector{D}\scount{so}$, we can already see from table~\ref{table:coefficients} that a measurement of~$\M$ is degenerate with a rescaling of~$\chi_\K$, since these quantities only appear in the combination~$\chi_\K/\M^4$. This should not prevent us from detecting the \emph{presence} of a disformal spin-orbit interaction, should one exist with a favourable value of~$\M$, but it is unlikely that the data can be used to pinpoint the exact value of~$\M$, unless it can separately pinpoint the values of~$\chi_\K$. (Note that thus far, all of the neutron stars that have been detected through gravitational waves have spins that are either unconstrained or only weakly bounded from above~\cite{LIGOScientific:2018hze, LIGOScientific:2020aai, LIGOScientific:2021qlt}.) On the whole, these limitations motivate undertaking a more robust, Bayesian analysis in the future.

% ============================================== %
% Section 7
% ============================================== %
\section{Discussion}
\label{sec:discussion}

\subsection{Summary of main results}

In this work, we used a generalisation of the NRGR approach~\cite{Goldberger:2004jt, Goldberger:2007hy, Goldberger:2009qd, Galley:2009px, Porto:2005ac, Porto:2006bt, Porto:2008tb, Porto:2008jj, Levi:2010zu, Levi:2015msa, Foffa:2013qca, Rothstein:2014sra, Porto:2016pyg, Levi:2018nxp} to compute the leading spin-orbit effects that would arise when a compact binary is coupled both conformally and disformally to a light scalar field. After reviewing the main conceptual ideas in section~\ref{sec:review}, we showed in section~\ref{sec:pp} how one could construct a general point-particle action for spinning bodies in this class of theories by utilising a two-step approach. Our first step was to perform a frame transformation on the corresponding action in general relativity as a means of systematically generating new terms that couple the effective point particle directly to the scalar field. The result of this procedure suffices as a model for how \emph{weakly} gravitating bodies behave (when viewed from large distances) in the Einstein frame, but for strongly gravitating objects like black holes and neutron stars, a second step where we generalised the action to allow for arbitrary values of the Wilson coefficients was required. This extra step accounts for the fact that when a body has a considerable amount of self-gravity, the strength with which it couples to the scalar is largely influenced by the microscopic details of its interior. The precise relation between a body's Wilson coefficients and its internal structure can be determined on a case-by-case basis by performing a number of matching calculations (see~section~\ref{sec:pp_stt} for examples), but for most of this paper, we have left these coefficients as free parameters, thereby rendering our results general enough to describe \emph{any} spinning body [up~to subleading effects that we neglect; see the discussion below~\eqref{eq:pp_formal_power_series}]. The~final result of this section is given in the form of a Routhian in~\eqref{eq:pp_Routhian}.

In section~\ref{sec:consv}, two distinct copies of this Routhian were combined with our fiducial field action $S_\text{fields}$ to produce the total effective action in~\eqref{eq:consv_S_eff}. With the help of Feynman diagrams, we then integrated out the potential modes of the scalar and gravitational field from this action to arrive at a perturbative result for the two-body potential~$V$, which we divided into four ``sectors'' according to~\eqref{eq:consv_decomposition}. The result itself is given in~\eqref{eq:consv_V}. A~mixture of Euler--Lagrange and Hamilton equations were then used to derive the conservative equations of motion for this system, which are given by~\eqref{eq:consv_eom_a} and~\eqref{eq:consv_eom_spin_cmv} when expressed in the centre-of-mass frame. Finally, we concluded our discussion of the conservative sector by specialising to circular nonprecessing binaries in section~\ref{sec:consv_circles}. The orbital binding energy for systems in this particular configuration can be found in~\eqref{eq:consv_circles_energy}.

Turning to the radiative sector, in section~\ref{sec:rad_scalar} we used a similar diagrammatic technique to compute the multipole moments $\mathcal Q^L$ [see~\eqref{eq:rad_Q_off_shell}] that are responsible for sourcing outgoing scalar waves. The amount of power~$P_\phi$ that is carried away by these waves is given in~\eqref{eq:rad_power_phi}, assuming again that the binary is in a circular nonprecessing orbit for simplicity. In~section~\ref{sec:rad_gw}, we then used power-counting arguments to show that the gravitational multipole moments $\mathcal I^L$~and~$\mathcal J^L$ are the same as they would be in general relativity (at the order to which we are working), although the presence of a scalar does contribute indirectly to the power~$P_g$ radiated into gravitational waves via the solution to the conservative equations of motion, which is needed to put the result ``on shell.'' For circular nonprecessing binaries, the result for~$P_g$ is given in~\eqref{eq:rad_power_gw}. Finally, in section~\ref{sec:rad_phase}, we combined these results for the total power with our earlier expression for the orbital binding energy to determine the evolution of the binary's orbital phase~$\psi$, and by extension, its gravitational-wave phase~$\Psi$ (up to nonlinear radiative corrections like tail terms, which we neglect).

Care was taken to distinguish between two possible scenarios: a dipole-driven inspiral, wherein the binary loses energy predominantly to scalar dipole radiation, and a quadrupole-driven inspiral, during which energy is radiated mostly into the quadrupolar modes of the scalar and gravitational field. The results for these two cases are given in \eqref{eq:rad_DD_phase} and~\eqref{eq:rad_QD_phase}, respectively, when expressed in the time domain, while the corresponding frequency-domain results can be found in \eqref{eq:app_gw_DD_phase} and~\eqref{eq:app_gw_QD_phase}. We should clarify that our results for the latter, quadrupole-driven regime will be the most relevant for establishing constraints if our default assumption is the validity of general relativity, but because current observations do not yet preclude the possibility of a binary that undergoes a (temporary) dipole-driven phase during the very early stages of its inspiral, we have made allowances for this more exotic~scenario~as~well.

At this stage, it is useful to mention a number of consistency checks that have been performed to validate the results of this paper. First, note that none of our results in the conformal spin-independent sector are new, as they have been lifted directly from refs.~\cite{Damour:1992we, Huang:2018pbu, Kuntz:2019zef}. Additionally, note that the leading effects (both conservative and radiative) from a disformal spin-independent coupling have previously been calculated in refs.~\cite{Brax:2018bow, Brax:2019tcy}, albeit only for the special case in which ${\alpha_1 = \alpha_2}$ and~${\beta_1 = \beta_2}$. Our results for this sector correctly reduce to those of refs.~\cite{Brax:2018bow, Brax:2019tcy} when we take this limit, but we think it is important to stress that our results are more general, as they also allow for body-dependent Wilson coefficients, which are necessary when modelling systems of strongly~gravitating~objects.

That said, the key novelty of this work is the calculation of the additional spin-orbit effects that are induced by a conformal and disformal coupling to the scalar field. Some preliminary results in this direction can be found in a previous paper of ours; namely ref.~\cite{Brax:2020vgg}, although we restricted ourselves to a conservative setting in that study and, moreover, focused only on weakly gravitating bodies. We have checked that the more general results of this present work correctly reduce to those of ref.~\cite{Brax:2020vgg} after setting ${\alpha_1 = \alpha_2}$ and ${\tilde\beta_1 = \tilde\beta_2}$,%
\footnote{To translate our results into the notation of ref.~\cite{Brax:2020vgg}, set ${\alpha_\K = -\bar c/2}$ and ${\beta_\K = \tilde\beta_\K = \bar d/2}$.}
but note that this only validates our results in the conservative sector. To the best of our knowledge, our results for the scalar-induced, spin-orbit parts of the radiated power and gravitational-wave phase have been calculated here for the very first time. (Of course, the conformal spin-orbit sector has a part that overlaps with general relativity, and we have verified that this subset of terms is consistent with the results in ref.~\cite{Blanchet:2013haa}.)

To illustrate that these effects are not merely theoretical curiosities, but are also relevant for phenomenological applications, we concluded this paper by assessing their observational prospects in current and future ground-based, gravitational-wave detectors. The results of section~\ref{sec:obsv} suggest that a future detector like the Einstein Telescope will be sensitive enough to probe disformal spin-orbit effects in double neutron star binaries for values of~$\M \tilde\beta_1\!{}^{-1/4}$ in the $\mathcal O(10)$ to $\mathcal O(100)$~keV range if at least one of the two bodies is sufficiently scalarised. This is encouraging, although as we highlighted towards the end of section~\ref{sec:obsv}, our rudimentary analysis leaves several important factors (like the various degeneracies between parameters) unaccounted for, and thus this preliminary result should be taken mostly as motivation for performing a more robust, Bayesian~analysis~in~the~future.

\subsection{Future directions}
\label{sec:discussion_future}

Additionally, some other key directions for future work are as follows.

\paragraph{Amplitude corrections.}
Because a matched-filtering analysis is much more sensitive to variations in the phase than in the amplitude of the gravitational wave, we have concentrated only on the former in this work. This should suffice for placing order-of-magnitude constraints on new physics~\cite{Tahura:2019dgr}, but if (or when) a deviation from general relativity is found, both amplitude and phase corrections should be included in the waveform to minimise systematic~errors.

\paragraph{Going beyond the ladder expansion.}
The inspiral phase of a binary system splits into a perturbative and nonperturbative part when the disformal coupling scale~$\M$ is small enough to be  of phenomenological interest [see the discussion around~\eqref{eq:obsv_N}]. Because the results of this paper were truncated to first order in the disformal coupling (i.e., to first order in~$1/\M^4$) for simplicity, its predictions are valid only in the former perturbative regime. Constructing a waveform that remains valid over the entire course of the inspiral would be highly desirable, but this rests on our ability to gain a good handle on the nonperturbative regime, which entails being able to resum a certain class of ladder diagrams. Fortunately, this has previously been shown to be possible for systems of nonspinning bodies~\cite{Davis:2019ltc, Davis:2021oce}, and it is likely that similar techniques could also be used to address the spinning case.

\paragraph{Other observational probes.}
As a precursor to more extensive studies in the future, our discussion in section~\ref{sec:obsv} focused solely on the constraining power of ground-based detectors. It~would be interesting to explore how much we could learn from future space-borne missions, like the Laser Interferometer Space Antenna (LISA)~\cite{Amaro-Seoane:2017lisa} or the Decihertz Interferometer Gravitational Wave Observatory (DECIGO)~\cite{Kawamura:2011zz}; and also, to examine what constraints can be placed by using observational data of binary pulsars~\cite{Damour:1992ppk, Stairs:2003eg, Shao:2014wja}. We expect that each of these experiments will have their uses, since they are sensitive to binaries within a different range of orbital frequencies. Accordingly, they will be sensitive to different values of the coupling scale~$\M$, and thus will cover complementary regions of parameter space. Note, however, that the results of this paper should be generalised to include a nonzero eccentricity before applying them to the above cases, since binary systems with lower orbital frequencies have yet~to~fully~circularise.

\paragraph{Orbital precession.}
Another generalisation we could make in the future would be to relax our assumption of a nonprecessing orbit. In particular, so-called ``up-down binaries'' are known to be unstable to small perturbations in the orientations of their spins~\cite{Gerosa:2015hba, Lousto:2016nlp, Mould:2020cgc, Varma:2020bon}, and it would be interesting to see what effect a disformal spin-orbit interaction might have on~this~instability.

\paragraph{Spinning black holes.}
While the tail end of this paper focused primarily on the prospect of observing disformal spin-orbit effects in double neutron stars, it is worth reiterating that the general results in sections~\ref{sec:pp} to~\ref{sec:rad} can also be used to describe systems of ``hairy'' black holes. Such objects would not be affected by a disformal coupling, but our results in the conformal spin-orbit sector still apply. With a minimal amount of effort (mostly needed to square differences in notation), these can be combined with the spin-independent results of refs.~\cite{Yagi:2011xp, Julie:2019sab, Shiralilou:2020gah, Shiralilou:2021mfl} to produce 1.5PN-accurate waveforms for spinning black hole binaries in Einstein-scalar-Gauss-Bonnet~theories.%
\footnote{Our field action in~\eqref{eq:consv_S_fields_raw} does not include the all-important term that couples the scalar to the Gauss-Bonnet invariant, but from power counting we know that we only need its contribution in the spin-independent sector (which is already given in refs.~\cite{Julie:2019sab, Shiralilou:2020gah, Shiralilou:2021mfl}) to obtain accurate spin-orbit results at~leading~order.}

\paragraph{Constraining gravity on multiple scales.}
By treating the masses, spins, and Wilson coefficients $\{ \alpha_\K, \tilde\beta_\K, \dots \}$ of the two bodies as free parameters, our results are general enough to describe how spinning compact binaries evolve in a broad class of scalar-tensor theories. There is, however, only so much information that can be gleaned from placing constraints on Wilson coefficients; hence, as we discussed at the end of section~\ref{sec:obsv}, it would also be interesting to apply our results to specific models. Performing the requisite matching calculations that link the two bodies' Wilson coefficients to the underlying parameters of the scalar-tensor theory would then allow us to establish more definitive constraints on those model parameters directly. This would also allow us to explore how the constraints on conformal and disformal couplings that derive from compact binaries compare with those that have already been established in the Solar System~\cite{Bertotti:2003rm, Hofmann:2018llr, Adelberger:2009zz, Burrage:2017qrf, Brax:2020vgg} and on much larger cosmological scales~\cite{Koyama:2015vza, Joyce:2016vqv, Ferreira:2019xrr, Creminelli:2017sry, Sakstein:2017xjx, Ezquiaga:2017ekz, Baker:2017hug, Langlois:2017dyl, Heisenberg:2017qka, Akrami:2018yjz, BeltranJimenez:2018ymu, Copeland:2018yuh}. 

\paragraph{Nonlinear interactions.}
It is important to appreciate, however, that viable cosmological models typically rely on nonlinear interaction terms in the field action~$S_\text{fields}$ to screen their effects on smaller scales (see refs.~\cite{Burrage:2017qrf, Sakstein:2018fwz, Baker:2019gxo} for modern reviews), but such terms were omitted in this work for simplicity. The impact of scalar self-interactions on the evolution of a binary is not fully understood, and remains an active area of research~\cite{deRham:2012fw, deRham:2012fg, Dar:2018dra, Kuntz:2019plo, Renevey:2021tcz, Bezares:2021dma}.

% ============================================== %
% Acknowledgments
% ============================================== %
\acknowledgments
This project has received funding/support from the European Union's Horizon 2020 research and innovation programme under the Marie Sk\l{}odowska-Curie grant agreement No.~860881-HIDDeN.
This work was partially supported by STFC consolidated grants ST/P000681/1 and ST/T000694/1.
S.M.~is supported by a UKRI Stephen Hawking Fellowship (EP/T017481/1).
The \emph{xAct}~package~\cite{xAct} for Mathematica was used to aid some of our calculations.

% ============================================== %
% Appendix A
% ============================================== %
\appendix
\section{Preserving the spin supplementary condition}
\label{app:ssc}

One of the key steps in our construction of the point-particle Routhian was the elimination of the Lagrange multipliers $\chi_a$ and~$\xi^a$. In section~\ref{sec:pp_Routhian}, we accomplished this by substituting in hand-picked solutions for these vectors, which we claimed would automatically preserve the covariant SSC and its conjugate constraint under time evolution. The goal of this appendix is to substantiate that claim by deriving the aforementioned solutions from first principles.

\paragraph{Equations of motion.}
Our first course of action is to write down the equations of motion that follow from the first-order Lagrangian in~\eqref{eq:pp_L_final}. To begin with, note that the variation of this Lagrangian with respect to $e$, $\chi_a$, and~$\xi^a$ leads to the set of constraints
\begin{align}
	C_0
	&\coloneq
	p^2 + 2\B (p\cdot\!\nabla\phi)^2 + \A^2 m^2 \approx 0,
\label{eq:app_ssc_cst_0}\\ 
	C_1^a
	&\coloneq
	\sqrt{-p^2}\Lambda^a{}_0 - p^a \approx 0,
\label{eq:app_ssc_cst_1}\\
	C_2^a
	&\coloneq
	S^{ab} p_b \approx 0.
\label{eq:app_ssc_cst_2}
\end{align}
The hypersurface in phase space along which all three of these conditions are satisfied is known as the \emph{constraint surface}, and we say that two quantities are ``weakly equal'' if they differ only by terms that vanish along this surface~\cite{Dirac:1964lqm}; i.e., ${F_1 \approx F_2}$ if and only if ${ F_2 - F_1 = \sum_n^\vph{n} b_n \cdot C_n }$, where $b_n$ are arbitrary functions of the phase-space variables.

Varying the Lagrangian with respect to the momentum~$p_\mu$ then gives us an equation that relates it to the tangent vector~$\dot x^\mu$, while varying with respect to $S_{ab}$ establishes an analogous relation between the spin tensor and the angular velocity~$\Omega_\Lambda^{ab}$. Explicitly, we~find
\looseness=-1
\begin{align}
	p_\mu - me^{-1}\dot x _\mu
	\approx &
	-
	2\B[(p-me^{-1}\dot x)\cdot\!\nabla\phi]\nabla_\mu\phi
	+
	m S_{\mu\nu}\xi^\nu
	+
	m P_{\mu\nu} \chi^\nu
	\nonumber\\&
	+
	\frac{ me^{-1}\dot x^\rho p^\sigma }{p^2}
	\big(
		\acute\B S_{\mu\nu} \nabla^\nu\phi \nabla_\rho\nabla_\sigma\phi
		+
		\grave\B S_{\mu\nu} \nabla^\nu\nabla_\rho\phi\nabla_\sigma\phi
	\big),
\label{eq:app_ssc_eom_xdot}
\allowdisplaybreaks\\[5pt]	
	\Omega_\Lambda^{ab} + \dot x^\mu\w_\mu^{ab}
	\approx &\:
	\frac{e\A^2}{m} \pdiff{m^2}{S^2} S^{ab}
	+
	2e \xi^{[a}p^{b]}
	+
	2 \dot x^{[a}\nabla^{b]}\log\widetilde\A
	+
	2\widetilde\B \dot x^\mu\nabla^{[a}\phi\nabla^{b]}\nabla_\mu\phi
	\nonumber\\
	&
	-
	\frac{ 2\dot x^\mu p^\nu }{p^2} p^{[a}
	\big(
		\acute\B \nabla^{b]}\phi \nabla_\mu\nabla_\nu\phi
		+
		\grave\B \nabla^{b]}\nabla_\mu\phi\nabla_\nu\phi
	\big).
\label{eq:app_ssc_eom_omega}
\end{align}
In writing~\eqref{eq:app_ssc_eom_xdot}, we have introduced ${ P^\mu{}_\nu \coloneq \delta^\mu_\nu - p^\mu p_\nu/p^2 }$ as the projection matrix onto the hypersurface orthogonal to~$p_\mu$. Note also that the first term on the rhs of~\eqref{eq:app_ssc_eom_omega} accounts for the fact that the total ADM mass~$m$ of a spinning body generally includes a contribution from the magnitude of its spin~\cite{Hanson:1974qy, Lorentsen:1997wt, Steinhoff:2015ksa}. Note, however, that because the quantity $\partial m^2/\partial S^2$ does not appear in any of the other equations of motion, and because we can describe the intrinsic rotation of a spinning body exclusively in terms of its spin tensor~$S_{ab}$ without making any reference to the angular velocity (on account of the angular coordinates in~$\Lambda^a{}_A$ being cyclic), knowledge of exactly how~$m$ depends on~$S^2$ will not be necessary here.%
\footnote{To clarify, this will be true when building gravitational-wave templates, as we do in this paper, but will not be true if we wanted to confront our predictions with, say, the timing data of pulsars, since in this case angular velocities are what we can actually measure.}

Moving on, the equation of motion for $S_{ab}$ follows from varying~\eqref{eq:pp_L_final} with respect to the Lorentz matrix~$\Lambda^a{}_A$. We have that ${ \Lambda^a{}_A \mapsto (\Lambda + \delta\Lambda)^a{}_A }$ under an infinitesimal variation, and since ${\Lambda + \delta\Lambda}$ must also be a Lorentz matrix, the variation $\delta\Lambda^a{}_A$ is constrained to be of the form~${\delta\Lambda^a{}_A = -\theta^a{}_b\Lambda^b{}_A}$, where $\theta_{ab}$ is an antisymmetric matrix. It~then~follows~that~\cite{Hanson:1974qy}
\begin{equation}
	\delta\Omega_\Lambda^{ab}
	=
	\dot\theta^{ab}
	+
	\Omega_\Lambda^{ac}\theta_c{}^{b}
	-
	\theta^a{}_c \Omega_\Lambda^{cb},
\end{equation}
and thus the equation of motion for the spin tensor in the locally flat frame~is
\begin{equation}
	\dot S^{ab}
	=
	2e \chi^{[a}p^{b]}
	-
	2\Omega_\Lambda^{c[a} S^{\hspace{0.5pt} b]}_\vph{\Lambda}{}_c.
\label{eq:app_ssc_eom_S_loc}
\end{equation}
The transformation rule in~\eqref{eq:pp_SOmega} can then be used to show that the corresponding result in a general coordinate frame reads
\begin{align}
	\covdiff{}{\lambda}S_{\mu\nu}
	=&\;
	2e \chi_{[\mu}p_{\nu]}
	-
	2\Omega^\alpha{}_{[\mu} S_{\nu]\alpha}
\label{eq:app_ssc_eom_S_xframe}\\
	\approx &
	\bigg[
		2 e \chi_\mu p_\nu
		+
		2S_\mu{}^\alpha
		\bigg(
			\frac{e\A^2}{m} \pdiff{m^2}{S^2} S_{\alpha\nu}
			+
			2e \xi_{[\alpha} p_{\nu]}
			+
			2 \dot x_{[\alpha}\nabla_{\nu]}\log\widetilde\A
			+
			2\tilde\B \dot x^\sigma
			\nabla_{[\alpha}\phi\nabla_{\nu]}\nabla_\sigma\phi
			\nonumber\\&\;
			-
			\frac{ 2\dot x^\rho p^\sigma }{p^2} p^{[\alpha}
			\big(
				\acute\B \nabla^{\nu]}\phi \nabla_\rho\nabla_\sigma\phi
				+
				\grave\B \nabla^{\nu]}\nabla_\rho\phi\nabla_\sigma\phi
			\big)
			\bigg)
	\bigg]_\text{asym.},
\label{eq:app_ssc_eom_S_asym}
\end{align}
where the second equality follows after using~\eqref{eq:app_ssc_eom_omega}. The subscript ``asym.''~instructs us to antisymmetrise over any free indices, and since ${ [S_\mu{}^\alpha S_{\alpha\nu}]_\text{asym.} = 0 }$, we see that the $\partial m^2/\partial S^2$ term drops out. It so happens that \eqref{eq:app_ssc_eom_xdot} can now be used to eliminate both $\chi_a$ and $\xi^a$ simultaneously from~\eqref{eq:app_ssc_eom_S_asym}, and thus our final result for the spin equation~of~motion~is\looseness=-1
\begin{align}
	\covdiff{}{\lambda}S_{\mu\nu}
	\approx\;&
	2 p_{[\mu} \dot x_{\nu]}
	-
	4em^{-1}\B
	[(p-me^{-1}\dot x)\cdot\!\nabla\phi]
	p_{[\mu}\nabla_{\nu]}\phi
	\nonumber\\
	&\!
	-
	4S^\rho{}_{[\mu}\delta^\sigma{}_{\nu]}
	\big(
		\dot x_{[\rho}\nabla_{\sigma]}\log\widetilde\A
		+
		\widetilde\B \dot x^\alpha
		\nabla_{[\rho}\phi\nabla_{\sigma]}\nabla_\alpha\phi
	\big).
\label{eq:app_ssc_eom_S}
\end{align}

We stated in the main text that this equation invariably conserves the magnitude of the spin tensor, and we are now in a position to verify this explicitly. Either~\eqref{eq:app_ssc_eom_S_loc} or~\eqref{eq:app_ssc_eom_S} can be used for this purpose, but working with the former turns out to be more instructive. Contracting it with~$S_{ab}$ gives us ${ S_{ab}\dot S^{ab} \equiv \dx S^2/\dx\lambda}$, and this can be seen to vanish since ${ \chi^{[a}p^{b]} S_{ab} \approx 0 }$ by virtue of the SSC in~\eqref{eq:app_ssc_cst_2}, while
${
	\Omega_\Lambda^{c[a} S^{b]}_\vph{\Lambda}{}_{c \vph{b}} S_{ab}
	=
	\Omega_\Lambda^{ab} S^{c}_\vph{\Lambda}{}_{a \vph{b}} S_{bc}
	= 0
}$
on account of symmetry. Crucially, this conclusion holds independently of how $\Omega_\Lambda^{ab}$ is related to~$S_{ab}$, and therefore applies to any Lagrangian of the form in~\eqref{eq:pp_general_action} whose Hamiltonian~$\mathcal H_\pp$ is independent of~$\Lambda^a{}_A$ (up to a constraint term of the form $\chi_a C_1^a$).

Finally, the equation of motion for~$p_\mu$ follows from varying~\eqref{eq:pp_L_final} with respect to~$x^\mu$. This is incredibly tedious to do in an arbitrary coordinate frame, but since the action is generally covariant, much of this difficulty can be sidestepped by working in normal coordinates around a point where the Christoffel symbols vanish (but their derivatives do not), and then covariantising the result. After what is still a lengthy calculation, we find
\begin{align}
	\covdiff{}{\lambda}p_\mu
	\approx &	
	-
	\frac{1}{2} R_{\mu\nu\rho\sigma} \dot x^\nu S^{\rho\sigma}
	-
	me\A^2\nabla_\mu\log\A
	-
	2\B(p^\rho\dot x^\sigma\nabla_\rho\nabla_\sigma\phi)\nabla_\mu\phi
	\nonumber\\&
	-
	2em^{-1}\B
	[(p - me^{-1}\dot x)\cdot\!\nabla\phi]
	p^\alpha\nabla_\alpha\nabla_\mu\phi
	-
	2\dot x^\rho S^\sigma_{[\mu}\nabla_{\rho]}
	\nabla_\sigma\log\widetilde\A
	\nonumber\\&
	+
	\covdiff{S_{\rho\sigma}}{\lambda}
	\big(
		\delta_\mu^\rho \nabla^\sigma\log\widetilde\A
		+
		\widetilde\B \nabla^\rho\nabla^\sigma\nabla_\mu\phi
	\big)
	-
	\widetilde\B S_{\rho\sigma} \dot x^\nu
	\nabla^\rho\nabla_\mu\phi \nabla^\sigma\nabla_\nu\phi
	\nonumber\\&
	+
	\B \nabla_\mu\phi\nabla^\alpha\phi
	R_{\alpha\nu\rho\sigma}\dot x^\nu S^{\rho\sigma}
	+
	\widetilde\B R_{\mu\nu\rho\alpha}
	\dot x^\nu S^{\rho\sigma} \nabla^\alpha\phi\nabla_\sigma\phi.
\label{eq:app_ssc_eom_p}
\end{align}
As a useful consistency check, we note that \eqref{eq:app_ssc_eom_S} and \eqref{eq:app_ssc_eom_p} correctly reduce to the Mathisson--Papapetrou--Dixon equations~\cite{Mathisson:1937zz, Papapetrou:1951pa, Dixon:1970zza, Dixon:1970zz, Dixon:1974} of general relativity in the limit where~${\phi \to \text{constant}}$.

\paragraph{Consistency conditions.}
The four equations of motion in \eqref{eq:app_ssc_eom_xdot}, \eqref{eq:app_ssc_eom_omega}, \eqref{eq:app_ssc_eom_S}, and \eqref{eq:app_ssc_eom_p}, along with the three constraints in \eqref{eq:app_ssc_cst_0}--\eqref{eq:app_ssc_cst_2}, already contain all of the information that can be extracted from varying the Lagrangian, but they still do not specify a unique solution for the evolution of this spinning body. This is not uncommon when studying systems with constraints~\cite{Henneaux:1992ig}, and~in our case, the underlying reason is that the four equations of motion for the phase-space variables $(x^\mu, p_\mu, \Lambda^a{}_A, S_{ab})$ also depend explicitly on the Lagrange multipliers~$(e, \chi_a, \xi^a)$, which do not have equations of motion of their own. How we deal with these extra, nondynamical degrees of freedom will depend on whether or not the Lagrange multiplier is associated with a gauge symmetry, and thus we will have to treat the einbein~$e$ differently from the two vectors~$\chi_a$ and~$\xi^a$.
\looseness=-1

The claim is that because the former is responsible for rendering the point-particle action invariant under reparametrisation, it can be eliminated systematically from all four equations of motion. To see this, first multiply \eqref{eq:app_ssc_eom_omega}, \eqref{eq:app_ssc_eom_S}, and \eqref{eq:app_ssc_eom_p} by a factor of~$e^{-1}$. It should then become apparent that the einbein appears in all four equations of motion only as part of the combination~$e^{-1} D/D\lambda$, and thus can be eliminated by simply defining a new parameter~${s}$ such that ${\dx{s} = e\,\dx\lambda}$. In general relativity, ${s}$~is nothing but the proper time, but this is no longer the case in a scalar-tensor theory. Instead, the relation between ${s}$ and the proper time~$\tau$ is determined by solving \eqref{eq:app_ssc_eom_xdot} for~$p_\mu$ and then substituting the result into~\eqref{eq:app_ssc_cst_0}. Working perturbatively up to first order in the spin and in the disformal coupling,%
\footnote{Naturally, any term proportional to~$S_{ab}$ is counted as being of first order in the spin, but additionally, we also count~$\chi_a$ as being of first order in the spin, since one does not have to include the constraint term $\chi_a C_1^a$ in the Lagrangian when modelling a nonspinning body.}
we find that the solution to \eqref{eq:app_ssc_eom_xdot}~is
\begin{align}
	p_\mu
	\approx &\;
	me^{-1}\dot x _\mu
	+
	m (S_{\mu\nu}\xi^\nu + U_{\mu\nu} \chi^\nu)
	-
	2\B m
	[(S_{\rho\sigma}\xi^\sigma + U_{\rho\sigma}\chi^\sigma)\nabla^\rho\phi]
	\nabla_\mu\phi
	\nonumber\\&
	+
	\frac{ \dot x^\rho \dot x^\sigma }{\dot x^2}
	\big(
		\acute\B S_{\mu\nu} \nabla^\nu\phi \nabla_\rho\nabla_\sigma\phi
		+
		\grave\B S_{\mu\nu} \nabla^\nu\nabla_\rho\phi\nabla_\sigma\phi
	\big),
\label{eq:app_ssc_p_x}
\end{align}
where ${ U^\mu{}_\nu \coloneq \delta^\mu_\nu - \dot x^\mu \dot x_\nu/\dot x^2 }$ is the projection matrix onto the hypersurface orthogonal to~$\dot x^\mu$. Substituting this result into~\eqref{eq:app_ssc_cst_0} then returns
\begin{equation}
	\bigg(\frac{\dx {s}}{\dx \lambda}\bigg)^2
	=
	e^2
	=
	- \A^{-2}[\dot x^2 + 2\B(\dot x \cdot\!\nabla\phi)^2],
\label{eq:app_ssc_sol_e}
\end{equation}
which is spin-independent at the order to which we are working. It follows immediately that
${
	(\dx {s}/\dx\tau)^2
	=
	\A^{-2}[1 - 2\B( u \cdot\!\nabla\phi)^2]
}$
when written in terms of the proper time~$\tau$, which is defined from the requirement that the 4-velocity~${u^\mu \coloneq \dx x^\mu/\dx\tau}$ satisfy the normalisation condition~${u_\mu u^\mu = -1}$.

These relations now offer us three equally viable methods for eliminating~$e$ from the equations of motion. We could reparametrise all four equations in terms of~${s}$ as described above; or, we could reparametrise in terms of~$\tau$; or, we could continue to use the parameter~$\lambda$ while using \eqref{eq:app_ssc_sol_e} to eliminate~$e$. This last option turns out to be the most convenient for our purposes, given that it does not impose any particular gauge choice on~$\lambda$.%
\footnote{It is also worth mentioning that this option is essentially identical to the process of integrating out $p_\mu$ and~$e$ from the Lagrangian, which is what we did when going from \eqref{eq:pp_L_final} to~\eqref{eq:pp_L_intermediate} in the main text.}
Accordingly, we can easily set ${\lambda \to \tau}$ and ${\dot x^\mu \to u^\mu}$ at a later stage should we find ourselves wanting to specialise to the proper-time gauge, or alternatively, we could set ${\lambda \to x^0}$ and ${\dot x^\mu \to v^\mu = (1, \bm v)}$ to establish a nonrelativistic expansion, as is done in the main text.

Having dealt with the einbein~$e$, we now turn to the issue of fixing $\chi_a$~and~$\xi^a$. These two Lagrange multipliers are not associated with any kind of gauge symmetry, and so cannot be eliminated from our system of equations. What we can do instead is tune them in such a way that the overall solution remains self-consistent. A~basic requirement for consistency is that the SSC and its conjugate constraint are preserved under time evolution, and indeed, demanding that ${D C_1^a/D\lambda \approx 0}$ and ${D C_2^a /D\lambda \approx 0}$ gives us two new equations (on top of the usual four equations of motion) that can be solved simultaneously~for~$\chi_a$~and~$\xi^a$.
\looseness=-1

To proceed, we first note that
\begin{equation}
	\text{if }
	C^a \approx 0
	\text{ and }
	\covdiff{C^a}{\lambda} \approx 0,
	\text{ then }
	\covdiff{}{\lambda}(C^a T_{a b_1 \cdots\, b_n})
	\approx 0
\label{eq:app_ssc_lemma}
\end{equation}
for any tensor~$T_{ab_1 \cdots\, b_n}$. This result is useful because the consistency condition ${D C_1^a/D\lambda \approx 0}$ is difficult to manipulate as is, but it is easy to show that
\begin{equation}
	\eta^{AB} e^\mu_A
	\covdiff{}{\lambda}
	\bigg(
		\frac{
			g_{\rho\sigma}^{} e^\rho_B
			e^{\sigma\vph{\rho}}_{a \vph{B}} C_1^a}%
		{\sqrt{-p^2}}
	\bigg)
	=
	-\frac{1}{\sqrt{-p^2}}
	\bigg(
		\Omega^{\mu\nu} p_\nu
		+
		P^{\mu\nu}
		\covdiff{p_\nu}{\lambda}
	\bigg).
\end{equation}
The above necessarily vanishes along the constraint surface on account of~\eqref{eq:app_ssc_lemma}; hence, the preservation of the constraint ${C_1^a \approx 0}$ under time evolution implies that
\begin{equation}
	\Omega^{\mu\nu} p_\nu \approx - P^{\mu\nu} \covdiff{p_\nu}{\lambda}.
\label{eq:app_ssc_consistency_1}
\end{equation}
Similarly, we can contract $C_2^a$ with $\eta_{ab} e^b_\mu$ and then use the result in~\eqref{eq:app_ssc_eom_S_xframe} to show that
\begin{align}
	\covdiff{}{\lambda}(S_{\mu\nu} p^\nu)
	&=
	\big(
		2e \chi_{[\mu}p_{\nu]}
		-
		2\Omega^\alpha{}_{[\mu}S_{\nu]\alpha}
	\big)
	p^\nu
	+
	S_{\mu\nu} \covdiff{p_\nu}{\lambda}
	\approx 0.
\label{eq:app_ssc_consistency_2}
\end{align}
Substituting \eqref{eq:app_ssc_consistency_1} into \eqref{eq:app_ssc_consistency_2} then reveals that
${ 2e \chi_{[\mu}p_{\nu]} p^\nu \approx 0 }$,
and thus we must have
\begin{equation}
	\chi_a \approx 0
\label{eq:app_ssc_sol_chi}
\end{equation}
for a consistent solution to the equations of motion. More precisely, we must have ${\chi_a \approx 0}$ up to an additive term proportional to~$p_a$, but since this Lagrange multiplier only ever appears in the equations of motion as part of the combinations $P_{\mu\nu}\chi^\nu$ or $\chi_{[\mu} p_{\nu]}$, the additive term always drops out, and thus the solution in~\eqref{eq:app_ssc_sol_chi} is sufficiently general.

What remains is to solve for~$\xi^a$. Rather than substitute \eqref{eq:app_ssc_consistency_1} into~\eqref{eq:app_ssc_consistency_2}, we now substitute \eqref{eq:app_ssc_eom_omega} and \eqref{eq:app_ssc_eom_p} into~\eqref{eq:app_ssc_consistency_2}. After also using \eqref{eq:app_ssc_eom_S}, \eqref{eq:app_ssc_sol_e}, and \eqref{eq:app_ssc_sol_chi} to simplify terms, we~find% 
\begin{equation}
	S_{\mu\nu} \bigg( m \xi^\nu 
	-
	\nabla^\nu\log\,(\widetilde\A/\A)
	-
	\frac{\dot x^\rho \dot x^\sigma}{\dot x^2}
	\big[
		(2\B - \widetilde\B - \acute\B)
		\nabla_\rho\nabla_\sigma\phi \nabla^\nu\phi
		+
		(\widetilde\B - \grave\B)
		\nabla_\rho\phi \nabla^\nu \nabla_\sigma\phi
	\big]
	\bigg)
	\approx 0
\end{equation}
at the order to which we are working. The factor of~$S_{\mu\nu}$ can easily be stripped off to yield
\begin{equation}
	m\xi^a \approx \nabla^a\log(\tilde\A/\A) 
	+
	\frac{\dot x^\rho \dot x^\sigma}{\dot x^2}
	\big[
		(2\B-\widetilde\B-\acute\B)
		\nabla_\rho\nabla_\sigma\phi\nabla^a\phi
		+
		(\widetilde\B-\grave\B)
		\nabla_\rho\phi\nabla^a\nabla_\sigma\phi
	\big].
\label{eq:app_ssc_sol_xi}
\end{equation} 
As was the case with~$\chi_a$, this solution for~$\xi^a$ is unique up to an additive term that is proportional to~$p^a$, but that consistently drops out of the equations of motion. The solution in \eqref{eq:app_ssc_sol_xi} is therefore sufficiently general.

\paragraph{Equivalence with the Routhian approach.}
The above solutions for $\chi_a$ and~$\xi^a$ are exactly what we substituted into~\eqref{eq:pp_L_intermediate} to obtain the point-particle Routhian in~\eqref{eq:pp_Routhian}. In general, one should be wary about substituting things back into the action,%
\footnote{The main exception being when one is ``integrating'' something out at the classical level.}
but it is possible to check that the first-order Lagrangian in~\eqref{eq:pp_L_final} and the Routhian in~\eqref{eq:pp_Routhian} correctly give rise to the same equations of motion.

In particular, we find that the Euler--Lagrange equation for~$\mathcal R_\pp$ [see~\eqref{eq:pp_eom}]~yields
\begin{align}
	m\A \covdiff{}{\tau} u^\mu
	\approx &
	-
	\frac{1}{2} R^\mu{}_{\nu\rho\sigma} u^\nu S^{\rho\sigma}
	-
	m\A\, U^{\mu\nu}\nabla_\nu\log\A
	-
	2m\A\B\, U^{\mu\nu} u^\rho u^\sigma
	\nabla_\rho\nabla_\sigma\phi\nabla_\nu\phi
\nonumber\\&
	+
	S^{\mu\rho}u^\sigma
	(\nabla_\rho\nabla_\sigma\log\A - \nabla_\rho\log\A\nabla_\sigma\log\A)
	+
	2\B S^{\mu\nu} u^\rho u^\sigma u^\alpha
	\nabla_\alpha(\nabla_\rho\nabla_\sigma\phi\nabla_\nu\phi)
\nonumber\\&
	-
	\widetilde\B S^{\rho\sigma} u^\alpha
	\nabla^\mu\nabla_\rho\phi \nabla_\sigma\nabla_\alpha\phi
	-
	2\widetilde\B S^{\mu\nu} u^\rho u^\sigma u^\alpha
	\nabla_{[\nu}\phi \nabla_{\rho]}
	\nabla_{\sigma \vph{]}}\nabla_{\alpha \vph{]}}\phi
\nonumber\\&
	+
	\frac{1}{2}\B R^\mu{}_{\nu\rho\sigma} u^\nu S^{\rho\sigma}
	(u\cdot\!\nabla\phi)^2
	-
	\B R_{\rho\sigma\alpha\beta} S^{\rho\sigma}
	U^{\mu\nu} u^\alpha \nabla^\beta\phi \nabla_\nu\phi
\nonumber\\&
	+
	\widetilde\B R^\mu{}_{\nu\rho\sigma} u^\nu S^{\rho\alpha}
	\nabla^\sigma\phi \nabla_\alpha\phi
	-
	\widetilde\B S^{\mu\nu} R_{\nu\rho\sigma\beta}
	u^\rho u^\sigma u^\alpha \nabla^\beta\phi \nabla_\alpha\phi
\end{align}
after specialising to the proper-time gauge for simplicity and then truncating to first order in the spin and the disformal coupling. This second-order equation of motion for the worldline is equivalent to what we would get from combining the first-order equations in \eqref{eq:app_ssc_eom_p} and~\eqref{eq:app_ssc_p_x}, along with our solutions for the Lagrange multipliers in~\eqref{eq:app_ssc_sol_e}, \eqref{eq:app_ssc_sol_chi}, and~\eqref{eq:app_ssc_sol_xi}.
\looseness=-1

Likewise, we can use Hamilton's equation on~$\mathcal R_\pp$ [see~\eqref{eq:pp_eom} and~\eqref{eq:pp_S_PB}] to obtain the equation of motion for the spin, which reads
\begin{align}
	\covdiff{}{\tau} S_{\mu\nu}
	\approx &\:
	2 u_{[\mu}S_{\nu]\alpha} \nabla^\alpha\log\A
	+
	4\B u_{[\mu}S_{\nu]\alpha} u^\rho u^\sigma
	\nabla_\rho\nabla_\sigma\phi \nabla^\alpha\phi
\nonumber\\&
	+
	4\widetilde\B 
	u_{[\mu}S_{\nu]}{}^\alpha u^\rho u^\sigma
	\nabla_{[\rho}\phi \nabla_{\alpha]}\nabla_{\sigma \vph{]}}\phi
	+
	4\widetilde\B
	u^\rho S^\sigma{}_{[\mu}\delta^\alpha{}_{\nu]}
	\nabla_{\rho \vph{]}}\nabla_{[\sigma}\phi\nabla_{\alpha]}\phi
\end{align}
when written in the proper-time gauge. Once again, this is exactly what we would obtain from combining \eqref{eq:app_ssc_eom_S} and~\eqref{eq:app_ssc_p_x}, along with our solutions for the Lagrange multipliers.

% ============================================== %
% Appendix B
% ============================================== %
\section{Evaluating Feynman diagrams}
\label{app:feyn}

Our results for the conservative potential~$V$ and the scalar source function~$J(x)$ have been pieced together by combining previously known results~\cite{Damour:1992we, Huang:2018pbu, Kuntz:2019zef, Porto:2005ac} with a number of calculations that are novel to this work. In particular, three new Feynman diagrams had to be evaluated, and the details of those calculations are presented here in this appendix.

\paragraph{1.}
We begin with the Feynman diagram in figure~\ref{fig:feyn_V_SO}(c), whose contribution to~$V$~reads
\begin{align}
	\text{figure~\ref*{fig:feyn_V_SO}(c)}
	=&
	-\! i\sumKK \int \dx^4x\,\dx^4x'\,
	\bigg(
		i\delta^{(3)}\bm( \bm x - \bm x_\K(t) \bm)
		\frac{\alpha_\K}{2\mpl}
		(S_\K^{i0} - S_\K^{ij}v_\K^j)
	\bigg)
	\nonumber\\[-5pt]&
	\times\!
	\pdiff{ \avg{T \hat\varphi(x)\hat\varphi(x')} }{x^i}
	\bigg(
		i\delta^{(3)}\bm( \bm x' - \bm x_{\K'}(t') \bm)
		\frac{\alpha_{\K'} m_{\K'}}{2\mpl}
	\bigg).
\end{align}
Enclosed in parentheses are explicit expressions for the two worldline vertices, which follow from the Feynman rules in table~\ref{table:feyn_1}, while the sum over $\K$ and $\K'$ ensures that both permutations of the worldlines are taken into account.%
\footnote{Requiring that ${\K' \!\neq \K}$ automatically restricts our attention only to those diagrams in which a scalar is being exchanged between the two worldlines. In general, the Feynman rules also allow for diagrams in which the scalar propagates to and from a single worldline (i.e.,~${\K' \!= \K}$), but these correspond to divergent self-energy corrections that are pure counterterm and have been shown to vanish when using dimensional regularisation~\cite{Goldberger:2004jt}. Accordingly, such diagrams have no physical effect and can safely~be~neglected.\looseness=-1}
Note also that we have chosen to multiply the entire diagram by an extra factor of $-i$, such that it constitutes a term in the effective action~$S_\text{eff}'$~[cf.~\eqref{eq:review_S_eff'_integral}], rather~than one in~$i S_\text{eff}'$.

Now using the explicit form for the potential-mode propagator in~\eqref{eq:consv_propagators_potential_modes}, we find~that%
\begin{align}
	\text{figure~\ref*{fig:feyn_V_SO}(c)}
	=&
	\sumKK \int \dx^4x\,\dx^4x'\,
	\delta^{(3)}\bm( \bm x - \bm x_\K(t) \bm)\,
	\delta^{(3)}\bm( \bm x' - \bm x_{\K'}(t') \bm)
	\nonumber\\[-7pt]&
	\times\!
	\frac{ \alpha_{\K\vph{'}} }{2\mpl}
	\frac{\alpha_{\K'} m_{\K'}}{2\mpl}
	(S_\K^{i0} - S_\K^{ij}v_\K^j)
	\pdiff{}{x^i}
	\bigg(
		\frac{\delta(t-t')}{4\pi|\bm x - \bm x'|}
	\bigg)
\nonumber\allowdisplaybreaks\\[-6pt]
	=&
	-\!\sumKK\int \dx t\,
	\frac{ \alpha_{\K\vph{'}} \alpha_{\K'} m_{\K'} }{16\pi\mpl^2}
	(S_\K^{i0} - S_\K^{ij}v_\K^j)
	\frac{ ( \bm x_{\K\vph{'}} - \bm x_{\K'} )^i }%
		{|\bm x_{\K\vph{'}} - \bm x_{\K'} |^3},
\end{align}
where the last line follows after integrating over~$\bm x$, $\bm x'$, and~$t'$. This is equivalent to~writing
\begin{equation}
	\text{figure~\ref*{fig:feyn_V_SO}(c)}
	=
	-\int\dx t\,
	\frac{2\GN \alpha_1\alpha_2 m_2}{r^2}
	n^i(S_1^{i0} - S_1^{ij}v_1^j)
	+
	(1\leftrightarrow 2).
\end{equation}
We may now read off this diagram's contribution to~$V$ by simply discarding the ``${-\!\int\!\dx t}$'' out front~[cf.~\eqref{eq:review_def_V}]. What remains is exactly the term in~\eqref{eq:consv_V_SO_C} that is proportional to~$\alpha_1\alpha_2$.\looseness=-1

\paragraph{2.}
Similar steps underpin the evaluation of figure~\ref{fig:feyn_V_SO}(d), which provides the leading disformal contribution to the spin-orbit potential. Explicitly, we have that
\begin{align}
	\text{figure~\ref*{fig:feyn_V_SO}(d)}
	=&
	-\!i\sumKK \int \dx^4x\,\dx^4x'\,\dx^4x''
	\bigg(
		i\delta^{(3)}\bm( \bm x - \bm x_\K(t) \bm)
		\frac{\tilde\beta_\K^\vph{j} S_\K^{ij}}{\M^4}
	\bigg)
	\nonumber\\[-3pt]&
	\times
	v_\K^\mu(t)
	\pdiff{^2\avg{T\hat\varphi(x)\hat\varphi(x')} }{x^\mu\partial x^i}
	\bigg(
		i\delta^{(3)}\bm( \bm x' - \bm x_{\K'}(t') \bm)
		\frac{\alpha_{\K'} m_{\K'}}{2\mpl}
	\bigg)
	\nonumber\\[-3pt]&\times
	\pdiff{ \avg{T\hat\varphi(x)\hat\varphi(x'')} }{x^j}
	\bigg(
		i\delta^{(3)}\bm( \bm x'' - \bm x_{\K'}(t'') \bm)
		\frac{\alpha_{\K'} m_{\K'}}{2\mpl}
	\bigg),
\end{align}
and this can be shown to simplify to
\begin{align}
	\text{figure~\ref*{fig:feyn_V_SO}(d)}
	=&\,
	\sumKK
	\int \dx t\,\dx t'\,
	\bigg( \frac{\alpha_{\K'} m_{\K'}}{2\mpl} \bigg)^{\!2}
	\frac{\tilde\beta_\K^\vph{j} S_\K^{ij}}{16\pi^2\M^4}
	\frac{( \bm x_{\K\vph{'}} - \bm x_{\K'} )^j}
		{|\bm x_{\K\vph{'}} - \bm x_{\K'}|^3}
	\nonumber\\[-5pt]&\times
	\bigg( 
		\pdiff{}{t}\delta(t-t')
		+
		\delta(t-t')\,
		\bm v_\K \cdot \pdiff{}{\bm x_\K}
	\bigg)
	\frac{[ \bm x_{\K\vph{'}} - \bm x_{\K'}(t') ]^i}
		{|\bm x_{\K\vph{'}} - \bm x_{\K'}(t')|^3}
\label{eq:app_feyn_2_intermediate_1}
\end{align}
after evaluating the derivatives with respect to~$x^i$ and $x^j$, and then performing the integrals over all variables except $t$ and~$t'$. Written in this way, $\bm x_\K$ and $\bm x_{\K'}$ are to be understood as being functions of~$t$, except where indicated otherwise. The identity ${\partial_t \delta(t-t') \equiv -\partial_{t'}\delta(t-t')}$ and a subsequent integration by parts can then be used to show~that
\begin{align}
	\text{figure~\ref*{fig:feyn_V_SO}(d)}
	=&
	\sumKK
	\int \dx t\,\dx t'\,
	\bigg( \frac{\alpha_{\K'} m_{\K'}}{2\mpl} \bigg)^{\!2}
	\frac{\tilde\beta_\K^\vph{j} S_\K^{ij}}{16\pi^2\M^4}
	\frac{( \bm x_{\K\vph{'}} - \bm x_{\K'} )^j}
		{|\bm x_{\K\vph{'}} - \bm x_{\K'}|^3}
	\nonumber\\[-8pt]&\times
	\delta(t-t')
	\bigg( 
		\pdiff{}{t'}
		+
		\bm v_\K \cdot \pdiff{}{\bm x_\K}
	\bigg)
	\frac{[ \bm x_{\K\vph{'}} - \bm x_{\K'}(t') ]^i}
		{|\bm x_{\K\vph{'}} - \bm x_{\K'}(t')|^3}
\nonumber\allowdisplaybreaks\\[-4pt]
	=&
	\sumKK
	\int \dx t\,
	\bigg( \frac{\alpha_{\K'} m_{\K'}}{2\mpl} \bigg)^{\!2}
	\frac{\tilde\beta_\K^\vph{j} S_\K^{ij}}{16\pi^2\M^4}
	\frac{( \bm x_{\K\vph{'}} - \bm x_{\K'} )^j}
		{|\bm x_{\K\vph{'}} - \bm x_{\K'}|^3}
	\diff{}{t}
	\bigg(
		\frac{(\bm x_{\K\vph{'}} - \bm x_{\K'})^i}
			{|\bm x_{\K\vph{'}} - \bm x_{\K'}(t')|^3}
	\bigg),
\label{eq:app_feyn_2_intermediate_2}
\end{align}
where the second line follows essentially by definition of the total derivative. This is equivalent to~writing
\begin{equation}
	\text{figure~\ref*{fig:feyn_V_SO}(d)}
	=
	-\int\dx t\,
	\frac{\GN \alpha_2^2 m_2^2}{2\pi\M^4 r^5}
	\tilde\beta_1^\vph{j} S_1^{ij} n^i v^j
	+
	(1\leftrightarrow 2),
\end{equation}
in agreement with~\eqref{eq:consv_V_SO_D}.

\paragraph{3.}
The last diagram that needs evaluating is shown in figure~\ref{fig:feyn_J_SO}(b), and is responsible for the leading disformal spin-orbit contribution to~$J(x)$. Unlike the previous two diagrams, which involve only the exchange of potential modes, this third diagram also contains a radiation mode that carries energy and momentum off to infinity. Thus, whereas we could previously read off the Feynman rules directly from table~\ref{table:feyn_1} after replacing $\varphi$ with $\hat\varphi$, we must now substitute ${\varphi = \bar\varphi + \hat\varphi}$ [see~\eqref{eq:review_potential_radiation_split}] into table~\ref{table:feyn_1} as a first step. The vertex in figure~\ref{fig:feyn_J_SO}(b) that involves both a potential mode and a radiation mode comes from making this substitution in the fourth row of table~\ref{table:feyn_1} and then keeping only the cross terms. Written out explicitly, they~read%
\begin{equation}
	\frac{\tilde\beta_\K}{\M^4}
	\int\dx t\, S^{ij}_\K
	\big[
		(v_\K^\mu\partial_\mu\partial_i\hat\varphi)\,\partial_j\bar\varphi
		+
		(v_\K^\mu\partial_\mu\partial_i\bar\varphi)\,\partial_j\hat\varphi
	\big].
\label{eq:app_feyn_3_vertex_raw}
\end{equation}
An integration by parts can now be used to move the derivative ${v_\K^\mu\partial_\mu~(\equiv \dx/\dx t)}$ in the second term off the radiation-mode scalar~$\bar\varphi$ and onto the potential-mode scalar~$\hat\varphi$. After also exploiting the antisymmetry of the spin tensor, we find that \eqref{eq:app_feyn_3_vertex_raw} is~equivalent~to
\begin{equation}
	\frac{2\tilde\beta_\K}{\M^4}
	\int\dx t\: S^{ij}_\K
	(v_\K^\mu\partial_\mu\partial_i\hat\varphi)\,\partial_j\bar\varphi.
\end{equation}

It now follows that figure~\ref{fig:feyn_J_SO}(b) is given by
\begin{align}
	\text{figure~\ref*{fig:feyn_J_SO}(b)}
	=&
	-\! i\sumKK
	\int \dx^4x\,\dx^4x'
	\bigg(
		i\delta^{(3)}\bm( \bm x' - \bm x_{\K'}(t') \bm)
		\frac{\alpha_{\K'}m_{\K'}}{2\mpl}
	\bigg)
	\nonumber\\[-6pt]&\times
	v_\K^\mu(t)
	\pdiff{^2 \avg{T\hat\varphi(x')\hat\varphi(x)} }{x^\mu\partial x^i}
	\bigg(
		i\delta^{(3)}\bm( \bm x - \bm x_\K(t) \bm)
		\frac{2\tilde\beta_\K^\vph{j} S_\K^{ij} }{\M^4}
		\pdiff{\bar\varphi(x)}{x^j}
	\bigg),
\end{align}
and this can be shown to simplify to
\begin{align}
	\text{figure~\ref*{fig:feyn_J_SO}(b)}
	=&
	-\!\sumKK \int \dx t'\,\dx^4x\,
	\delta^{(3)}\bm( \bm x - \bm x_\K(t) \bm)\,
	\frac{\alpha_{\K'}m_{\K'}}{2\mpl}
	\frac{\tilde\beta_\K^\vph{j} S_\K^{ij} }{2\pi\M^4}
	\pdiff{\bar\varphi(x)}{x^j}
	\nonumber\\[-8pt]&
	\times\!
	\bigg(
		\pdiff{}{t}\delta(t-t')
		+
		\delta(t-t')\,\bm v_\K\cdot\pdiff{}{\bm x}
	\bigg)
	\frac{[ \bm x - \bm x_{\K'}(t') ]^i}%
		{|\bm x - \bm x_{\K'}(t')|^3}
\end{align}
after evaluating the derivative with respect to~$x^i$ and then performing the integral over~$\bm x'$. The derivative acting on the delta function can then be dealt with in a similar way to how we went from \eqref{eq:app_feyn_2_intermediate_1} to \eqref{eq:app_feyn_2_intermediate_2}, and we ultimately find that
\begin{equation}
	\text{figure~\ref*{fig:feyn_J_SO}(b)}
	=
	\sumKK \int \dx^4x\,
	\delta^{(3)}\bm( \bm x - \bm x_\K(t) \bm)\,
	\frac{\alpha_{\K'}m_{\K'}}{2\mpl}
	\frac{\tilde\beta_\K^\vph{j} S_\K^{ij} }{2\pi\M^4}
	\pdiff{\bar\varphi(x)}{x^i}
	\diff{}{t}
	\frac{(\bm x_{\K\vph{'}} - \bm x_{\K'})^j}%
		{|\bm x_{\K\vph{'}} - \bm x_{\K'}|^3}.
\end{equation}
An overall minus sign has disappeared because we have relabelled the indices ${i \leftrightarrow j}$ while using the antisymmetry of the spin tensor. Finally, integrating by parts to move the spatial derivative $\partial/\partial x^i$ off the radiation-mode scalar~$\bar\varphi$, we get
\begin{equation}
	\text{figure~\ref*{fig:feyn_J_SO}(b)}
	=
	-\!\sum_{\K\vph{'}} \!\int \dx^4x
	\bigg(
		\sum_{\K'\neq\K}
		\frac{\alpha_{\K'}m_{\K'}}{2\mpl}
		\frac{\tilde\beta_\K^\vph{j} S_\K^{ij} }{2\pi\M^4}
		\diff{}{t}
		\frac{(\bm x_{\K\vph{'}} - \bm x_{\K'})^j}%
			{|\bm x_{\K\vph{'}} - \bm x_{\K'}|^3}
	\bigg)
	\partial_i\delta^{(3)}\bm( \bm x - \bm x_\K(t) \bm)\,\bar\varphi(x).
\end{equation}
The quantity in parentheses is exactly the dipole moment~$q_\K^i(t)$ in~\eqref{eq:rad_J_SO_dipole}.

This concludes our discussion of the three diagrams most relevant to this work. Several other diagrams have been included in the main text for illustrative purposes, and while there is no need for us to evaluate any of them in exhaustive detail here, the Feynman rules that would be needed to do so are provided in table~\ref{table:feyn_2} for the sake of completeness. Also included in this table is a description of how each vertex scales with the EFT's expansion parameters, which does get used in the main text when we make power-counting arguments.

\begin{table}
\centering
\begin{tabular}{|@{\hspace{10pt}} c @{\hspace{30pt}} l @{\hspace{25pt}} l @{\hspace{10pt}}|}
\hline
Diagram & Interaction vertex & Scaling \\
\hline
% ------ PURE GRAVITY COUPLINGS ----- %
% h
\includegraphics[valign=c]{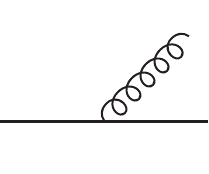}
&
$\tablestyle
\frac{m_\K}{2\mpl} \int\dx t\: h_{00}$
&
$\sqrt{L}$
\\
% hv
\includegraphics[valign=c]{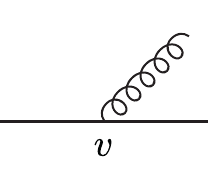}
&
$\tablestyle
\frac{m_\K}{\mpl}
\int\dx t\: h_{0i}^\vph{i} v^i_\K$
&
$\sqrt{L}v$
\\
% hv^2
\includegraphics[valign=c]{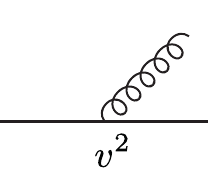}
&
$\tablestyle
\frac{m_\K}{2\mpl}
\int\dx t\,
\bigg(
	h_{ij}^\vph{i} v^{i \vph{j}}_\K v^j_\K
	+
	\frac{1}{2}h_{00}^\vph{2} \bm v_\K^2
\bigg)$
&
$\sqrt{L}v^2$
\\
% hS
\includegraphics[valign=c]{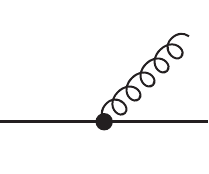}
&
$\tablestyle
 \frac{1}{2\mpl}
\int\dx t\: S_\K^{ij} \partial_i h_{j0}$
&
$\sqrt{L}v\epsSpin$
\\
% hSv
\includegraphics[valign=c]{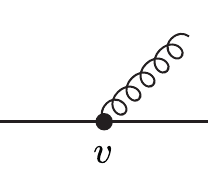}
&
$\tablestyle
\frac{1}{2\mpl} \int\dx t\:
\big(
	S_\K^{ij} v^m_\K \partial_i h_{jm}
	+
	S_\K^{i0}\partial_i h_{00}
\big)$
&
$\sqrt{L}v^2\epsSpin$
\\
% ------ QUADRATIC-IN-FIELDS COUPLINGS ----- %
% Conformal [O]'
\includegraphics[valign=c]{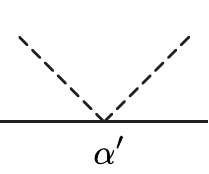}
&
$\tablestyle
\frac{\alphaprime_\K m_\K}{8\mpl^2}
\int\dx t\:\varphi^2$
&
$v^2$
\\
% h-phi
\includegraphics[valign=c]{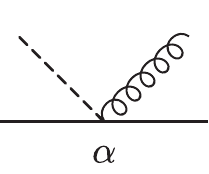}
&
$\tablestyle
-\frac{\alpha_\K m_\K}{8\mpl^2}
\int\dx t\: h_{00} \varphi$
&
$v^2$
\\
% h-phi S
\includegraphics[valign=c]{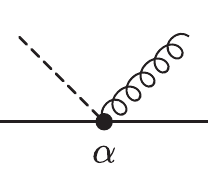}
&
$\tablestyle
\frac{\alpha_\K}{8\mpl^2}
\int\dx t\: S_\K^{ij} h_{0i} \partial_j \varphi$
&
$v^3\epsSpin$
\\
% h-phi Sv
\includegraphics[valign=c]{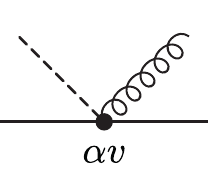}
&
$\tablestyle
\frac{\alpha_\K}{8\mpl^2} \int\dx t\:
\big(
	S_\K^{ij} v^m_\K h_{mi}\partial_j\varphi
	-
	S_\K^{i0} h_{00}\partial_i\varphi
\big)
$
&
$v^4\epsSpin$
\\[25pt]
\hline
\end{tabular}
\caption{Feynman rules for a number of worldline vertices. The graviton ${h_{\mu\nu}\equiv h_{\mu\nu}(t,\bm x_\K)}$ and the scalar ${\varphi \equiv \varphi(t,\bm x_\K)}$ are represented by helical and dashed lines, respectively, while the worldline itself is drawn as a solid line. Vertices without black dots are spin-independent, while those with a black dot are coupled to one power of the spin. The rightmost column lists the power-counting rules for how each vertex scales with the EFT's expansion parameters, assuming all factors of $h_{\mu\nu}$ and~$\varphi$ are taken to be potential modes. For each factor of $h_{\mu\nu}$ or $\varphi$ that is instead taken to be a radiation mode, simply include an extra factor of $\sqrt{v}$.}
\label{table:feyn_2}
\end{table}

% ============================================== %
% Appendix C
% ============================================== %
\section{Computing the scalar flux at leading order}
\label{app:Q}

The goal of this appendix is to derive the result in~\eqref{eq:rad_power_scalar_on_shell} for the power radiated into scalar waves. For circular nonprecessing binaries, we saw in the main text that the scalar multipole moments could all be cast into the general form~[cf.~\eqref{eq:rad_Q_on_shell}]
\begin{equation}
	\mathcal Q^L
	=
	\frac{m}{2\mpl}
	\bigg( \frac{\Geff m}{\x} \bigg)^{\!\ell}
	Q_\ell(\x)\,
	n^\avg{L}
\label{eq:app_Q_Q_on_shell}
\end{equation}
at the order to which we are working. What we have to do now is substitute this result into the master formula in~\eqref{eq:review_rad_power_scalar}. Because the circular nature of the orbit ensures that the overall magnitudes of these quantities remain constant, the time derivatives that act on $\mathcal Q^L$ in~\eqref{eq:review_rad_power_scalar} end up acting only on the product of unit vectors~$n^\avg{L}$. All that remains is for us to evaluate these derivatives.

It turns out that we can do this for arbitrary~$\ell$ by making use of the general properties of symmetric and trace-free tensors. We start by decomposing~$n^\avg{L}$ into its components along the directions spanned by a set of basis vectors~$\mathcal Y^{\ell\ym}_L$, which we shall take to correspond to the generators of the spherical harmonics. Accordingly, these basis vectors  are defined by the relation~${ Y_{\ell \ym}(\bm n) = \mathcal Y^{\ell \ym}_L n^L }$ and can be shown to satisfy the orthogonality condition~\cite{Thorne:1980ru}\looseness=-1
\begin{equation}
	(\mathcal Y^{\ell\ym^\vph{'}}_L)^*_{} \mathcal Y^{\ell\ym'}_L
	=
	\frac{(2\ell+1)!!}{4\pi\ell!}\delta^{\ym^\vph{'}\ym'}.
\label{eq:app_Q_Ylm_orthogonal}
\end{equation}

Combining this harmonic decomposition with our freedom to choose coordinates such that ${\bm n = (\cos\Omega t,\sin\Omega t,0)}$ now allows us to write~\cite{Wong:2019kru}%
\begin{equation}
	n^\avg{L}
	=
	\frac{4\pi\ell!}{(2\ell+1)!!}
	\sum_{\ym=-\ell}^\ell Y^*_{\ell \ym}(\bm d)
	\mathcal Y^{\ell \ym}_L e^{-i\ym\Omega t},
\label{eq:app_Q_nL}
\end{equation}
where ${\bm d = (1,0,0)}$ points in the direction of~$\bm n$ at the reference time~${t=0}$. This expression makes it easy for us to differentiate~$\mathcal Q^L$ as many times as needed, and indeed we find that
\begin{equation}
	{}^{(\ell+1)}\!{\mathcal Q}^L
	=
	\frac{m}{2\mpl}
	\bigg( \frac{\Geff m}{\x} \bigg)^{\!\ell} Q_\ell(\x) \!
	\sum_{\ym=-\ell}^\ell Y^*_{\ell \ym}(\bm d)
	\mathcal Y^{\ell \ym}_L (-i\ym\Omega)^{\ell+1}
	e^{-i\ym\Omega t}.
\end{equation}
Taking the inner product of this result with itself now gives us
\begin{align}
	\frac{\langle ({}^{(\ell+1)}\!\mathcal Q^L)^2 \rangle}%
		{4\pi\ell!(2\ell+1)!!}
	&=
	\bigg( \frac{m}{2\mpl} \bigg)^2
	\bigg( \frac{\Geff m}{\x} \bigg)^{\!2\ell}
	\Omega^{2\ell+2}
	N_\ell^\vph{1} Q_\ell^2(\x)
	=
	\frac{8\pi N_\ell}{\Geff (1+2\alpha_1\alpha_2)}
	Q_\ell^2(\x) \, \x^{\ell+3},
\label{eq:app_Q_summand}
\end{align}
where the second equality follows after using \eqref{eq:consv_def_x} to eliminate $\Omega$ in favour of~$\x$. Meanwhile, the overall numerical coefficient is~given~by
\begin{equation}
	N_\ell
	=
	\frac{1}{4\pi\ell!(2\ell+1)!!}
	\bigg|
		\frac{4\pi\ell!}{(2\ell+1)!!}
		\sum_{\ym=-\ell}^\ell Y^*_{\ell\ym}(\bm d)
		\mathcal Y^{\ell\ym}_L
		(-i\ym)^{\ell+1}	e^{-i\ym\Omega t}\,
	\bigg|^2,
\end{equation}
which further simplifies to
\begin{equation}
	N_\ell
	=
	\sum_{\ym=-\ell}^\ell
	\bigg|\frac{Y_{\ell\ym}(\bm d)}{(2\ell+1)!!}\bigg|^2
	\ym^{2\ell+2}
\label{eq:app_Q_Nl}
\end{equation}
after using the orthogonality condition in~\eqref{eq:app_Q_Ylm_orthogonal}. The sum over $\ym$ is straightforward to evaluate with an algebraic package like Mathematica, and we obtain ${N_0 = 0}$, ${N_1 = 1/12\pi}$, and ${N_2 = 4/15\pi}$ for the first few values of~$\ell$. Finally, taking the sum of \eqref{eq:app_Q_summand} over all integer values of ${\ell \geq 0}$ returns the desired result in~\eqref{eq:rad_power_scalar_on_shell}.

% ============================================== %
% Appendix D
% ============================================== %
\section{Gravitational-wave phase in the frequency domain}
\label{app:gw}

Compiled in this appendix are some additional formulae that supplement the discussion in section~\ref{sec:rad_phase}. In particular, our results for the gravitational-wave phase in the frequency domain are presented in~\eqref{eq:app_gw_DD_phase} and~\eqref{eq:app_gw_QD_phase}. 

\paragraph{Dipole-driven regime.}
For dipole-driven systems, the solution to~\eqref{eq:rad_dtdx} reads
\begin{equation}
	t^\vph{()}_\DD
	=
	t_0^\vph{()}
	+
	t_\DD^\vph{()}\scount{o}
	+
	t_\DD^\sector{C}\scount{so}
	+
	t_\DD^\sector{D}\scount{so},
\end{equation}
where $t_0$ is an integration constant to be fixed by initial conditions, and
\begin{subequations}
\begin{align}
	\frac{ t_\DD^\vph{()}\scount{o} }{\Geff m}
	&=
	-\frac{1+2\alpha_1\alpha_2}{2\Delta\alpha^2\nu}
	\frac{1}{2}\x^{-3},
	\allowdisplaybreaks\\
	\frac{ t_\DD^\sector{C}\scount{so} }{\Geff m}
	&=
	-\frac{2}{3}\mathcal{S}_\DD^\sector{C} \,\x^{-3/2},
	\\
	\frac{ t_\DD^\sector{D}\scount{so} }{\Geff m}
	&=
	\frac{2}{3}
	\mathcal{S}_\DD^\sector{D}
	\,\x^{3/2},
\end{align}
\end{subequations}
with the result for each constituent part given to leading order in~$\x$. Explicit expressions for all of the spin-orbit coefficients~$\mathcal{S}$ are listed in table~\ref{table:coefficients}. The above can now be substituted into \eqref{eq:rad_gw_phase_Fourier} along with~\eqref{eq:rad_DD_phase} to give us the gravitational-wave phase~$\tilde\Psi_{\yl\ym}(f)$. The result in the dipole-driven regime is schematically of the form
\begin{equation}
	\tilde\Psi^\vph{()}_\DD
	=
	(\ym \psi_0^\vph{()} + 2\pi f t_0^\vph{()})
	+
	\tilde\Psi_\DD^\vph{()}\scount{o}
	+
	\tilde\Psi_\DD^\sector{C}\scount{so}
	+
	\tilde\Psi_\DD^\sector{D}\scount{so},
\end{equation}
where we have dropped the labels $(\yl,\ym)$ on~$\tilde\Psi$ to declutter our notation. Written in terms of the dimensionless frequency variable ${ u_\ym \coloneq 2\pi\Geff m f/\ym }$, we find
\begin{subequations}
\label{eq:app_gw_DD_phase}
\begin{align}
	\tilde\Psi_\DD^\vph{()}\scount{o}
	&=
	- \frac{1+2\alpha_1\alpha_2}{2\Delta\alpha^2\nu}
	\frac{1}{2}\ym u_\ym^{-1},
	\allowdisplaybreaks\\
	\tilde\Psi_\DD^\sector{C}\scount{so}
	&=
	\frac{2}{3}\mathcal{S}_\DD^\sector{C}
	\,\ym (1+\log u_\ym),
	\\
	\tilde\Psi_\DD^\sector{D}\scount{so}
	&=
	-\frac{1}{3}
	\mathcal{S}_\DD^\sector{D}
	\,\ym u_\ym^2
\end{align}
\end{subequations}
at leading order in each sector. If desired, the phase for the dominant ${(2,2)}$~mode of the gravitational-wave signal can be obtained by simply setting~${\ym = 2}$.

\paragraph{Quadrupole-driven regime.}
For quadrupole-driven systems, integrating~\eqref{eq:rad_dtdx}~yields
\begin{align}
	t^\vph{()}_\QD
	=
	t_0^\vph{()}
	+
	\big(
		&
		t_{\QD,\nondip}^\vph{()}\scount{o}
		+
		t_{\QD,\nondip}^\sector{C}\scount{so}
		+
		t_{\QD,\nondip}^\sector{D}\scount{so}
	\big)
	\nonumber\\
	+\:
	\big(
		&
		t_{\QD,\dip}^\vph{()}\scount{o}
		\:+\:
		t_{\QD,\dip}^\sector{C}\scount{so}
		\:+\:
		t_{\QD,\dip}^\sector{D}\scount{so}
	\big),
\end{align}
where $t_0$ is the integration constant to be fixed by initial conditions, while
\begin{subequations}
\begin{align}
% [O]	
	&&
	\frac{ t_{\QD,\nondip}^\vph{()}\scount{o} }{\Geff m}
	&=
	-\frac{1+2\alpha_1\alpha_2}{32\zeta\nu}
	\frac{5}{8}\x^{-4},
&\qquad
	\frac{ t_{\QD,\dip}^\vph{()}\scount{o} }{\Geff m}
	&=
	\frac{1+2\alpha_1\alpha_2}{32\zeta\nu}
	\frac{25\Delta\alpha^2}{336\zeta}
	\frac{7}{10}\x^{-5},
&
\allowdisplaybreaks\\
% Conformal [SO]	
	&&
	\frac{ t_{\QD,\nondip}^\sector{C}\scount{so} }{\Geff m}
	&=
	-\frac{2}{5}\mathcal{S}_{\QD,\nondip}^\sector{C}
	\,\x^{-5/2},
	&
	\frac{ t_{\QD,\dip}^\sector{C}\scount{so} }{\Geff m}
	&=
	-\frac{2}{7}\mathcal{S}_{\QD,\dip}^\sector{C}
	\,\x^{-7/2},
&
\\
% Disformal [SO]	
	&&
	\frac{ t_{\QD,\nondip}^\sector{D}\scount{so} }{\Geff m}
	&=
	2\mathcal{S}_{\QD,\nondip}^\sector{D}
	\,\x^{1/2},
	&
	\frac{ t_{\QD,\dip}^\sector{D}\scount{so} }{\Geff m}
	&=
	-2\mathcal{S}_{\QD,\dip}^\sector{D}
	\,\x^{-1/2}.
&
\end{align}
\end{subequations}
As before, the result in each sector is given to leading order in~$\x$. Substituting this into~\eqref{eq:rad_gw_phase_Fourier} alongside~\eqref{eq:rad_QD_phase} returns the gravitational-wave~phase
\begin{align}
	\tilde\Psi^\vph{()}_\QD
	=
	(\ym \psi_0^\vph{()} + 2\pi f t_0^\vph{()})
	+
	\big(
		&
		\tilde\Psi_{\QD,\nondip}^\vph{()}\scount{o}
		+
		\tilde\Psi_{\QD,\nondip}^\sector{C}\scount{so}
		+
		\tilde\Psi_{\QD,\nondip}^\sector{D}\scount{so}
	\big)
	\nonumber\\
	+\:
	\big(
		&
		\tilde\Psi_{\QD,\dip}^\vph{()}\scount{o}
		\:+\:
		\tilde\Psi_{\QD,\dip}^\sector{C}\scount{so}
		\:+\:
		\tilde\Psi_{\QD,\dip}^\sector{D}\scount{so}
	\big),
\end{align}
and we have once again dropped the labels~${(\yl,\ym)}$ for notational convenience. At leading order, the six constituent parts of this phase are given by
\begin{subequations}
\label{eq:app_gw_QD_phase}
\begin{align}
% [O]	
	\tilde\Psi_{\QD,\nondip}^\vph{()}\scount{o}
	&=
	-\frac{1+2\alpha_1\alpha_2}{32\zeta\nu}
	\frac{3}{8}\ym u_\ym^{-5/3},
&\qquad
	\tilde\Psi_{\QD,\dip}^\vph{()}\scount{o}
	&=
	\frac{1+2\alpha_1\alpha_2}{32\zeta\nu}
	\frac{25\Delta\alpha^2}{336\zeta}
	\frac{3}{10}\ym u_\ym^{-7/3},
\allowdisplaybreaks\\
% Conformal [SO]	
	\tilde\Psi_{\QD,\nondip}^\sector{C}\scount{so}
	&=
	-\frac{3}{5}\mathcal{S}_{\QD,\nondip}^\sector{C}
	\,\ym u_\ym^{-2/3},
	&
	\tilde\Psi_{\QD,\dip}^\sector{C}\scount{so}
	&=
	-\frac{3}{14}\mathcal{S}_{\QD,\dip}^\sector{C}
	\,\ym u_\ym^{-4/3},
\\[3pt]
% Disformal [SO]	
	\tilde\Psi_{\QD,\nondip}^\sector{D}\scount{so}
	&=
	-\frac{3}{2}\mathcal{S}_{\QD,\nondip}^\sector{D}
	\,\ym u_\ym^{4/3},
	&
	\tilde\Psi_{\QD,\dip}^\sector{D}\scount{so}
	&=
	3\mathcal{S}_{\QD,\dip}^\sector{D}
	\,\ym u_\ym^{2/3}
\end{align}
\end{subequations}
when written in terms of the dimensionless variable ${ u_\ym \coloneq 2\pi\Geff m f/\ym }$. The phase for the dominant ${(2,2)}$~mode of the gravitational-wave signal simply follows from setting~${\ym = 2}$.

\bibliography{so}

\providecommand{\href}[2]{#2}\begingroup\raggedright\begin{thebibliography}{100}

\bibitem{Abbott:2016blz}
B.~P.~Abbott et~al. {(LIGO Scientific and Virgo Collaborations)},
  \emph{Observation of gravitational waves from a binary black hole merger},
  \href{https://doi.org/10.1103/PhysRevLett.116.061102}{Phys. Rev. Lett.
  {\bfseries 116} (2016) 061102}
  [\href{https://arxiv.org/abs/1602.03837}{{\ttfamily 1602.03837}}].

\bibitem{TheLIGOScientific:2016src}
B.~P.~Abbott et~al. {(LIGO Scientific and Virgo Collaborations)}, \emph{{Tests
  of general relativity with GW150914}},
  \href{https://doi.org/10.1103/PhysRevLett.116.221101}{Phys. Rev. Lett.
  {\bfseries 116} (2016) 221101}{ [Erratum ibid. \textbf{121} (2018) 129902]}
  [\href{https://arxiv.org/abs/1602.03841}{{\ttfamily 1602.03841}}].

\bibitem{LIGOScientific:2018dkp}
B.~P.~Abbott et~al. {(LIGO Scientific and Virgo Collaborations)}, \emph{Tests
  of general relativity with {GW170817}},
  \href{https://doi.org/10.1103/PhysRevLett.123.011102}{Phys. Rev. Lett.
  {\bfseries 123} (2019) 011102}
  [\href{https://arxiv.org/abs/1811.00364}{{\ttfamily 1811.00364}}].

\bibitem{LIGOScientific:2019fpa}
B.~P.~Abbott et~al. {(LIGO Scientific and Virgo Collaborations)}, \emph{Tests
  of general relativity with the binary black hole signals from the {LIGO-Virgo
  catalog GWTC-1}}, \href{https://doi.org/10.1103/PhysRevD.100.104036}{Phys.
  Rev. D {\bfseries 100} (2019) 104036}
  [\href{https://arxiv.org/abs/1903.04467}{{\ttfamily 1903.04467}}].

\bibitem{LIGOScientific:2020tif}
R.~Abbott et~al. {(LIGO Scientific and Virgo Collaborations)}, \emph{Tests of
  general relativity with binary black holes from the second {LIGO-Virgo
  gravitational-wave transient catalog}},
  \href{https://doi.org/10.1103/PhysRevD.103.122002}{Phys. Rev. D {\bfseries
  103} (2021) 122002} [\href{https://arxiv.org/abs/2010.14529}{{\ttfamily
  2010.14529}}].

\bibitem{Capozziello:2007ec}
S.~Capozziello and M.~Francaviglia, \emph{Extended theories of gravity and
  their cosmological and astrophysical applications},
  \href{https://doi.org/10.1007/s10714-007-0551-y}{Gen. Relativ. Gravit.
  {\bfseries 40} (2008) 357} [\href{https://arxiv.org/abs/0706.1146}{{\ttfamily
  0706.1146}}].

\bibitem{Capozziello:2011et}
S.~Capozziello and M.~De~Laurentis, \emph{Extended theories of gravity},
  \href{https://doi.org/10.1016/j.physrep.2011.09.003}{Phys. Rep. {\bfseries
  509} (2011) 167} [\href{https://arxiv.org/abs/1108.6266}{{\ttfamily
  1108.6266}}].

\bibitem{Clifton:2011jh}
T.~Clifton, P.~G.~Ferreira, A.~Padilla and C.~Skordis, \emph{Modified gravity
  and cosmology}, \href{https://doi.org/10.1016/j.physrep.2012.01.001}{Phys.
  Rep. {\bfseries 513} (2012) 1}
  [\href{https://arxiv.org/abs/1106.2476}{{\ttfamily 1106.2476}}].

\bibitem{Joyce:2014kja}
A.~Joyce, B.~Jain, J.~Khoury and M.~Trodden, \emph{Beyond the cosmological
  standard model}, \href{https://doi.org/10.1016/j.physrep.2014.12.002}{Phys.
  Rep. {\bfseries 568} (2015) 1}
  [\href{https://arxiv.org/abs/1407.0059}{{\ttfamily 1407.0059}}].

\bibitem{Bull:2015stt}
P.~{Bull}, Y.~{Akrami}, J.~{Adamek}, T.~{Baker}, E.~{Bellini}, J.~{Beltr{\'a}n
  Jim{\'e}nez} et~al., \emph{{Beyond $\Lambda\mathrm{CDM}$: Problems,
  solutions, and the road ahead}},
  \href{https://doi.org/10.1016/j.dark.2016.02.001}{Phys. Dark Univ. {\bfseries
  12} (2016) 56} [\href{https://arxiv.org/abs/1512.05356}{{\ttfamily
  1512.05356}}].

\bibitem{CANTATA:2021ktz}
E.~N.~{Saridakis}, R.~{Lazkoz}, V.~{Salzano}, P.~{Vargas Moniz},
  S.~{Capozziello}, J.~{Beltr{\'a}n Jim{\'e}nez} et~al. {(CANTATA
  Collaboration)}, \emph{Modified gravity and cosmology: {An} update by the
  {CANTATA} network},  \href{https://arxiv.org/abs/2105.12582}{{\ttfamily
  2105.12582}}.

\bibitem{Riess:1998cb}
A.~G.~{Riess}, A.~V.~{Filippenko}, P.~{Challis}, A.~{Clocchiatti},
  A.~{Diercks}, P.~M.~{Garnavich} et~al., \emph{Observational evidence from
  supernovae for an accelerating universe and a cosmological constant},
  \href{https://doi.org/10.1086/300499}{Astron. J. {\bfseries 116} (1998) 1009}
  [\href{https://arxiv.org/abs/astro-ph/9805201}{{\ttfamily
  astro-ph/9805201}}].

\bibitem{Perlmutter:1998np}
S.~{Perlmutter}, G.~{Aldering}, G.~{Goldhaber}, R.~A.~{Knop}, P.~{Nugent},
  P.~G.~{Castro} et~al. {(The Supernova Cosmology Project)},
  \emph{{Measurements of $\Omega$ and $\Lambda$ from 42 high-redshift
  supernovae}}, \href{https://doi.org/10.1086/307221}{Astrophys. J. {\bfseries
  517} (1999) 565} [\href{https://arxiv.org/abs/astro-ph/9812133}{{\ttfamily
  astro-ph/9812133}}].

\bibitem{Weinberg:1988cp}
S.~Weinberg, \emph{The cosmological constant problem},
  \href{https://doi.org/10.1103/RevModPhys.61.1}{Rev. Mod. Phys. {\bfseries 61}
  (1989) 1}.

\bibitem{Chiba:1999ka}
T.~Chiba, T.~Okabe and M.~Yamaguchi, \emph{{Kinetically driven quintessence}},
  \href{https://doi.org/10.1103/PhysRevD.62.023511}{Phys. Rev. D {\bfseries 62}
  (2000) 023511} [\href{https://arxiv.org/abs/astro-ph/9912463}{{\ttfamily
  astro-ph/9912463}}].

\bibitem{ArmendarizPicon:2000dh}
C.~Armendariz-Picon, V.~F.~Mukhanov and P.~J.~Steinhardt, \emph{Dynamical
  solution to the problem of a small cosmological constant and late-time cosmic
  acceleration}, \href{https://doi.org/10.1103/PhysRevLett.85.4438}{Phys. Rev.
  Lett. {\bfseries 85} (2000) 4438}
  [\href{https://arxiv.org/abs/astro-ph/0004134}{{\ttfamily
  astro-ph/0004134}}].

\bibitem{ArmendarizPicon:2000ah}
C.~Armendariz-Picon, V.~F.~Mukhanov and P.~J.~Steinhardt, \emph{{Essentials of
  $k$-essence}}, \href{https://doi.org/10.1103/PhysRevD.63.103510}{Phys. Rev. D
  {\bfseries 63} (2001) 103510}
  [\href{https://arxiv.org/abs/astro-ph/0006373}{{\ttfamily
  astro-ph/0006373}}].

\bibitem{Boisseau:2000pr}
B.~Boisseau, G.~Esposito-Far\`{e}se, D.~Polarski and A.~A.~Starobinsky,
  \emph{Reconstruction of a scalar-tensor theory of gravity in an accelerating
  universe}, \href{https://doi.org/10.1103/PhysRevLett.85.2236}{Phys. Rev.
  Lett. {\bfseries 85} (2000) 2236}
  [\href{https://arxiv.org/abs/gr-qc/0001066}{{\ttfamily gr-qc/0001066}}].

\bibitem{Copeland:2006wr}
E.~J.~Copeland, M.~Sami and S.~Tsujikawa, \emph{{Dynamics of dark energy}},
  \href{https://doi.org/10.1142/S021827180600942X}{Int. J. Mod. Phys. D
  {\bfseries 15} (2006) 1753}
  [\href{https://arxiv.org/abs/hep-th/0603057}{{\ttfamily hep-th/0603057}}].

\bibitem{Bamba:2012cp}
K.~Bamba, S.~Capozziello, S.~Nojiri and S.~D.~Odintsov, \emph{{Dark energy
  cosmology: The equivalent description via different theoretical models and
  cosmography tests}},
  \href{https://doi.org/10.1007/s10509-012-1181-8}{Astrophys. Space Sci.
  {\bfseries 342} (2012) 155}
  [\href{https://arxiv.org/abs/1205.3421}{{\ttfamily 1205.3421}}].

\bibitem{Gubitosi:2012hu}
G.~Gubitosi, F.~Piazza and F.~Vernizzi, \emph{The effective field theory of
  dark energy}, \href{https://doi.org/10.1088/1475-7516/2013/02/032}{JCAP
  {\bfseries 02} (2013) 032} [\href{https://arxiv.org/abs/1210.0201}{{\ttfamily
  1210.0201}}].

\bibitem{Bloomfield:2012ff}
J.~K.~Bloomfield, E.~E.~Flanagan, M.~Park and S.~Watson, \emph{{Dark energy or
  modified gravity? An effective field theory approach}},
  \href{https://doi.org/10.1088/1475-7516/2013/08/010}{JCAP {\bfseries 08}
  (2013) 010} [\href{https://arxiv.org/abs/1211.7054}{{\ttfamily 1211.7054}}].

\bibitem{Gleyzes:2014rba}
J.~Gleyzes, D.~Langlois and F.~Vernizzi, \emph{{A unifying description of dark
  energy}}, \href{https://doi.org/10.1142/S021827181443010X}{Int. J. Mod. Phys.
  D {\bfseries 23} (2015) 1443010}
  [\href{https://arxiv.org/abs/1411.3712}{{\ttfamily 1411.3712}}].

\bibitem{Bellini:2014fua}
E.~Bellini and I.~Sawicki, \emph{{Maximal freedom at minimum cost: Linear
  large-scale structure in general modifications of gravity}},
  \href{https://doi.org/10.1088/1475-7516/2014/07/050}{JCAP {\bfseries 07}
  (2014) 050} [\href{https://arxiv.org/abs/1404.3713}{{\ttfamily 1404.3713}}].

\bibitem{Sin:1992bg}
S.-J.~Sin, \emph{{Late time cosmological phase transition and galactic halo as
  Bose liquid}}, \href{https://doi.org/10.1103/PhysRevD.50.3650}{Phys. Rev. D
  {\bfseries 50} (1994) 3650}
  [\href{https://arxiv.org/abs/hep-ph/9205208}{{\ttfamily hep-ph/9205208}}].

\bibitem{Hu:2000ke}
W.~Hu, R.~Barkana and A.~Gruzinov, \emph{{Cold and fuzzy dark matter}},
  \href{https://doi.org/10.1103/PhysRevLett.85.1158}{Phys. Rev. Lett.
  {\bfseries 85} (2000) 1158}
  [\href{https://arxiv.org/abs/astro-ph/0003365}{{\ttfamily
  astro-ph/0003365}}].

\bibitem{Burgess:2000yq}
C.~P.~Burgess, M.~Pospelov and T.~ter Veldhuis, \emph{{The minimal model of
  nonbaryonic dark matter: A singlet scalar}},
  \href{https://doi.org/10.1016/S0550-3213(01)00513-2}{Nucl. Phys. B {\bfseries
  619} (2001) 709} [\href{https://arxiv.org/abs/hep-ph/0011335}{{\ttfamily
  hep-ph/0011335}}].

\bibitem{Bekenstein:2004ne}
J.~D.~Bekenstein, \emph{{Relativistic gravitation theory for the MOND
  paradigm}}, \href{https://doi.org/10.1103/PhysRevD.70.083509}{Phys. Rev. D
  {\bfseries 70} (2004) 083509}[Erratum: Phys.Rev.D 71, 069901 (2005)]
  [\href{https://arxiv.org/abs/astro-ph/0403694}{{\ttfamily
  astro-ph/0403694}}].

\bibitem{Hui:2016ltb}
L.~Hui, J.~P.~Ostriker, S.~Tremaine and E.~Witten, \emph{{Ultralight scalars as
  cosmological dark matter}},
  \href{https://doi.org/10.1103/PhysRevD.95.043541}{Phys. Rev. D {\bfseries 95}
  (2017) 043541} [\href{https://arxiv.org/abs/1610.08297}{{\ttfamily
  1610.08297}}].

\bibitem{Urena-Lopez:2019kud}
L.~A.~Ureña-López, \emph{Brief review on scalar field dark matter models},
  \href{https://doi.org/10.3389/fspas.2019.00047}{Front. Astron. Space Sci.
  {\bfseries 6} (2019) 47}.

\bibitem{Hui:2021tkt}
L.~Hui, \emph{Wave dark matter},
  \href{https://arxiv.org/abs/2101.11735}{{\ttfamily 2101.11735}}.

\bibitem{Burrage:2018zuj}
C.~Burrage, E.~J.~Copeland, C.~K\"ading and P.~Millington, \emph{{Symmetron
  scalar fields: Modified gravity, dark matter, or both?}},
  \href{https://doi.org/10.1103/PhysRevD.99.043539}{Phys. Rev. D {\bfseries 99}
  (2019) 043539} [\href{https://arxiv.org/abs/1811.12301}{{\ttfamily
  1811.12301}}].

\bibitem{Brax:2020gqg}
P.~Brax, K.~Kaneta, Y.~Mambrini and M.~Pierre, \emph{{Disformal dark matter}},
  \href{https://doi.org/10.1103/PhysRevD.103.015028}{Phys. Rev. D {\bfseries
  103} (2021) 015028} [\href{https://arxiv.org/abs/2011.11647}{{\ttfamily
  2011.11647}}].

\bibitem{Damour:1993hw}
T.~Damour and G.~Esposito-Far{\`e}se, \emph{{Nonperturbative strong-field
  effects in tensor-scalar theories of gravitation}},
  \href{https://doi.org/10.1103/PhysRevLett.70.2220}{Phys. Rev. Lett.
  {\bfseries 70} (1993) 2220}.

\bibitem{Damour:1996ke}
T.~Damour and G.~Esposito-Far{\`e}se, \emph{{Tensor-scalar gravity and
  binary-pulsar experiments}},
  \href{https://doi.org/10.1103/PhysRevD.54.1474}{Phys. Rev. D {\bfseries 54}
  (1996) 1474} [\href{https://arxiv.org/abs/gr-qc/9602056}{{\ttfamily
  gr-qc/9602056}}].

\bibitem{Minamitsuji:2016hkk}
M.~Minamitsuji and H.~O.~Silva, \emph{{Relativistic stars in scalar-tensor
  theories with disformal coupling}},
  \href{https://doi.org/10.1103/PhysRevD.93.124041}{Phys. Rev. D {\bfseries 93}
  (2016) 124041} [\href{https://arxiv.org/abs/1604.07742}{{\ttfamily
  1604.07742}}].

\bibitem{Silva:2017uqg}
H.~O.~Silva, J.~Sakstein, L.~Gualtieri, T.~P.~Sotiriou and E.~Berti,
  \emph{{Spontaneous scalarization of black holes and compact stars from a
  Gauss-Bonnet coupling}},
  \href{https://doi.org/10.1103/PhysRevLett.120.131104}{Phys. Rev. Lett.
  {\bfseries 120} (2018) 131104}
  [\href{https://arxiv.org/abs/1711.02080}{{\ttfamily 1711.02080}}].

\bibitem{Doneva:2017duq}
D.~D.~Doneva and S.~S.~Yazadjiev, \emph{{Neutron star solutions with curvature
  induced scalarization in the extended Gauss-Bonnet scalar-tensor theories}},
  \href{https://doi.org/10.1088/1475-7516/2018/04/011}{JCAP {\bfseries 04}
  (2018) 011} [\href{https://arxiv.org/abs/1712.03715}{{\ttfamily
  1712.03715}}].

\bibitem{Andreou:2019ikc}
N.~Andreou, N.~Franchini, G.~Ventagli and T.~P.~Sotiriou, \emph{{Spontaneous
  scalarization in generalised scalar-tensor theory}},
  \href{https://doi.org/10.1103/PhysRevD.99.124022}{Phys. Rev. D {\bfseries 99}
  (2019) 124022}{ [Erratum ibid. \textbf{101} (2020) 109903]}
  [\href{https://arxiv.org/abs/1904.06365}{{\ttfamily 1904.06365}}].

\bibitem{Ventagli:2020rnx}
G.~Ventagli, A.~Leh\'ebel and T.~P.~Sotiriou, \emph{{Onset of spontaneous
  scalarization in generalized scalar-tensor theories}},
  \href{https://doi.org/10.1103/PhysRevD.102.024050}{Phys. Rev. D {\bfseries
  102} (2020) 024050} [\href{https://arxiv.org/abs/2006.01153}{{\ttfamily
  2006.01153}}].

\bibitem{Shao:2017gwu}
L.~Shao, N.~Sennett, A.~Buonanno, M.~Kramer and N.~Wex, \emph{{Constraining
  nonperturbative strong-field effects in scalar-tensor gravity by combining
  pulsar timing and laser-interferometer gravitational-wave detectors}},
  \href{https://doi.org/10.1103/PhysRevX.7.041025}{Phys. Rev. X {\bfseries 7}
  (2017) 041025} [\href{https://arxiv.org/abs/1704.07561}{{\ttfamily
  1704.07561}}].

\bibitem{Zhao:2019suc}
J.~Zhao, L.~Shao, Z.~Cao and B.-Q.~Ma, \emph{{Reduced-order surrogate models
  for scalar-tensor gravity in the strong field regime and applications to
  binary pulsars and GW170817}},
  \href{https://doi.org/10.1103/PhysRevD.100.064034}{Phys. Rev. D {\bfseries
  100} (2019) 064034} [\href{https://arxiv.org/abs/1907.00780}{{\ttfamily
  1907.00780}}].

\bibitem{Guo:2021leu}
M.~Guo, J.~Zhao and L.~Shao, \emph{{Extended reduced-order surrogate models for
  scalar-tensor gravity in the strong field and applications to binary pulsars
  and gravitational waves}},
  \href{https://arxiv.org/abs/2106.01622}{{\ttfamily 2106.01622}}.

\bibitem{Yagi:2021loe}
K.~Yagi and M.~Stepniczka, \emph{Neutron stars in scalar-tensor theories:
  {Analytic} scalar charges and universal relation},
  \href{https://arxiv.org/abs/2105.01614}{{\ttfamily 2105.01614}}.

\bibitem{Kanti:1995vq}
P.~Kanti, N.~E.~Mavromatos, J.~Rizos, K.~Tamvakis and E.~Winstanley,
  \emph{{Dilatonic black holes in higher curvature string gravity}},
  \href{https://doi.org/10.1103/PhysRevD.54.5049}{Phys. Rev. D {\bfseries 54}
  (1996) 5049} [\href{https://arxiv.org/abs/hep-th/9511071}{{\ttfamily
  hep-th/9511071}}].

\bibitem{Kleihaus:2011tg}
B.~Kleihaus, J.~Kunz and E.~Radu, \emph{Rotating black holes in dilatonic
  {Einstein-Gauss-Bonnet} theory},
  \href{https://doi.org/10.1103/PhysRevLett.106.151104}{Phys. Rev. Lett.
  {\bfseries 106} (2011) 151104}
  [\href{https://arxiv.org/abs/1101.2868}{{\ttfamily 1101.2868}}].

\bibitem{Pani:2011gy}
P.~Pani, C.~F.~B.~Macedo, L.~C.~B.~Crispino and V.~Cardoso, \emph{{Slowly
  rotating black holes in alternative theories of gravity}},
  \href{https://doi.org/10.1103/PhysRevD.84.087501}{Phys. Rev. D {\bfseries 84}
  (2011) 087501} [\href{https://arxiv.org/abs/1109.3996}{{\ttfamily
  1109.3996}}].

\bibitem{Yunes:2011we}
N.~Yunes and L.~C.~Stein, \emph{{Nonspinning black holes in alternative
  theories of gravity}},
  \href{https://doi.org/10.1103/PhysRevD.83.104002}{Phys. Rev. D {\bfseries 83}
  (2011) 104002} [\href{https://arxiv.org/abs/1101.2921}{{\ttfamily
  1101.2921}}].

\bibitem{Maselli:2015tta}
A.~Maselli, P.~Pani, L.~Gualtieri and V.~Ferrari, \emph{Rotating black holes in
  {Einstein-dilaton-Gauss-Bonnet} gravity with finite coupling},
  \href{https://doi.org/10.1103/PhysRevD.92.083014}{Phys. Rev. D {\bfseries 92}
  (2015) 083014} [\href{https://arxiv.org/abs/1507.00680}{{\ttfamily
  1507.00680}}].

\bibitem{Sotiriou:2014pfa}
T.~P.~Sotiriou and S.-Y.~Zhou, \emph{{Black hole hair in generalized
  scalar-tensor gravity: An explicit example}},
  \href{https://doi.org/10.1103/PhysRevD.90.124063}{Phys. Rev. D {\bfseries 90}
  (2014) 124063} [\href{https://arxiv.org/abs/1408.1698}{{\ttfamily
  1408.1698}}].

\bibitem{Antoniou:2017hxj}
G.~Antoniou, A.~Bakopoulos and P.~Kanti, \emph{{Black-hole solutions with
  scalar hair in Einstein-scalar-Gauss-Bonnet theories}},
  \href{https://doi.org/10.1103/PhysRevD.97.084037}{Phys. Rev. D {\bfseries 97}
  (2018) 084037} [\href{https://arxiv.org/abs/1711.07431}{{\ttfamily
  1711.07431}}].

\bibitem{Delgado:2020rev}
J.~F.~M.~Delgado, C.~A.~R.~Herdeiro and E.~Radu, \emph{{Spinning black holes in
  shift-symmetric Horndeski theory}},
  \href{https://doi.org/10.1007/JHEP04(2020)180}{JHEP {\bfseries 04} (2020)
  180} [\href{https://arxiv.org/abs/2002.05012}{{\ttfamily 2002.05012}}].

\bibitem{Sullivan:2020zpf}
A.~Sullivan, N.~Yunes and T.~P.~Sotiriou, \emph{{Numerical black hole solutions
  in modified gravity theories: Axial symmetry case}},
  \href{https://doi.org/10.1103/PhysRevD.103.124058}{Phys. Rev. D {\bfseries
  103} (2021) 124058} [\href{https://arxiv.org/abs/2009.10614}{{\ttfamily
  2009.10614}}].

\bibitem{Doneva:2017bvd}
D.~D.~Doneva and S.~S.~Yazadjiev, \emph{New {Gauss-Bonnet} black holes with
  curvature-induced scalarization in extended scalar-tensor theories},
  \href{https://doi.org/10.1103/PhysRevLett.120.131103}{Phys. Rev. Lett.
  {\bfseries 120} (2018) 131103}
  [\href{https://arxiv.org/abs/1711.01187}{{\ttfamily 1711.01187}}].

\bibitem{Cunha:2019dwb}
P.~V.~P.~Cunha, C.~A.~R.~Herdeiro and E.~Radu, \emph{Spontaneously scalarized
  {Kerr} black holes in extended scalar-tensor-{Gauss-Bonnet} gravity},
  \href{https://doi.org/10.1103/PhysRevLett.123.011101}{Phys. Rev. Lett.
  {\bfseries 123} (2019) 011101}
  [\href{https://arxiv.org/abs/1904.09997}{{\ttfamily 1904.09997}}].

\bibitem{Minamitsuji:2019iwp}
M.~Minamitsuji and T.~Ikeda, \emph{{Spontaneous scalarization of black holes in
  the Horndeski theory}},
  \href{https://doi.org/10.1103/PhysRevD.99.104069}{Phys. Rev. D {\bfseries 99}
  (2019) 104069} [\href{https://arxiv.org/abs/1904.06572}{{\ttfamily
  1904.06572}}].

\bibitem{Herdeiro:2020wei}
C.~A.~R.~Herdeiro, E.~Radu, H.~O.~Silva, T.~P.~Sotiriou and N.~Yunes,
  \emph{{Spin-induced scalarized black holes}},
  \href{https://doi.org/10.1103/PhysRevLett.126.011103}{Phys. Rev. Lett.
  {\bfseries 126} (2021) 011103}
  [\href{https://arxiv.org/abs/2009.03904}{{\ttfamily 2009.03904}}].

\bibitem{Berti:2020kgk}
E.~Berti, L.~G.~Collodel, B.~Kleihaus and J.~Kunz, \emph{{Spin-induced
  black-hole scalarization in Einstein-scalar-Gauss-Bonnet theory}},
  \href{https://doi.org/10.1103/PhysRevLett.126.011104}{Phys. Rev. Lett.
  {\bfseries 126} (2021) 011104}
  [\href{https://arxiv.org/abs/2009.03905}{{\ttfamily 2009.03905}}].

\bibitem{East:2021bqk}
W.~E.~East and J.~L.~Ripley, \emph{{Dynamics of spontaneous black hole
  scalarization and mergers in Einstein-scalar-Gauss-Bonnet gravity}},
  \href{https://arxiv.org/abs/2105.08571}{{\ttfamily 2105.08571}}.

\bibitem{Yagi:2011xp}
K.~Yagi, L.~C.~Stein, N.~Yunes and T.~Tanaka, \emph{{Post-Newtonian,
  quasicircular binary inspirals in quadratic modified gravity}},
  \href{https://doi.org/10.1103/PhysRevD.85.064022}{Phys. Rev. D {\bfseries 85}
  (2012) 064022}{ [Erratum ibid. \textbf{93} (2016) 029902]}
  [\href{https://arxiv.org/abs/1110.5950}{{\ttfamily 1110.5950}}].

\bibitem{Julie:2019sab}
F.-L.~Juli\'e and E.~Berti, \emph{{Post-Newtonian dynamics and black hole
  thermodynamics in Einstein-scalar-Gauss-Bonnet gravity}},
  \href{https://doi.org/10.1103/PhysRevD.100.104061}{Phys. Rev. D {\bfseries
  100} (2019) 104061} [\href{https://arxiv.org/abs/1909.05258}{{\ttfamily
  1909.05258}}].

\bibitem{Shiralilou:2020gah}
B.~Shiralilou, T.~Hinderer, S.~Nissanke, N.~Ortiz and H.~Witek,
  \emph{{Nonlinear curvature effects in gravitational waves from inspiralling
  black hole binaries}},
  \href{https://doi.org/10.1103/PhysRevD.103.L121503}{Phys. Rev. D {\bfseries
  103} (2021) L121503} [\href{https://arxiv.org/abs/2012.09162}{{\ttfamily
  2012.09162}}].

\bibitem{Shiralilou:2021mfl}
B.~Shiralilou, T.~Hinderer, S.~Nissanke, N.~Ortiz and H.~Witek,
  \emph{{Post-Newtonian gravitational and scalar waves in scalar-Gauss-Bonnet
  gravity}},  \href{https://arxiv.org/abs/2105.13972}{{\ttfamily 2105.13972}}.

\bibitem{Barausse:2012da}
E.~Barausse, C.~Palenzuela, M.~Ponce and L.~Lehner, \emph{{Neutron-star mergers
  in scalar-tensor theories of gravity}},
  \href{https://doi.org/10.1103/PhysRevD.87.081506}{Phys. Rev. D {\bfseries 87}
  (2013) 081506} [\href{https://arxiv.org/abs/1212.5053}{{\ttfamily
  1212.5053}}].

\bibitem{Palenzuela:2013hsa}
C.~Palenzuela, E.~Barausse, M.~Ponce and L.~Lehner, \emph{{Dynamical
  scalarization of neutron stars in scalar-tensor gravity theories}},
  \href{https://doi.org/10.1103/PhysRevD.89.044024}{Phys. Rev. D {\bfseries 89}
  (2014) 044024} [\href{https://arxiv.org/abs/1310.4481}{{\ttfamily
  1310.4481}}].

\bibitem{Shibata:2013pra}
M.~Shibata, K.~Taniguchi, H.~Okawa and A.~Buonanno, \emph{{Coalescence of
  binary neutron stars in a scalar-tensor theory of gravity}},
  \href{https://doi.org/10.1103/PhysRevD.89.084005}{Phys. Rev. D {\bfseries 89}
  (2014) 084005} [\href{https://arxiv.org/abs/1310.0627}{{\ttfamily
  1310.0627}}].

\bibitem{Taniguchi:2014fqa}
K.~Taniguchi, M.~Shibata and A.~Buonanno, \emph{{Quasiequilibrium sequences of
  binary neutron stars undergoing dynamical scalarization}},
  \href{https://doi.org/10.1103/PhysRevD.91.024033}{Phys. Rev. D {\bfseries 91}
  (2015) 024033} [\href{https://arxiv.org/abs/1410.0738}{{\ttfamily
  1410.0738}}].

\bibitem{Sennett:2016rwa}
N.~Sennett and A.~Buonanno, \emph{{Modeling dynamical scalarization with a
  resummed post-Newtonian expansion}},
  \href{https://doi.org/10.1103/PhysRevD.93.124004}{Phys. Rev. D {\bfseries 93}
  (2016) 124004} [\href{https://arxiv.org/abs/1603.03300}{{\ttfamily
  1603.03300}}].

\bibitem{Sennett:2017lcx}
N.~Sennett, L.~Shao and J.~Steinhoff, \emph{{Effective action model of
  dynamically scalarizing binary neutron stars}},
  \href{https://doi.org/10.1103/PhysRevD.96.084019}{Phys. Rev. D {\bfseries 96}
  (2017) 084019} [\href{https://arxiv.org/abs/1708.08285}{{\ttfamily
  1708.08285}}].

\bibitem{Khalil:2019wyy}
M.~Khalil, N.~Sennett, J.~Steinhoff and A.~Buonanno, \emph{{Theory-agnostic
  framework for dynamical scalarization of compact binaries}},
  \href{https://doi.org/10.1103/PhysRevD.100.124013}{Phys. Rev. D {\bfseries
  100} (2019) 124013} [\href{https://arxiv.org/abs/1906.08161}{{\ttfamily
  1906.08161}}].

\bibitem{Silva:2020omi}
H.~O.~Silva, H.~Witek, M.~Elley and N.~Yunes, \emph{Dynamical descalarization
  in binary black hole mergers},
  \href{https://doi.org/10.1103/PhysRevLett.127.031101}{Phys. Rev. Lett.
  {\bfseries 127} (2021) 031101}
  [\href{https://arxiv.org/abs/2012.10436}{{\ttfamily 2012.10436}}].

\bibitem{Bekenstein:1992pj}
J.~D.~Bekenstein, \emph{Relation between physical and gravitational geometry},
  \href{https://doi.org/10.1103/PhysRevD.48.3641}{Phys. Rev. D {\bfseries 48}
  (1993) 3641} [\href{https://arxiv.org/abs/gr-qc/9211017}{{\ttfamily
  gr-qc/9211017}}].

\bibitem{Bertotti:2003rm}
B.~{Bertotti}, L.~{Iess} and P.~{Tortora}, \emph{{A test of general relativity
  using radio links with the Cassini spacecraft}},
  \href{https://doi.org/10.1038/nature01997}{Nature {\bfseries 425} (2003)
  374}.

\bibitem{Hofmann:2018llr}
F.~{Hofmann} and J.~{M{\"u}ller}, \emph{{Relativistic tests with lunar laser
  ranging}}, \href{https://doi.org/10.1088/1361-6382/aa8f7a}{Class. Quantum
  Gravity {\bfseries 35} (2018) 035015}.

\bibitem{Adelberger:2009zz}
E.~G.~Adelberger, J.~H.~Gundlach, B.~R.~Heckel, S.~Hoedl and S.~Schlamminger,
  \emph{{Torsion balance experiments: A low-energy frontier of particle
  physics}}, \href{https://doi.org/10.1016/j.ppnp.2008.08.002}{Prog. Part.
  Nucl. Phys. {\bfseries 62} (2009) 102}.

\bibitem{Burrage:2017qrf}
C.~Burrage and J.~Sakstein, \emph{Tests of chameleon gravity},
  \href{https://doi.org/10.1007/s41114-018-0011-x}{Living Rev. Relativity
  {\bfseries 21} (2018) 1} [\href{https://arxiv.org/abs/1709.09071}{{\ttfamily
  1709.09071}}].

\bibitem{Brax:2018iyo}
P.~Brax, C.~Burrage and A.-C.~Davis, \emph{{Laboratory constraints}},
  \href{https://doi.org/10.1142/S0218271818480097}{Int. J. Mod. Phys. D
  {\bfseries 27} (2018) 1848009}.

\bibitem{Brax:2018zfb}
P.~Brax, A.-C.~Davis, B.~Elder and L.~K.~Wong, \emph{{Constraining screened
  fifth forces with the electron magnetic moment}},
  \href{https://doi.org/10.1103/PhysRevD.97.084050}{Phys. Rev. D {\bfseries 97}
  (2018) 084050} [\href{https://arxiv.org/abs/1802.05545}{{\ttfamily
  1802.05545}}].

\bibitem{Berge:2017ovy}
J.~Berg\'e, P.~Brax, G.~M\'etris, M.~Pernot-Borr\`as, P.~Touboul and
  J.-P.~Uzan, \emph{{MICROSCOPE} mission: {F}irst constraints on the violation
  of the weak equivalence principle by a light scalar dilaton},
  \href{https://doi.org/10.1103/PhysRevLett.120.141101}{Phys. Rev. Lett.
  {\bfseries 120} (2018) 141101}
  [\href{https://arxiv.org/abs/1712.00483}{{\ttfamily 1712.00483}}].

\bibitem{Sakstein:2018fwz}
J.~Sakstein, \emph{{Astrophysical tests of screened modified gravity}},
  \href{https://doi.org/10.1142/S0218271818480085}{Int. J. Mod. Phys. D
  {\bfseries 27} (2018) 1848008}
  [\href{https://arxiv.org/abs/2002.04194}{{\ttfamily 2002.04194}}].

\bibitem{Naik:2019moz}
A.~P.~Naik, E.~Puchwein, A.-C.~Davis, D.~Sijacki and H.~Desmond,
  \emph{Constraints on chameleon {$f(R)$}-gravity from galaxy rotation curves
  of the {SPARC} sample}, \href{https://doi.org/10.1093/mnras/stz2131}{Mon.
  Not. R. Astron. Soc. {\bfseries 489} (2019) 771}
  [\href{https://arxiv.org/abs/1905.13330}{{\ttfamily 1905.13330}}].

\bibitem{Desmond:2020gzn}
H.~Desmond and P.~G.~Ferreira, \emph{{Galaxy morphology rules out
  astrophysically relevant Hu-Sawicki $f(R)$ gravity}},
  \href{https://doi.org/10.1103/PhysRevD.102.104060}{Phys. Rev. D {\bfseries
  102} (2020) 104060} [\href{https://arxiv.org/abs/2009.08743}{{\ttfamily
  2009.08743}}].

\bibitem{Hees:2017aal}
A.~Hees, T.~Do, A.~M.~Ghez, G.~D.~Martinez, S.~Naoz, E.~E.~Becklin et~al.,
  \emph{Testing general relativity with stellar orbits around the supermassive
  black hole in our {Galactic} center},
  \href{https://doi.org/10.1103/PhysRevLett.118.211101}{Phys. Rev. Lett.
  {\bfseries 118} (2017) 211101}
  [\href{https://arxiv.org/abs/1705.07902}{{\ttfamily 1705.07902}}].

\bibitem{Kramer:2006nb}
M.~{Kramer}, I.~H.~{Stairs}, R.~N.~{Manchester}, M.~A.~{McLaughlin},
  A.~G.~{Lyne}, R.~D.~{Ferdman} et~al., \emph{{Tests of general relativity from
  timing the double pulsar}},
  \href{https://doi.org/10.1126/science.1132305}{Science {\bfseries 314} (2006)
  97} [\href{https://arxiv.org/abs/astro-ph/0609417}{{\ttfamily
  astro-ph/0609417}}].

\bibitem{Freire:2012mg}
P.~C.~C.~Freire, N.~Wex, G.~Esposito-Far\`{e}se, J.~P.~W.~Verbiest, M.~Bailes,
  B.~A.~Jacoby et~al., \emph{{The relativistic pulsar-white dwarf binary PSR
  J1738+0333 II. The most stringent test of scalar-tensor gravity}},
  \href{https://doi.org/10.1111/j.1365-2966.2012.21253.x}{Mon. Not. R. Astron.
  Soc. {\bfseries 423} (2012) 3328}
  [\href{https://arxiv.org/abs/1205.1450}{{\ttfamily 1205.1450}}].

\bibitem{Seymour:2019tir}
B.~C.~Seymour and K.~Yagi, \emph{Probing massive scalar fields from a pulsar in
  a stellar triple system},
  \href{https://doi.org/10.1088/1361-6382/ab9933}{Class. Quantum Gravity
  {\bfseries 37} (2020) 145008}
  [\href{https://arxiv.org/abs/1908.03353}{{\ttfamily 1908.03353}}].

\bibitem{Koyama:2015vza}
K.~Koyama, \emph{Cosmological tests of modified gravity},
  \href{https://doi.org/10.1088/0034-4885/79/4/046902}{Rept. Prog. Phys.
  {\bfseries 79} (2016) 046902}
  [\href{https://arxiv.org/abs/1504.04623}{{\ttfamily 1504.04623}}].

\bibitem{Joyce:2016vqv}
A.~Joyce, L.~Lombriser and F.~Schmidt, \emph{Dark energy versus modified
  gravity}, \href{https://doi.org/10.1146/annurev-nucl-102115-044553}{Annu.
  Rev. Nucl. Part. Sci. {\bfseries 66} (2016) 95}
  [\href{https://arxiv.org/abs/1601.06133}{{\ttfamily 1601.06133}}].

\bibitem{Ferreira:2019xrr}
P.~G.~Ferreira, \emph{Cosmological tests of gravity},
  \href{https://doi.org/10.1146/annurev-astro-091918-104423}{Ann. Rev. Astron.
  Astrophys. {\bfseries 57} (2019) 335}
  [\href{https://arxiv.org/abs/1902.10503}{{\ttfamily 1902.10503}}].

\bibitem{Will:2014kxa}
C.~M.~Will, \emph{The confrontation between general relativity and experiment},
  \href{https://doi.org/10.12942/lrr-2014-4}{Living Rev. Relativity {\bfseries
  17} (2014) 4} [\href{https://arxiv.org/abs/1403.7377}{{\ttfamily
  1403.7377}}].

\bibitem{Berti:2015itd}
E.~{Berti}, E.~{Barausse}, V.~{Cardoso}, L.~{Gualtieri}, P.~{Pani},
  U.~{Sperhake} et~al., \emph{Testing general relativity with present and
  future astrophysical observations},
  \href{https://doi.org/10.1088/0264-9381/32/24/243001}{Class. Quantum Gravity
  {\bfseries 32} (2015) 243001}
  [\href{https://arxiv.org/abs/1501.07274}{{\ttfamily 1501.07274}}].

\bibitem{Damour:PDGReview}
T.~Damour, \emph{Experimental tests of gravitational theory},
\newblock {in \emph{Review of Particle Physics}, P. A. Zyla et al. (Particle
  Data Group), \href{https://doi.org/10.1093/ptep/ptaa104}{ Prog. Theor. Exp.
  Phys. \textbf{2020} (2020) 083C01}.}

\bibitem{Horndeski:1974wa}
G.~W.~Horndeski, \emph{{Second-order scalar-tensor field equations in a
  four-dimensional space}}, \href{https://doi.org/10.1007/BF01807638}{Int. J.
  Theor. Phys. {\bfseries 10} (1974) 363}.

\bibitem{Deffayet:2009wt}
C.~Deffayet, G.~Esposito-Far\`{e}se and A.~Vikman, \emph{{Covariant Galileon}},
  \href{https://doi.org/10.1103/PhysRevD.79.084003}{Phys. Rev. D {\bfseries 79}
  (2009) 084003} [\href{https://arxiv.org/abs/0901.1314}{{\ttfamily
  0901.1314}}].

\bibitem{Zumalacarregui:2012us}
M.~Zumalac\'arregui, T.~S.~Koivisto and D.~F.~Mota, \emph{{DBI Galileons in the
  Einstein frame: Local gravity and cosmology}},
  \href{https://doi.org/10.1103/PhysRevD.87.083010}{Phys. Rev. D {\bfseries 87}
  (2013) 083010} [\href{https://arxiv.org/abs/1210.8016}{{\ttfamily
  1210.8016}}].

\bibitem{Bettoni:2013diz}
D.~Bettoni and S.~Liberati, \emph{{Disformal invariance of second order
  scalar-tensor theories: Framing the Horndeski action}},
  \href{https://doi.org/10.1103/PhysRevD.88.084020}{Phys. Rev. D {\bfseries 88}
  (2013) 084020} [\href{https://arxiv.org/abs/1306.6724}{{\ttfamily
  1306.6724}}].

\bibitem{Zumalacarregui:2013pma}
M.~Zumalac\'arregui and J.~Garc\'\i{}a-Bellido, \emph{{Transforming gravity:
  From derivative couplings to matter to second-order scalar-tensor theories
  beyond the Horndeski Lagrangian}},
  \href{https://doi.org/10.1103/PhysRevD.89.064046}{Phys. Rev. D {\bfseries 89}
  (2014) 064046} [\href{https://arxiv.org/abs/1308.4685}{{\ttfamily
  1308.4685}}].

\bibitem{Gleyzes:2014dya}
J.~Gleyzes, D.~Langlois, F.~Piazza and F.~Vernizzi, \emph{{Healthy theories
  beyond Horndeski}},
  \href{https://doi.org/10.1103/PhysRevLett.114.211101}{Phys. Rev. Lett.
  {\bfseries 114} (2015) 211101}
  [\href{https://arxiv.org/abs/1404.6495}{{\ttfamily 1404.6495}}].

\bibitem{Gleyzes:2014qga}
J.~Gleyzes, D.~Langlois, F.~Piazza and F.~Vernizzi, \emph{{Exploring
  gravitational theories beyond Horndeski}},
  \href{https://doi.org/10.1088/1475-7516/2015/02/018}{JCAP {\bfseries 02}
  (2015) 018} [\href{https://arxiv.org/abs/1408.1952}{{\ttfamily 1408.1952}}].

\bibitem{Langlois:2015cwa}
D.~Langlois and K.~Noui, \emph{{Degenerate higher derivative theories beyond
  Horndeski: Evading the Ostrogradski instability}},
  \href{https://doi.org/10.1088/1475-7516/2016/02/034}{JCAP {\bfseries 02}
  (2016) 034} [\href{https://arxiv.org/abs/1510.06930}{{\ttfamily
  1510.06930}}].

\bibitem{Crisostomi:2016tcp}
M.~Crisostomi, M.~Hull, K.~Koyama and G.~Tasinato, \emph{{Horndeski: Beyond, or
  not beyond?}}, \href{https://doi.org/10.1088/1475-7516/2016/03/038}{JCAP
  {\bfseries 03} (2016) 038}
  [\href{https://arxiv.org/abs/1601.04658}{{\ttfamily 1601.04658}}].

\bibitem{Crisostomi:2016czh}
M.~Crisostomi, K.~Koyama and G.~Tasinato, \emph{Extended scalar-tensor theories
  of gravity}, \href{https://doi.org/10.1088/1475-7516/2016/04/044}{JCAP
  {\bfseries 04} (2016) 044}
  [\href{https://arxiv.org/abs/1602.03119}{{\ttfamily 1602.03119}}].

\bibitem{BenAchour:2016cay}
J.~Ben~Achour, D.~Langlois and K.~Noui, \emph{{Degenerate higher order
  scalar-tensor theories beyond Horndeski and disformal transformations}},
  \href{https://doi.org/10.1103/PhysRevD.93.124005}{Phys. Rev. D {\bfseries 93}
  (2016) 124005} [\href{https://arxiv.org/abs/1602.08398}{{\ttfamily
  1602.08398}}].

\bibitem{BenAchour:2016fzp}
J.~Ben~Achour, M.~Crisostomi, K.~Koyama, D.~Langlois, K.~Noui and G.~Tasinato,
  \emph{{Degenerate higher order scalar-tensor theories beyond Horndeski up to
  cubic order}}, \href{https://doi.org/10.1007/JHEP12(2016)100}{JHEP {\bfseries
  12} (2016) 100} [\href{https://arxiv.org/abs/1608.08135}{{\ttfamily
  1608.08135}}].

\bibitem{deRham:2010ik}
C.~de~Rham and G.~Gabadadze, \emph{{Generalization of the Fierz-Pauli action}},
  \href{https://doi.org/10.1103/PhysRevD.82.044020}{Phys. Rev. D {\bfseries 82}
  (2010) 044020} [\href{https://arxiv.org/abs/1007.0443}{{\ttfamily
  1007.0443}}].

\bibitem{deRham:2010kj}
C.~de~Rham, G.~Gabadadze and A.~J.~Tolley, \emph{Resummation of massive
  gravity}, \href{https://doi.org/10.1103/PhysRevLett.106.231101}{Phys. Rev.
  Lett. {\bfseries 106} (2011) 231101}
  [\href{https://arxiv.org/abs/1011.1232}{{\ttfamily 1011.1232}}].

\bibitem{deRham:2014zqa}
C.~de~Rham, \emph{Massive gravity},
  \href{https://doi.org/10.12942/lrr-2014-7}{Living Rev. Relativity {\bfseries
  17} (2014) 7} [\href{https://arxiv.org/abs/1401.4173}{{\ttfamily
  1401.4173}}].

\bibitem{Alcaraz:2002iu}
J.~Alcaraz, J.~A.~R.~Cembranos, A.~Dobado and A.~L.~Maroto, \emph{{Limits on
  the brane fluctuations mass and on the brane tension scale from
  electron-positron colliders}},
  \href{https://doi.org/10.1103/PhysRevD.67.075010}{Phys. Rev. D {\bfseries 67}
  (2003) 075010} [\href{https://arxiv.org/abs/hep-ph/0212269}{{\ttfamily
  hep-ph/0212269}}].

\bibitem{Cembranos:2004jp}
J.~A.~R.~Cembranos, A.~Dobado and A.~L.~Maroto, \emph{{Branon search in
  hadronic colliders}}, \href{https://doi.org/10.1103/PhysRevD.70.096001}{Phys.
  Rev. D {\bfseries 70} (2004) 096001}
  [\href{https://arxiv.org/abs/hep-ph/0405286}{{\ttfamily hep-ph/0405286}}].

\bibitem{deRham:2010eu}
C.~de~Rham and A.~J.~Tolley, \emph{{DBI and the Galileon reunited}},
  \href{https://doi.org/10.1088/1475-7516/2010/05/015}{JCAP {\bfseries 05}
  (2010) 015} [\href{https://arxiv.org/abs/1003.5917}{{\ttfamily 1003.5917}}].

\bibitem{Koivisto:2013fta}
T.~Koivisto, D.~Wills and I.~Zavala, \emph{{Dark D-brane cosmology}},
  \href{https://doi.org/10.1088/1475-7516/2014/06/036}{JCAP {\bfseries 06}
  (2014) 036} [\href{https://arxiv.org/abs/1312.2597}{{\ttfamily 1312.2597}}].

\bibitem{Kaloper:2003yf}
N.~Kaloper, \emph{{Disformal inflation}},
  \href{https://doi.org/10.1016/j.physletb.2004.01.005}{Phys. Lett. B
  {\bfseries 583} (2004) 1}
  [\href{https://arxiv.org/abs/hep-ph/0312002}{{\ttfamily hep-ph/0312002}}].

\bibitem{Koivisto:2008ak}
T.~S.~Koivisto, \emph{{Disformal quintessence}},
  \href{https://arxiv.org/abs/0811.1957}{{\ttfamily 0811.1957}}.

\bibitem{Zumalacarregui:2010wj}
M.~Zumalac\'arregui, T.~S.~Koivisto, D.~F.~Mota and P.~Ruiz-Lapuente,
  \emph{Disformal scalar fields and the dark sector of the universe},
  \href{https://doi.org/10.1088/1475-7516/2010/05/038}{JCAP {\bfseries 05}
  (2010) 038} [\href{https://arxiv.org/abs/1004.2684}{{\ttfamily 1004.2684}}].

\bibitem{Koivisto:2012za}
T.~S.~Koivisto, D.~F.~Mota and M.~Zumalac\'arregui, \emph{Screening
  modifications of gravity through disformally coupled fields},
  \href{https://doi.org/10.1103/PhysRevLett.109.241102}{Phys. Rev. Lett.
  {\bfseries 109} (2012) 241102}
  [\href{https://arxiv.org/abs/1205.3167}{{\ttfamily 1205.3167}}].

\bibitem{Brax:2012ie}
P.~Brax, C.~Burrage and A.-C.~Davis, \emph{Shining light on modifications of
  gravity}, \href{https://doi.org/10.1088/1475-7516/2012/10/016}{JCAP
  {\bfseries 10} (2012) 016} [\href{https://arxiv.org/abs/1206.1809}{{\ttfamily
  1206.1809}}].

\bibitem{vandeBruck:2012vq}
C.~van~de Bruck and G.~Sculthorpe, \emph{Modified gravity and the radiation
  dominated epoch}, \href{https://doi.org/10.1103/PhysRevD.87.044004}{Phys.
  Rev. D {\bfseries 87} (2013) 044004}
  [\href{https://arxiv.org/abs/1210.2168}{{\ttfamily 1210.2168}}].

\bibitem{vandeBruck:2013yxa}
C.~van~de Bruck, J.~Morrice and S.~Vu, \emph{Constraints on nonconformal
  couplings from the properties of the cosmic microwave background radiation},
  \href{https://doi.org/10.1103/PhysRevLett.111.161302}{Phys. Rev. Lett.
  {\bfseries 111} (2013) 161302}
  [\href{https://arxiv.org/abs/1303.1773}{{\ttfamily 1303.1773}}].

\bibitem{Brax:2013nsa}
P.~Brax, C.~Burrage, A.-C.~Davis and G.~Gubitosi, \emph{Cosmological tests of
  the disformal coupling to radiation},
  \href{https://doi.org/10.1088/1475-7516/2013/11/001}{JCAP {\bfseries 11}
  (2013) 001} [\href{https://arxiv.org/abs/1306.4168}{{\ttfamily 1306.4168}}].

\bibitem{Neveu:2014vua}
J.~Neveu, V.~Ruhlmann-Kleider, P.~Astier, M.~Besan\c{c}on, A.~Conley, J.~Guy
  et~al., \emph{{First experimental constraints on the disformally coupled
  Galileon model}}, \href{https://doi.org/10.1051/0004-6361/201423758}{Astron.
  Astrophys. {\bfseries 569} (2014) A90}
  [\href{https://arxiv.org/abs/1403.0854}{{\ttfamily 1403.0854}}].

\bibitem{Sakstein:2014isa}
J.~Sakstein, \emph{Disformal theories of gravity: {F}rom the solar system to
  cosmology}, \href{https://doi.org/10.1088/1475-7516/2014/12/012}{JCAP
  {\bfseries 12} (2014) 012} [\href{https://arxiv.org/abs/1409.1734}{{\ttfamily
  1409.1734}}].

\bibitem{Sakstein:2014aca}
J.~Sakstein, \emph{Towards viable cosmological models of disformal theories of
  gravity}, \href{https://doi.org/10.1103/PhysRevD.91.024036}{Phys. Rev. D
  {\bfseries 91} (2015) 024036}
  [\href{https://arxiv.org/abs/1409.7296}{{\ttfamily 1409.7296}}].

\bibitem{Brax:2014vva}
P.~Brax and C.~Burrage, \emph{Constraining disformally coupled scalar fields},
  \href{https://doi.org/10.1103/PhysRevD.90.104009}{Phys. Rev. D {\bfseries 90}
  (2014) 104009} [\href{https://arxiv.org/abs/1407.1861}{{\ttfamily
  1407.1861}}].

\bibitem{Brax:2015hma}
P.~Brax, C.~Burrage and C.~Englert, \emph{{Disformal dark energy at
  colliders}}, \href{https://doi.org/10.1103/PhysRevD.92.044036}{Phys. Rev. D
  {\bfseries 92} (2015) 044036}
  [\href{https://arxiv.org/abs/1506.04057}{{\ttfamily 1506.04057}}].

\bibitem{Ip:2015qsa}
H.~Y.~Ip, J.~Sakstein and F.~Schmidt, \emph{Solar system constraints on
  disformal gravity theories},
  \href{https://doi.org/10.1088/1475-7516/2015/10/051}{JCAP {\bfseries 10}
  (2015) 051} [\href{https://arxiv.org/abs/1507.00568}{{\ttfamily
  1507.00568}}].

\bibitem{Sakstein:2015jca}
J.~Sakstein and S.~Verner, \emph{Disformal gravity theories: {A Jordan} frame
  analysis}, \href{https://doi.org/10.1103/PhysRevD.92.123005}{Phys. Rev. D
  {\bfseries 92} (2015) 123005}
  [\href{https://arxiv.org/abs/1509.05679}{{\ttfamily 1509.05679}}].

\bibitem{vandeBruck:2015ida}
C.~van~de Bruck and J.~Morrice, \emph{{Disformal couplings and the dark sector
  of the universe}}, \href{https://doi.org/10.1088/1475-7516/2015/04/036}{JCAP
  {\bfseries 04} (2015) 036}
  [\href{https://arxiv.org/abs/1501.03073}{{\ttfamily 1501.03073}}].

\bibitem{vandeBruck:2016cnh}
C.~van~de Bruck, C.~Burrage and J.~Morrice, \emph{{Vacuum Cherenkov radiation
  and bremsstrahlung from disformal couplings}},
  \href{https://doi.org/10.1088/1475-7516/2016/08/003}{JCAP {\bfseries 08}
  (2016) 003} [\href{https://arxiv.org/abs/1605.03567}{{\ttfamily
  1605.03567}}].

\bibitem{TheLIGOScientific:2017qsa}
B.~P.~Abbott et~al. {(LIGO Scientific and Virgo Collaborations)},
  \emph{{GW170817: Observation} of gravitational waves from a binary neutron
  star inspiral}, \href{https://doi.org/10.1103/PhysRevLett.119.161101}{Phys.
  Rev. Lett. {\bfseries 119} (2017) 161101}
  [\href{https://arxiv.org/abs/1710.05832}{{\ttfamily 1710.05832}}].

\bibitem{Goldstein:2017mmi}
A.~Goldstein et~al., \emph{An ordinary short gamma-ray burst with extraordinary
  implications: {Fermi-GBM} detection of {GRB 170817A}},
  \href{https://doi.org/10.3847/2041-8213/aa8f41}{Astrophys. J. {\bfseries 848}
  (2017) L14} [\href{https://arxiv.org/abs/1710.05446}{{\ttfamily
  1710.05446}}].

\bibitem{Monitor:2017mdv}
B.~P.~Abbott et~al. {({LIGO Scientific and Virgo Collaborations, Fermi
  Gamma-ray Burst Monitor, and INTEGRAL})}, \emph{Gravitational waves and
  gamma-rays from a binary neutron star merger: {GW170817 and GRB 170817A}},
  \href{https://doi.org/10.3847/2041-8213/aa920c}{Astrophys. J. Lett.
  {\bfseries 848} (2017) L13}
  [\href{https://arxiv.org/abs/1710.05834}{{\ttfamily 1710.05834}}].

\bibitem{GBM:2017lvd}
B.~P.~Abbott et~al., \emph{Multi-messenger observations of a binary neutron
  star merger}, \href{https://doi.org/10.3847/2041-8213/aa91c9}{Astrophys. J.
  Lett. {\bfseries 848} (2017) L12}
  [\href{https://arxiv.org/abs/1710.05833}{{\ttfamily 1710.05833}}].

\bibitem{deRham:2018red}
C.~de~Rham and S.~Melville, \emph{Gravitational rainbows: {LIGO} and dark
  energy at its cutoff},
  \href{https://doi.org/10.1103/PhysRevLett.121.221101}{Phys. Rev. Lett.
  {\bfseries 121} (2018) 221101}
  [\href{https://arxiv.org/abs/1806.09417}{{\ttfamily 1806.09417}}].

\bibitem{Creminelli:2017sry}
P.~Creminelli and F.~Vernizzi, \emph{{Dark energy after GW170817 and
  GRB170817A}}, \href{https://doi.org/10.1103/PhysRevLett.119.251302}{Phys.
  Rev. Lett. {\bfseries 119} (2017) 251302}
  [\href{https://arxiv.org/abs/1710.05877}{{\ttfamily 1710.05877}}].

\bibitem{Sakstein:2017xjx}
J.~Sakstein and B.~Jain, \emph{Implications of the neutron star merger
  {GW170817} for cosmological scalar-tensor theories},
  \href{https://doi.org/10.1103/PhysRevLett.119.251303}{Phys. Rev. Lett.
  {\bfseries 119} (2017) 251303}
  [\href{https://arxiv.org/abs/1710.05893}{{\ttfamily 1710.05893}}].

\bibitem{Ezquiaga:2017ekz}
J.~M.~Ezquiaga and M.~Zumalac\'arregui, \emph{Dark energy after {GW170817:
  Dead} ends and the road ahead},
  \href{https://doi.org/10.1103/PhysRevLett.119.251304}{Phys. Rev. Lett.
  {\bfseries 119} (2017) 251304}
  [\href{https://arxiv.org/abs/1710.05901}{{\ttfamily 1710.05901}}].

\bibitem{Baker:2017hug}
T.~Baker, E.~Bellini, P.~G.~Ferreira, M.~Lagos, J.~Noller and I.~Sawicki,
  \emph{{Strong constraints on cosmological gravity from GW170817 and GRB
  170817A}}, \href{https://doi.org/10.1103/PhysRevLett.119.251301}{Phys. Rev.
  Lett. {\bfseries 119} (2017) 251301}
  [\href{https://arxiv.org/abs/1710.06394}{{\ttfamily 1710.06394}}].

\bibitem{Langlois:2017dyl}
D.~Langlois, R.~Saito, D.~Yamauchi and K.~Noui, \emph{{Scalar-tensor theories
  and modified gravity in the wake of GW170817}},
  \href{https://doi.org/10.1103/PhysRevD.97.061501}{Phys. Rev. D {\bfseries 97}
  (2018) 061501} [\href{https://arxiv.org/abs/1711.07403}{{\ttfamily
  1711.07403}}].

\bibitem{Heisenberg:2017qka}
L.~Heisenberg and S.~Tsujikawa, \emph{{Dark energy survivals in massive gravity
  after GW170817: SO(3) invariant}},
  \href{https://doi.org/10.1088/1475-7516/2018/01/044}{JCAP {\bfseries 01}
  (2018) 044} [\href{https://arxiv.org/abs/1711.09430}{{\ttfamily
  1711.09430}}].

\bibitem{Akrami:2018yjz}
Y.~Akrami, P.~Brax, A.-C.~Davis and V.~Vardanyan, \emph{{Neutron star merger
  GW170817 strongly constrains doubly coupled bigravity}},
  \href{https://doi.org/10.1103/PhysRevD.97.124010}{Phys. Rev. D {\bfseries 97}
  (2018) 124010} [\href{https://arxiv.org/abs/1803.09726}{{\ttfamily
  1803.09726}}].

\bibitem{BeltranJimenez:2018ymu}
J.~Beltr\'an~Jim\'enez and L.~Heisenberg, \emph{{Non-trivial gravitational
  waves and structure formation phenomenology from dark energy}},
  \href{https://doi.org/10.1088/1475-7516/2018/09/035}{JCAP {\bfseries 09}
  (2018) 035} [\href{https://arxiv.org/abs/1806.01753}{{\ttfamily
  1806.01753}}].

\bibitem{Copeland:2018yuh}
E.~J.~Copeland, M.~Kopp, A.~Padilla, P.~M.~Saffin and C.~Skordis, \emph{{Dark
  energy after GW170817 revisited}},
  \href{https://doi.org/10.1103/PhysRevLett.122.061301}{Phys. Rev. Lett.
  {\bfseries 122} (2019) 061301}
  [\href{https://arxiv.org/abs/1810.08239}{{\ttfamily 1810.08239}}].

\bibitem{Will:1994fb}
C.~M.~Will, \emph{{Testing scalar-tensor gravity with gravitational-wave
  observations of inspiralling compact binaries}},
  \href{https://doi.org/10.1103/PhysRevD.50.6058}{Phys. Rev. D {\bfseries 50}
  (1994) 6058} [\href{https://arxiv.org/abs/gr-qc/9406022}{{\ttfamily
  gr-qc/9406022}}].

\bibitem{Damour:1998jk}
T.~Damour and G.~Esposito-Far{\`e}se, \emph{{Gravitational-wave versus
  binary-pulsar tests of strong-field gravity}},
  \href{https://doi.org/10.1103/PhysRevD.58.042001}{Phys. Rev. D {\bfseries 58}
  (1998) 042001} [\href{https://arxiv.org/abs/gr-qc/9803031}{{\ttfamily
  gr-qc/9803031}}].

\bibitem{Berti:2012bp}
E.~Berti, L.~Gualtieri, M.~Horbatsch and J.~Alsing, \emph{{Light scalar field
  constraints from gravitational-wave observations of compact binaries}},
  \href{https://doi.org/10.1103/PhysRevD.85.122005}{Phys. Rev. D {\bfseries 85}
  (2012) 122005} [\href{https://arxiv.org/abs/1204.4340}{{\ttfamily
  1204.4340}}].

\bibitem{Yunes:2013dva}
N.~Yunes and X.~Siemens, \emph{Gravitational-wave tests of general relativity
  with ground-based detectors and pulsar timing-arrays},
  \href{https://doi.org/10.12942/lrr-2013-9}{Living Rev. Relativity {\bfseries
  16} (2013) 9} [\href{https://arxiv.org/abs/1304.3473}{{\ttfamily
  1304.3473}}].

\bibitem{Yunes:2016jcc}
N.~Yunes, K.~Yagi and F.~Pretorius, \emph{Theoretical physics implications of
  the binary black-hole mergers {GW150914 and GW151226}},
  \href{https://doi.org/10.1103/PhysRevD.94.084002}{Phys. Rev. D {\bfseries 94}
  (2016) 084002} [\href{https://arxiv.org/abs/1603.08955}{{\ttfamily
  1603.08955}}].

\bibitem{Chamberlain:2017fjl}
K.~Chamberlain and N.~Yunes, \emph{Theoretical physics implications of
  gravitational wave observation with future detectors},
  \href{https://doi.org/10.1103/PhysRevD.96.084039}{Phys. Rev. D {\bfseries 96}
  (2017) 084039} [\href{https://arxiv.org/abs/1704.08268}{{\ttfamily
  1704.08268}}].

\bibitem{Barack:2018yly}
L.~Barack, V.~Cardoso, S.~Nissanke, T.~P.~Sotiriou, A.~Askar, C.~Belczynski
  et~al., \emph{{Black holes, gravitational waves and fundamental physics: A
  roadmap}}, \href{https://doi.org/10.1088/1361-6382/ab0587}{Class. Quantum
  Gravity {\bfseries 36} (2019) 143001}
  [\href{https://arxiv.org/abs/1806.05195}{{\ttfamily 1806.05195}}].

\bibitem{Carson:2019fxr}
Z.~Carson, B.~C.~Seymour and K.~Yagi, \emph{{Future prospects for probing
  scalar--tensor theories with gravitational waves from mixed binaries}},
  \href{https://doi.org/10.1088/1361-6382/ab6a1f}{Class. Quantum Gravity
  {\bfseries 37} (2020) 065008}
  [\href{https://arxiv.org/abs/1907.03897}{{\ttfamily 1907.03897}}].

\bibitem{Perkins:2020tra}
S.~E.~Perkins, N.~Yunes and E.~Berti, \emph{{Probing fundamental physics with
  gravitational waves: The next generation}},
  \href{https://doi.org/10.1103/PhysRevD.103.044024}{Phys. Rev. D {\bfseries
  103} (2021) 044024} [\href{https://arxiv.org/abs/2010.09010}{{\ttfamily
  2010.09010}}].

\bibitem{Thorne:1971}
K.~S.~{Thorne} and J.~J.~{Dykla}, \emph{{Black holes in the Dicke-Brans-Jordan
  theory of gravity}}, \href{https://doi.org/10.1086/180734}{Astrophys. J.
  {\bfseries 166} (1971) L35}.

\bibitem{Bekenstein:1971hc}
J.~D.~Bekenstein, \emph{{Nonexistence of Baryon number for static black
  holes}}, \href{https://doi.org/10.1103/PhysRevD.5.1239}{Phys. Rev. D
  {\bfseries 5} (1972) 1239}.

\bibitem{Adler:1978dp}
S.~L.~Adler and R.~B.~Pearson, \emph{{``No-hair'' theorems for the Abelian
  Higgs and Goldstone models}},
  \href{https://doi.org/10.1103/PhysRevD.18.2798}{Phys. Rev. D {\bfseries 18}
  (1978) 2798}.

\bibitem{Hawking:1972qk}
S.~W.~Hawking, \emph{{Black holes in the Brans-Dicke theory of gravitation}},
  \href{https://doi.org/10.1007/BF01877518}{Commun. Math. Phys. {\bfseries 25}
  (1972) 167}.

\bibitem{Zannias:1994jf}
T.~Zannias, \emph{{Black holes cannot support conformal scalar hair}},
  \href{https://doi.org/10.1063/1.531201}{J. Math. Phys. (N.Y.) {\bfseries 36}
  (1995) 6970} [\href{https://arxiv.org/abs/gr-qc/9409030}{{\ttfamily
  gr-qc/9409030}}].

\bibitem{Bekenstein:1995un}
J.~D.~Bekenstein, \emph{{Novel ``no-scalar-hair'' theorem for black holes}},
  \href{https://doi.org/10.1103/PhysRevD.51.R6608}{Phys. Rev. D {\bfseries 51}
  (1995) R6608}.

\bibitem{Saa:1996aw}
A.~Saa, \emph{{New no-scalar-hair theorem for black holes}},
  \href{https://doi.org/10.1063/1.531513}{J. Math. Phys. (N.Y.) {\bfseries 37}
  (1996) 2346} [\href{https://arxiv.org/abs/gr-qc/9601021}{{\ttfamily
  gr-qc/9601021}}].

\bibitem{Sotiriou:2011dz}
T.~P.~Sotiriou and V.~Faraoni, \emph{Black holes in scalar-tensor gravity},
  \href{https://doi.org/10.1103/PhysRevLett.108.081103}{Phys. Rev. Lett.
  {\bfseries 108} (2012) 081103}
  [\href{https://arxiv.org/abs/1109.6324}{{\ttfamily 1109.6324}}].

\bibitem{Hui:2012qt}
L.~Hui and A.~Nicolis, \emph{No-hair theorem for the galileon},
  \href{https://doi.org/10.1103/PhysRevLett.110.241104}{Phys. Rev. Lett.
  {\bfseries 110} (2013) 241104}
  [\href{https://arxiv.org/abs/1202.1296}{{\ttfamily 1202.1296}}].

\bibitem{Graham:2014mda}
A.~A.~H.~Graham and R.~Jha, \emph{{Nonexistence of black holes with
  noncanonical scalar fields}},
  \href{https://doi.org/10.1103/PhysRevD.89.084056}{Phys. Rev. D {\bfseries 89}
  (2014) 084056}{ [Erratum ibid. \textbf{92} (2015) 069901]}
  [\href{https://arxiv.org/abs/1401.8203}{{\ttfamily 1401.8203}}].

\bibitem{Chrusciel:2012jk}
P.~T.~Chru{\'{s}}ciel, J.~L.~Costa and M.~Heusler, \emph{{Stationary black
  holes: Uniqueness and beyond}},
  \href{https://doi.org/10.12942/lrr-2012-7}{Living Rev. Relativity {\bfseries
  15} (2012) 7} [\href{https://arxiv.org/abs/1205.6112}{{\ttfamily
  1205.6112}}].

\bibitem{Berti:2004bd}
E.~Berti, A.~Buonanno and C.~M.~Will, \emph{{Estimating spinning binary
  parameters and testing alternative theories of gravity with LISA}},
  \href{https://doi.org/10.1103/PhysRevD.71.084025}{Phys. Rev. D {\bfseries 71}
  (2005) 084025} [\href{https://arxiv.org/abs/gr-qc/0411129}{{\ttfamily
  gr-qc/0411129}}].

\bibitem{Jacobson:1999vr}
T.~Jacobson, \emph{{Primordial black hole evolution in tensor-scalar
  cosmology}}, \href{https://doi.org/10.1103/PhysRevLett.83.2699}{Phys. Rev.
  Lett. {\bfseries 83} (1999) 2699}
  [\href{https://arxiv.org/abs/astro-ph/9905303}{{\ttfamily
  astro-ph/9905303}}].

\bibitem{Horbatsch:2011ye}
M.~W.~Horbatsch and C.~P.~Burgess, \emph{Cosmic black-hole hair growth and
  quasar {OJ287}}, \href{https://doi.org/10.1088/1475-7516/2012/05/010}{JCAP
  {\bfseries 05} (2012) 010} [\href{https://arxiv.org/abs/1111.4009}{{\ttfamily
  1111.4009}}].

\bibitem{Berti:2013gfa}
E.~Berti, V.~Cardoso, L.~Gualtieri, M.~Horbatsch and U.~Sperhake,
  \emph{{Numerical simulations of single and binary black holes in
  scalar-tensor theories: Circumventing the no-hair theorem}},
  \href{https://doi.org/10.1103/PhysRevD.87.124020}{Phys. Rev. D {\bfseries 87}
  (2013) 124020} [\href{https://arxiv.org/abs/1304.2836}{{\ttfamily
  1304.2836}}].

\bibitem{Wong:2019yoc}
L.~K.~Wong, A.-C.~Davis and R.~Gregory, \emph{{Effective field theory for black
  holes with induced scalar charges}},
  \href{https://doi.org/10.1103/PhysRevD.100.024010}{Phys. Rev. D {\bfseries
  100} (2019) 024010} [\href{https://arxiv.org/abs/1903.07080}{{\ttfamily
  1903.07080}}].

\bibitem{Clough:2019jpm}
K.~Clough, P.~G.~Ferreira and M.~Lagos, \emph{{Growth of massive scalar hair
  around a Schwarzschild black hole}},
  \href{https://doi.org/10.1103/PhysRevD.100.063014}{Phys. Rev. D {\bfseries
  100} (2019) 063014} [\href{https://arxiv.org/abs/1904.12783}{{\ttfamily
  1904.12783}}].

\bibitem{Hui:2019aqm}
L.~Hui, D.~Kabat, X.~Li, L.~Santoni and S.~S.~C.~Wong, \emph{Black hole hair
  from scalar dark matter},
  \href{https://doi.org/10.1088/1475-7516/2019/06/038}{JCAP {\bfseries 06}
  (2019) 038} [\href{https://arxiv.org/abs/1904.12803}{{\ttfamily
  1904.12803}}].

\bibitem{Bamber:2020bpu}
J.~Bamber, K.~Clough, P.~G.~Ferreira, L.~Hui and M.~Lagos, \emph{{Growth of
  accretion driven scalar hair around Kerr black holes}},
  \href{https://doi.org/10.1103/PhysRevD.103.044059}{Phys. Rev. D {\bfseries
  103} (2021) 044059} [\href{https://arxiv.org/abs/2011.07870}{{\ttfamily
  2011.07870}}].

\bibitem{Babichev:2013cya}
E.~Babichev and C.~Charmousis, \emph{{Dressing a black hole with a
  time-dependent Galileon}},
  \href{https://doi.org/10.1007/JHEP08(2014)106}{JHEP {\bfseries 08} (2014)
  106} [\href{https://arxiv.org/abs/1312.3204}{{\ttfamily 1312.3204}}].

\bibitem{Kobayashi:2014eva}
T.~Kobayashi and N.~Tanahashi, \emph{{Exact black hole solutions in shift
  symmetric scalar-tensor theories}},
  \href{https://doi.org/10.1093/ptep/ptu096}{Prog. Theor. Exp. Phys. {\bfseries
  2014} (2014) 073E02} [\href{https://arxiv.org/abs/1403.4364}{{\ttfamily
  1403.4364}}].

\bibitem{Herdeiro:2014goa}
C.~A.~R.~Herdeiro and E.~Radu, \emph{{Kerr black holes with scalar hair}},
  \href{https://doi.org/10.1103/PhysRevLett.112.221101}{Phys. Rev. Lett.
  {\bfseries 112} (2014) 221101}
  [\href{https://arxiv.org/abs/1403.2757}{{\ttfamily 1403.2757}}].

\bibitem{Degollado:2018ypf}
J.~C.~Degollado, C.~A.~R.~Herdeiro and E.~Radu, \emph{{Effective stability
  against superradiance of Kerr black holes with synchronised hair}},
  \href{https://doi.org/10.1016/j.physletb.2018.04.052}{Phys. Lett. B
  {\bfseries 781} (2018) 651}
  [\href{https://arxiv.org/abs/1802.07266}{{\ttfamily 1802.07266}}].

\bibitem{Alexander:2009tp}
S.~Alexander and N.~Yunes, \emph{{Chern-Simons} modified general relativity},
  \href{https://doi.org/10.1016/j.physrep.2009.07.002}{Phys. Rep. {\bfseries
  480} (2009) 1} [\href{https://arxiv.org/abs/0907.2562}{{\ttfamily
  0907.2562}}].

\bibitem{Yunes:2009hc}
N.~Yunes and F.~Pretorius, \emph{{Dynamical Chern-Simons modified gravity:
  Spinning black holes in the slow-rotation approximation}},
  \href{https://doi.org/10.1103/PhysRevD.79.084043}{Phys. Rev. D {\bfseries 79}
  (2009) 084043} [\href{https://arxiv.org/abs/0902.4669}{{\ttfamily
  0902.4669}}].

\bibitem{Delsate:2018ome}
T.~Delsate, C.~Herdeiro and E.~Radu, \emph{{Non-perturbative spinning black
  holes in dynamical Chern--Simons gravity}},
  \href{https://doi.org/10.1016/j.physletb.2018.09.060}{Phys. Lett. B
  {\bfseries 787} (2018) 8} [\href{https://arxiv.org/abs/1806.06700}{{\ttfamily
  1806.06700}}].

\bibitem{Yagi:2012vf}
K.~Yagi, N.~Yunes and T.~Tanaka, \emph{Gravitational waves from quasi-circular
  black hole binaries in dynamical {Chern-Simons} gravity},
  \href{https://doi.org/10.1103/PhysRevLett.116.169902}{Phys. Rev. Lett.
  {\bfseries 109} (2012) 251105}{ [Errata ibid. \textbf{116} (2016) 169902 and
  ibid. \textbf{124} (2020) 029901]}
  [\href{https://arxiv.org/abs/1208.5102}{{\ttfamily 1208.5102}}].

\bibitem{Loutrel:2018ydv}
N.~Loutrel, T.~Tanaka and N.~Yunes, \emph{{Spin-precessing black hole binaries
  in dynamical Chern-Simons gravity}},
  \href{https://doi.org/10.1103/PhysRevD.98.064020}{Phys. Rev. D {\bfseries 98}
  (2018) 064020} [\href{https://arxiv.org/abs/1806.07431}{{\ttfamily
  1806.07431}}].

\bibitem{Loutrel:2018rxs}
N.~Loutrel, T.~Tanaka and N.~Yunes, \emph{Scalar tops and perturbed
  quadrupoles: {P}robing fundamental physics with spin-precessing binaries},
  \href{https://doi.org/10.1088/1361-6382/ab15fa}{Class. Quantum Gravity
  {\bfseries 36} (2019) 10LT02}
  [\href{https://arxiv.org/abs/1806.07425}{{\ttfamily 1806.07425}}].

\bibitem{Brax:2018bow}
P.~Brax and A.-C.~Davis, \emph{{Gravitational effects of disformal couplings}},
  \href{https://doi.org/10.1103/PhysRevD.98.063531}{Phys. Rev. D {\bfseries 98}
  (2018) 063531} [\href{https://arxiv.org/abs/1809.09844}{{\ttfamily
  1809.09844}}].

\bibitem{Brax:2019tcy}
P.~Brax, A.-C.~Davis and A.~Kuntz, \emph{Disformally coupled scalar fields and
  inspiralling trajectories},
  \href{https://doi.org/10.1103/PhysRevD.99.124034}{Phys. Rev. D {\bfseries 99}
  (2019) 124034} [\href{https://arxiv.org/abs/1903.03842}{{\ttfamily
  1903.03842}}].

\bibitem{Goldberger:2004jt}
W.~D.~Goldberger and I.~Z.~Rothstein, \emph{{Effective field theory of gravity
  for extended objects}},
  \href{https://doi.org/10.1103/PhysRevD.73.104029}{Phys. Rev. D {\bfseries 73}
  (2006) 104029} [\href{https://arxiv.org/abs/hep-th/0409156}{{\ttfamily
  hep-th/0409156}}].

\bibitem{Goldberger:2007hy}
W.~D.~Goldberger, \emph{{Effective field theories and gravitational
  radiation}}, {}in
  \href{https://www.sciencedirect.com/bookseries/les-houches/vol/86/suppl/C}{\emph{{Particle
  Physics and Cosmology: The Fabric of Spacetime}}}, F.~Bernardeau, C.~Grojean
  and J.~Dalibard, eds., pp.~351--396{, Lecture Notes of the Les Houches Summer
  School 2006} [\href{https://arxiv.org/abs/hep-ph/0701129}{{\ttfamily
  hep-ph/0701129}}].

\bibitem{Goldberger:2009qd}
W.~D.~Goldberger and A.~Ross, \emph{{Gravitational radiative corrections from
  effective field theory}},
  \href{https://doi.org/10.1103/PhysRevD.81.124015}{Phys. Rev. D {\bfseries 81}
  (2010) 124015} [\href{https://arxiv.org/abs/0912.4254}{{\ttfamily
  0912.4254}}].

\bibitem{Galley:2009px}
C.~R.~Galley and M.~Tiglio, \emph{{Radiation reaction and gravitational waves
  in the effective field theory approach}},
  \href{https://doi.org/10.1103/PhysRevD.79.124027}{Phys. Rev. D {\bfseries 79}
  (2009) 124027} [\href{https://arxiv.org/abs/0903.1122}{{\ttfamily
  0903.1122}}].

\bibitem{Porto:2005ac}
R.~A.~Porto, \emph{{Post-Newtonian corrections to the motion of spinning bodies
  in nonrelativistic general relativity}},
  \href{https://doi.org/10.1103/PhysRevD.73.104031}{Phys. Rev. D {\bfseries 73}
  (2006) 104031} [\href{https://arxiv.org/abs/gr-qc/0511061}{{\ttfamily
  gr-qc/0511061}}].

\bibitem{Porto:2006bt}
R.~A.~Porto and I.~Z.~Rothstein, \emph{The hyperfine {Einstein-Infeld-Hoffmann}
  potential}, \href{https://doi.org/10.1103/PhysRevLett.97.021101}{Phys. Rev.
  Lett. {\bfseries 97} (2006) 021101}
  [\href{https://arxiv.org/abs/gr-qc/0604099}{{\ttfamily gr-qc/0604099}}].

\bibitem{Porto:2008tb}
R.~A.~Porto and I.~Z.~Rothstein, \emph{Spin(1)spin(2) effects in the motion of
  inspiralling compact binaries at third order in the post-{N}ewtonian
  expansion}, \href{https://doi.org/10.1103/PhysRevD.78.044012}{Phys. Rev. D
  {\bfseries 78} (2008) 044012}{ [Erratum ibid. \textbf{81} (2010) 029904]}
  [\href{https://arxiv.org/abs/0802.0720}{{\ttfamily 0802.0720}}].

\bibitem{Porto:2008jj}
R.~A.~Porto and I.~Z.~Rothstein, \emph{Next to leading order spin(1)spin(1)
  effects in the motion of inspiralling compact binaries},
  \href{https://doi.org/10.1103/PhysRevD.78.044013}{Phys. Rev. D {\bfseries 78}
  (2008) 044013}{ [Erratum ibid. \textbf{81} (2010) 029905]}
  [\href{https://arxiv.org/abs/0804.0260}{{\ttfamily 0804.0260}}].

\bibitem{Levi:2010zu}
M.~Levi, \emph{Next-to-leading order gravitational spin-orbit coupling in an
  effective field theory approach},
  \href{https://doi.org/10.1103/PhysRevD.82.104004}{Phys. Rev. D {\bfseries 82}
  (2010) 104004} [\href{https://arxiv.org/abs/1006.4139}{{\ttfamily
  1006.4139}}].

\bibitem{Levi:2015msa}
M.~Levi and J.~Steinhoff, \emph{{Spinning gravitating objects in the effective
  field theory in the post-Newtonian scheme}},
  \href{https://doi.org/10.1007/JHEP09(2015)219}{JHEP {\bfseries 09} (2015)
  219} [\href{https://arxiv.org/abs/1501.04956}{{\ttfamily 1501.04956}}].

\bibitem{Foffa:2013qca}
S.~Foffa and R.~Sturani, \emph{{Effective field theory methods to model compact
  binaries}}, \href{https://doi.org/10.1088/0264-9381/31/4/043001}{Class.
  Quantum Gravity {\bfseries 31} (2014) 043001}
  [\href{https://arxiv.org/abs/1309.3474}{{\ttfamily 1309.3474}}].

\bibitem{Rothstein:2014sra}
I.~Z.~Rothstein, \emph{{Progress in effective field theory approach to the
  binary inspiral problem}},
  \href{https://doi.org/10.1007/s10714-014-1726-y}{Gen. Relativ. Gravit.
  {\bfseries 46} (2014) 1726}.

\bibitem{Porto:2016pyg}
R.~A.~Porto, \emph{{The effective field theorist's approach to gravitational
  dynamics}}, \href{https://doi.org/10.1016/j.physrep.2016.04.003}{Phys. Rep.
  {\bfseries 633} (2016) 1} [\href{https://arxiv.org/abs/1601.04914}{{\ttfamily
  1601.04914}}].

\bibitem{Levi:2018nxp}
M.~Levi, \emph{{Effective field theories of post-Newtonian gravity: A
  comprehensive review}}, \href{https://doi.org/10.1088/1361-6633/ab12bc}{Rept.
  Prog. Phys. {\bfseries 83} (2020) 075901}
  [\href{https://arxiv.org/abs/1807.01699}{{\ttfamily 1807.01699}}].

\bibitem{Damour:1992we}
T.~Damour and G.~Esposito-Far{\`e}se, \emph{{Tensor-multi-scalar theories of
  gravitation}}, \href{https://doi.org/10.1088/0264-9381/9/9/015}{Class.
  Quantum Gravity {\bfseries 9} (1992) 2093}.

\bibitem{Huang:2018pbu}
J.~Huang, M.~C.~Johnson, L.~Sagunski, M.~Sakellariadou and J.~Zhang,
  \emph{{Prospects for axion searches with Advanced LIGO through binary
  mergers}}, \href{https://doi.org/10.1103/PhysRevD.99.063013}{Phys. Rev. D
  {\bfseries 99} (2019) 063013}
  [\href{https://arxiv.org/abs/1807.02133}{{\ttfamily 1807.02133}}].

\bibitem{Kuntz:2019zef}
A.~Kuntz, F.~Piazza and F.~Vernizzi, \emph{{Effective field theory for
  gravitational radiation in scalar-tensor gravity}},
  \href{https://doi.org/10.1088/1475-7516/2019/05/052}{JCAP {\bfseries 05}
  (2019) 052} [\href{https://arxiv.org/abs/1902.04941}{{\ttfamily
  1902.04941}}].

\bibitem{Mirshekari:2013vb}
S.~Mirshekari and C.~M.~Will, \emph{{Compact binary systems in scalar-tensor
  gravity: Equations of motion to 2.5 post-Newtonian order}},
  \href{https://doi.org/10.1103/PhysRevD.87.084070}{Phys. Rev. D {\bfseries 87}
  (2013) 084070} [\href{https://arxiv.org/abs/1301.4680}{{\ttfamily
  1301.4680}}].

\bibitem{Lang:2013fna}
R.~N.~Lang, \emph{{Compact binary systems in scalar-tensor gravity. II. Tensor
  gravitational waves to second post-Newtonian order}},
  \href{https://doi.org/10.1103/PhysRevD.89.084014}{Phys. Rev. D {\bfseries 89}
  (2014) 084014} [\href{https://arxiv.org/abs/1310.3320}{{\ttfamily
  1310.3320}}].

\bibitem{Lang:2014osa}
R.~N.~Lang, \emph{{Compact binary systems in scalar-tensor gravity. III. Scalar
  waves and energy flux}},
  \href{https://doi.org/10.1103/PhysRevD.91.084027}{Phys. Rev. D {\bfseries 91}
  (2015) 084027} [\href{https://arxiv.org/abs/1411.3073}{{\ttfamily
  1411.3073}}].

\bibitem{Sennett:2016klh}
N.~Sennett, S.~Marsat and A.~Buonanno, \emph{{Gravitational waveforms in
  scalar-tensor gravity at 2PN relative order}},
  \href{https://doi.org/10.1103/PhysRevD.94.084003}{Phys. Rev. D {\bfseries 94}
  (2016) 084003} [\href{https://arxiv.org/abs/1607.01420}{{\ttfamily
  1607.01420}}].

\bibitem{Bernard:2018hta}
L.~Bernard, \emph{{Dynamics of compact binary systems in scalar-tensor
  theories: Equations of motion to the third post-Newtonian order}},
  \href{https://doi.org/10.1103/PhysRevD.98.044004}{Phys. Rev. D {\bfseries 98}
  (2018) 044004} [\href{https://arxiv.org/abs/1802.10201}{{\ttfamily
  1802.10201}}].

\bibitem{Bernard:2018ivi}
L.~Bernard, \emph{{Dynamics of compact binary systems in scalar-tensor
  theories: II. Center-of-mass and conserved quantities to 3PN order}},
  \href{https://doi.org/10.1103/PhysRevD.99.044047}{Phys. Rev. D {\bfseries 99}
  (2019) 044047} [\href{https://arxiv.org/abs/1812.04169}{{\ttfamily
  1812.04169}}].

\bibitem{Brax:2020vgg}
P.~Brax, A.-C.~Davis, S.~Melville and L.~K.~Wong, \emph{{Spin precession as a
  new window into disformal scalar fields}},
  \href{https://doi.org/10.1088/1475-7516/2021/03/001}{JCAP {\bfseries 03}
  (2021) 001} [\href{https://arxiv.org/abs/2011.01213}{{\ttfamily
  2011.01213}}].

\bibitem{Nordtvedt:1968qs}
K.~Nordtvedt, \emph{Equivalence principle for massive bodies. {II. Theory}},
  \href{https://doi.org/10.1103/PhysRev.169.1017}{Phys. Rev. {\bfseries 169}
  (1968) 1017}.

\bibitem{Khoury:2003aq}
J.~Khoury and A.~Weltman, \emph{{Chameleon fields: Awaiting surprises for tests
  of gravity in space}},
  \href{https://doi.org/10.1103/PhysRevLett.93.171104}{Phys. Rev. Lett.
  {\bfseries 93} (2004) 171104}
  [\href{https://arxiv.org/abs/astro-ph/0309300}{{\ttfamily
  astro-ph/0309300}}].

\bibitem{Khoury:2003rn}
J.~Khoury and A.~Weltman, \emph{{Chameleon cosmology}},
  \href{https://doi.org/10.1103/PhysRevD.69.044026}{Phys. Rev. D {\bfseries 69}
  (2004) 044026} [\href{https://arxiv.org/abs/astro-ph/0309411}{{\ttfamily
  astro-ph/0309411}}].

\bibitem{Hinterbichler:2010es}
K.~Hinterbichler and J.~Khoury, \emph{Symmetron fields: {S}creening long-range
  forces through local symmetry restoration},
  \href{https://doi.org/10.1103/PhysRevLett.104.231301}{Phys. Rev. Lett.
  {\bfseries 104} (2010) 231301}
  [\href{https://arxiv.org/abs/1001.4525}{{\ttfamily 1001.4525}}].

\bibitem{Hinterbichler:2011ca}
K.~Hinterbichler, J.~Khoury, A.~Levy and A.~Matas, \emph{{Symmetron
  cosmology}}, \href{https://doi.org/10.1103/PhysRevD.84.103521}{Phys. Rev. D
  {\bfseries 84} (2011) 103521}
  [\href{https://arxiv.org/abs/1107.2112}{{\ttfamily 1107.2112}}].

\bibitem{Brax:2010gi}
P.~Brax, C.~van~de Bruck, A.-C.~Davis and D.~Shaw, \emph{The dilaton and
  modified gravity}, \href{https://doi.org/10.1103/PhysRevD.82.063519}{Phys.
  Rev. D {\bfseries 82} (2010) 063519}
  [\href{https://arxiv.org/abs/1005.3735}{{\ttfamily 1005.3735}}].

\bibitem{Zhang:2017srh}
X.~Zhang, T.~Liu and W.~Zhao, \emph{Gravitational radiation from compact binary
  systems in screened modified gravity},
  \href{https://doi.org/10.1103/PhysRevD.95.104027}{Phys. Rev. D {\bfseries 95}
  (2017) 104027} [\href{https://arxiv.org/abs/1702.08752}{{\ttfamily
  1702.08752}}].

\bibitem{Liu:2018sia}
T.~Liu, X.~Zhang, W.~Zhao, K.~Lin, C.~Zhang, S.~Zhang et~al., \emph{{Waveforms
  of compact binary inspiral gravitational radiation in screened modified
  gravity}}, \href{https://doi.org/10.1103/PhysRevD.98.083023}{Phys. Rev. D
  {\bfseries 98} (2018) 083023}
  [\href{https://arxiv.org/abs/1806.05674}{{\ttfamily 1806.05674}}].

\bibitem{Nicolis:2008in}
A.~Nicolis, R.~Rattazzi and E.~Trincherini, \emph{{The Galileon as a local
  modification of gravity}},
  \href{https://doi.org/10.1103/PhysRevD.79.064036}{Phys. Rev. D {\bfseries 79}
  (2009) 064036} [\href{https://arxiv.org/abs/0811.2197}{{\ttfamily
  0811.2197}}].

\bibitem{Babichev:2009ee}
E.~Babichev, C.~Deffayet and R.~Ziour, \emph{{k-Mouflage gravity}},
  \href{https://doi.org/10.1142/S0218271809016107}{Int. J. Mod. Phys. D
  {\bfseries 18} (2009) 2147}
  [\href{https://arxiv.org/abs/0905.2943}{{\ttfamily 0905.2943}}].

\bibitem{Deffayet:2011gz}
C.~Deffayet, X.~Gao, D.~A.~Steer and G.~Zahariade, \emph{From k-essence to
  generalised galileons},
  \href{https://doi.org/10.1103/PhysRevD.84.064039}{Phys. Rev. D {\bfseries 84}
  (2011) 064039} [\href{https://arxiv.org/abs/1103.3260}{{\ttfamily
  1103.3260}}].

\bibitem{Babichev:2013usa}
E.~Babichev and C.~Deffayet, \emph{{An introduction to the Vainshtein
  mechanism}}, \href{https://doi.org/10.1088/0264-9381/30/18/184001}{Class.
  Quantum Gravity {\bfseries 30} (2013) 184001}
  [\href{https://arxiv.org/abs/1304.7240}{{\ttfamily 1304.7240}}].

\bibitem{deRham:2012fw}
C.~de~Rham, A.~J.~Tolley and D.~H.~Wesley, \emph{Vainshtein mechanism in binary
  pulsars}, \href{https://doi.org/10.1103/PhysRevD.87.044025}{Phys. Rev. D
  {\bfseries 87} (2013) 044025}
  [\href{https://arxiv.org/abs/1208.0580}{{\ttfamily 1208.0580}}].

\bibitem{deRham:2012fg}
C.~de~Rham, A.~Matas and A.~J.~Tolley, \emph{Galileon radiation from binary
  systems}, \href{https://doi.org/10.1103/PhysRevD.87.064024}{Phys. Rev. D
  {\bfseries 87} (2013) 064024}
  [\href{https://arxiv.org/abs/1212.5212}{{\ttfamily 1212.5212}}].

\bibitem{Dar:2018dra}
F.~Dar, C.~De~Rham, J.~T.~Deskins, J.~T.~Giblin and A.~J.~Tolley, \emph{Scalar
  gravitational radiation from binaries: {Vainshtein} mechanism in
  time-dependent systems},
  \href{https://doi.org/10.1088/1361-6382/aaf5e8}{Class. Quantum Gravity
  {\bfseries 36} (2019) 025008}
  [\href{https://arxiv.org/abs/1808.02165}{{\ttfamily 1808.02165}}].

\bibitem{Kuntz:2019plo}
A.~Kuntz, \emph{{Two-body potential of Vainshtein screened theories}},
  \href{https://doi.org/10.1103/PhysRevD.100.024024}{Phys. Rev. D {\bfseries
  100} (2019) 024024} [\href{https://arxiv.org/abs/1905.07340}{{\ttfamily
  1905.07340}}].

\bibitem{Renevey:2021tcz}
C.~Renevey, R.~McManus, C.~Dalang and L.~Lombriser, \emph{{The effect of
  screening mechanisms on black hole binary inspiral waveforms}},
  \href{https://arxiv.org/abs/2106.05678}{{\ttfamily 2106.05678}}.

\bibitem{Bezares:2021dma}
M.~Bezares, R.~Aguilera-Miret, L.~ter Haar, M.~Crisostomi, C.~Palenzuela and
  E.~Barausse, \emph{{No evidence of kinetic screening in merging binary
  neutron stars}},  \href{https://arxiv.org/abs/2107.05648}{{\ttfamily
  2107.05648}}.

\bibitem{Yee:1993ya}
K.~Yee and M.~Bander, \emph{{Equations of motion for spinning particles in
  external electromagnetic and gravitational fields}},
  \href{https://doi.org/10.1103/PhysRevD.48.2797}{Phys. Rev. D {\bfseries 48}
  (1993) 2797} [\href{https://arxiv.org/abs/hep-th/9302117}{{\ttfamily
  hep-th/9302117}}].

\bibitem{Blas:2016ddr}
D.~Blas, D.~L.~Nacir and S.~Sibiryakov, \emph{Ultralight dark matter resonates
  with binary pulsars},
  \href{https://doi.org/10.1103/PhysRevLett.118.261102}{Phys. Rev. Lett.
  {\bfseries 118} (2017) 261102}
  [\href{https://arxiv.org/abs/1612.06789}{{\ttfamily 1612.06789}}].

\bibitem{Blas:2019hxz}
D.~Blas, D.~L\'opez~Nacir and S.~Sibiryakov, \emph{Secular effects of
  ultralight dark matter on binary pulsars},
  \href{https://doi.org/10.1103/PhysRevD.101.063016}{Phys. Rev. D {\bfseries
  101} (2020) 063016} [\href{https://arxiv.org/abs/1910.08544}{{\ttfamily
  1910.08544}}].

\bibitem{Damour:1985mt}
T.~Damour and G.~{Sch\"{a}fer}, \emph{{Lagrangians for $n$ point masses at the
  second post-Newtonian approximation of general relativity}},
  \href{https://doi.org/10.1007/BF00773685}{Gen. Relativ. Gravit. {\bfseries
  17} (1985) 879}.

\bibitem{Damour:1990jh}
T.~Damour and G.~{Sch\"{a}fer}, \emph{{Redefinition of position variables and
  the reduction of higher-order Lagrangians}},
  \href{https://doi.org/10.1063/1.529135}{J. Math. Phys. {\bfseries 32} (1991)
  127}.

\bibitem{Levi:2014sba}
M.~Levi and J.~Steinhoff, \emph{{Equivalence of ADM Hamiltonian and Effective
  Field Theory approaches at next-to-next-to-leading order spin1-spin2 coupling
  of binary inspirals}},
  \href{https://doi.org/10.1088/1475-7516/2014/12/003}{JCAP {\bfseries 12}
  (2014) 003} [\href{https://arxiv.org/abs/1408.5762}{{\ttfamily 1408.5762}}].

\bibitem{Ross:2012fc}
A.~Ross, \emph{{Multipole expansion at the level of the action}},
  \href{https://doi.org/10.1103/PhysRevD.85.125033}{Phys. Rev. D {\bfseries 85}
  (2012) 125033} [\href{https://arxiv.org/abs/1202.4750}{{\ttfamily
  1202.4750}}].

\bibitem{Thorne:1980ru}
K.~S.~Thorne, \emph{{Multipole expansions of gravitational radiation}},
  \href{https://doi.org/10.1103/RevModPhys.52.299}{Rev. Mod. Phys. {\bfseries
  52} (1980) 299}.

\bibitem{Blanchet:2013haa}
L.~Blanchet, \emph{Gravitational radiation from post-{N}ewtonian sources and
  inspiralling compact binaries},
  \href{https://doi.org/10.12942/lrr-2014-2}{Living Rev. Relativity {\bfseries
  17} (2014) 2} [\href{https://arxiv.org/abs/1310.1528}{{\ttfamily
  1310.1528}}].

\bibitem{Steinhoff:2009ei}
J.~Steinhoff and G.~Sch{\"a}fer, \emph{{Canonical formulation of
  self-gravitating spinning-object systems}},
  \href{https://doi.org/10.1209/0295-5075/87/50004}{EPL {\bfseries 87} (2009)
  50004} [\href{https://arxiv.org/abs/0907.1967}{{\ttfamily 0907.1967}}].

\bibitem{Tulczyjew:1959ssc}
W.~Tulczyjew, \emph{Motion of multipole particles in general relativity
  theory}, {Acta Phys. Pol. {\bfseries 18} (1959) 37}.

\bibitem{Dirac:1964lqm}
P.~A.~M.~Dirac{}, \emph{{Lectures on Quantum Mechanics}}, Yeshiva University,
  New York (1964).

\bibitem{Hanson:1974qy}
A.~J.~Hanson and T.~Regge, \emph{The relativistic spherical top},
  \href{https://doi.org/10.1016/0003-4916(74)90046-3}{Annals Phys. {\bfseries
  87} (1974) 498}.

\bibitem{Pryce:1948pf}
M.~H.~L.~Pryce, \emph{{The mass-centre in the restricted theory of relativity
  and its connexion with the quantum theory of elementary particles}},
  \href{https://doi.org/10.1098/rspa.1948.0103}{Proc. Roy. Soc. Lond. A
  {\bfseries 195} (1948) 62}.

\bibitem{Lorentsen:1997wt}
J.~H.~Lorentsen and N.~K.~Nielsen, \emph{Gauge theory of a massive relativistic
  spinning point particle},
  \href{https://arxiv.org/abs/hep-th/9705081}{{\ttfamily hep-th/9705081}}.

\bibitem{Steinhoff:2015ksa}
J.~Steinhoff, \emph{{Spin gauge symmetry in the action principle for classical
  relativistic particles}},  \href{https://arxiv.org/abs/1501.04951}{{\ttfamily
  1501.04951}}.

\bibitem{Goldberger:2020fot}
W.~D.~Goldberger, J.~Li and I.~Z.~Rothstein, \emph{{Non-conservative effects on
  spinning black holes from world-line effective field theory}},
  \href{https://doi.org/10.1007/JHEP06(2021)053}{JHEP {\bfseries 06} (2021)
  053} [\href{https://arxiv.org/abs/2012.14869}{{\ttfamily 2012.14869}}].

\bibitem{Henneaux:1992ig}
M.~Henneaux and C.~Teitelboim{}, \emph{{Quantization of Gauge Systems}},
  Princeton University Press (1992).

\bibitem{Bini:2012gu}
D.~Bini, T.~Damour and G.~Faye, \emph{{Effective action approach to
  higher-order relativistic tidal interactions in binary systems and their
  effective one body description}},
  \href{https://doi.org/10.1103/PhysRevD.85.124034}{Phys. Rev. D {\bfseries 85}
  (2012) 124034} [\href{https://arxiv.org/abs/1202.3565}{{\ttfamily
  1202.3565}}].

\bibitem{Endlich:2015mke}
S.~Endlich and R.~Penco, \emph{{Effective field theory approach to tidal
  dynamics of spinning astrophysical systems}},
  \href{https://doi.org/10.1103/PhysRevD.93.064021}{Phys. Rev. D {\bfseries 93}
  (2016) 064021} [\href{https://arxiv.org/abs/1510.08889}{{\ttfamily
  1510.08889}}].

\bibitem{Steinhoff:2016rfi}
J.~Steinhoff, T.~Hinderer, A.~Buonanno and A.~Taracchini, \emph{{Dynamical
  tides in general relativity: Effective action and effective-one-body
  Hamiltonian}}, \href{https://doi.org/10.1103/PhysRevD.94.104028}{Phys. Rev. D
  {\bfseries 94} (2016) 104028}
  [\href{https://arxiv.org/abs/1608.01907}{{\ttfamily 1608.01907}}].

\bibitem{Kol:2011vg}
B.~Kol and M.~Smolkin, \emph{{Black hole stereotyping: Induced gravito-static
  polarization}}, \href{https://doi.org/10.1007/JHEP02(2012)010}{JHEP
  {\bfseries 02} (2012) 010} [\href{https://arxiv.org/abs/1110.3764}{{\ttfamily
  1110.3764}}].

\bibitem{Hui:2020xxx}
L.~Hui, A.~Joyce, R.~Penco, L.~Santoni and A.~R.~Solomon, \emph{{Static
  response and Love numbers of Schwarzschild black holes}},
  \href{https://arxiv.org/abs/2010.00593}{{\ttfamily 2010.00593}}.

\bibitem{Goldberger:2005cd}
W.~D.~Goldberger and I.~Z.~Rothstein, \emph{{Dissipative effects in the
  worldline approach to black hole dynamics}},
  \href{https://doi.org/10.1103/PhysRevD.73.104030}{Phys. Rev. D {\bfseries 73}
  (2006) 104030} [\href{https://arxiv.org/abs/hep-th/0511133}{{\ttfamily
  hep-th/0511133}}].

\bibitem{Mathisson:1937zz}
M.~Mathisson, \emph{Neue mechanik materieller systemes \emph{(in German)}},
  {Acta Phys. Polon. {\bfseries 6} (1937) 163}{
  [\href{https://doi.org/10.1007/s10714-010-0939-y}{Gen. Relativ. Gravit.
  \textbf{42} (2010) 11011}]}.

\bibitem{Papapetrou:1951pa}
A.~Papapetrou, \emph{{Spinning test-particles in general relativity. I}},
  \href{https://doi.org/10.1098/rspa.1951.0200}{Proc. Roy. Soc. Lond. A
  {\bfseries 209} (1951) 248}.

\bibitem{Dixon:1970zza}
W.~G.~Dixon, \emph{{Dynamics of extended bodies in general relativity. I.
  Momentum and angular momentum}},
  \href{https://doi.org/10.1098/rspa.1970.0020}{Proc. Roy. Soc. Lond. A
  {\bfseries 314} (1970) 499}.

\bibitem{Dixon:1970zz}
W.~G.~Dixon, \emph{{Dynamics of extended bodies in general relativity. II.
  Moments of the charge-current vector}},
  \href{https://doi.org/10.1098/rspa.1970.0191}{Proc. Roy. Soc. Lond. A
  {\bfseries 319} (1970) 509}.

\bibitem{Dixon:1974}
W.~G.~Dixon, \emph{{Dynamics of extended bodies in general relativity. III.
  Equations of motion}}, \href{https://doi.org/10.1098/rsta.1974.0046}{Proc.
  Roy. Soc. Lond. A {\bfseries 277} (1974) 59}.

\bibitem{Bernard:2019yfz}
L.~Bernard, \emph{{Dipolar tidal effects in scalar-tensor theories}},
  \href{https://doi.org/10.1103/PhysRevD.101.021501}{Phys. Rev. D {\bfseries
  101} (2020) 021501} [\href{https://arxiv.org/abs/1906.10735}{{\ttfamily
  1906.10735}}].

\bibitem{Damour:1992kf}
T.~Damour and K.~Nordtvedt, \emph{{General relativity as a cosmological
  attractor of tensor-scalar theories}},
  \href{https://doi.org/10.1103/PhysRevLett.70.2217}{Phys. Rev. Lett.
  {\bfseries 70} (1993) 2217}.

\bibitem{Damour:1993id}
T.~Damour and K.~Nordtvedt, \emph{{Tensor-scalar cosmological models and their
  relaxation toward general relativity}},
  \href{https://doi.org/10.1103/PhysRevD.48.3436}{Phys. Rev. D {\bfseries 48}
  (1993) 3436}.

\bibitem{Anderson:2016aoi}
D.~Anderson, N.~Yunes and E.~Barausse, \emph{{Effect of cosmological evolution
  on Solar System constraints and on the scalarization of neutron stars in
  massless scalar-tensor theories}},
  \href{https://doi.org/10.1103/PhysRevD.94.104064}{Phys. Rev. D {\bfseries 94}
  (2016) 104064} [\href{https://arxiv.org/abs/1607.08888}{{\ttfamily
  1607.08888}}].

\bibitem{Alby:2017dzl}
T.~A.~de~Pirey Saint~Alby and N.~Yunes, \emph{{Cosmological evolution and Solar
  System consistency of massive scalar-tensor gravity}},
  \href{https://doi.org/10.1103/PhysRevD.96.064040}{Phys. Rev. D {\bfseries 96}
  (2017) 064040} [\href{https://arxiv.org/abs/1703.06341}{{\ttfamily
  1703.06341}}].

\bibitem{Anson:2019uto}
T.~Anson, E.~Babichev, C.~Charmousis and S.~Ramazanov, \emph{{Cosmological
  instability of scalar-Gauss-Bonnet theories exhibiting scalarization}},
  \href{https://doi.org/10.1088/1475-7516/2019/06/023}{JCAP {\bfseries 06}
  (2019) 023} [\href{https://arxiv.org/abs/1903.02399}{{\ttfamily
  1903.02399}}].

\bibitem{Franchini:2019npi}
N.~Franchini and T.~P.~Sotiriou, \emph{{Cosmology with subdominant Horndeski
  scalar field}}, \href{https://doi.org/10.1103/PhysRevD.101.064068}{Phys. Rev.
  D {\bfseries 101} (2020) 064068}
  [\href{https://arxiv.org/abs/1903.05427}{{\ttfamily 1903.05427}}].

\bibitem{Anson:2019ebp}
T.~Anson, E.~Babichev and S.~Ramazanov, \emph{{Reconciling spontaneous
  scalarization with cosmology}},
  \href{https://doi.org/10.1103/PhysRevD.100.104051}{Phys. Rev. D {\bfseries
  100} (2019) 104051} [\href{https://arxiv.org/abs/1905.10393}{{\ttfamily
  1905.10393}}].

\bibitem{Antoniou:2020nax}
G.~Antoniou, L.~Bordin and T.~P.~Sotiriou, \emph{{Compact object scalarization
  with general relativity as a cosmic attractor}},
  \href{https://doi.org/10.1103/PhysRevD.103.024012}{Phys. Rev. D {\bfseries
  103} (2021) 024012} [\href{https://arxiv.org/abs/2004.14985}{{\ttfamily
  2004.14985}}].

\bibitem{Galley:2015kus}
C.~R.~Galley, A.~K.~Leibovich, R.~A.~Porto and A.~Ross, \emph{{Tail effect in
  gravitational radiation reaction: Time nonlocality and renormalization group
  evolution}}, \href{https://doi.org/10.1103/PhysRevD.93.124010}{Phys. Rev. D
  {\bfseries 93} (2016) 124010}
  [\href{https://arxiv.org/abs/1511.07379}{{\ttfamily 1511.07379}}].

\bibitem{Foffa:2021pkg}
S.~Foffa and R.~Sturani, \emph{{Near and far zone in two-body dynamics: An
  effective field theory perspective}},
  \href{https://arxiv.org/abs/2103.03190}{{\ttfamily 2103.03190}}.

\bibitem{Davis:2019ltc}
A.-C.~Davis and S.~Melville, \emph{Novel screening with two bodies: {S}umming
  the ladder in disformal scalar-tensor theories},
  \href{https://doi.org/10.1088/1475-7516/2020/09/013}{JCAP {\bfseries 09}
  (2020) 013} [\href{https://arxiv.org/abs/1910.08831}{{\ttfamily
  1910.08831}}].

\bibitem{Davis:2021oce}
A.-C.~Davis and S.~Melville, \emph{Scalar fields near compact objects:
  {Resummation versus UV completion}},
  \href{https://arxiv.org/abs/2107.00010}{{\ttfamily 2107.00010}}.

\bibitem{Kidder:1992fr}
L.~E.~Kidder, C.~M.~Will and A.~G.~Wiseman, \emph{{Spin effects in the inspiral
  of coalescing compact binaries}},
  \href{https://doi.org/10.1103/PhysRevD.47.R4183}{Phys. Rev. D {\bfseries 47}
  (1993) R4183} [\href{https://arxiv.org/abs/gr-qc/9211025}{{\ttfamily
  gr-qc/9211025}}].

\bibitem{Kidder:1995zr}
L.~E.~Kidder, \emph{{Coalescing binary systems of compact objects to
  (post)${}^{5/2}$-Newtonian order. V. Spin effects}},
  \href{https://doi.org/10.1103/PhysRevD.52.821}{Phys. Rev. D {\bfseries 52}
  (1995) 821} [\href{https://arxiv.org/abs/gr-qc/9506022}{{\ttfamily
  gr-qc/9506022}}].

\bibitem{Faye:2006gx}
G.~Faye, L.~Blanchet and A.~Buonanno, \emph{{Higher-order spin effects in the
  dynamics of compact binaries. I. Equations of motion}},
  \href{https://doi.org/10.1103/PhysRevD.74.104033}{Phys. Rev. D {\bfseries 74}
  (2006) 104033} [\href{https://arxiv.org/abs/gr-qc/0605139}{{\ttfamily
  gr-qc/0605139}}].

\bibitem{deAndrade:2000gf}
V.~C.~de~Andrade, L.~Blanchet and G.~Faye, \emph{{Third post-Newtonian dynamics
  of compact binaries: Noetherian conserved quantities and equivalence between
  the harmonic-coordinate and ADM-Hamiltonian formalisms}},
  \href{https://doi.org/10.1088/0264-9381/18/5/301}{Class. Quantum Gravity
  {\bfseries 18} (2001) 753}
  [\href{https://arxiv.org/abs/gr-qc/0011063}{{\ttfamily gr-qc/0011063}}].

\bibitem{Newton:1949ssc}
T.~D.~Newton and E.~P.~Wigner, \emph{Localized states for elementary systems},
  \href{https://doi.org/10.1103/RevModPhys.21.400}{Rev. Mod. Phys. {\bfseries
  21} (1949) 400}.

\bibitem{Buonanno:2009zt}
A.~Buonanno, B.~Iyer, E.~Ochsner, Y.~Pan and B.~S.~Sathyaprakash,
  \emph{{Comparison of post-Newtonian templates for compact binary inspiral
  signals in gravitational-wave detectors}},
  \href{https://doi.org/10.1103/PhysRevD.80.084043}{Phys. Rev. D {\bfseries 80}
  (2009) 084043} [\href{https://arxiv.org/abs/0907.0700}{{\ttfamily
  0907.0700}}].

\bibitem{Yunes:2009ke}
N.~Yunes and F.~Pretorius, \emph{{Fundamental theoretical bias in gravitational
  wave astrophysics and the parametrized post-Einsteinian framework}},
  \href{https://doi.org/10.1103/PhysRevD.80.122003}{Phys. Rev. D {\bfseries 80}
  (2009) 122003} [\href{https://arxiv.org/abs/0909.3328}{{\ttfamily
  0909.3328}}].

\bibitem{Tahura:2018zuq}
S.~Tahura and K.~Yagi, \emph{Parameterized post-{E}insteinian gravitational
  waveforms in various modified theories of gravity},
  \href{https://doi.org/10.1103/PhysRevD.98.084042}{Phys. Rev. D {\bfseries 98}
  (2018) 084042}{ [Erratum ibid. \textbf{101} (2020) 109902]}
  [\href{https://arxiv.org/abs/1809.00259}{{\ttfamily 1809.00259}}].

\bibitem{Peters:1964:ee}
P.~C.~Peters, \emph{Gravitational radiation and the motion of two point
  masses}, \href{https://doi.org/10.1103/PhysRev.136.B1224}{Phys. Rev.
  {\bfseries 136} (1964) B1224}.

\bibitem{Krolak:1987ee}
A.~{Krolak} and B.~F.~{Schutz}, \emph{{Coalescing binaries---Probe of the
  universe}}, \href{https://doi.org/10.1007/BF00759095}{Gen. Relativ. Gravit.
  {\bfseries 19} (1987) 1163}.

\bibitem{Cardoso:2020iji}
V.~Cardoso, C.~F.~B.~Macedo and R.~Vicente, \emph{{Eccentricity evolution of
  compact binaries and applications to gravitational-wave physics}},
  \href{https://doi.org/10.1103/PhysRevD.103.023015}{Phys. Rev. D {\bfseries
  103} (2021) 023015} [\href{https://arxiv.org/abs/2010.15151}{{\ttfamily
  2010.15151}}].

\bibitem{Radice:2020ddv}
D.~Radice, S.~Bernuzzi and A.~Perego, \emph{The dynamics of binary neutron star
  mergers and {GW170817}},
  \href{https://doi.org/10.1146/annurev-nucl-013120-114541}{Ann. Rev. Nucl.
  Part. Sci. {\bfseries 70} (2020) 95}
  [\href{https://arxiv.org/abs/2002.03863}{{\ttfamily 2002.03863}}].

\bibitem{Baiotti:2016qnr}
L.~Baiotti and L.~Rezzolla, \emph{{Binary neutron star mergers: A review of
  Einstein's richest laboratory}},
  \href{https://doi.org/10.1088/1361-6633/aa67bb}{Rept. Prog. Phys. {\bfseries
  80} (2017) 096901} [\href{https://arxiv.org/abs/1607.03540}{{\ttfamily
  1607.03540}}].

\bibitem{Faber:2012rw}
J.~A.~Faber and F.~A.~Rasio, \emph{Binary neutron star mergers},
  \href{https://doi.org/10.12942/lrr-2012-8}{Living Rev. Relativity {\bfseries
  15} (2012) 8} [\href{https://arxiv.org/abs/1204.3858}{{\ttfamily
  1204.3858}}].

\bibitem{Martynov:2016fzi}
B.~P.~Abbott et~al., \emph{{Sensitivity of the Advanced LIGO detectors at the
  beginning of gravitational wave astronomy}},
  \href{https://doi.org/10.1103/PhysRevD.93.112004}{Phys. Rev. D {\bfseries 93}
  (2016) 112004}{ [Erratum ibid. \textbf{97} (2018) 059901]}
  [\href{https://arxiv.org/abs/1604.00439}{{\ttfamily 1604.00439}}].

\bibitem{Hild:2010id}
S.~Hild et~al., \emph{Sensitivity studies for third-generation gravitational
  wave observatories},
  \href{https://doi.org/10.1088/0264-9381/28/9/094013}{Class. Quantum Gravity
  {\bfseries 28} (2011) 094013}
  [\href{https://arxiv.org/abs/1012.0908}{{\ttfamily 1012.0908}}].

\bibitem{Lorimer:2008se}
D.~R.~Lorimer, \emph{Binary and millisecond pulsars},
  \href{https://doi.org/10.12942/lrr-2008-8}{Living Rev. Relativity {\bfseries
  11} (2008) 8} [\href{https://arxiv.org/abs/0811.0762}{{\ttfamily
  0811.0762}}].

\bibitem{LIGOScientific:2018hze}
B.~P.~Abbott et~al. {(LIGO Scientific and Virgo Collaborations)},
  \emph{{Properties of the binary neutron star merger GW170817}},
  \href{https://doi.org/10.1103/PhysRevX.9.011001}{Phys. Rev. X {\bfseries 9}
  (2019) 011001} [\href{https://arxiv.org/abs/1805.11579}{{\ttfamily
  1805.11579}}].

\bibitem{LIGOScientific:2020aai}
B.~P.~Abbott et~al. {(LIGO Scientific and Virgo Collaborations)},
  \emph{{GW190425}: {Observation} of a compact binary coalescence with total
  mass {$\sim 3.4 M_{\odot}$}},
  \href{https://doi.org/10.3847/2041-8213/ab75f5}{Astrophys. J. Lett.
  {\bfseries 892} (2020) L3}
  [\href{https://arxiv.org/abs/2001.01761}{{\ttfamily 2001.01761}}].

\bibitem{LIGOScientific:2021qlt}
R.~Abbott et~al. {(LIGO Scientific, Virgo, and KAGRA Collaborations)},
  \emph{Observation of gravitational waves from two neutron
  star\textendash{}black hole coalescences},
  \href{https://doi.org/10.3847/2041-8213/ac082e}{Astrophys. J. Lett.
  {\bfseries 915} (2021) L5}
  [\href{https://arxiv.org/abs/2106.15163}{{\ttfamily 2106.15163}}].

\bibitem{Tahura:2019dgr}
S.~Tahura, K.~Yagi and Z.~Carson, \emph{{Testing gravity with gravitational
  waves from binary black hole mergers: Contributions from amplitude
  corrections}}, \href{https://doi.org/10.1103/PhysRevD.100.104001}{Phys. Rev.
  D {\bfseries 100} (2019) 104001}
  [\href{https://arxiv.org/abs/1907.10059}{{\ttfamily 1907.10059}}].

\bibitem{Amaro-Seoane:2017lisa}
P.~{Amaro-Seoane}, H.~{Audley}, S.~{Babak}, J.~{Baker}, E.~{Barausse},
  P.~{Bender} et~al., \emph{{Laser Interferometer Space Antenna}},
  \href{https://arxiv.org/abs/1702.00786}{{\ttfamily 1702.00786}}.

\bibitem{Kawamura:2011zz}
S.~Kawamura, M.~Ando, N.~Seto, S.~Sato, T.~Nakamura, K.~Tsubono et~al.,
  \emph{{The Japanese space gravitational wave antenna: DECIGO}},
  \href{https://doi.org/10.1088/0264-9381/28/9/094011}{Class. Quantum Gravity
  {\bfseries 28} (2011) 094011}.

\bibitem{Damour:1992ppk}
T.~Damour and J.~H.~Taylor, \emph{Strong-field tests of relativistic gravity
  and binary pulsars}, \href{https://doi.org/10.1103/PhysRevD.45.1840}{Phys.
  Rev. D {\bfseries 45} (1992) 1840}.

\bibitem{Stairs:2003eg}
I.~H.~Stairs, \emph{{Testing general relativity with pulsar timing}},
  \href{https://doi.org/10.12942/lrr-2003-5}{Living Rev. Relativity {\bfseries
  6} (2003) 5} [\href{https://arxiv.org/abs/astro-ph/0307536}{{\ttfamily
  astro-ph/0307536}}].

\bibitem{Shao:2014wja}
L.~Shao, I.~Stairs, J.~Antoniadis, A.~Deller, P.~Freire, J.~Hessels et~al.,
  \emph{Testing gravity with pulsars in the {SKA} era},
  \href{https://doi.org/10.22323/1.215.0042}{PoS Proc. Sci. (AASKA14) 042}
  [\href{https://arxiv.org/abs/1501.00058}{{\ttfamily 1501.00058}}].

\bibitem{Gerosa:2015hba}
D.~Gerosa, M.~Kesden, R.~O'Shaughnessy, A.~Klein, E.~Berti, U.~Sperhake et~al.,
  \emph{{Precessional instability in binary black holes with aligned spins}},
  \href{https://doi.org/10.1103/PhysRevLett.115.141102}{Phys. Rev. Lett.
  {\bfseries 115} (2015) 141102}
  [\href{https://arxiv.org/abs/1506.09116}{{\ttfamily 1506.09116}}].

\bibitem{Lousto:2016nlp}
C.~O.~Lousto and J.~Healy, \emph{{Unstable flip-flopping spinning binary black
  holes}}, \href{https://doi.org/10.1103/PhysRevD.93.124074}{Phys. Rev. D
  {\bfseries 93} (2016) 124074}
  [\href{https://arxiv.org/abs/1601.05086}{{\ttfamily 1601.05086}}].

\bibitem{Mould:2020cgc}
M.~Mould and D.~Gerosa, \emph{{Endpoint of the up-down instability in
  precessing binary black holes}},
  \href{https://doi.org/10.1103/PhysRevD.101.124037}{Phys. Rev. D {\bfseries
  101} (2020) 124037} [\href{https://arxiv.org/abs/2003.02281}{{\ttfamily
  2003.02281}}].

\bibitem{Varma:2020bon}
V.~Varma, M.~Mould, D.~Gerosa, M.~A.~Scheel, L.~E.~Kidder and H.~P.~Pfeiffer,
  \emph{{Up-down instability of binary black holes in numerical relativity}},
  \href{https://doi.org/10.1103/PhysRevD.103.064003}{Phys. Rev. D {\bfseries
  103} (2021) 064003} [\href{https://arxiv.org/abs/2012.07147}{{\ttfamily
  2012.07147}}].

\bibitem{Baker:2019gxo}
T.~Baker, A.~Barreira, H.~Desmond, P.~Ferreira, B.~Jain, K.~Koyama et~al.,
  \emph{{Novel Probes Project: Tests of gravity on astrophysical scales}},
  \href{https://doi.org/10.1103/RevModPhys.93.015003}{Rev. Mod. Phys.
  {\bfseries 93} (2021) 015003}
  [\href{https://arxiv.org/abs/1908.03430}{{\ttfamily 1908.03430}}].

\bibitem{xAct}
J.~M.~Mart\'in-Garc\'ia, \emph{{xAct: Efficient tensor computer algebra for the
  Wolfram Language}},
\newblock {\url{http://www.xact.es/}}.

\bibitem{Wong:2019kru}
L.~K.~Wong, \emph{{Superradiant scattering by a black hole binary}},
  \href{https://doi.org/10.1103/PhysRevD.100.044051}{Phys. Rev. D {\bfseries
  100} (2019) 044051} [\href{https://arxiv.org/abs/1905.08543}{{\ttfamily
  1905.08543}}].

\end{thebibliography}\endgroup
\end{document}